\documentclass[12pt,a4paper,titlepage,twoside]{scrreprt}
\usepackage{graphicx, epsfig,amssymb} %include figure files
\usepackage{epstopdf}
\usepackage{amsmath, amsfonts}
\usepackage{bm} %include bold math: \bm{} creates bold letters in math mode
\usepackage{color}
\usepackage[sort&compress,numbers]{natbib}
\usepackage{epigraph}
\usepackage{enumitem}
\usepackage[hypertexnames=false,breaklinks]{hyperref}
\usepackage{comment}

\newcommand{\be}{\begin{equation}}
\newcommand{\ee}{\end{equation}}
\def\beq{\begin{eqnarray}}
\def\eeq{\end{eqnarray}}
\newcommand{\nn}{\nonumber}

\def\bf{\normalfont\bfseries}
\def\it{\normalfont\itshape}
\def\tt{\normalfont\ttfamily}
\def\rm{\mathrm}

\numberwithin{equation}{chapter} %For numbering equations according to section
\numberwithin{figure}{chapter} %For numbering figures according to section
\numberwithin{table}{chapter} %For numbering tables according to section

\graphicspath{{plots/}}

\begin{document}
%%%%%%%%%%%%%%%%%%%%%%%%%%%%%%%%%%%%%%%%%%%%%%%%%%%%%%%%%%%%%%%%%%%%%%%%
\begin{titlepage}

\includegraphics[viewport=1.86cm 0cm 10cm 3.16cm,scale=0.29]{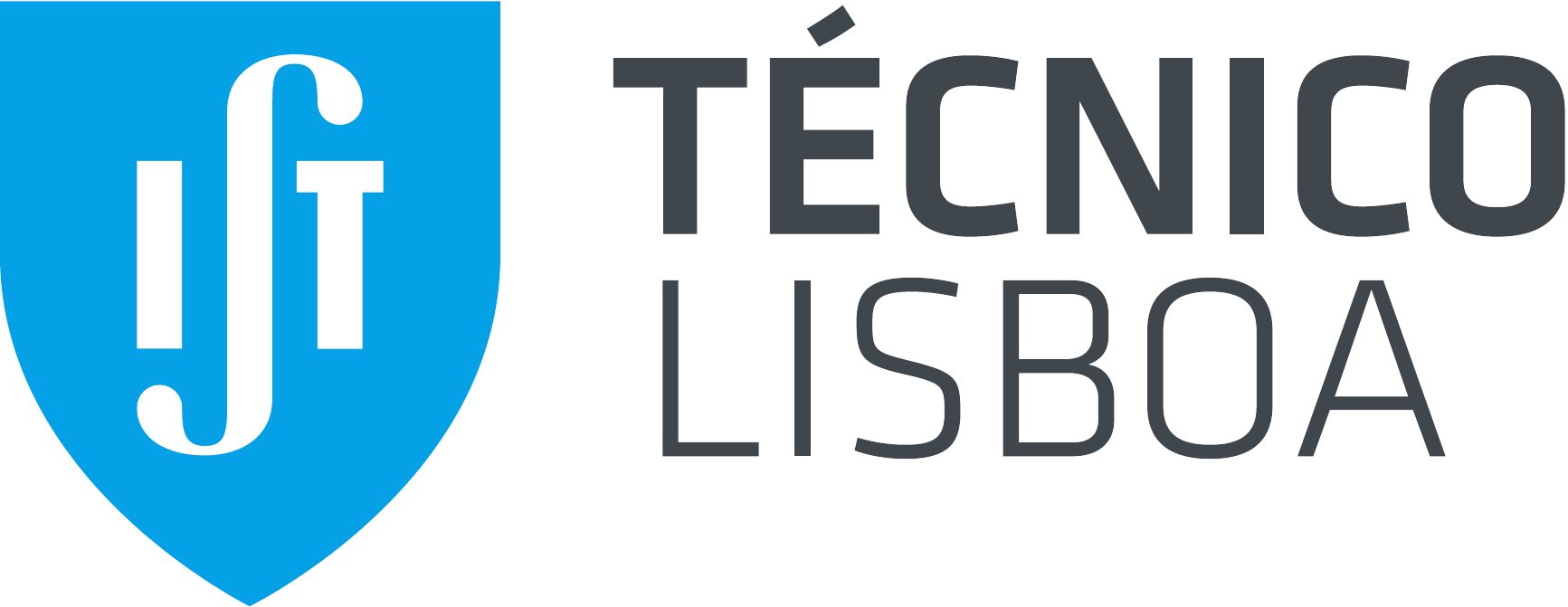}

\thispagestyle {empty}

\vspace{0.0cm}
\begin{center}
    %begin{Spacing}{2}
          \textsc{{\large UNIVERSIDADE DE LISBOA}}\\    \vspace{.25cm}
          \textsc{{\large INSTITUTO SUPERIOR T\'{E}CNICO}}
    %\end{Spacing}
\end{center}
%\vspace{0.5cm}

%\begin{center}
%    \includegraphics[width=1\textwidth]{./pics/Auger.jpg}
%\end{center}

%%%%%%%%%%%%%%%%%%%%%%%%%%%%%%%%%%%%%%
%% Title 
%   \vspace{2cm}
\begin{center}
   %{\upshape \huge Doctoral Thesis }\\
   
  % \hrule % Horizontal line
	\vspace{ 2cm}
    \textbf{{\Large Fundamental fields around compact objects:
		%\vspace{0.2cm}
		Massive spin-2 fields, Superradiant instabilities and Stars with dark matter cores}}
		%\vspace{ 0.2cm}
%    \hrule  % Horizontal line
\end{center}
%%%%%%%%%%%%%%%%%%%%%%%%%%%%%%%%%%%%%%%%
\begin{center}
    	\vspace{0.8cm}
    \textbf{\large Richard Pires Brito}\\[0.5cm]
    \par
    %{\Large Dissertation to obtain the PhD Degree in Physics}
\end{center}
\vspace{0.8cm}

%%%%%%%%%%%%%%%%%%%%%%%%%%%%%%%%%%%%%%%%%%%%
%%__ Supervisor ______
\hspace{0ex}
  \parbox[t]{\textwidth}{
  \textbf{Supervisor:} Doctor V\'{i}tor Manuel dos Santos Cardoso
	
	\textbf{Co-Supervisor:} Doctor Paolo Pani
  }

%%%%%%%%%%%%%%%%%%%%%%%%%%%%%%%%%%%%%%%%
 \vspace{0.8cm}
 \begin{center}
    \textbf{\normalsize Thesis approved in public session to obtain the PhD Degree in Physics}\\
    \vspace{0.1cm}
        \textbf{\normalsize Jury final classification: Pass with Distinction and Honour}
\end{center}

%%%%%%%%%%%%%%%%%%%%%%%%%%%%%%%%%%%%%%%%%%%%%
%%_____  Juri _________
%\begin{comment}
%\hspace{-10ex}
\vspace{0.8cm}
   \begin{center}
       \textbf{Jury}
   \end{center} 
   \hspace{1cm}
   \begin{minipage}[h!]{0.9\textwidth}
       \large \normalsize
       \textbf{Chairperson:} Chairman of the IST Scientific Board  \newline
       \textbf{Members of the Committee:} \newline
  
   \begin{minipage}{1\textwidth}
      \normalsize \small
      \leftskip2em
        Doctor Jos\'{e} Pizarro de Sande e Lemos \vspace{0.2cm} \newline 
        Doctor Eugeny Babichev \vspace{0.2cm} \newline 
        Doctor V\'{i}tor Manuel dos Santos Cardoso\vspace{0.2cm} \newline 
        Doctor Jos\'{e} Ant\'{o}nio Maciel Nat\'{a}rio\vspace{0.2cm} \newline 
				Doctor Carlos Alberto Ruivo Herdeiro\vspace{0.2cm} \newline 
   \end{minipage}
\end{minipage}
%\end{comment}
%%%%%%%%%%%%%%%%%%%%%%
%%%%%%%%%%%%%%%%%%%%%%%%%% %%%%%%%%%%%%%%%%%%%%%%%%%%
%%%%%%%%%%%%%%%%%%%%%%%%%% %%%%%%%%%%%%%%%%%%%%%%%%%%
\newline
 \vspace{1cm} 
 %%%%%%%%%%%%%%%%%%%%%%%%%%
 %%__ Foot institutions
%\begin{center}
%       \textbf{Funding Institution:}\\ 
%   \begin{minipage}[h!]{0.9\textwidth}
%      \textbf{Funding Institution:}\\
%\hspace{0.01\textwidth}
%        Grant SFRH/BD/52047/2012 from Funda\c{c}\~{a}o para a Ci\^{e}ncia e Tecnologia (FCT).
%\end{center}
 %  \end{minipage}
%   \hspace{0.05\textwidth}  
  
 \vspace{0.05cm}
\begin{center}
\textbf{\large 2016}
\end{center}
 \vspace{-2.5cm}

\newpage
\cleardoublepage

\thispagestyle {empty}

\includegraphics[viewport=1.86cm 0cm 10cm 3.16cm,scale=0.29]{LOGOist}

\vspace{0.0cm}
\begin{center}
    %begin{Spacing}{2}
          \textsc{{\large UNIVERSIDADE DE LISBOA}}\\    \vspace{.25cm}
          \textsc{{\large INSTITUTO SUPERIOR T\'{E}CNICO}}
    %\end{Spacing}
\end{center}
%\vspace{0.5cm}

%\begin{center}
%    \includegraphics[width=1\textwidth]{./pics/Auger.jpg}
%\end{center}

%%%%%%%%%%%%%%%%%%%%%%%%%%%%%%%%%%%%%%
%% Title 
%   \vspace{2cm}
\begin{center}
   %{\upshape \huge Doctoral Thesis }\\
   
  % \hrule % Horizontal line
	\vspace{0.6cm}
    \textbf{{\Large Fundamental fields around compact objects:
		%\vspace{0.2cm}
		Massive spin-2 fields, Superradiant instabilities and Stars with dark matter cores}}
		%\vspace{ 0.2cm}
%    \hrule  % Horizontal line
\end{center}
\vspace{ 0.cm}

%%%%%%%%%%%%%%%%%%%%%%%%%%%%%%%%%%%%%%%%
\begin{center}
    	\vspace{0.05cm}
    \textbf{\large Richard Pires Brito}\\[0.5cm]
    \par
    %{\Large Dissertation to obtain the PhD Degree in Physics}
\end{center}
\vspace{1cm}

%%%%%%%%%%%%%%%%%%%%%%%%%%%%%%%%%%%%%%%%%%%%
%%__ Supervisor ______
\hspace{0ex}
  \parbox[t]{\textwidth}{
  \textbf{Supervisor:} Doctor V\'{i}tor Manuel dos Santos Cardoso
	
	\textbf{Co-Supervisor:} Doctor Paolo Pani
  }

%%%%%%%%%%%%%%%%%%%%%%%%%%%%%%%%%%%%%%%%
 %\vspace{1cm}
 \begin{center}
    \textbf{\normalsize Thesis approved in public session to obtain the PhD Degree in Physics}\\
    \vspace{0.1cm}
        \textbf{\normalsize Jury final classification: Pass with Distinction and Honour}
\end{center}

%%%%%%%%%%%%%%%%%%%%%%%%%%%%%%%%%%%%%%%%%%%%%
%%_____  Juri _________
%\begin{comment}
%\hspace{-10ex}
\vspace{.1cm}
   \begin{center}
       \textbf{Jury}
   \end{center} 
   \hspace{1cm}
   \begin{minipage}[h!]{0.9\textwidth}
       \large \normalsize
       \textbf{Chairperson:} Chairman of the IST Scientific Board  \newline
       \textbf{Members of the Committee:} \newline
  
   \begin{minipage}{1\textwidth}
      \normalsize \small
      \leftskip2em
        Doctor Jos\'{e} Pizarro de Sande e Lemos, {\small Professor Catedr\'{a}tico do Instituto Superior T\'{e}cnico da Universidade de Lisboa.} \vspace{0.2cm} \newline 
        Doctor Eugeny Babichev, {\small  Charg\'{e} de Recherche, Laboratoire de Physique Th\'{e}orique, Centre National de la Recherche Scientifique, Universit\'{e} Paris-Sud, Universit\'{e} Paris-Saclay.} \vspace{0.2cm} \newline 
        Doctor V\'{i}tor Manuel dos Santos Cardoso, {\small  Professor Associado (com Agrega\c c\~{a}o) do Instituto Superior T\'{e}cnico da Universidade de Lisboa.}\vspace{0.2cm} \newline 
        Doctor Jos\'{e} Ant\'{o}nio Maciel Nat\'{a}rio, {\small  Professor Associado do Instituto Superior T\'{e}cnico da Universidade de Lisboa.}\vspace{0.2cm} \newline 
				Doctor Carlos Alberto Ruivo Herdeiro, {\small  Equiparado a Investigador Principal (com Agrega\c c\~{a}o) da Universidade de Aveiro.}\vspace{0.2cm} \newline 
   \end{minipage}
\end{minipage}
%\end{comment}
%%%%%%%%%%%%%%%%%%%%%%
%%%%%%%%%%%%%%%%%%%%%%%%%% %%%%%%%%%%%%%%%%%%%%%%%%%%
%%%%%%%%%%%%%%%%%%%%%%%%%% %%%%%%%%%%%%%%%%%%%%%%%%%%
\newline
 \vspace{.25cm} 
 %%%%%%%%%%%%%%%%%%%%%%%%%%
 %%__ Foot institutions
\begin{center}
       \textbf{Funding Institution:}\\ 
%   \begin{minipage}[h!]{0.9\textwidth}
%      \textbf{Funding Institution:}\\
%\hspace{0.01\textwidth}
        Grant SFRH/BD/52047/2012 from Funda\c{c}\~{a}o para a Ci\^{e}ncia e Tecnologia (FCT).
\end{center}
 %  \end{minipage}
%   \hspace{0.05\textwidth}  
  
 \vspace{0.05cm}
\begin{center}
\textbf{\large 2016}
\end{center}
 \vspace{-2.5cm}

\end{titlepage}
%%%%%%%%%%%%%%%%%%%%%%%%%%%%%%%%%%%%%%%%%%%%%%%%%%%%%%%%%%%%%%%%%%%%%%%%
\cleardoublepage
\newpage
\pagenumbering{Roman}
%===================================================================================
\chapter*{Resumo}
%\subsection*{T\'itulo: Campos fundamentais em torno de objectos compactos: Campos massivos de spin-2, instabilidades de superradi\^{a}ncia e estrelas com núcleos de mat\'{e}ria escura}
%\subsection*{Nome: Richard Pires Brito}
%\subsection*{Doutoramento em F\'isica}
%\subsection*{Orientador: Doutor V\'itor Manuel dos Santos Cardoso}
%\subsection*{Co-orientador: Doutor Paolo Pani}
%\subsection*{Resumo:}

Campos bos\'{o}nicos fundamentais com spin arbitr\'{a}rio s\~{a}o genericamente previstos em extens\~{o}es do Modelo Standard e da Relatividade Geral, e s\~{a}o fortes candidatos para explicar as componentes de energia e mat\'{e}ria escura do Universo. Um dos canais mais promissores para detectar a sua presen\c ca \'e atrav\'es da sua intera\c c\~{a}o gravitacional com objectos compactos. Neste contexto, esta tese dedica-se ao estudo de diferentes mecanismos nos quais campos bos\'{o}nicos afectam a din\^{a}mica e a estrura de buracos negros e estrelas de neutr\~{o}es.
 
A primeira parte da tese \'{e} dedicada ao estudo de campos massivos de spin-2 em torno de buracos negros esfericamente sim\'etricos. Campos massivos de spin-2 podem ser consistentemente descritos em teorias de gravidade massiva, tornando poss\'ivel um estudo sistem\'atico da propaga\c c\~ao destes campos em espa\c cos-tempo curvos. Em particular, mostramos que devido \`{a} presen\c ca de graus de liberdade adicionais nestas teorias, a estrutura das solu\c c\~{o}es descrevendo buracos negros \'e mais complexa do que na Relatividade Geral. 

Na segunda parte desta tese, discutimos em detalhe instabilidades de superradi\^{a}ncia no contexto da f\'isica de buracos negros. Mostramos que diferentes mecanismos, tais como campos bos\'{o}nicos massivos e campos magn\'eticos, podem tornar buracos negros em rota\c c\~ao inst\'aveis contra modos superradiantes, o que tem importantes implica\c c\~oes para a astrof\'isica e para a f\'isica para al\'em do Modelo Standard.  

Na \'{u}ltima parte da tese apresentamos um estudo sobre a intera\c c\~ao gravitational entre condensados de mat\'eria escura bos\'onica e estrelas compactas. 
Em particular, mostramos que configura\c c\~{o}es estelares est\'{a}veis compostas por um fluido perfeito e por um condensado bos\'{o}nico existem e podem descrever as \'{u}ltimas fases da acre\c c\~{a}o de mat\'{e}ria escura por estrelas, em ambientes ricos em mat\'eria escura.

\paragraph{Palavras-chave:} 
Objectos compactos,
campos bos\'onicos fundamentais,
campos massivos de spin-2,
instabilidades de superradi\^{a}ncia,
estrelas bos\'{o}nicas.

%===================================================================================
\newpage\mbox{}
\chapter*{Abstract}
%
%\subsection*{Title: Fundamental fields around compact objects: Massive spin-2 fields, Superradiant instabilities\\and\\Stars with dark matter cores}
%\subsection*{Name: Richard Brito}
%\subsection*{PhD in Physics}
%\subsection*{Supervisor: Doutor V\'itor Manuel dos Santos Cardoso}
%\subsection*{Abstract:}

Fundamental bosonic fields of arbitrary spin are predicted by generic extensions of the Standard Model and of General Relativity, and are well-motivated candidates to explain the dark components of the Universe. One of most promising channels to look for their presence is through their gravitational interaction with compact objects. Within this context, in this thesis I study several mechanisms by which bosonic fields may affect the dynamics and structure of black holes and neutron stars. 

The first part of the thesis is devoted to the study of massive spin-2 fields around spherically symmetric black-hole spacetimes. Massive spin-2 fields can be consistently described within theories of massive gravity, making it possible to perform a systematic study of the propagation of these fields in curved spacetimes. In particular, I show that due to the presence of additional degrees of freedom in these theories, the structure of black-hole solutions is richer than in General Relativity.

In the second part of the thesis, I discuss in detail superradiant instabilities in the context of black-hole physics. I show that several mechanisms, such as massive bosonic fields and magnetic fields, can turn spinning black holes unstable against superradiant modes, which has important implications for astrophysics and for physics beyond the Standard Model.

In the last part of this thesis, I present a study of how bosonic dark matter condensates interact gravitationally with compact stars. In particular, I show that stable stellar configurations formed by a perfect fluid and a bosonic condensate exist and can describe the late stages of dark matter accretion onto stars, in dark matter rich environments. 

\paragraph{Keywords:} 
Compact objects,
fundamental bosonic fields, 
massive spin-2 fields,
superradiant instabilities,
bosonic stars.
%===================================================================================

\newpage\mbox{}
%%%%%%%%%%%%%%%%%%%%%%%%%%%%%%%%%%%%%%%%%%%%%%%%%%%%%%%%%%%%%%%%%%%%%%%%%%%%%%%
\chapter*{Acknowledgments}

I am extremely grateful to my supervisor Prof. Vitor Cardoso, and my co-supervisor Dr. Paolo Pani for giving me the opportunity to spend four years of my life thinking about physics. I am grateful for all the discussions and all the physics that I have learned from them in the past few years. A great amount of these discussions ended up in successful collaborations (including the publication of a book!). I could not have asked for better supervisors, and I am truly grateful to have had the chance to work with them. Most of the work here presented could not have been done without their constant support and always valuable and crucial input. Thanks Vitor and Paolo! It was a pleasure to work with you and I hope that we will continue to do so in the future.

I am indebted to Emanuele Berti, Eugeny Babichev, Helvi Witek, Hirotada Okawa, Matthew Johnson, Alexandra Terrana, Carlos Palenzuela, Caio Macedo, Jorge Rocha, Carlos Herdeiro, Eugen Radu, Jo\~{a}o Lu\'is Rosa and Miguel Duarte for very fruitful collaborations, in which part of this work was done.

I am also grateful to all the members of CENTRA and especially the gravity group and my office mates, for the joyful and exceptional working environment that they provided me. I hope our paths will cross again in the future! 

I acknowledge the hospitality of the Perimeter Institute for Theoretical Physics and the gravitational physics group at the University of Mississippi, where parts of this work have been done. I also want to thank Lu\'is Crispino and its group in Universidade Federal do Par\'a for their hospitality when I visited Bel\'em.

I am indebted to Paolo Pani and Guilherme Sanches for carefully proof-reading this thesis and for useful suggestions.

I am grateful to the Funda\c c\~{a}o Calouste Gulbenkian for awarding me the ``Est\'{i}mulo \`{a} Investiga\c c\~{a}o'' Prize, to support part of the work here presented. Finally, I am indebted to the IDPASC program for awarding me the FCT fellowship, which made it possible for me to complete this thesis.

I am grateful to the friends that made my stay in Canada a wonderful experience, with a special mention to Farbod Kamiab, Ravi Kunjwal, Mansour Karami, Heidar Moradi and Kelly Leonard.

Last, but certainly not least, I want to thank all my family and friends for their constant support through these years. I am grateful to all the friends that I have made in Lisbon in these last nine years. I wouldn't be writing a PhD thesis if it wasn't for their presence. A special mention goes to all my incredible housemates that turned the painful ride of doing a PhD into a memorable couple of years.

%%%%%%%%%%%%%%%%%%%%%%%%%%%%%%%%%%%%%%%%%%%%%%%%%%%%%%%%%%%%%%%%%%%%%%%%%%%%%%%
%\newpage

%This work was supported by Funda\c c\~ao para a Ci\^encia e a Tecnologia, under the grant SFRH/BD/52047/2012.

%\cleardoublepage

%===================================================================================
\newpage\mbox{}
\tableofcontents 

\clearpage
\newpage
\section*{Acronyms}
%%%%%%%%%%%%%%%%%%%%%%%%%%%%%%%%%%%%%%%%%%%%%%%%%%%%%%%%%%%%%%%%%%%%%%%%%%%%%%%%%%%%%
% \begin{table}[ht]
\begin{tabular}{ll}
ADM    & Arnowitt-Deser-Misner                                \\
%AGN    & Active Galactic Nucleus				\\
AdS    & Anti-de Sitter                                       \\
BH     & Black hole                                           \\
%CFT    & Conformal field theory                                 \\
DM     & Dark Matter \\
% EoS    & Equation of state                                    \\
% ER     & Ergoregion                                           \\
GR     & General Relativity                                   \\ 
%GW     & Gravitational Wave                                   \\
ISCO   & Innermost stable circular orbit  \\
% HEP    & High Energy Physics                                  \\
% LHC    & Large Hadron Collider                               \\
LIGO   & Laser Interferometric Gravitational Wave Observatory \\
ODE    & Ordinary differential equation                       \\
% NR     & Numerical Relativity                                 \\
%NS     & Neutron star                                        \\
PDE    & Partial differential equation                        \\
%QCD    & Quantum Chromodynamics                               \\
% PN     & Post-Newtonian                                       \\
QNM    & Quasinormal mode                                     \\
%RN     & Reissner-Nordstr\"om				      \\
% RHIC   & Relativistic Heavy Ion Collider                      \\
WIMP   & Weakly interacting massive particles \\
ZAMO   & Zero Angular Momentum Observer
\end{tabular}
% \end{table}
%
% \bigskip

\clearpage
\newpage

%\newpage
%%%%%%%%%%%%%%%%%%%%%%%%%%%%%%%%%%%%%%%%%%%%%%%%%%%%%%%%%%%%%%%%%%%%
\chapter*{Preamble}\label{chapter:Preamble}
The research presented in this thesis has been carried out at the 
Centro Multidisciplinar de Astrof\'isica (CENTRA) at the Instituto Superior T\'ecnico / 
Universidade de Lisboa. 

I declare that this thesis is not substantially the same as any that I have submitted for a 
degree, diploma or other qualification at any other university and that no part of it has 
already been or is concurrently submitted for any such degree, diploma or other qualification.

Most of the work presented in Part~\ref{part:massive2} and~\ref{part:super} was done in collaboration with Professor Vitor Cardoso and Dr. Paolo Pani. Chapter~\ref{chapter:nonbi} was done in collaboration with Dr. Paolo Pani and Dr. Eugeny Babichev. Part~\ref{part:BFstar} is the outcome of a collaboration with Professor Vitor Cardoso, Dr. Hirotada Okawa, Dr. Caio Macedo and Dr. Carlos Palenzuela. 
Most of the chapters of this thesis have been published. The publications here presented are included below:
\begin{enumerate}[leftmargin=*]
\item
\underline{R.~Brito}, V.~Cardoso, P.~Pani,
``Massive spin-2 fields on black hole spacetimes: Instability of the Schwarzschild and Kerr solutions and bounds on graviton mass'',
Phys.\ Rev.\ D {\textbf 88} (2013) 023514, [arXiv:1304.6725[gr-qc]] 
(Chapter~\ref{chapter:massive2} and Chapter~\ref{chapter:Kerr});
\item
\underline{R.~Brito}, V.~Cardoso, P.~Pani,
``Partially massless gravitons do not destroy general relativity black holes'',
Phys.\ Rev.\ D {\textbf 87} (2013) 124024, [arXiv:1306.0908[gr-qc]] 
(Chapter~\ref{chapter:partial});
\item
\underline{R.~Brito}, V.~Cardoso, P.~Pani,
``Black holes with massive graviton hair'',
Phys.\ Rev.\ D {\textbf 88} (2013) 064006, [arXiv:1309.0818[gr-qc]] 
(Chapter~\ref{chapter:BHhair});
\item
E.~Babichev, \underline{R.~Brito}, P.~Pani, ``Linear stability of nonbidiagonal black holes in massive gravity '',
Phys.\ Rev.\ D {\textbf 93} (2016) 044041, [arXiv:1512.04058 [gr-qc]]
(Chapter~\ref{chapter:nonbi});
\item
\underline{R.~Brito}, V.~Cardoso, P.~Pani,
``Superradiant instability of black holes immersed in a magnetic field'',
Phys.\ Rev.\ D {\textbf 89} (2014) 104045, [arXiv:1405.2098[gr-qc]] 
(Chapter~\ref{chapter:magnetic});
\item
\underline{R.~Brito}, V.~Cardoso, P.~Pani, ``Black holes as particle detectors: evolution of superradiant instabilities'',
Class.\ Quant.\ Grav. 32 (2015) 13, 134001, [arXiv:1411.0686 [gr-qc]];
(Chapter~\ref{chapter:detectors});
\item
\underline{R.~Brito}, V.~Cardoso, H.~Okawa, ``Accretion of dark matter by stars'',
Phys.\ Rev.\ Lett. 115 (2015) 11, 111301, [arXiv:1508.04773 [gr-qc]];
(Part~\ref{part:BFstar})
\item
\underline{R.~Brito}, V.~Cardoso, C.~Macedo, H.~Okawa, C.~Palenzuela, ``Interaction between bosonic dark matter and stars'',
Phys.\ Rev.\ D {\textbf 93} (2016) 044045, [arXiv:1512.00466 [astro-ph.SR]].
(Part~\ref{part:BFstar})
\end{enumerate}

\bigskip
Part~\ref{part:massive2} is also partially based on a review written in collaboration with Dr. Eugeny Babichev:
\begin{itemize}
\item E.~Babichev, \underline{R.~Brito}, ``Black holes in massive gravity'',
Class.\ Quant.\ Grav. 32 (2015) 15, 154001, [arXiv:1503.07529 [gr-qc]].
\end{itemize}
Part~\ref{part:super} is also partially based on a book co-authored by the author of this thesis, and done in collaboration with Professor Vitor Cardoso and Dr. Paolo Pani:
\begin{itemize}
\item \underline{R.~Brito}, V.~Cardoso, P.~Pani, ``Superradiance: Energy Extraction, Black-Hole Bombs and Implications for Astrophysics and
Particle Physics'', Lecture Notes in Physics 906 (2015), Springer International Publishing, Switzerland.
\end{itemize}

\bigskip
 Further publications by the author written during the development of the thesis but not discussed here:
\begin{enumerate}[leftmargin=*]
\item
\underline{R.~Brito}, A.~Terrana, M.~C.~Johnson, V.~Cardoso,
``Nonlinear dynamical stability of infrared modifications of gravity'',
Phys.\ Rev.\ D {\textbf 90} (2014) 124035, \newline[arXiv:1409.0886[hep-th]];
\item
E.~Berti, \underline{R.~Brito}, V. Cardoso, ``Ultrahigh-energy debris from the collisional Penrose process'',
Phys.\ Rev.\ Lett. 114 (2015) 25, 251103, [arXiv:1410.8534[gr-qc]];
\item
V.~Cardoso, \underline{R.~Brito}, J.~L.~Rosa, ``Superradiance in stars'',
Phys.\ Rev.\ D {\textbf 91} (2015) 12, 124026, [arXiv:1505.05509 [gr-qc]];
\item
\underline{R.~Brito}, V.~Cardoso, C.~A.~R.~Herdeiro, E.~Radu, ``Proca Stars: gravitating Bose-Einstein condensates of massive spin 1 particles '',
Phys.\ Lett.\ B752 (2016) 291-295, [arXiv:1508.05395 [gr-qc]];
%
%\item 
%\underline{R.~Brito}, V.~Cardoso, J.~V.~Rocha, ``Two worlds collide: Interacting shells in AdS spacetime and chaos'',
%[arXiv:1602.03535 [hep-th]].
\end{enumerate}

%%%%%%%%%%%%%%%%%%%%%%%%%%%%%%%%%%%%%%%%%%%%%%%%%%%%%%%%%%%%%%%%%%%%

% arabische Seitenzahlen im Hauptteil 
\clearpage
\pagenumbering{arabic}

%%%%%%%%%%%%%%%%%%%%%%%%%%%%%%%%%%%%%%%%%%%%%%%%%%%%%%%%%%%%%%%%%%%%
\chapter{General introduction}\label{part:Intro}
%%%%%%%%%%%%%%%%%%%%%%%%%%%%%%%%%%%%%%%%%%%%%%%%%%%%%%%%%%%%%%%%%%%%%%%%%%%%%%
%\epigraph{}{}

Einstein's theory of General Relativity (GR) is a singular achievement in mankind's history. The theory describes how matter interacts gravitationally and has far-reaching implications. Among its many successes, one can name a few that drastically changed the way we understand our Universe. In particular, it gave us a new picture of the origin and evolution of the Universe; predicted the existence of new exotic astrophysical objects, such as black holes (BHs); predicted the existence of gravitational waves; and at the more fundamental level, changed the way we think about space and time.

No less impressive has been the quite accurate picture of the world at very small scales that the Standard Model of particle physics has given us. Although a description of gravity at small scales is still missing, there is no doubt that we now have an incredibly precise understanding of the fundamental building blocks of matter.

However, from rotation curves of galaxies~\cite{Bertone:2004pz} to Supernovae observations~\cite{Riess:1998cb}, it has gradually become clear that our Universe is not compatible with either GR on large scales, the Standard Model of particle physics, or both. 
The confirmation of these observations by a number of other experiments~\cite{Bertone:2004pz,Ade:2013sjv}, gives now compelling evidence that only $5\%$ of our Universe is composed of baryonic matter~\cite{Ade:2013sjv}, the kind of matter which forms the basis of everything we know.
The remaining $95$\% are poorly understood and constitute one the biggest unresolved mysteries of contemporary physics. 
Dark matter (DM), constituting roughly $27\%$ of the Universe, is necessary to explain the apparent existence of more matter than what is actually seen~\cite{Bertone:2004pz}, while dark energy makes up $68\%$ of the Universe, and is a key ingredient to explain the cosmological acceleration~\cite{Riess:1998cb}. Both these constituents are a mystery, in the sense that a) DM has never been detected in any Earth-bound experiment and b) the magnitude of the cosmological constant necessary to explain dark energy is 120 orders of magnitude {\it smaller} than that predicted by quantum field theory, if one believes in it up to the Planck scale~\cite{Carroll:2000fy}. Different proposals have been put forward to solve these problems. The difficulty lies not only in the fact that one must necessarily modify or even abandon some pillars of XXth century physics, but also in the intricate task of devising theoretically viable models that pass all  experimental tests at hand.

Either motivated to solve the DM problem or to explain the accelerated expansion of the Universe, a generic aspect of most of the proposals to solve these problems is the prediction of new fundamental degrees of freedom. In particular, fundamental bosonic fields stand out as a quite generic well-motivated feature of extensions of the Standard Model~\cite{Arvanitaki:2009fg,Goodsell:2009xc,Jaeckel:2010ni} and modified theories of gravity~\cite{Clifton:2011jh,Berti:2015itd}.

The feebleness with which these fundamental fields couple to ordinary matter lies at the heart of the difficulty to detect them. Extra fundamental fields may couple to Standard Model particles in various ways, which makes it challenging to exclude, or possibly detect, new effects. Fortunately, the equivalence principle guarantees that gravity is universal for all forms of matter and energy. Although gravity is way too weak for us to hope to detect the presence of these fields here on Earth through their gravitational interaction, one can expect that strongly gravitating objects, such as BHs and neutron stars, might be ideal candidates to look for smoking gun effects of the existence of new fundamental degrees of freedom. We are then offered with the intriguing possibility of using the growing wealth of observations in high-energy astrophysics~\cite{Narayan:2005ie,Brenneman:2011wz,2013Natur.494..449R} and gravitational-wave astronomy~\cite{Aasi:2013wya} to put physics beyond GR and the Standard Model to the test. 

Within this context, this thesis is devoted to the study of fundamental bosonic fields around compact objects. All the works that I here present involve classical bosonic fields propagating on curved spacetimes and are part of a broader program aiming to fully understand the physics of fundamental fields when coupled to gravity.

\vspace{0.5cm}

\noindent\textbf{Fundamental bosonic fields}

\vspace{0.5cm}

All observed elementary bosons are all either massless or very massive, such as the $W$ and $Z$ bosons and the recently-discovered Higgs boson, whose masses are of the order $m\sim 100~{\rm{GeV}}$~\cite{PDG}. 
For a compact object with mass $M$, the Compton wavelength of the bosonic field $\lambda_c$ is comparable to its size when $M/\lambda_c\sim 1$ (in units $G=c=1$). This sets the range of masses $\hbar\lambda_c^{-1}$ which are phenomenologically relevant for a given $M$. A hypothetical boson with mass in the electronvolt range would have a Compton wavelength comparable to objects with masses $M\sim 10^{20}{\rm{kg}}$. Although this kind of compact object could exist, in particular ``primordial'' BHs~\cite{1971MNRAS.152...75H,1966AZh....43..758Z,1974MNRAS.168..399C} formed in the early Universe, I will mostly focus on massive compact objects, i.e. those with masses ranging from a few solar masses to billions of solar masses. To expect any significant impact on the dynamics and structure of these objects, we must then rely on the existence of ultralight particles with masses from $\sim 10^{-25}$ eV up to $\sim 10^{-10}$ eV.

One of the most promising candidates to fall within this mass range, is the Peccei-Quinn axion~\cite{PecceiQuinn,1978PhRvL..40..223W,Wilczek:1977pj}, a pseudo-scalar field with a mass theoretically predicted to be below the electronvolt scale~\cite{Marsh:2015wka}, and introduced as a possible resolution for the strong CP problem in QCD, i.e. the observed suppression of CP violations in the Standard Model despite the fact that, in principle, the nontrivial vacuum structure of QCD allows for large CP violations. 
In addition to solve the strong CP problem, light axions are also interesting candidates for cold DM~\cite{Fairbairn:2014zta,Marsh:2014qoa}.
Furthermore, a plenitude of ultralight bosons might arise from moduli compactification in string theory. In the ``axiverse'' scenario, multiples of light axion-like fields can populate the mass spectrum down to the Planck mass, $M_P \sim10^{-33}~{\mathrm{eV}}$, and can provide interesting phenomenology at astrophysical and cosmological scales~\cite{Arvanitaki:2009fg}.

In addition to these beyond-the-Standard-Model particles, effective scalar degrees of freedom also arise in several modified theories of gravity~\cite{Berti:2015itd}.
For example, in so-called scalar-tensor theories, the gravitational interaction is mediated by a scalar field in addition to the standard massless graviton~\cite{Fujii:2003pa}. Due to a correspondence between scalar-tensor theories and theories which replace the Einstein-Hilbert term by a generic function of the Ricci curvature (so-called $f(R)$ gravity~\cite{Sotiriou:2008rp}), effective massive scalar degrees of freedom are also present in these theories.

Bosonic fields with spin are also a generic feature of extensions of the Standard Model and of GR, but have received much less attention, mainly due to the complexity of their field equations. For example, massive vector fields (``dark photons''~\cite{Ackerman:mha}) arise in the so-called hidden $U(1)$ sector~\cite{Goodsell:2009xc,Jaeckel:2010ni,Camara:2011jg,Goldhaber:2008xy,Hewett:2012ns}. In fact, several proposals have advocated massive spin vector fields as a DM ingredient~\cite{Holdom:1985ag,ArkaniHamed:2008qn,Pospelov:2008jd,Goodsell:2009xc}, making the study of these fields of special importance.

On the other hand, massive tensor fields are a much more involved problem from a theoretical standpoint, but progress in describing consistently the gravitational interaction of massive tensor fields with gravity has been recently done in the context of nonlinear massive gravity and bimetric theories~\cite{deRham:2010kj,Hassan:2011hr} (see also Refs.~\cite{Hinterbichler:2011tt,deRham:2014zqa,Schmidt-May:2015vnx} for reviews). As pointed out in Refs.~\cite{Hassan:2012wr,Schmidt-May:2015vnx,Aoki:2016zgp,Babichev:2016hir}, massive bimetric theories can, in certain limits, consistently describe the coupling of massive spin-2 fields to gravity. These theories were originally motivated by the hope of solving the cosmological constant problem~\cite{deRham:2014zqa}, but the possibility that they could also mimic the presence of DM was also recently considered in Refs.~\cite{Bernard:2014psa,Blanchet:2015bia,Blanchet:2015sra,Aoki:2016zgp,Babichev:2016hir}. 

Massive gravity theories can also describe a putative massive graviton. A non-zero mass for the graviton would have potential impacts in gravitational-wave searches. In fact, as I will discuss in this thesis, the recent first direct detection of gravitational waves emitted by a compact-binary coalescence~\cite{Abbott:2016blz} already constrains the mass of the graviton to a range where interesting effects might occur around supermassive BHs with mass of the order of $M\sim 10^8 M_{\odot}$.

\vspace{0.5cm}

\noindent\textbf{Bosonic fields and black holes}

\vspace{0.5cm}

Most of the interesting phenomena resulting from the interaction between bosonic fields with BHs are associated to BH superradiance~\cite{Brito:2015oca}. Under certain conditions, bosonic waves scattering off spinning BHs can be amplified at the expense of the BH rotational energy. In confined systems, this superradiant scattering can lead to strong instabilities, with applications to high-energy physics, astrophysics and to physics beyond the Standard Model.
The mass term of a massive bosonic field provides the necessary confinement for low-frequency waves to trigger superradiant instabilities around Kerr BHs~\cite{Brito:2015oca}. This is of particular interest to probe new fundamental degrees of freedom. In fact, efforts have already been made in order to use BHs as particle-physics laboratories, through which one can constrain the mass of the QCD axion~\cite{Arvanitaki:2014wva}, of stringy pseudoscalars populating the so-called axiverse~\cite{Arvanitaki:2009fg,Arvanitaki:2010sy}, and the hidden $U(1)$ sector~\cite{Goodsell:2009xc,Jaeckel:2010ni,Pani:2012bp,Pani:2012vp}. In addition to their phenomenological relevance, such studies have revealed unexpected aspects related to the dynamics of these fields in curved spacetime. 

One of the main purposes of this thesis is to explore the physics behind these superradiant instabilities. In particular, I will show that massive spin-2 fields, including a putative massive graviton, can also render Kerr BHs superradiantly unstable, which has strong implications for the existence of ultra-light spin-2 fields.

As already mentioned, these particles can be described within a class of alternative theories of gravity known as massive gravity and massive bimetric gravity. If one wishes to study the phenomenology of these theories or any other modified theory of gravity, it is also crucial that they pass theoretical tests. 
These include internal theoretical consistency, absence of pathologies, and existence of stable gravitational solutions describing physical systems. In this context, BH solutions are the ideal test bed to probe the strong-curvature regime of any relativistic (classical) theory of gravity~\cite{Berti:2015itd}. Viable candidates of modified-gravity theories should possess BH solutions and the latter should (presumably) be dynamically stable, at least over the typical observation time scale of astrophysical compact objects. 
Therefore, I will also provide a detailed study on BH solutions in massive gravity, with particular emphasis on their stability properties under small perturbations.

As a by-product of understanding how massive spin-2 fluctuations behave in BH spacetimes, we will also start to understand how gravitational waveforms might differ from GR if the graviton has a small but non-vanishing mass. In fact, the extra gravitational polarizations and the nontrivial dispersion introduced by a putative small graviton mass may leave important imprints on gravitational waveforms. Given that advanced gravitational-wave detectors~\cite{LIGO,VIRGO} are now starting to detect their first sources~\cite{Abbott:2016blz}, an accurate description of these effects is of the utmost importance. 

\vspace{0.5cm}

\noindent\textbf{Bosonic fields and compact stars}

\vspace{0.5cm}

Over the past three decades, several studies on the interaction of DM and compact stars have concluded that, for old enough stars, the accumulation of DM inside the star can lead, quite generically, to the formation of a BH, which eventually destroys the whole star (see e.g., Refs.~\cite{Goldman:1989nd,Kouvaris:2011fi,Bramante:2014zca,Bramante:2015cua,Kurita:2015vga}). These strong claims, which have been used to constrain DM particles, lack of a rigorous proof, making it crucial to better understand these processes. In particular, situations leading to BH formation can only be accessed within full GR. 
Here I present the first steps taken in this direction, by considering DM as being composed of coherent massive bosonic fields minimally coupled to GR. This simple but rigorous model allow us to start to understand how DM might affect the global structure of compact stars. One of the main results here presented, is that stable stellar configurations with DM cores can, in principle, form and avoid collapse to a BH if DM is composed of bosonic fields, \emph{independently} of the mass of the field.

An important side-product of these studies will be the construction of novel self-gravitating compact solutions for massive vector fields. In fact, massive bosonic fields coupled to gravity have the tendency to form stellar-like structures.
These objects, known as boson stars~\cite{Kaup:1968zz,Ruffini:1969qy,Brito:2015pxa} or oscillatons~\cite{Seidel:1991zh,Brito:2015yfh} depending on the nature of the field, have been proposed to solve long-standing problems in DM physics, in particular if DM is composed of ultra-light bosonic fields with masses $\ll$ eV~\cite{Suarez:2013iw,Li:2013nal}. 

In fact, a popular viewpoint is that DM consists of weakly interacting massive particles (WIMPs), with masses $\gtrsim$ GeV~\cite{Bertone:2004pz}. Despite its success in modeling structure formation, WIMPs DM models have been shown to form too much structure at small scales, leading to a mismatch with observations. Examples such as the ``missing satellite''~\cite{Moore:1999nt,Klypin:1999uc}  and the ``cuspy core''~\cite{deBlok:2009sp} problem are an evidence that our current best models cannot be complete. Although these issues may eventually be solved within the WIMP paradigm, for example by taking properly into account the effect of baryonic processes~\cite{Governato:2012fa,Brooks:2012vi}\footnote{Another possibility to solve the small-scale problems is the introduction of particles with strong self-interactions~\cite{Spergel:1999mh}.}, an interesting possibility is that the tendency for ultra-light bosonic fields to form gravitationally-bound macroscopic condensates may hamper the formation of structure at small scales, thus effectively solving these problems. Studying the physics of self-gravitating bosonic structures is thus timely and relevant in the context of DM searches.

\vspace{0.5cm}

\noindent\textbf{Structure of the thesis}

\vspace{0.5cm}

For the reader's convenience, I summarize here the structure of the thesis.
Part of it will be dedicated to the study of the impact of massive spin-2 fields on the dynamics of BH spacetimes. I will show in Part~\ref{part:massive2}, that due to the complex structure of theories describing massive spin-2 fields, BHs have a much richer structure than in GR. I will study generic massive spin-2 fields around spherically symmetric BHs and show how this can be consistently described within theories of massive gravity. I will also show that massive spin-2 perturbations can render the Schwarzchild BH unstable and that, due to this instability, these theories allow for non-Schwarzschild BH solutions. For completeness, I will also study perturbations around Schwarzschild-de Sitter spacetimes in a specific limit of linear massive gravity known as partially massless gravity, and will also discuss generic perturbations of another class of solutions of massive gravity theories (known as non-bidiagonal solutions).

In Part~\ref{part:super}, I will show how rotating Kerr BHs can be rendered superradiantly unstable in the presence of bosonic fields. I will set the stage by studying different scenarios where such instability occurs, namely for a Kerr BH enclosed by a reflecting mirror and for a magnetized Kerr BH. I will then show that, under certain conditions, Kerr BHs are unstable against massive spin-2 fields, similarly to massive scalar and vector fields~\cite{Brito:2015oca}. Finally, I study the evolution of this instability within an adiabatic approximation and show how these results can be used to put strong constraints on ultralight bosonic particles. In particular, these constraints can be used to place a bound on the graviton mass $m_g\lesssim 5\times 10^{-23}$ eV, which is one order of magnitude stronger than the constraints imposed by the gravitational-wave observation GW150914 by Advanced LIGO~\cite{Abbott:2016blz}.

Finally, Part~\ref{part:BFstar} will be dedicated to the study of bosonic fields around compact stars. I will show that massive bosonic fields minimally coupled to gravity generically form compact structures, and argue that if DM is composed of massive bosonic fields, then stable stellar configurations formed by a perfect fluid and a bosonic condensate can form and avoid collapse to a BH. 

Unless otherwise stated, I use geometrized units where $G=c=1$, so that energy and time have units of length. I also adopt the
$(-+++)$ convention for the metric.  
%%%%%%%%%%%%%%%%%%%%%%%%%%%%%%%%%%%%%%%%%%%%%%%%%%%%%%%%%%%%%%%%%%%%

%%%%%%%%%%%%%%%%%%%%%%%%%%%%%%%%%%%%%%%%%%%%%%%%%%%%%%%%%%%%%%%%%%%%
\part{Massive spin-2 fields and black-hole spacetimes}\label{part:massive2}
%%%%%%%%%%%%%%%%%%%%%%%%%%%%%%%%%%%%%%%%%%%%%%%%%%%%%%%%%%%%%
\chapter{Massive spin-2 fields and strong gravity}\label{sec:longintro}
\emph{This part is based on Refs.~\cite{Brito:2013wya,Brito:2013yxa,Babichev:2015zub,Brito:2013xaa,Babichev:2015xha}}.
%%%%%%%%%%%%%%%%%%%%%%%%%%%%%%%%%%%%%%%%%%%%%%%%%%%%%%%%%%%%%
\section{Massive spin-2 fields?}
%%%%%%%%%%%%%%%%%%%%%%%%%%%%%%%%%%%%%%%%%%%%%%%%%%%%%%%%
Higher-spin fields are predicted to arise in several contexts~\cite{Bouatta:2004kk,Sorokin:2004ie,Sagnotti:2013qp}. The motivation to investigate their gravitational dynamics is twofold.
The first reason is conceptual and is tied to a renewed interest in theories of massive gravity.
Massive gravity is a modification of GR based on the idea of equipping the graviton with mass. 
A model of non self-interacting massive gravitons was first suggested by Fierz and Pauli in the early beginnings of field theory~\cite{Fierz:1939ix}. They showed that at the linear level there is only one ghost- and tachyon-free, Lorentz-invariant mass term that describes the five polarizations of a massive spin-2 field on a flat background.
However, in the zero-mass limit, the Fierz-Pauli theory does not recover linear GR due to the existence of extra degrees of freedom introduced by the graviton mass. In the massless limit, the helicity-0 state maintains a finite coupling to the trace of the source stress-energy tensor, modifying the Newtonian potential and hence yielding predictions
which differ from the massless graviton theory, rendering the theory inconsistent with observations~\cite{vanDam:1970vg,Zakharov:1970cc,Iwasaki:1971uz,VanNieuwenhuizen:1973qf,Carrera:2001pj}. 

To overcome this difficulty, Vainshtein~\cite{Vainshtein:1972sx} argued that the discontinuity present in the Fierz-Pauli theory is an artifact of the linear theory, and that the full nonlinear theory has a smooth limit for $m_g\to 0$, where $m_g$ is the graviton mass. He found that around any massive object of mass $M$, for generic nonlinear massive gravity theories there is a new length scale known as the Vainshtein radius, $r_V$, where the actual form of the Vainsthein radius depends on the nonlinear theory. For example, in ghost-free theories, it is given by $r_V\sim\left(r_S \lambda_g^2\right)^{1/3}$~\cite{Babichev:2013usa}, where $r_S=2M$ is the gravitational radius of the object and $\lambda_g$ is the Compton wavelength of the massive graviton.
Nonlinearities begin to dominate at $r\lesssim r_V$ invalidating the predictions made by the linear theory. This is due to the fact that at high energies the helicity-$0$ mode of the graviton, responsible for the discontinuity, becomes strongly coupled to itself, screening its presence near the source. Much later, rigorous studies of several non-linear completions of massive gravity showed that, in general, one could indeed recover GR solutions at small length scales through the Vainshtein mechanism (for a review on the Vainshtein mechanism see Ref.~\cite{Babichev:2013usa}).
However, it was believed until recently that Lorentz-invariant nonlinear massive gravity theories were doomed to fail due to the (re)appearance of a ghost-like sixth degree of freedom~\cite{Boulware:1973my}. This was studied by Boulware and Deser who showed that in nontrivial backgrounds there are 6 degrees of freedom, where the extra degree of freedom was shown to be a ghost scalar, known as the Boulware-Deser ghost. 

This problem has only recently been solved in a series of works~\cite{deRham:2010kj,deRham:2010ik,deRham:2011rn,Hassan:2011tf,Hassan:2012qv,Hassan:2011hr,Hassan:2011ea}, 
where it was shown that for a subclass of nonlinear massive gravity theories, the Boulware-Deser ghost does not appear, both in massive gravity --- a theory with one dynamical and one fixed metric, the so-called de Rham, Gabadadze and Tolley (dRGT) model --- and 
its bimetric extension (see Refs.~\cite{Hinterbichler:2011tt,deRham:2014zqa,Schmidt-May:2015vnx} for recent reviews on massive gravity and bimetric theories)\footnote{A settled term often used in the literature to refer to dRGT theory and its extensions is the ``ghost-free massive gravity''. This name should be used with care, since dRGT theory is free from the Boulware-Deser ghost, but the other degrees of freedom are not necessary healthy for particular solutions. Therefore the right term should rather be ``Boulware-Deser ghost free massive gravity''. We will sometimes refer to dRGT theory as ``ghost free massive gravity'', keeping in mind the above reservation.}.
Bimetric theories contain two dynamical metrics, which interact with each other via non-derivative terms. If ordinary matter only couples to one of the metrics, then one can interpret the theory as an extension of Einstein's gravity with an extra spin-2 field coupled to gravity in a particular non-minimal way.

 %%%%%%%%%%%%%%%%%%%%%%%%%%%%%%%%%%%%%%%%%%%%%%%%%%%%%%%%
\section{Gravitational-wave searches and astrophysics}
%%%%%%%%%%%%%%%%%%%%%%%%%%%%%%%%%%%%%%%%%%%%%%%%%%%%%%%%

The second motivation to investigate massive spin-2 fields is of a more practical and phenomenological nature. We now have strong evidences that gravitational waves exist. The LIGO experiment recently reported the first direct observation of gravitational waves emitted by two merging BHs~\cite{Abbott:2016blz}.
Before this detection, the best dynamical bound on the mass of the graviton came from binary-pulsar observations~\cite{Finn:2001qi}, which provided the first compelling evidence for gravitational waves through the Hulse-Taylor pulsar~\cite{1975ApJ}, given by $m_g\lesssim 7.6\times 10^{-20} {\rm{eV}}$. This bound comes from the fact that a hypothetical massive graviton would affect the decay rate of an orbiting pulsar~\cite{Damour:1990wz,Goldhaber:2008xy}\footnote{Note however that the theory considered in Ref~\cite{Finn:2001qi} does not satisfy the Fierz-Pauli tuning and hence it contains a ghost. It would be interesting to repeat such calculation for viable theories. In this case however, the Vainshtein mechanism discussed in the main text may prevent a consistent linear analysis.}.
On the other hand, a Yukawa-like potential of a hypothetical graviton mass would also be responsible for a modified dispersion relation and consequent deformation of the gravitational-wave signal during its journey from the source to the observer. In other words, a small graviton mass {\it may} not affect the inspiral of a binary system to a significant extent (including the changes in period of the binary pulsar), but introduces nontrivial dispersion which acts over several Compton wavelengths, $\sim \hbar m_g^{-1}$. This peculiar effect can leave a signature in the gravitational waveform, and in fact the gravitational-wave observation GW150914 by Advanced LIGO already allows to put strong constraints on the mass of the graviton $m_g\lesssim 1.2\times 10^{-22} {\rm{eV}}$~\cite{Abbott:2016blz,TheLIGOScientific:2016src}, precisely due to this effect\footnote{I will show in Chapter~\ref{chapter:Kerr} that a putative massive graviton would also turn Kerr BHs unstable against superradiant instabilities, and this can be used to place even stronger constraints on the graviton mass $m_g\lesssim 5\times 10^{-23}$ eV.}.

However, even with these tight constraints in place and because any gravitational wave detection will occur with very low signal-to-noise-ratio, an accurate knowledge of these effects may be important, in the sense that accurate templates are required to detect extra polarizations without introducing bias~\cite{Will:1997bb,Chatziioannou:2012rf} [see Ref.~\cite{Yunes:2013dva} for a recent review]. 

In summary, gravitational waveforms for inspiralling objects emitting massive gravitons are necessary. There are several ways to deal with this problem, e.g., full nonlinear simulations, slow-motion expansions or perturbative expansions around some background. We will initiate here the latter, by understanding how small vacuum fluctuations behave in bimetric theories and massive gravity. As a by-product, we are able to understand stability properties of BHs in these theories and begin to understand
how gravitational waveforms differ from GR [see also Ref.~\cite{DeFelice:2013nba} for a recent attempt].

%%%%%%%%%%%%%%%%%%%%%%%%%%%%%%%%%%%%%%%%%%%%%%%%%%%%%%%%%%%%%%%%%%%%%%%%%%%
\section{Massive Gravity}\label{sec:theory}
%%%%%%%%%%%%%%%%%%%%%%%%%%%%%%%%%%%%%%%%%%%%%%%%%%%%%%%%%%%%%%%%%%%%%%%%%%%
\subsection{The Fierz-Pauli tuning in flat spacetime}
%%%%%%%%%%%%%%%%%%%%%%%%%%%%%%%%%%%%%%%%%%%%%%%%%%%%%%%%%%%%%%%%%%%%%%%%%%%
Let us start by reviewing the classical Fierz-Pauli theory describing a linear massive spin-2 field propagating in a four-dimensional flat spacetime~\cite{Fierz:1939ix}. 
The linear theory can also be viewed as an expansion of the full non-linear massive gravity model around a Minkowski background. 
Expanding the Einstein-Hilbert action in metric perturbations $h_{\mu\nu}$ as $g_{\mu\nu}=\eta_{\mu\nu}+h_{\mu\nu}$, 
where $\eta_{\mu\nu}$ is the Minkowski metric $\eta_{\mu\nu}= \rm{diag}(-1,+1,+1,+1)$ and the indices of $h_{\mu \nu}$ are moved up and down with the metric $\eta_{\mu \nu}$, and keeping only quadratic terms in the action we obtain the linear approximation of GR,
\begin{equation}\label{GRq}
	S_{GR} = M_P^2\int d^4 x \sqrt{-g}R = M_P^2 \int d^{4}x
\left( -\frac12 h^{\mu\nu}\mathcal{E}^{\alpha\beta}_{\mu\nu} h_{\alpha\beta}\right) +\mathcal{O} (h^3),
\end{equation}
where $M_P$ is the Planck mass and 
\begin{equation}
{\cal E}_{\mu \nu}  \equiv \mathcal{E}^{\alpha\beta}_{\mu\nu}h_{\alpha\beta} 
	= -\frac{1}{2} \partial_\mu \partial_\nu h - \frac{1}{2} \Box h_{\mu \nu} + \frac{1}{2} \partial_\rho \partial_\mu h^\rho_\nu 
		+ \frac{1}{2}\partial_\rho \partial_\nu h^\rho_\mu  - \frac{1}{2} \eta_{\mu \nu}(\partial^\rho \partial^\sigma h_{\rho \sigma} - \Box h),
\end{equation}
is the linearized Einstein tensor $G_{\mu\nu} = {\cal E}_{\mu \nu}  +\mathcal{O} (h^2) $. 
When matter is present, the metric perturbation $h_{\mu\nu}$ is also coupled to the energy-momentum tensor $T_{\mu\nu}$, via the interaction term $h_{\mu\nu}T^{\mu\nu}$, 
but since here we will mostly consider vacuum solutions, we omit this term.  

The action (\ref{GRq}) contains only derivative terms and is the unique Lorentz-invariant and ghost-free action that one can write for a massless spin-2 field in Minkowski spacetime~\cite{Fierz:1939ix}. In fact, starting from~\eqref{GRq} one can show that Einstein-Hilbert's action is the only consistent nonlinear generalization of~\eqref{GRq}, when one requires that the massless spin-2 field gravitates, i.e., couples to its own stress-energy tensor~\cite{Weinberg:1965rz,Gupta:1954zz}\footnote{The uniqueness of Einstein's equations was proven by Lovelock~\cite{Lovelock:1971yv,Lovelock:1972vz}, assuming four-dimensional spacetimes, no extra degrees of freedom besides the massless spin-2 field, diffeomorphism invariance, and linear coupling to the stress-energy tensor of the matter sector. See also Ref.~\cite{Berti:2015itd}.}.

On the other hand, to describe a massive spin-2 field one can start by adding non-derivative quadratic terms $h^2$ (where $h=h_{\mu \nu}\eta^{\mu \nu}$) and $h_{\mu\nu}h^{\mu\nu}$ to the action (\ref{GRq}).
The most generic action that one can write is~\cite{Fierz:1939ix}
\begin{equation}\label{FPaction}
S_{FP} = M_P^2 \int d^4x   \left[  -\frac12 h^{\mu\nu}\mathcal{E}^{\alpha\beta}_{\mu\nu} h_{\alpha\beta}
- \frac{1}{4} \mu^2 \left( h_{\mu\nu}h^{\mu\nu}-\kappa h^2\right) \right],
\end{equation}
where $\kappa$ is an arbitrary constant, and $m_g=\mu\hbar$ is the graviton mass. When $\mu=0$, the action reduces to the linearized Einstein-Hilbert action. When $\mu\neq0$, the mass term violates the diffeomorphism invariance of GR, i.e., this action is not invariant under infinitesimal transformations of the form
\be
\delta h_{\mu\nu}=\partial_{\mu}\xi_{\nu}(x)+\partial_{\nu}\xi_{\mu}(x)\,.
\ee

The equations of motion are given by
\begin{equation}\label{eqmotion}
\Box h_{\mu\nu}-\partial_{\lambda}\partial_{\mu}h^{\lambda}_{\nu}-\partial_{\lambda}\partial_{\nu}h^{\lambda}_{\mu}+
\eta_{\mu\nu}\partial_{\lambda}\partial_{\sigma}h^{\lambda\sigma}
+\partial_{\mu}\partial_{\nu}h-\eta_{\mu\nu}\Box h-\mu^2\left(h_{\mu\nu}-\kappa \eta_{\mu\nu}h\right)=0\,. 
\end{equation}
Acting with $\partial^{\mu}$ on \eqref{eqmotion} we find the constraint
\be
\label{harmonic}
\partial^{\nu}h_{\nu\mu}-\kappa\partial_{\mu}h=0\,.
\ee
Note that for $\kappa={1}/{2}$ this corresponds to the harmonic gauge in linearized GR. Plugging this back into the field equations and taking the trace, we find
\be
\label{trace}
2(1-\kappa)\Box h+(1-4\kappa)\mu^2 h=0\,.
\ee
Substituting the trace condition, Eq.~\eqref{eqmotion} reads
\be
\label{eqmotion2}
(\Box-\mu^2)h_{\mu\nu}=(2\kappa-1)\left[\partial_{\mu}\partial_{\nu}h+\frac{1}{2}\eta_{\mu\nu}\mu^2 h\right]\,.
\ee
For massive spin-2 particles we must have $2s+1=5$ degrees of freedom. The only choice for the constant $\kappa$ that describes a single massive graviton is the Fierz-Pauli tuning, $\kappa=1$~\cite{Fierz:1939ix}.
In this case, the full set of linearized equations reads:
\be
\label{eqmotion3}
(\Box-\mu^2)h_{\mu\nu}=0\,,\qquad \partial^{\mu}h_{\mu\nu}=0\,,\qquad h=0\,.
\ee
On the other hand, for $\kappa\neq 1$ the theory propagates 6 degrees of freedom. The extra polarization comes from a scalar ghost (a scalar with negative kinetic energy) of mass $m_{{\rm{ghost}}}^2=-\frac{1-4\kappa}{2(1-\kappa)}\mu^2$, which arises from the trace equation~\eqref{trace}. The ghost mass approaches infinity as the Fierz-Pauli tuning is approached, so that the ghost decouples in this limit. 

%%%%%%%%%%%%%%%%%%%%%%%%%%%%%%%%%%%%%%%%%%%%%%%%%%%%%%%%%%
\subsection{Non-linear massive gravity}
%%%%%%%%%%%%%%%%%%%%%%%%%%%%%%%%%%%%%%%%%%%%%%%%%%%%%%%%%%

Starting from the quadratic action \eqref{FPaction}, one can try to guess a non-linear generalization of the theory.
Obviously, the first term in~\eqref{FPaction} constructed solely out of the metric $g$ should be the Einstein-Hilbert term~\footnote{In principle, since Lovelock's theorem is not valid in massive gravity, other kinetic terms could be possible. However it was shown in Ref.~\cite{deRham:2013tfa} that any new kinetic term would generically lead to ghost instabilities.}. 
However, it is not possible to get a non-linear massive term using only the metric $g_{\mu\nu}$, 
since the only nontrivial term corresponds to a Lagrangian density proportional to $\sqrt{-g}$, which acts as a cosmological constant. 
Thus, the only way to have a non-trivial mass term is to add a second metric, say $f_{\mu\nu}$, which can be chosen to be fixed 
(in this case the theory has a preferred background, i.e. ``aether'') or dynamical, in which case the theory is called massive bimetric gravity. We will sometimes simply call it massive bigravity for short. 
The metric $g_{\mu\nu}$ can be non-derivatively coupled to the second metric $f_{\mu\nu}$, in order to form a non-trivial mass term. 
Non-linear mass terms should be chosen such that: (i) the action is invariant under a coordinate change common to both metrics; (ii) there is a flat solution for $g_{\mu \nu}$; (iii) in the limit where $g_{\mu \nu}=\eta_{\mu\nu}+h_{\mu\nu}$ and  $f_{\mu\nu}=\eta_{\mu\nu}$ the potential at quadratic order for $h_{\mu \nu}$ takes the Fierz-Pauli form~\eqref{FPaction}. In spite of these restrictions, there is a huge freedom in choosing the mass term. In fact, one can choose the interaction term in a class of functions satisfying these conditions (see e.g. Ref.~\cite{Damour:2002ws}). 
The term  
$$
\sqrt{-g}\;(g_{\mu \nu} - f_{\mu \nu}) (g_{\sigma \tau} - f_{\sigma \tau}) \left(g^{\mu\sigma}g^{\nu\tau}-g^{\mu\nu}g^{\sigma\tau}\right),
$$
considered in \cite{ArkaniHamed:2002sp}, is an example of such an interaction. 
As it has been proposed by Vainshtein~\cite{Vainshtein:1972sx}, the inclusion of non-linear terms in the equations of motion of massive gravity is essential for this theory to have a smooth GR limit, when the graviton mass is continuously taken to zero (for a recent review see~\cite{Babichev:2013usa}).
However, as shown by Boulware and Deser~\cite{Boulware:1973my}, generic non-linear interaction terms lead to ghost instabilities (appearing at the non-linear level).

%%%%%%%%%%%%%%%%%%%%%%%%%
\subsection{dRGT massive gravity and bimetric theories}
\label{dRGT}
%%%%%%%%%%%%%%%%%%%%%%%

Although,  in general,  the Boulware-Deser ghost persists in non-linear massive and bimetric theories, it has been found by de Rham, Gabadadze and Tolley (dRGT)  that 
in the so-called ``decoupling limit'' --- a limit where the degrees of freedom of the theory (almost) decouple --- there is a restricted subclass of potential terms, for which the Boulware-Deser ghost is absent even at the non-linear level~\cite{deRham:2010kj,deRham:2010ik,deRham:2011rn,Hassan:2012qv}. A full Hamiltonian analysis confirmed the absence of the Boulware-Deser ghost in this model~\cite{Hassan:2011tf,Hassan:2011hr,Hassan:2011ea}, while fully covariant proofs were given in Ref.~\cite{Deffayet:2012nr} for a subset of possible massive terms and for generic mass terms in~\cite{Kugo:2014hja} (see also the review~\cite{deRham:2014zqa}).

The Lagrangian of this most general bimetric theory without the Boulware-Deser ghost reads~\cite{deRham:2010ik,Hassan:2011zd} (dRGT massive gravity is recovered when one of the metrics is taken to be non-dynamical),
\be
{\cal L}=\sqrt{|g|}\left[M_g^2R_g+M_f^2\sqrt{{f}/{g}}\, R_f-2M_v^4\, V\left(g,f\right)\right]\,. \label{biaction}
\ee
Here $R_g$ and $R_f$ are the Ricci scalars corresponding to the two metrics $g_{\mu\nu}$ and $f_{\mu\nu}$, respectively; $M_g^{-2}={16\pi G}$, $M_f^{-2}={16\pi \mathcal{G}}$ are the corresponding gravitational couplings; and $M_v$ is a constant related to the graviton mass. The quantities $f,g$ denote the determinants of the corresponding metric. The potential can be written as
\be
\label{potential}
V:=\sum_{n=0}^4\,\beta_n V_n\left(\gamma\right)\,, \quad \gamma^{\mu}\,_{\nu}:=\left(\sqrt{g^{-1}f}\right)^{\mu}\,_{\nu}\,,
\ee
where $\beta_n$ are real parameters,
%%%%%
\beq
V_0&=&1\,,\quad
V_1=[\gamma]\,,\quad
V_2=\frac{1}{2}\left([\gamma]^2-[\gamma^2]\right)\,,\nn\\
V_3&=&\frac{1}{6}\left([\gamma]^3-3[\gamma][\gamma^2]+2[\gamma^3]\right)\,,\quad
V_4=\det(\gamma)\,,
\eeq
and the square brackets denote the matrix trace.

The Lagrangian~\eqref{biaction} gives rise to two sets of modified Einstein's equations for $g_{\mu\nu}$ and $f_{\mu\nu}$, 
\beq
G_{\mu\nu} +\frac{M_v^4}{M_g^2}T_{\mu\nu}(\gamma)&=&0\,, \label{field_eqs1} \\
\mathcal{G}_{\mu\nu} +\frac{M_v^4}{M_f^2}\mathcal{T}_{\mu\nu}(\gamma)&=&0\,, \label{field_eqs2}
\eeq
where $G_{\mu\nu}$ and $\mathcal{G}_{\mu\nu}$ are the corresponding Einstein tensors for the two  metrics $g_{\mu\nu}$ and $f_{\mu\nu}$, and
\be
T_{\mu\nu}=\sum_{n=0}^3(-1)^n\beta_n g_{\mu\lambda}Y^{\lambda}_{\nu}(\gamma)\,,\qquad
\mathcal{T}_{\mu\nu}=\sum_{n=0}^3(-1)^n\beta_{4-n} f_{\mu\lambda}Y^{\lambda}_{\nu}(\gamma^{-1})\,,
\ee
with $Y(\gamma)=\sum_{r=0}^n (-1)^r\gamma^{n-r}V_r(\gamma)$~\cite{Hassan:2011zd}.
The Bianchi identity implies the conservation conditions
\be
\nabla_g^{\mu}T_{\mu\nu}(\gamma)=0\,,\quad \nabla_f^{\mu}\mathcal{T}_{\mu\nu}(\gamma)=0\,, \label{bianchi1}
\ee
where $\nabla_g$ and $\nabla_f$ are the covariant derivatives with respect to $g_{\mu\nu}$ and $f_{\mu\nu}$, respectively. In fact, these two conditions are not independent from each other due to the diffeomorphism invariance of the interaction term in \eqref{biaction}, which is a general property of the ``Fierz-Pauli like'' interaction terms~\cite{Damour:2002ws}.  

%%%%%%%%%%%%%%%%%%%%%%%%%%%%%%%%%%%%%%%%%%%%%%%%%%%%%%%%
\section{Outline of Part I}
%%%%%%%%%%%%%%%%%%%%%%%%%%%%%%%%%%%%%%%%%%%%%%%%%%%%%%%%
%
In Chapter~\ref{chapter:massive2} we obtain the Fierz-Pauli theory~\cite{Fierz:1939ix} for a linearized massive spin-2 field propagating on \emph{curved} backgrounds and show how it can be consistently obtained in bimetric and massive gravity (or massive (bi)gravity for short). Within this context we perform  a complete analysis of the linear dynamics of massive spin-2 fields on a Schwarzschild BH. We first focus on the monopole mode that corresponds to the scalar polarization of a massive graviton. We find a strongly unstable, spherically symmetric mode, which was also discussed in Ref.~\cite{Babichev:2013una}. We also show that the inclusion of a cosmological constant makes the Schwarzschild-de Sitter BHs even more unstable.
We then study non-spherically symmetric massive spin-2 perturbations around the Schwarzschild geometry, and show that the spectrum supports a rich structure of quasinormal modes (QNMs) and quasibound, long-lived states.

In Chapter~\ref{chapter:partial} we show that for a particular tuning of the cosmological constant with the mass of the spin-2 field, known as partially massless gravity, the spherically symmetric instability disappears. Remarkably, for this particular case the spectrum of massive gravitational perturbations is isospectral, i.e., perturbations with opposite parity have the same quasi-normal spectrum.

We complete these results in Chapter~\ref{chapter:nonbi} where we study generic linear perturbations of another class of spherically symmetric solutions of massive (bi)gravity, namely the nonbidiagonal solutions (this terminology will be explained in the next chapter). 
We show that the quasinormal spectrum of these solutions coincides with that of a Schwarzschild BH in GR, thus proving that these solutions are mode stable. This is in contrast to the case of bidiagonal BH solutions studied in Chapter~\ref{chapter:massive2}  which are affected by a radial instability.
On the other hand, we show that the full set of perturbation equations is generically richer than that of a Schwarzschild BH in GR, and this affects the linear response of the BH to external perturbations.

The instability of the Schwarzshild solutions against massive spin-2 perturbations suggests the existence of new static solutions in massive bi(gravity). We end the first part of this thesis in Chapter~\ref{chapter:BHhair}, by confirming that novel BH solutions indeed exist in these theories. 

%%%%%%%%%%%%%%%%%%%%%%%%%%%%%%%%%%%%%%%%%%%%%%%%%%%%%%%%
\chapter{Massive spin-2 fields around Schwarzschild black holes}\label{chapter:massive2}
%%%%%%%%%%%%%%%%%%%%%%%%%%%%%%%%%%%%%%%%%%%%%%%%%%%%%%%%

%%%%%%%%%%%%%%%%%%%%%%%%%%%%%%%%%%%%%%%%%%%%%%%%%%%%%%%%
\section{Introduction}
%%%%%%%%%%%%%%%%%%%%%%%%%%%%%%%%%%%%%%%%%%%%%%%%%%%%%%%%

We wish to describe two different cases: i) the interaction of a generic massive spin-2 field with standard gravity, that is, we consider the massive tensor as a probe field propagating on a geometry which solves Einstein equations; ii) the linearized dynamics of a massive graviton as it emerges in nonlinear massive gravity. It turns out that both cases can be described consistently within a common framework.

More specifically, we consider the action for two tensor fields, $g_{\mu\nu}$ and $f_{\mu\nu}$, with a ghost-free nonlinear interaction between them (cf. Eq.~\eqref{biaction}).
The fluctuations of the two dynamical metrics can be separated and describe two interacting gravitons, one massive and one massless.

A crucial point is to identify the background solution over which the massive tensor perturbations propagate. Linearization of massive gravity is typically considered around a flat, Minkowski background. Here instead we wish to describe the linearized dynamics around a nonlinear vacuum solution, i.e. a BH geometry. Regular, nonlinear, solutions in bimetric and massive gravity are challenging to find and they might exhibit a rich structure~\cite{Deffayet:2011rh,Berezhiani:2011mt,Banados:2011hk,Cai:2012db,Volkov:2012wp,Volkov:2014ooa,Babichev:2015xha}. It was shown that the only way to avoid a singular horizon is to either require both metrics to have coincident horizons, or that at least one of the metrics is non-diagonal (or non-stationary and axisymmetric) when written in the same coordinate patch~\cite{Deffayet:2011rh,Banados:2011hk}.

In this Chapter we consider the special case in which the background solutions are proportional, $f_{\mu\nu}=C^2g_{\mu\nu}$. This choice also avoids the singular horizon problem, as the two metrics have the same horizon. The linearized equations describing the fluctuations of the two metrics can be easily decoupled and they describe one massless graviton (which is described by usual linearized Einstein dynamics), and a massive graviton which is described by the Fierz-Pauli theory on a curved background~\cite{Hassan:2012wr,Volkov:2013roa}.

On the other hand, in the limit of massive gravity this is equivalent of taking the nondynamical metric as being the BH spacetime instead of the usual flat spacetime. Although perfectly consistent with the field equations, this choice seems somewhat unnatural and other nonlinear background metrics can be considered.
The fluctuations of the physical metric $g_{\mu\nu}$ propagate on a nonlinear BH background $f_{\mu\nu}$ and they are also described by Fierz-Pauli theory. 

In Ref.~\cite{Babichev:2013una}, Babichev and Fabbri showed that around these solutions, the mass term for the graviton can be interpreted as a Kaluza-Klein
momentum of a four-dimensional Schwarzschild BH extended into a flat higher dimensional spacetime.
Such ``black string'' spacetimes are known to be unstable against long-wavelength perturbations,
or in other words, against low-mass perturbations, which are spherically symmetric on the four-dimensional subspace. 
This is known as the Gregory-Laflamme instability \cite{Gregory:1993vy,Kudoh:2006bp}, which in turn is the analog of a Rayleigh-Plateau instability of fluids~\cite{Cardoso:2006ks,Camps:2010br}. Based on these results, Ref.~\cite{Babichev:2013una} pointed out that massive tensor perturbations on a Schwarzschild BH in massive gravity and bimetric theories would generically give rise to a (spherically symmetric) instability. In this Chapter we will confirm these results within a more generic framework and extend them to generic modes and to the case of Schwarzschild-de Sitter BHs.

One of the important open questions, that we will partially address in Chapter~\ref{chapter:BHhair}, is the end-state of the instability. For black strings, there is reasonable evidence that break-up occurs \cite{Lehner:2010pn}. But the spacetimes we deal with are spherically symmetric, and so is the unstable mode. 
A possible end-state is a spherically symmetric BH endowed with a graviton cloud. An analysis of the nonlinear equations will be presented in Chapter~\ref{chapter:BHhair}, where we will show that such solutions exist. However, whether they can be the end-state of the instability is still unclear.

%%%%%%%%%%%%%%%%%%%%%%%%%%%%%%%%%%%%%%%%%%%%%%%%%%%%%
\section{Black holes in massive (bi)gravity theories}
%%%%%%%%%%%%%%%%%%%%%%%%%%%%%%%%%%%%%%%%%%%%%%%%%%%%%
\label{BHSOLUTIONS}

The structure of the solutions in massive (bi)gravity is more complex that in GR, mainly due to the fact that this theory has two metrics (see e.g. Ref.~\cite{Blas:2005yk} to see how the global structure of these solutions is affected by the co-existence of two metrics). In particular, the well-known Birkhoff's theorem for spherically symmetric solutions does not apply. This suggests that in massive (bi)gravity the classes of BH solutions are richer than in GR. 

Indeed, the spherically symmetric BH solutions in bimetric theories can be divided into two classes. 
The first class corresponds to the case for which the metrics cannot be brought simultaneously to a bidiagonal form. Said differently, in this class, 
if one metric in some coordinates is diagonal, the other metric is not. BHs of the second class have two metrics which can be both written in the diagonal form, but not necessarily equal. 

The first BH solutions for a nonlinear massive gravity theory (suffering from the Boulware-Deser ghost) were constructed in Ref.~\cite{Salam:1976as}. 
Much later, spherically symmetric solutions for other classes of pathological massive bigravity theories were found and classified in detail in Refs.~\cite{Blas:2005yk,Berezhiani:2008nr}.
In the framework of the original dRGT model a class of non-bidiagonal Schwarzschild-de-Sitter solutions was found in Ref.~\cite{Koyama:2011xz}.
In Refs.~\cite{Comelli:2011wq,Volkov:2012wp}, spherically symmetric BH solutions were found for the bi-metric extension of dRGT theory, while 
spherically symmetric (charged and uncharged) solutions in the dRGT model for a special choice of the parameters of the action were presented in Refs.~\cite{Nieuwenhuizen:2011sq,Berezhiani:2011mt}. 
More recently, Ref.~\cite{Babichev:2014fka} found a more general class of charged BH solutions (in both the dRGT model and its bimetric extension). Finally, Ref.~\cite{Babichev:2014tfa} generalized these findings by including rotation in the geometry. This last class of solutions, jointly with the charged solutions of Ref.~\cite{Babichev:2014fka},  includes as particular cases most of the previously found spherically symmetric solutions\footnote{With the exception of Schwarzschild non-bidiagonal solutions, presented in~\cite{Koyama:2011xz}, where an extra constant of integration
appears in the solution. However, in~\cite{Blas:2005yk} it has been argued (for similar BH solutions in a ghostly massive gravity) that the extra parameter should be set to a specific value for the solutions to be physical. In this case the solutions of~\cite{Koyama:2011xz} are a particular subclass of the solutions found in~\cite{Babichev:2014fka,Babichev:2014tfa}.
In~\cite{Volkov:2014ooa}, a method was presented to construct more general spherically symmetric non-bidiagonal solutions. These solutions are implicitly written in terms of one function (of two coordinates), which must satisfy a particular partial differential equation (PDE), and thus describe an infinite-dimensional family of solutions (a similar technique has been used in~\cite{Mazuet:2015pea} to find de Sitter solutions in dRGT massive gravity). }.

Interestingly, spherically symmetric BH solutions with hair --- solutions differing from the Schwarzschild family --- also exist in bimetric gravity, both with Anti-de Sitter~\cite{Volkov:2012wp} and flat asymptotics~\cite{Brito:2013xaa} (we will discuss these solutions in Chapter~\ref{chapter:BHhair}).
For reviews on solutions of BHs in massive gravity we refer the reader to Refs.~\cite{Volkov:2013roa,Tasinato:2013rza,Volkov:2014ooa,Babichev:2015xha}.

%%%%%%%%%%%%%%%%%%%%%%%%%%%%%%%%%%%%%%%%%%%%%%%%%%%%%%%%%%%%%%%%%%%%%%%%%%%%%%%%%%%%%%%%%%%%%%%%%%%%%%%%%%%%%%
\subsection{Bidiagonal black-hole solutions}
%%%%%%%%%%%%%%%%%%%%%%%%%%%%%%%%%%%%%%%%%%%%%%%%%%%%%%%%%%%%%%%%%%%%%%%%%%%%%%%%%%%%%%%%%%%%%%%%%%%%%%%%%%%%%%

Let us first discuss the simplest BH solutions of the theory. We leave the discussion on the non-bidiagonal class of solutions to Chapter~\ref{chapter:nonbi}.

The simplest bidiagonal BH solutions can be obtained by taking two proportional metrics $\bar{f}_{\mu\nu}=C^2\bar{g}_{\mu\nu}$ (we use the bar notation to denote background quantities later on).
Remarkably, in this case the solutions coincide with those of GR. Indeed, Eqs.~\eqref{field_eqs1} and \eqref{field_eqs2} reduce to~\cite{Hassan:2012wr} 
\be
\label{eqs_pro}
\bar{R}_{\mu\nu}-\frac{1}{2}\bar{g}_{\mu\nu}\bar{R}+\Lambda_g \bar{g}_{\mu\nu}=0\,,\qquad
\bar{R}_{\mu\nu}-\frac{1}{2}\bar{g}_{\mu\nu}\bar{R}+\Lambda_f \bar{g}_{\mu\nu}=0\,,
\ee
which are just two copies of Einstein's equations with two different cosmological constants given by~\cite{Hassan:2012wr}:
\be\label{cosmo_bigrav}
\Lambda_g=\frac{M_v^4\left(\beta_0+3C\beta_1+3 C^2\beta_2+C^3\beta_3\right)}{M_g^2}\,,\qquad
\Lambda_f=\frac{M_v^4\left(C^3\beta_4+\beta_1+3C\beta_2+3C^2\beta_3\right)}{C M_f^2}\,,\\
\ee
Furthermore, consistency of the background equations requires $\bar{\Lambda}_g=\bar{\Lambda}_f$, which translates into a quartic algebraic equation for the constant $C$. 
Classical no-hair theorems of GR guarantee that the most general stationary BH solution in vacuum and with a cosmological constant is the Kerr-(Anti) de Sitter metric. Therefore, when $\Lambda_g=\Lambda_f=\Lambda>0$ the fields $g_{\mu\nu}$ and $f_{\mu\nu}$ describe two identical Kerr-de Sitter BHs. For non-rotating BHs, these solutions reduce to two diagonal Schwarzschild-de Sitter geometries.

%%%%%%%%%%%%%%%%%%%%%%%%%%%%%%%%%%%%%%%%%%%%%%%%%%%%%%%%%%%%%%%%%%%%%%%%%%%%%%%%%%%%%%%%%%%%%%%%%%%%%%%%%%%%%%
\section{Linear spin-2 field in a curved background}\label{sec:linearspin2}
%%%%%%%%%%%%%%%%%%%%%%%%%%%%%%%%%%%%%%%%%%%%%%%%%%%%%%%%%%%%%%%%%%%%%%%%%%%%%%%%%%%%%%%%%%%%%%%%%%%%%%%%%%%%%%

Starting from the action~\eqref{biaction} let us write down the field equations describing a linear massive spin-2 field propagating in a curved spacetime.
We consider fluctuations around the background metrics:
\begin{equation}\label{pert}
g_{\mu\nu}=\bar{g}_{\mu\nu}+\frac{1}{M_g}\delta g_{\mu\nu}\,,\qquad
f_{\mu\nu}=C^2\bar{g}_{\mu\nu}+\frac{C}{M_f}\delta f_{\mu\nu}\,.
\end{equation}
Note that the perturbations are generically independent, $\delta g_{\mu\nu}\neq\delta f_{\mu\nu}$. 
From Eqs.~\eqref{field_eqs1} and~\eqref{field_eqs2}, the linearized field equations read
\begin{align}
\label{eq_g}
&\bar{\mathcal{E}}^{\rho\sigma}_{\mu\nu}\delta g_{\rho\sigma}+\Lambda \delta g_{\mu\nu}-\frac{M_v^4B}{M_g}\bar{g}_{\mu\rho}\left(\delta S^{\rho}\,_{\nu}-\delta_{\nu}^{\rho}\delta S^{\sigma}\,_{\sigma}\right)=0\,,\\
\label{eq_f}
&\bar{\mathcal{E}}^{\rho\sigma}_{\mu\nu}\delta f_{\rho\sigma}+\Lambda \delta f_{\mu\nu}+\frac{M_v^4B}{C M_f}\bar{g}_{\mu\rho}\left(\delta S^{\rho}\,_{\nu}-\delta_{\nu}^{\rho}\delta S^{\sigma}\,_{\sigma}\right)=0\,,
\end{align}
where $B$ is a constant~\cite{Hassan:2012wr},
\be
\delta S^{\rho}\,_{\nu}=\frac{\bar{g}^{\rho\mu}}{2M_f}\left(\delta f_{\mu\nu}-C\frac{M_f}{M_g}\delta g_{\mu\nu}\right)\,,
\ee
and $\bar{\mathcal{E}}^{\rho\sigma}_{\mu\nu}$ is the operator representing the linearized Einstein equations in curved spacetimes:
\beq
&&\bar{\mathcal{E}}^{\rho\sigma}_{\mu\nu}=-\frac{1}{2}\left[\delta^{\rho}_{\mu}\delta^{\sigma}_{\nu}\bar{\Box}+\bar{g}^{\rho\sigma}\bar{\nabla}_{\mu}\bar{\nabla}_{\nu}
-\delta^{\rho}_{\mu}\bar{\nabla}^{\sigma}\bar{\nabla}_{\nu}
-\delta^{\rho}_{\nu}\bar{\nabla}^{\sigma}\bar{\nabla}_{\mu}-\bar{g}_{\mu\nu}\bar{g}^{\sigma\rho}\bar{\Box}+\bar{g}_{\mu\nu}\bar{\nabla}^{\rho}\bar{\nabla}^{\sigma}\right.\nn\\
&&\left.-\bar{g}_{\mu\nu}\bar{R}^{\rho\sigma}+\delta^{\rho}_{\mu}\delta^{\sigma}_{\nu}\bar{R}\right]\,.
\label{operator}
\eeq

Taking appropriate linear combinations of the metric fluctuations,
\begin{equation}
h^{(0)}_{\mu\nu}=\frac{M_g\delta g_{\mu\nu}+C\,M_f\delta f_{\mu\nu}}{\sqrt{C^2M^2_f+M^2_g}}\,,\qquad
h^{(m)}_{\mu\nu}=\frac{M_g\delta f_{\mu\nu}-C\,M_f\delta g_{\mu\nu}}{\sqrt{C^2M^2_f+M^2_g}}\,,
\end{equation}
the linear equations decouple:
\begin{align}
\label{eq_0}
&\bar{\mathcal{E}}^{\rho\sigma}_{\mu\nu}h^{(0)}_{\rho\sigma}+\Lambda h^{(0)}_{\mu\nu}=0\,,\\
\label{eq_m}
&\bar{\mathcal{E}}^{\rho\sigma}_{\mu\nu}h^{(m)}_{\rho\sigma}+\Lambda h^{(m)}_{\mu\nu}+\frac{\mu^2}{2}\left(h^{(m)}_{\mu\nu}-\bar{g}_{\mu\nu}h^{(m)}\right)=0\,.
\end{align}
From the equations above, it is clear that the theory describes two spin-2 fields, $h^{(0)}_{\mu\nu}$ and $h^{(m)}_{\mu\nu}$. The former is massless and it is described by the linearized Einstein-Hilbert action, whereas the latter has a Fierz-Pauli mass term defined as
\be\label{mass_g}
\mu^2=M_v^4(C\beta_1+2C^2\beta_2+C^3\beta_3)\left(\frac{1}{C^2M_f^2}+\frac{1}{M_g^2}\right)\,.
\ee

What we have discussed so far is valid for bimetric theories~\eqref{biaction}. It is worth stressing that linearized massive gravity can be recovered taking the limit $\delta f_{\mu\nu}\to 0$ and $M_f\to \infty$ in Eq.~\eqref{pert} such that $\delta f_{\mu\nu}/M_f\to 0$. In this limit only Eq.~\eqref{eq_g} survives as a dynamical equation. In the massive gravity limit, this equation can be written in the same form as in Eq.~\eqref{eq_m} for the perturbation $\delta g_{\mu\nu}$, but with a mass term 
\be
\mu=\sqrt{BC}M_v^2/M_g\,.
\ee
Therefore, also in this case the theory describes a massive graviton propagating in the curved background $\bar{g}_{\mu\nu}\equiv \bar{f}_{\mu\nu}/C^2$.

We have just proved that in both cases (bimetric theories and massive gravity) the linearized equations describing a massive spin-2 field on a curved spacetime are described by an equation of the form~\eqref{eq_m}. In the case of bimetric theory one also has Eq.~\eqref{eq_0}, which we ignore since it describes a standard massless graviton and it is decoupled.

In flat spacetime, the equations of motion~\eqref{eq_m} reduce to Eq.~\eqref{eqmotion} whereas, on curved background they reduce to the system (after taking the divergence and trace of Eq.~\eqref{eq_m}):
\begin{eqnarray}
&& \bar\Box h_{\mu\nu}+2 \bar R_{\alpha\mu\beta\nu} h^{\alpha\beta}-\mu^2 h_{\mu\nu}=0 \label{eqmotioncurved}\,,\\
&& \mu^2\bar\nabla^{\mu}h_{\mu\nu}=0\,,\label{constraint1}\\
&& \left(\mu^2-{2\Lambda}/{3}\right)h=0\,.\label{constraint2}
\end{eqnarray}
where, here and in the following, we have suppressed the superscript ``$(m)$'' for simplicity, and we used the tensorial relation
\be
(\bar \nabla_c\bar \nabla_d - \bar \nabla_d\bar \nabla_c)h_{ab} = \bar R_{aecd}h^e\,_b + \bar R_{becd}h_a\,^e\,.  \label{commutator}
\ee
This set of equations can be shown to be the only one that consistently describes a massive spin-2 with five degrees of freedom and coupled to gravity in generic backgrounds~\cite{Buchbinder:1999ar}. In fact, in the limit $M_f\ll M_g$, interactions of the massive mode with the massless mode and matter fields are suppressed (if the matter fields only couple minimally to $g_{\mu\nu}$), and thus in this limit, and at the linear level, one can interpret these theories as a model of a massive spin-2 fields coupled to gravity~\cite{Hassan:2012wr}.
In the rest of this Chapter we will investigate the system~\eqref{eqmotioncurved}--\eqref{constraint2} on a Schwarzschild background. We leave the study of this system on top of a Kerr BH to Chapter~\ref{chapter:Kerr}.

%%%%%%%%%%%%%%%%%%%%%%%%%%%%%%%%%%%%%%%%%%%%%%%%%%%%%%%%%
\section{Setup}
%%%%%%%%%%%%%%%%%%%%%%%%%%%%%%%%%%%%%%%%%%%%%%%%%%%%%%%%%
%%%%%%%%%%%%%%%%%%%%%%%%%%%%%%%%%%%%%%%%%%%%%%%%%%%%%%%%%
\subsection{The Schwarzschild(-de Sitter) geometry}
%%%%%%%%%%%%%%%%%%%%%%%%%%%%%%%%%%%%%%%%%%%%%%%%%%%%%%%%%
The most general static solution of eqs.~\eqref{eqs_pro} is the Schwarzschild-de Sitter spacetime, described by
\be\label{eq:SdS}
d\bar{s}^2 = -f(r)\, dt^2 + f(r)^{-1}\, dr^2 + r^2 d\Omega^2,
\ee
where
\begin{equation}
f=1-\frac{2M}{r}-\frac{\Lambda}{3} r^2 \equiv \frac{\Lambda}{3 r}\, (r-r_b)(r_c-r)(r-r_0)\,,
\label{eq:f_def}
\end{equation}
with $r_0 = -(r_b + r_c)$, $r_b$ and $r_c>r_b$ being the BH horizon and the cosmological horizon, respectively.
The cosmological constant can be expressed as $3/\Lambda={r_b}^2 + r_b r_c + {r_c}^2$
and the spacetime has mass $M= \Lambda r_b r_c (r_b + r_c)/6$. 
The Schwarzschild geometry is recovered when $\Lambda=0$. In this case, there is only one horizon given by $r_b=2M$, while $r_c\to \infty$.
Finally, another quantity that will be useful later on, is the surface gravity $\kappa_b$ associated with the BH horizon, given by
\begin{equation}
\kappa_b\equiv\frac{f'(r_b)}{2} = \frac{ \Lambda(r_c-r_b)(r_b-r_0) }{ 6 r_b }.
\label{surface}
\end{equation}
%

%%%%%%%%%%%%%%%%%%%%%%%%%%%%%%%%%%%%%%%%%%%%%%%%%%%%%%%%%
\subsection{Tensor spherical harmonic decomposition of spin-2 fields}
%%%%%%%%%%%%%%%%%%%%%%%%%%%%%%%%%%%%%%%%%%%%%%%%%%%%%%%%%
To lay down the necessary framework, consider a generic tensor field $h_{\mu\nu}$ in a spherically symmetric background. Due to spherical symmetry, the tensor field $h_{\mu\nu}$ can be conveniently decomposed in a complete basis of tensor spherical harmonics~\cite{Regge:1957td,1970JMP....11.2203Z}. Furthermore, the perturbation variables are classified as ``polar'' or ``axial'' depending on how they transform under parity inversion ($\theta\to \pi-\theta$, $\phi\to \phi+\pi$). Polar perturbations are multiplied by $(-1)^l$ whereas axial perturbations pick up the opposite sign $(-1)^{l+1}$. We refer the reader to Refs.~\cite{Berti:2009kk,Zerilli:1971wd} for further terminology used in the literature. 

We decompose the spin-2 perturbation in Fourier space as follows:
\be
\label{decom}
h_{\mu\nu}(t,r,\theta,\phi)=\sum_{l,m}\int_{-\infty}^{+\infty}e^{-i\omega t}\left[h^{{\mathrm{axial}},lm}_{\mu\nu}(\omega,r,\theta,\phi)
+h^{{\mathrm{polar}},lm}_{\mu\nu}(\omega,r,\theta,\phi)\right]d\omega\,.
\ee
The tensorial quantities $h^{{\mathrm{axial}},lm}_{\mu\nu}$ and $h^{{\mathrm{polar}},lm}_{\mu\nu}$ are explicitly given by:
%
%\begin{widetext}
% \begin{center}
% \textbf{Axial sector:}
% \end{center}
\begin{equation}\label{oddpart}
h^{{\mathrm{axial}},lm}_{\mu\nu}(\omega,r,\theta,\phi) =
 \begin{pmatrix}
  0 & 0 & h^{lm}_0(\omega,r)\csc\theta\partial_{\phi}Y_{lm}(\theta,\phi) & -h^{lm}_0(\omega,r)\sin\theta\partial_{\theta}Y_{lm}(\theta,\phi) \\
  * & 0 & h^{lm}_1(\omega,r)\csc\theta\partial_{\phi}Y_{lm}(\theta,\phi) & -h^{lm}_1(\omega,r)\sin\theta\partial_{\theta}Y_{lm}(\theta,\phi) \\
  *  & *  & -h^{lm}_2(\omega,r)\frac{X_{lm}(\theta,\phi)}{\sin\theta} & h^{lm}_2(\omega,r)\sin\theta W_{lm}(\theta,\phi)  \\
  * & * & * & h^{lm}_2(\omega,r)\sin\theta X_{lm}(\theta,\phi)
 \end{pmatrix}\,,
\end{equation}
%
%
% \begin{center}
% \textbf{Polar sector:}
% \end{center}
{\footnotesize
\begin{align}\label{evenpart}
&h^{{\mathrm{polar}},lm}_{\mu\nu}(\omega,r,\theta,\phi)=\nn\\
&\begin{pmatrix}
f(r)H_0^{lm}(\omega,r)Y_{lm} & H_1^{lm}(\omega,r)Y_{lm} & \eta^{lm}_0(\omega,r)\partial_{\theta}Y_{lm}& \eta^{lm}_0(\omega,r)\partial_{\phi}Y_{lm}\\
  * & f(r)^{-1} H_2^{lm}(\omega,r)Y_{lm} & \eta^{lm}_1(\omega,r)\partial_{\theta}Y_{lm}& \eta^{lm}_1(\omega,r)\partial_{\phi}Y_{lm}\\
  *  & *  & \begin{array}{c}r^2\left[K^{lm}(\omega,r)Y_{lm}\right.\\\left. +G^{lm}(\omega,r)W_{lm}\right]\end{array} & r^2  G^{lm}(\omega,r)X_{lm}  \\
  * & * & * & \begin{array}{c}r^2\sin^2\theta\left[K^{lm}(\omega,r)Y_{lm}\right.\\\left.-G^{lm}(\omega,r)W_{lm}\right]\end{array}
\end{pmatrix}\,,
\end{align}
%\end{widetext}
%
}
where asterisks represent symmetric components, $Y_{lm}\equiv Y_{lm}(\theta,\phi)$ are the scalar spherical harmonics and
\beq
X_{lm}(\theta,\phi)&=&2\partial_{\phi}\left[\partial_{\theta}Y_{lm}-\cot\theta Y_{lm}\right]\,,\\
W_{lm}(\theta,\phi)&=&\partial^2_{\theta}Y_{lm}-\cot\theta\partial_{\theta}Y_{lm}-\csc^2\theta\partial^2_{\phi}Y_{lm}\,.
\eeq

In a spherically symmetric background, the field equations do not depend on the azimuthal number $m$ and they are also decoupled for each harmonic index $l$. In addition, perturbations with opposite parity decouple from each other.

%%%%%%%%%%%%%%%%%%%%%%%%%%%%%%%%%%%%%%%%%%%%%%%%%%%%%%%%%%%%%%%%%%%%%%%%%%%%%%%
\section{Field equations for massive spin-2 fields on a Schwarzschild background}\label{sec:schwar}
%%%%%%%%%%%%%%%%%%%%%%%%%%%%%%%%%%%%%%%%%%%%%%%%%%%%%%%%%%%%%%%%%%%%%%%%%%%%%%%

In this Section we write down the main field equations for a massive spin-2 field in a Schwarzschild background. 
Since we are interested in local physics near massive BHs, we focus on the case where $\Lambda_g\approx0\approx\Lambda_f$. This condition can be satisfied exactly by requiring a fine tuning of the interaction couplings, as can be seen from Eq.~\eqref{cosmo_bigrav}. Alternatively, even without fine tuning, realistic values of the cosmological constant should not play any role in describing local physics at the scale of astrophysical compact objects. For completeness, and because it is the most interesting case, in Section~\ref{sec:monopoleinstability}, we will also consider spherical perturbations for $\Lambda_g\neq 0 \neq \Lambda_f$. 

%%%%%%%%%%%%%%%%%%%%%%%%%%%%%%%%%%%%%%%%%%%%%%%%%%%%%%%%%%%%%%%%%%
\subsection{Axial equations}
%%%%%%%%%%%%%%%%%%%%%%%%%%%%%%%%%%%%%%%%%%%%%%%%%%%%%%%%%%%%%%%%%%
The field equations for the axial sector are obtained
by using the decomposition \eqref{oddpart} in Eq.~\eqref{eqmotioncurved}. Substituting into the linearized field equations, we obtain:
\begin{align}
&f^2 h''_0+\left[\omega^2-f\left(\mu^2+\frac{\lambda}{r^2}-\frac{4M}{r^3}\right)\right]h_0-\frac{2Mi\omega f}{r^2} h_1=0\,,\label{odd1}\\
&f^2 h''_1+\frac{4 M f}{r^2}h'_1+\left[\omega^2-f\left(\mu^2+\frac{\lambda+4}{r^2}-\frac{8M}{r^3}\right)\right]h_1
-\frac{2Mi\omega}{r(r-2M)}h_0\nn\\
&+\frac{2(2-\lambda)f}{r^3}h_2=0\,,\label{odd2}\\
&f^2 h''_2-\frac{2 f(r-3M)}{r^2}h'_2-\frac{2f^2}{r}h_1
+\left[\omega^2-f\left(\mu^2+\frac{\lambda-4}{r^2}+\frac{8M}{r^3}\right)\right]h_2=0\,,\label{odd3}
\end{align}
where $\lambda=l(l+1)$ and $f\equiv f(r)$. Equations \eqref{odd1} and \eqref{odd2} correspond to the $(t\theta)$ and the $(r\theta)$
component of the field equations respectively, and \eqref{odd3} corresponds to the $(\theta\theta)$ component.
The transverse constraint \eqref{constraint1} leads to the radial equation
\begin{equation}
\label{harmonicodd}
f h'_1-\frac{2 (M-r)}{r^2}h_1+\frac{i\omega}{f}h_0+\frac{\lambda-2}{r^2}h_2=0\,,
\end{equation}
which can be obtained either from the $\theta$ or the $\phi$ component.
For the axial terms the trace~\eqref{constraint2} vanishes identically,
\begin{equation}
\label{traceodd}
h^{\mathrm{axial}}=0\,.
\end{equation}
%

%%%%%%%%%%%%%%%%%%%%%%%%%%
\subsubsection{Axial sector: master equations for $l\geq 2$} 
%%%%%%%%%%%%%%%%%%%%%%%%%%
Using the constraint \eqref{harmonicodd} we can reduce the system to a pair of coupled differential equations. Eliminating $h_0$, and defining the functions $Q(r)\equiv f(r)h_1$ and $Z(r)\equiv h_2/r$, we finally obtain the following system:
\beq\label{axial_bi1}
&&\frac{d^2}{dr_*^2}Q+\left[\omega^2-f\left(\mu^2+\frac{\lambda+4}{r^2}-\frac{16M}{r^3}\right)\right]Q=S_Q\,,\\
\label{axial_bi2}
&&\frac{d^2}{dr_*^2}Z+\left[\omega^2-f\left(\mu^2+\frac{\lambda-2}{r^2}+\frac{2M}{r^3}\right)\right]Z=S_Z\,,
\eeq
where $\lambda=l(l+1)$ and we have defined the tortoise coordinate $r_*$ via $dr/dr_*=f\equiv1-2M/r$. The source terms are given by
\be
S_Q= (\lambda-2)\frac{2f(r-3M)}{r^3}Z\,,\qquad
S_Z= \frac{2}{r^2}f\,Q\,.
\ee
%
%%%%%%%%%%%%%%%%%%%%%%%%%%%%%%%%%
\subsubsection{Axial dipole mode}
%%%%%%%%%%%%%%%%%%%%%%%%%%%%%%%%%
The $l=0$ monopole mode does not exist in the axial sector since the angular part of the axial perturbations \eqref{oddpart} vanishes for $l=0$. For the dipole mode ($l=1$ or equivalently $\lambda=2$), the angular functions $W_{lm}$ and $X_{lm}$ vanish and one is left with a single decoupled equation:
\begin{equation}
\label{oddl1}
\frac{d^2}{dr_*^2} Q + \left[\omega^2-f\left(\mu^2+\frac{6}{r^2}-\frac{16M}{r^3}\right)\right]Q=0\,.
\end{equation}
%
%%%%%%%%%%%%%%%%%%%%%%%%%%%%%%%%%%%%
\subsubsection{Axial massless limit}
%%%%%%%%%%%%%%%%%%%%%%%%%%%%%%%%%%%%
It is interesting to note that in the massless limit we can use the transformations
\begin{align}
&h_0=\frac{1}{i\omega}\left[\varphi_1+\frac{\lambda-2}{3}\varphi_2\right]\,,\nonumber\\
&h_1=\frac{1}{(i\omega)^2}\left[\frac{2}{r}\varphi_1+\frac{2-\lambda}{3r}\varphi_2-\frac{d\varphi_1}{dr}+\frac{2-\lambda}{3}\frac{d\varphi_2}{dr}\right]\,,\nonumber\\
&h_2=\frac{1}{(i\omega)^2}\left[\varphi_1+\frac{(\lambda+1)r-6M}{3r}\varphi_2+(r-2M)\frac{d\varphi_2}{dr}\right]\,,\nn
\end{align}
to reduce the system to a pair of decoupled equations, given by a ``vectorial'' and a ``tensorial'' Regge-Wheeler equation
\be
\frac{d^2}{dr_*^2} \varphi_s + \left[\omega^2-f\left(\frac{\lambda}{r^2}+(1-s^2)\frac{2M}{r^3}\right)\right]\varphi_s=0\,,\label{RW}
\ee
where $s=0,1,2$ for scalar, vectorial, or tensorial perturbations. These transformations were first found by Berndtson~\cite{Berndtson:2009hp} when studying the massless graviton perturbations of the Schwarzschild metric in the harmonic gauge. In the massless limit the vectorial degree of freedom can be removed by a gauge transformation, but for $\mu\neq 0$ it becomes a physical mode.
Note that the wave equation \eqref{RW} for $s=1$ is identical to that describing electromagnetic perturbations of Schwarzschild BHs \cite{Berti:2009kk}; thus the axial spectrum of massive spin-2 perturbations should include a mode which approaches that of an electromagnetic mode in the low-mass limit.

%%%%%%%%%%%%%%%%%%%%%%%%%%%%%%
\subsection{Polar equations} 
%%%%%%%%%%%%%%%%%%%%%%%%%%%%%%
Using the decomposition \eqref{evenpart} in Eq.~\eqref{eqmotioncurved} and substituting into the linearized field equations, we obtain:
%
%\begin{widetext}
\begin{align}
&f^2 H''_0+\frac{2 f(r-M)}{r^2}H'_0+\left[\omega^2-\frac{2M^2}{r^4}-f\left(\mu^2+\frac{\lambda}{r^2}\right)\right]H_0\nn\\
&-\frac{4iM\omega}{r^2}H_1-\frac{2M(2r-3M)}{r^4}H_2+\frac{4M f}{r^3}K=0\,,\label{even1}\\
&f^2 H''_1+\frac{2 f(r-M)}{r^2}H'_1+\left[\omega^2-\frac{4M^2}{r^4}-f\left(\mu^2+\frac{\lambda+2}{r^2}\right)\right]H_1\nn\\
&-\frac{2iM\omega}{r^2}(H_0+H_2)+\frac{2\lambda f}{r^3}\eta_0=0\,,\label{even2}\\
&f^2 H''_2+\frac{2 f(r-M)}{r^2}H'_2+\left[\omega^2-\frac{2M^2}{r^4}-f\left(\mu^2+\frac{\lambda+4}{r^2}-\frac{8M}{r^3}\right)\right]H_2\nn\\
&-\frac{2M(2r-3M)}{r^4}H_0-\frac{4iM\omega}{r^2}H_1+\frac{4(r-3M)f}{r^3}K+\frac{4\lambda f^2}{r^3}\eta_1=0\,,\label{even3}\\
&f^2 \eta''_0+\left[\omega^2-f\left(\mu^2+\frac{\lambda}{r^2}-\frac{4M}{r^3}\right)\right]\eta_0-\frac{2M i\omega f}{r^2}\eta_1+\frac{2 f^2}{r}H_1=0\,,\label{even4}
\\
&f^2 \eta''_1+\frac{4M f}{r^2}\eta'_1+\left[\omega^2-f\left(\mu^2+\frac{\lambda+4}{r^2}-\frac{8M}{r^3}\right)\right]\eta_1-\frac{2M i\omega}{r(r-2M)}\eta_0\nn\\
&+\frac{2 f}{r}\left[H_2-K+(\lambda-2)G\right]=0\,,\label{even5}\\
&f^2 G''+\frac{2(r-M) f}{r^2}G'+\left[\omega^2-f\left(\mu^2+\frac{\lambda-2}{r^2}\right)\right]G+\frac{2f^2}{r^3}\eta_1=0\,,\label{even6}\\
&f^2 K''+\frac{2(r-M) f}{r^2}K'+\left[\omega^2-f\left(\mu^2+\frac{\lambda+2}{r^2}-\frac{8M}{r^3}\right)\right]K+\frac{2M f}{r^3}H_0\nn\\
&+\frac{2(r-3M) f}{r^3}H_2-\frac{2\lambda f^2}{r^3}\eta_1=0\label{even7}\,,
\end{align}
Equations \eqref{even1}-\eqref{even5} correspond to the $(tt),\,(tr),\,(rr),\,(t\theta)$ and $(r\theta)$ components of the field equations, respectively. From the $(\theta\phi)$ component we get Eq.~\eqref{even6}, which combined with the $(\theta\theta)$
component yields Eq.~\eqref{even7}.

The transverse constraint \eqref{constraint1} leads to the following radial equations
\beq
&&f H'_1-\frac{2 (M-r)}{r^2}H_1+i\omega H_0-\frac{\lambda}{r^2}\eta_0=0\,,\label{harmoniceven1}\\
&&f H'_2+\frac{2r-3M}{r^2}H_2+i\omega H_1+\frac{M}{r^2}H_0-\frac{2f}{r}K-\frac{f\lambda}{r^2}\eta_1=0\,,\label{harmoniceven2}\\
&&f \eta'_1-\frac{2(M-r)}{r^2}\eta_1+\frac{i\omega}{f} \eta_0+K-(\lambda-2)G=0\label{harmoniceven3}\,,
\eeq
for the $t,r$ and $\theta$ component of the constraint, respectively.
Finally, in the polar case the traceless constraint~\eqref{constraint2} yields
\be
H_0=H_2+2 K\,.\label{traceeven}
\ee
%

%%%%%%%%%%%%%%%%%%%%%%%%%%%%%%%
\subsubsection{Polar sector: master equations for $l\geq 2$} 
%%%%%%%%%%%%%%%%%%%%%%%%%%%%%%%
Unlike the axial sector, the polar equations are not so straightforward to further reduce. For $l\geq 2$ one could use the constraint equations to eliminate $\eta_0$, $\eta_1$, $H_0$ and $G$ and obtain three second-order equations for $K$, $H_1$ and $H_2$. However, this choice is not particularly  useful, because the system does not directly contain the monopole and dipole cases ($l=0,1$). For this reason we chose to work with $K$, $\eta_1$ and $G$ as dynamical variables instead.

After some tedious algebra, we obtain that the polar sector is fully described by a system of three coupled ordinary differential equations:
\beq
\label{polar_eq1}
f^2\frac{d^2 K}{dr^2}+\hat\alpha_1 \frac{d K}{dr}+\hat\beta_1 K &=& S_K\,,\\
\label{polar_eq2}
f^2\frac{d^2 \eta_1}{dr^2}+\hat\alpha_2 \frac{d \eta_1}{dr}+\hat\beta_2 \eta_1 &=& S_{\eta_1}\,,\\
\label{polar_eq3}
f^2\frac{d^2 G}{dr^2}+\hat\alpha_3 \frac{d G}{dr}+\hat\beta_3 G &=& S_G\,,
\eeq
where the source terms are given by
\begin{align}
 S_K&= \lambda\hat\gamma_1 \frac{d \eta_1}{dr}+\hat\delta_1\lambda \eta_1+\lambda(\lambda-2)\hat\sigma_1 \frac{d G}{dr}+\lambda(\lambda-2)\hat\rho_1 G\,,\\
S_{\eta_1}&= \hat\gamma_2 \frac{d K}{dr}+\hat\delta_2 K+\lambda(\lambda-2)\hat\sigma_2 \frac{d G}{dr}+\lambda(\lambda-2)\hat\rho_2 G\,,\\
 S_G &= \hat\gamma_3 \frac{d K}{dr}+\hat\delta_3 K+\hat\sigma_3 \frac{d \eta_1}{dr}+\hat\rho_3 \eta_1\,.
\end{align}
%%%
The coefficients $\hat\alpha_i,\,\hat\beta_i,\,\hat\gamma_i,\,\hat\delta_i,\,\hat\sigma_i,\,\hat\rho_i$ are radial functions which also depend on $\omega$ and $l$. These equations are rather lengthy and since their explicit form is not fundamental here, we made them available online in {\scshape Mathematica} notebooks~\cite{webpage}.

%%%%%%%%%%%%%%%%%%%%%%%%%%%%%%%%%
\subsubsection{Polar dipole mode}
%%%%%%%%%%%%%%%%%%%%%%%%%%%%%%%%%
In the dipole case, $l=1$, $\lambda=2$, the radial function $G$ identically vanishes and we are left with a pair of coupled equations satisfying the following system:
\beq
\label{polar_dipole1}
f^2\frac{d^2 K}{dr^2}+\hat\alpha_1 \frac{d K}{dr}+\hat\beta_1 K &=& 2(\hat\gamma_1 \frac{d \eta_1}{dr}+\hat\delta_1 \eta_1)\,,\\
\label{polar_dipole2}
f^2\frac{d^2 \eta_1}{dr^2}+\hat\alpha_2 \frac{d \eta_1}{dr}+\hat\beta_2 \eta_1 &=& \hat\gamma_2 \frac{d K}{dr}+\hat\delta_2 K\,.
\eeq
%

%%%%%%%%%%%%%%%%%%%%%%%%%%%%%%%%%
\subsubsection{Polar monopole mode}
%%%%%%%%%%%%%%%%%%%%%%%%%%%%%%%%%
For the $l=0$ polar sector, the perturbations $G$, $\eta_0$ and $\eta_1$ as given in Eq.~\eqref{evenpart} are not defined because their angular dependence vanishes. The remaining dynamical variables can be recast into a simple monopole equation. First, we use the constraints \eqref{traceeven} and \eqref{harmoniceven1} to eliminate $H_0$ and $H_2$. Then, we use a generalization of the Berndtson-Zerilli transformations~\cite{Berndtson:2009hp}:
\begin{align}
&\frac{H_1}{2}=\left[\frac{i\omega(M-r)}{fr^3}+\mu ^2\frac{3 i r  \omega }{2 M+r^3 \mu ^2}\right]\varphi_0+\frac{i\omega}{r}\frac{d\varphi_0}{dr}\,,\nonumber\\
&\frac{K}{2}=\left[\frac{f}{r^3}-\mu ^2\frac{ 6 r+r^3 \mu ^2-10 M}{2\left(2 M r+r^4 \mu ^2\right)}\right]\varphi_0-\frac{f}{r^2}\frac{d\varphi_0}{dr}\,.\nn
\end{align}
After substituting these transformations into the system of equations we arrive at a single wave equation of the form: 
\begin{equation}
\label{evenl0}
\frac{d^2}{dr_*^2} \varphi_0 + \left[\omega^2-V_0(r)\right]\varphi_0=0\,,
\end{equation}
with
\begin{equation}
V_0=f\left[\frac{2M}{r^3}+\mu^2+\frac{24M(M-r)\mu^2+6 r^3 (r-4 M) \mu ^4}{\left(2 M+r^3 \mu ^2\right)^2}\right]\,.\nn
\end{equation}
In this form it is clear that in the massless limit the monopole reduces to the scalar-field wave equation with $l=0$ \cite{Berti:2009kk}.

%%%%%%%%%%%%%%%%%%%%%%%%%%%%%%%%%%%%
\subsubsection{Polar massless limit}
%%%%%%%%%%%%%%%%%%%%%%%%%%%%%%%%%%%%
In the massless limit we can use the argument presented by Berndtson in Ref.~\cite{Berndtson:2009hp} to reduce the system to three decoupled equations, one ``scalar'', one ``vectorial''~\eqref{RW} and one ``tensorial'' equation described by Zerilli's equation~\cite{Zerilli:1970se}
\footnote{Note that in these transformations there are four functions. One tensorial, one vectorial, and two scalars. However one of the scalar functions is simply the trace of $h_{\mu\nu}$, which vanishes in our case (in their notation is the scalar function $\varphi_0$, not to be confused with the scalar function used here). We stress again the importance of having a vanishing trace in order to have a correct number of degrees of freedom.}.
In the massless limit the scalar and the vectorial degrees of freedom can be removed by a gauge transformation but, for $\mu\neq 0$, they become physical. Thus, we expect that the small-mass limit of the massive gravity spectrum includes a family of modes which
are identical to that of a scalar and an electromagnetic mode (these modes are discussed in Ref.~\cite{Berti:2009kk} and available online at 
\cite{webpage}).

%%%%%%%%%%%%%%%%%%%%%%%%%%%%%%%%%%%%%%
\section{Results}
%%%%%%%%%%%%%%%%%%%%%%%%%%%%%%%%%%%%%%
We have solved the previous systems of equations subjected to appropriate boundary conditions,
which defines an eigenvalue problem for the complex frequency $\omega\equiv\omega_R+i\omega_I$; this problem can be solved using 
several different techniques~\cite{Berti:2009kk,Pani:2013pma} which we detail in Appendix \ref{app:modes}. 

At the BH horizon, $r=2M$, regular boundary conditions imply that all perturbation functions behave as an ingoing wave, $\Phi_j(r)\sim e^{-i\omega r_*}$, where $\Phi_j$ describes generically the perturbation functions.
At infinity, the asymptotic behavior of the solution is given by
\be\label{BC_inf_2}
\Phi_j(r)\sim B_j e^{-k_{\infty} r}r^{-\frac{M(\mu^2-2\omega^2)}{k_{\infty}}}+C_j e^{k_{\infty} r}r^{\frac{M(\mu^2-2\omega^2)}{k_{\infty}}}\,,
\ee
where $k_{\infty}=\sqrt{\mu^2-\omega^2}$ and, without loss of generality, we assume Re$(k_{\infty})>0$. The spectrum of massive perturbations admits two different families of physically motivated modes, which are distinguished according to how they behave at spatial infinity. The first family includes the standard QNMs, which corresponds to purely outgoing waves at infinity, i.e., they are defined by $B_j=0$~\cite{Berti:2009kk}. The second family includes quasibound states, defined by $C_j=0$. The latter correspond to modes spatially localized within the vicinity of the BH and that decay exponentially at spatial infinity~\cite{Dolan:2007mj,Rosa:2011my,Pani:2012bp,Pani:2013pma}. 
On the other hand, for modes with purely imaginary frequencies, regularity requires that they must satisfy the bound-state condition $C_j=0$.

%%%%%%%%%%%%%%%%%%%%%%%%%%%%%%%%%%%%%%%%%%%%%%%%%%%%%%%%%%%%%%%%%%%%%%%%%%%%%%%%%%%%%%%%%%%%%%%%%%%%%%%%%%%%%%
\subsection{Instability of black holes against spherically symmetric fluctuations}\label{sec:monopoleinstability}
%%%%%%%%%%%%%%%%%%%%%%%%%%%%%%%%%%%%%%%%%%%%%%%%%%%%%%%%%%%%%%%%%%%%%%%%%%%%%%%%%%%%%%%%%%%%%%%%%%%%%%%%%%%%%%

\begin{figure*}[htb]
\begin{center}
\begin{tabular}{cc}
\epsfig{file=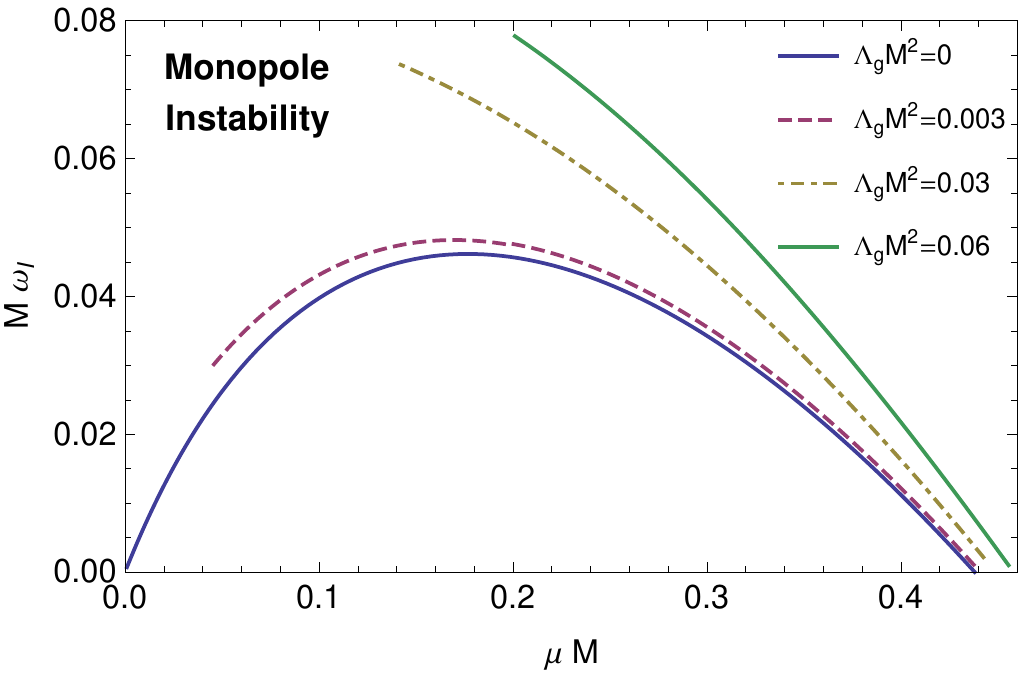,width=7.5cm,angle=0,clip=true}
\epsfig{file=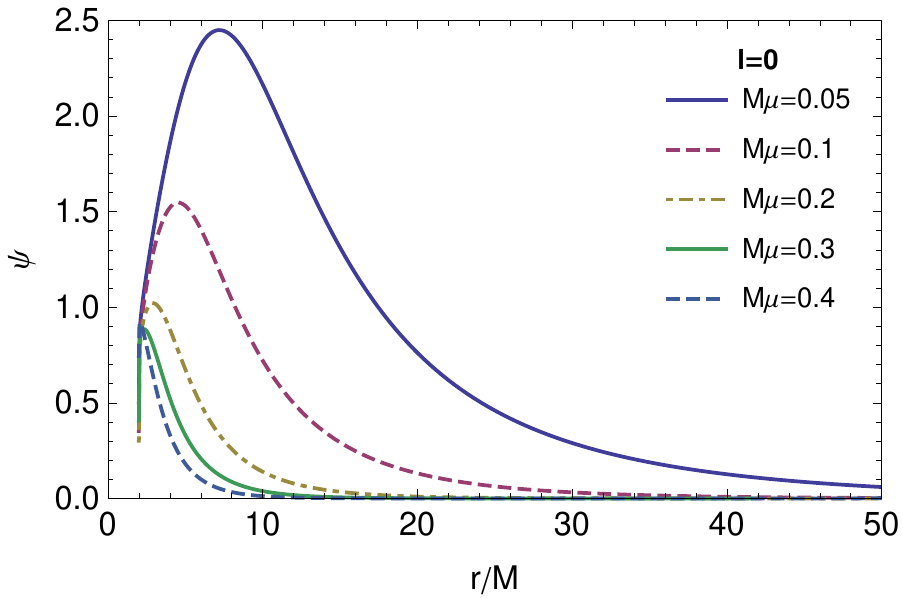,width=7.5cm,angle=0,clip=true}\\
\end{tabular}
\caption{Details of the instability of Schwarzschild (de Sitter) BHs against spherically symmetric polar modes of a massive spin-2 field. The left panel shows the inverse of the instability timescale $\omega_I=1/\tau$ as a function of the graviton mass $\mu$ for different values of the cosmological constant $\Lambda_g=\Lambda_f$, including the asymptotically flat case $\Lambda_g=0$. Curves are truncated when the Higuchi bound is reached $\mu^2=2\Lambda_g/3$~\cite{Higuchi:1986py}. For any value of $\Lambda_g$, unstable modes exist in the range $0<M\mu\lesssim 0.47$, the upper bound being only mildly sensitive to $\Lambda_g$. The right panel shows some eigenfunctions in the asymptotically flat case. The eigenfunctions decay exponentially at spatial infinity and are
progressively peaked closer and closer to the BH horizon for masses close to the threshold mass $M\mu\sim 0.43$.
\label{fig:GL}}
\end{center}
\end{figure*}

We start by showing that Schwarzschild BHs are generically unstable against spherically symmetric perturbations~\cite{Babichev:2013una}.
This is a generic and strong instability, as we will show.

We have solved Eq.~\eqref{evenl0} subjected to the appropriate boundary conditions stated above, by direct integration, looking for eigenvalues $\omega=\omega_R+i\omega_I$ (see Appendix~\ref{app:modes} for more details on the numerical methods). Given the time dependence \eqref{decom}, stable modes are characterized by $\omega_I<0$
and unstable modes by $\omega_I>0$. We found one unstable mode, detailed in Fig.~\ref{fig:GL} and characterized by a purely imaginary, positive
component. This is a low-mass instability which disappears for $M\mu\geq 0.43$ and has a minimum growth timescale of around $M\omega_I\sim 0.046$.
In fact, as recognized very recently~\cite{Babichev:2013una}, the linearized equations~\eqref{eqmotioncurved}--\eqref{constraint2} are equivalent to those describing four-dimensional perturbations of a five-dimensional black string after a Kaluza-Klein reduction of the extra dimension. Therefore, the system is affected by Gregory-Laflamme instability~\cite{Gregory:1993vy,Kudoh:2006bp} that manifests itself in the spherically symmetric, monopole mode.
One interesting aspect of our own formulation is that we are able to reduce this instability to the study of a very simple wave equation,
described by~\eqref{evenl0}.

To summarize, in this setup Schwarzschild BHs are unstable. The instability timescale depends strongly on the mass scale $\mu$.
For low masses, we find numerically that $\omega_I\sim 0.7\mu$, in good agreement with analytic calculation by Camps and Emparan~\cite{Camps:2010br}. 
The Gregory-Laflamme instability only affects spherically-symmetric ($l=0$) modes~\cite{Kudoh:2006bp}, so we expect the rest of the sector to be stable. We confirm this result in the next subsections, where we derive the complete linear dynamics on a Schwarzschild metric. 

A more relevant question is related to the role of a cosmological constant. When the background metrics are two copies of Schwarzschild-de Sitter solutions, the field equations~\eqref{eqmotioncurved} do not arise from a Kaluza-Klein decomposition of a five-dimensional black string. Thus, it is not obvious a priori if the monopole instability discussed above survives when $\Lambda_g=\Lambda_f\neq0$. 

From the system~\eqref{eqmotioncurved}--\eqref{constraint2}, it is straightforward to obtain a master equation for spherical perturbations of Schwarzschild-de Sitter BHs. The monopole is described by an equation of the same form as Eq.~\eqref{evenl0}, but where the potential now reads:
%%%%
\beq
&& V_0^{\Lambda_g}=\frac{1-2M/r-\Lambda_g /3\,r^2}{r^3 \left[2 M+r^3 \left(\mu ^2-2 {\Lambda_g/3}\right)\right]^2}
\times\left\{8 M^3+12 M^2 r^3 \left(3 \mu ^2-8 {\Lambda_g/3}\right)\right.\nn\\
&&\left. +r^7 \left(\mu ^2-2 {\Lambda_g/3}\right)^2 \left[6+r^2 \left(\mu ^2-2 {\Lambda_g/3}\right)\right]\right.\nn\\
&& \left.-6 M r^4 \left(\mu ^2-2 {\Lambda_g/3}\right) \left[4+r^2 \left(3 \mu ^2-10 {\Lambda_g/3}\right)\right]\right\}\,.\label{V0dS}
\eeq
%%%
Using the same technique as before, we have integrated Eq.~\eqref{evenl0} with the potential~\eqref{V0dS}. The results are shown in Fig.~\ref{fig:GL} for various values of $\Lambda_g=\Lambda_f$. Note that massive spin-2 perturbations propagating in an asymptotically de Sitter spacetime are subjected to the bound $\mu^2>2\Lambda_g/3$~\cite{Higuchi:1986py}. Below such bound, the helicity-0 component of the massive graviton becomes a ghost.
%%%
When the bound is saturated, $\mu^2=2\Lambda_g/3$, the helicity-0 mode becomes pure gauge and the instability disappears. Theories with such fine-tuning are called ``partially massless gravities''~\cite{Deser:1983mm,Deser:2001pe} [see also Refs.~\cite{Hassan:2012gz,Deser:2012qg,Deser:2013uy,Hassan:2012rq,Hassan:2013pca}] and they are not affected by the monopole instability discussed above.
%%%
Finally, as shown in Fig.~\ref{fig:GL}, the instability is even more effective for Schwarzschild-de Sitter BHs and it exists roughly in the same range of graviton mass.
%%%%

%%%%%%%%%%%%%%%%%%%%%%%%%%%%%%%%%%%%%%%%%%%%

For both Schwarzschild and Schwarzschild-de Sitter BHs, the instability timescale is of the order of the Hubble time when $m_g=\hbar\mu\sim 2\times 10^{-33} {\mathrm{eV}}$~\cite{Babichev:2013una}.
This of course, does {\it not} mean that the observation of compact objects imposes constraints on the graviton mass \footnote{The monopole instability does not impose limits on the
graviton mass, but the observation of rotating compact BHs, discussed in Chapter~\ref{chapter:Kerr}, does impose strict limits on the graviton mass.}.
Rather, it suggests that the background solution used to describe these geometries is likely not the physical one.
It would seem that a suitable background geometry is given by the end-state of this monopole instability.

Our linear analysis cannot handle the nonlinear development of the instability, nor the nonlinear final state.
However, from the mode profile in Fig.~\ref{fig:GL}, it is tempting to conjecture that a Schwarzschild BH surrounded by a graviton
cloud could be a possible solution of the field equations. We will confirm in Chapter~\ref{chapter:BHhair} that such solutions indeed exist.
We note that this possible endstate is completely different, as it must be, from the standard Gregory-Laflamme instability which acts to fragment black strings \cite{Cardoso:2006ks,Lehner:2010pn}.

%%%%%%%%%%%%%%%%%%%%%%%%%%%%%%%%%%%%%%
\subsection{Quasinormal modes}
%%%%%%%%%%%%%%%%%%%%%%%%%%%%%%%%%%%%%%
%
\begin{figure}[htb]
\begin{center}
% \begin{tabular}{c}
\epsfig{file=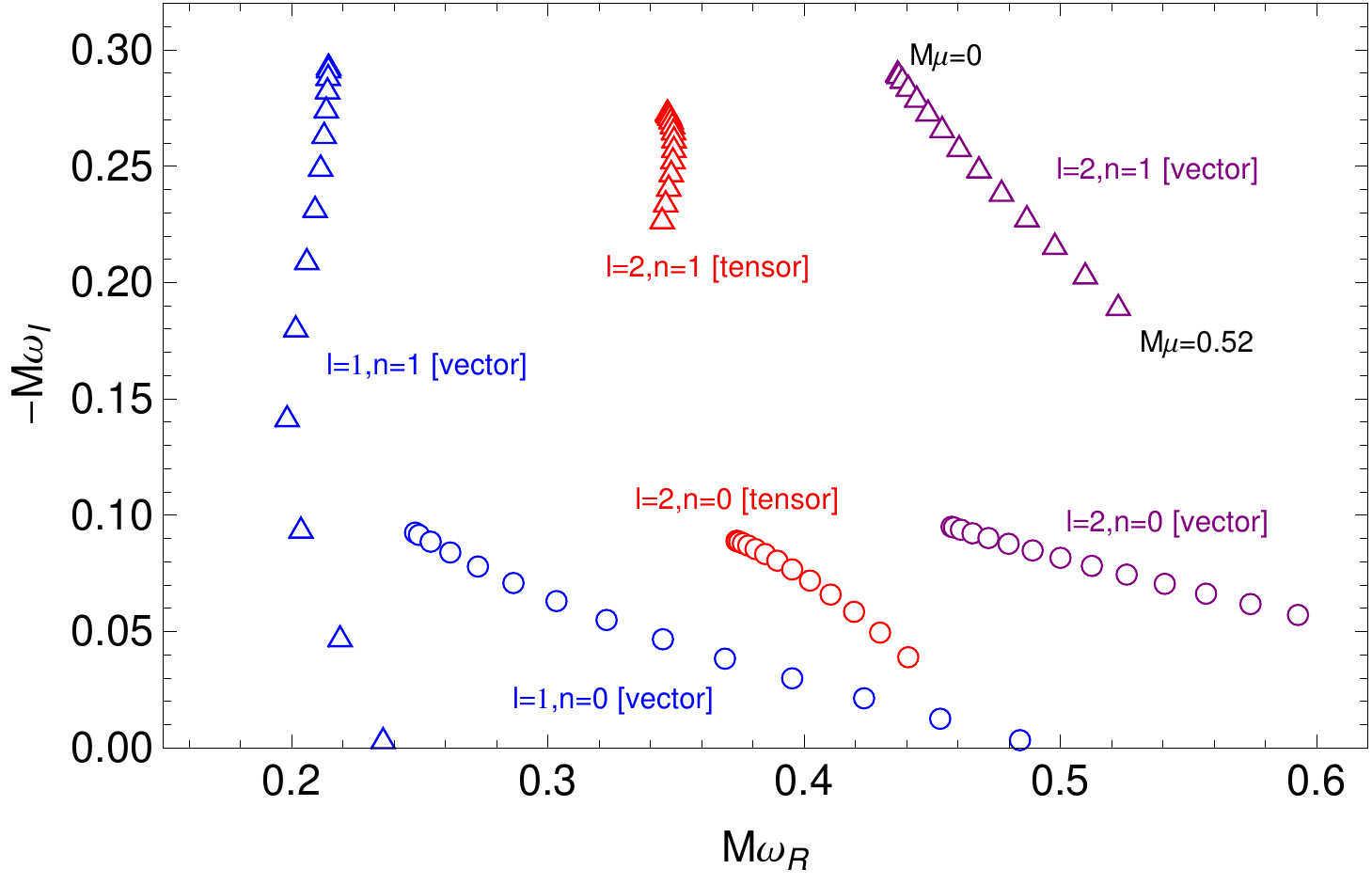,width=10cm,angle=0,clip=true}
% \end{tabular}
\caption{QNM frequencies for axial $l=1,2$ modes, for a range of field masses $M\mu=0,0.04,\ldots,0.52$. Points with largest $|\omega_I|$ correspond to $\mu\to0$. The fundamental mode ($n=0$, circles) and the first overtones ($n=1$, triangles) are shown. In the massless limit the ``vector'' modes have the same QNM frequency as the electromagnetic field, and the ``tensor'' modes have the same QNM frequency as the massless gravity perturbations. \label{fig:QNM}}
\end{center}
\end{figure}
Let us now turn to non-spherically symmetric perturbations. We have computed the axial QNM frequencies using a continued-fraction method that we outline in Appendix~\ref{app:modes}.
In Figure~\ref{fig:QNM} we show the axial QNM frequencies for different values of the spin-2 mass.
As expected, for $l\geq 2$ one can sensibly group the modes in two families for any given $l$ and $n$. They can be distinguished by their behavior in the massless limit, the spectrum of the ``vector'' modes reduces to the spectrum of the photon, while the ``tensor'' modes, which are the only physical modes in the massless limit, approaches the spectrum of the massless gravity perturbations.   
For the lowest overtones, as the mass increases the decay rate decreases to zero, reaching a limit where the QNM disappears. This is linked with the decreasing height of the effective potential barrier as was previously discussed in Ref.~\cite{Ohashi:2004wr}. The limiting behavior, when the damping rate reaches zero are the so-called \emph{quasiresonant} modes, which were already shown to occur for massive scalar~\cite{Ohashi:2004wr,Konoplya:2004wg} and massive vector~\cite{Konoplya:2005hr} fields.     

Polar QNMs are more challenging to compute, because the perturbation equations are lengthy and translate into higher-term recurrence relations in a matrix-valued continued-fraction method~\cite{Pani:2013pma}. On the other hand, due to the well-known divergent nature of the QNM eigenfunctions~\cite{Berti:2009kk}, a direct integration is not well suited to compute these modes precisely. Instead of computing these modes, in the following we shall rather focus on quasibound states --~both in the axial and polar sector~-- which are easier to compute.

%%%%%%%%%%%%%%%%%%%%%%%%%%%%%%%%%
\subsection{Quasibound states}\label{subsec:quasibound}
%%%%%%%%%%%%%%%%%%%%%%%%%%%%%%%%%

Besides the QNM spectrum, massive fields can also be localized in the vicinity of the BH, showing a rich spectrum of quasibound states with complex frequencies. Here the terminology `quasi' stands for the fact that these states decay due to the absorption by the BH, hence the complex frequencies. Bound states were already considered for massive scalar~\cite{Dolan:2007mj}, Dirac~\cite{Gal'tsov:1983wq,Lasenby:2002mc} and Proca~\cite{Gal'tsov:1984nb,Rosa:2011my} fields. In the small-mass limit $M\mu\ll l$, it was shown that for these fields the spectrum resembles that of the hydrogen atom: 
\be
\omega_R/\mu \sim 1-\frac{(M\mu)^2}{2(j+1+n)^2}\,, \label{hydrogenic}
\ee
where $j=l+S$ is the total angular momentum of the state with spin projections $S=-s,-s+1,\ldots,s-1,s$. Here $s$ is the spin of the field. For a given $l$ and $n$, the total angular momentum $j$ satisfies the quantum mechanical rules for addition of angular momenta, $|l-s|\leq j\leq l+s$.
\begin{figure*}[htb]
\begin{center}
\begin{tabular}{cc}
\epsfig{file=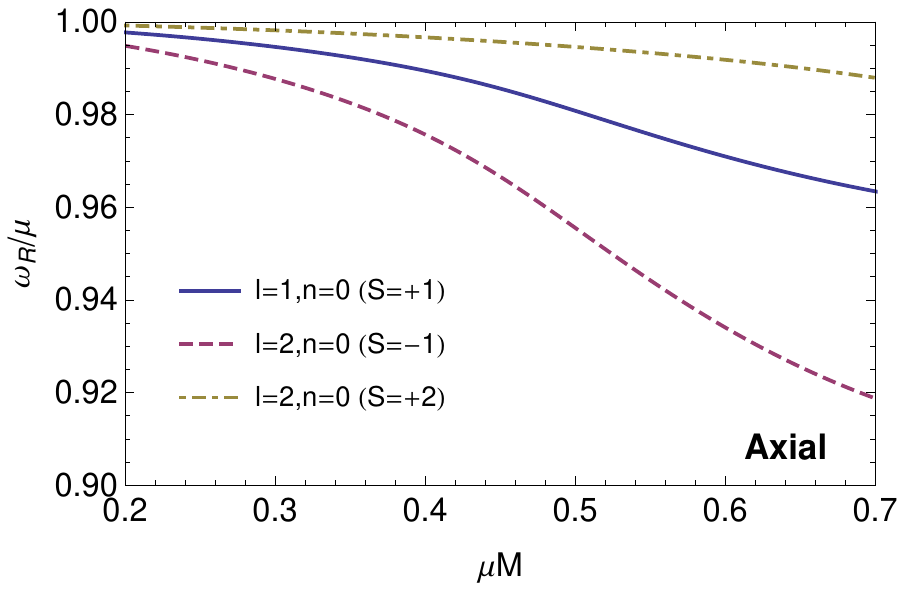,width=7.3cm,angle=0,clip=true}
\epsfig{file=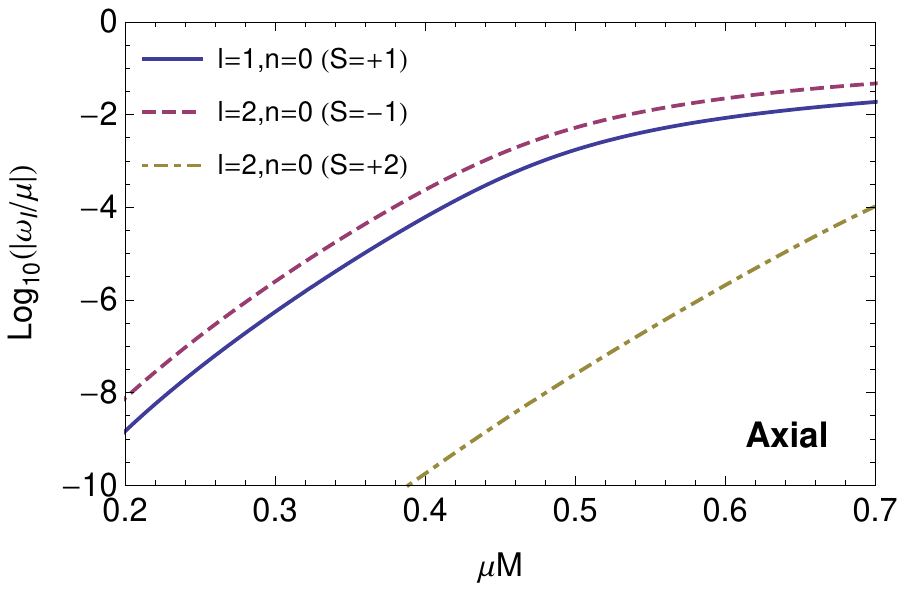,width=7.3cm,angle=0,clip=true}\\
\epsfig{file=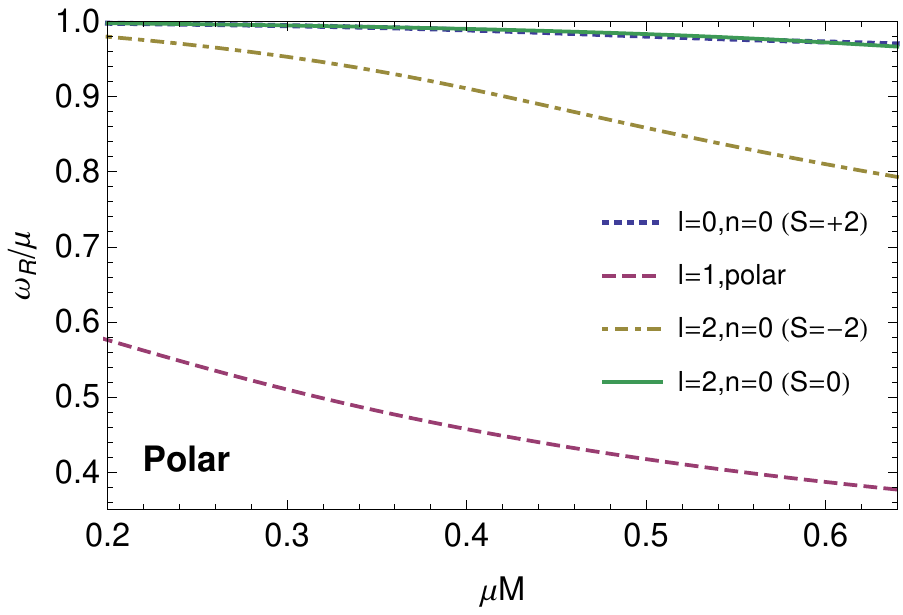,width=7.3cm,angle=0,clip=true}
\epsfig{file=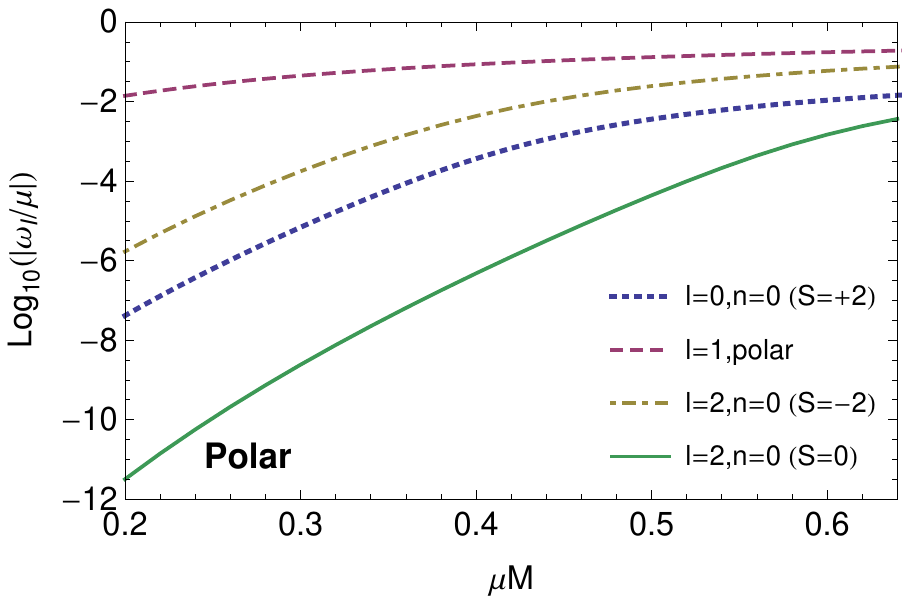,width=7.3cm,angle=0,clip=true}
\end{tabular}
\caption{Axial (Top) and polar (bottom) quasibound state levels of the massive spin-2 field. The left and right panels show the real part, $\omega_R/\mu$, and the imaginary part, $\omega_I/\mu$, of the mode as a function of the mass coupling $M\mu$, respectively. We label the modes by their angular momentum $l$, overtone number $n$ and spin projection $S$. Except for the polar dipole $l=1$, the spectrum is hydrogenic in the massless limit. \label{fig:BS}}
\end{center}
\end{figure*}

Our results show that the spectrum \eqref{hydrogenic} also describes massive spin-2 perturbations which is also confirmed analytically for the axial mode $l=1$ (see Eq.~\eqref{ana_real} of Appendix~\ref{app:ana}).
In Fig.~\ref{fig:BS} we show the quasibound-state frequency spectrum for the lowest modes. Apart from the polar dipole (we discuss this in detail below), all other modes follow a hydrogenic spectrum as predicted by Eq.~\eqref{hydrogenic}. The monopole $l=0$ [which belongs to a different family than the unstable monopole mode discussed in Sec.~\ref{sec:monopoleinstability}] is fully consistent with $S=+2$ which is in agreement with the rules for the sum of angular momenta, $|l-s|\leq j\leq l+s\implies j=2$. For each pair $l\geq 2$ and $n$ there are five kinds of modes, characterized by their spin projections. Here we do not show the mode $l=2$, $n=0$, $S=1$, which is very difficult to find numerically due the complicated form of the polar equations and his tiny imaginary part. Besides that, the existence of the mode $l=2$, $n=1$, $S=0$ with approximately the same real frequency makes it even more challenging to evaluate the $l=2$, $n=0$, $S=1$ mode with sufficient precision.  

\begin{figure}[htb]
\begin{center}
% \begin{tabular}{c}
\epsfig{file=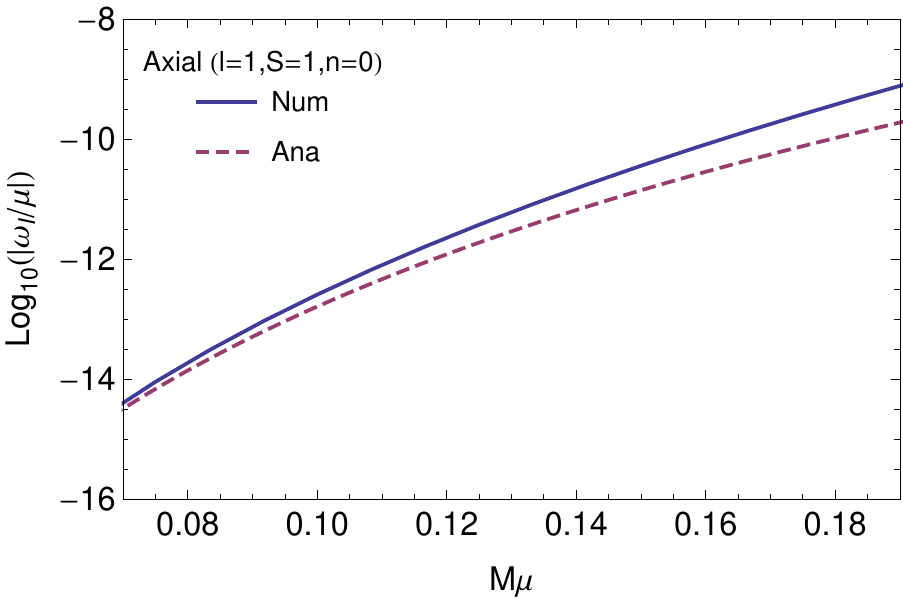,width=10cm,angle=0,clip=true}
% \end{tabular}
\caption{Comparison between the numerical and analytical results for the the axial mode $l=1,\,S=1,\,n=0$ as a function of the mass coupling $M\mu$. The solid line shows the numerical data and the dashed shows the analytical formula~\eqref{wI_ana}.\label{fig:axial_ana}}
\end{center}
\end{figure}

Evaluating the dependence of $\omega_I(\mu)$ in the small-$M\mu$ limit turns out to be extremely challenging, due to the fact that
$\omega_I$ is extremely small in this regime. Our results indicate a power-law dependence of the kind found previously for other massive fields~\cite{Rosa:2011my}, $\omega_I/\mu\propto -(M\mu)^{\eta}$, with
\be
\eta=4l+2S+5\,. \label{wIslope}
\ee
The fact that the modes $l=L$, $S=S_1$ and $l=L+S_1$, $S=-S_1$ have the same exponent is a further confirmation of this scaling. Note that only the constant of proportionality depends on the overtone number $n$ and it also generically depends on $l$ and $S$. This is confirmed analytically for the axial mode $l=1,\,S=1\,,n=0$, as shown in Fig.~\ref{fig:axial_ana}, where we see that in the low-mass limit the numerical results approaches the analytical formula derived in Appendix~\ref{app:ana}, given by
\be
\omega_I/\mu\approx -\frac{320}{19683}(M\mu)^{11}\,. \label{wI_ana}
\ee
%
%%%%%%%%%%%%%%%%%%%%%%%%%
%%%%%%%%%%%%%%%%%%%%%%%%%

The quasibound state found for the polar dipole is clearly the more interesting. This mode appears to be isolated from the rest of the modes and it does not follow the small-mass behavior predicted by Eqs.~\eqref{hydrogenic} and~\eqref{wIslope}. Furthermore, we have found only a single fundamental mode for this state, and no overtones. For this mode, the real part is much smaller than the mass of the spin-2 field. 

%%%%%%%% COMMENTED %%%%%%%%%%%%%%%%%%%%%%%
% This is further confirmed by Fig.~\ref{fig:polar_dipole_BW} where we plot the determinant of the matrix $\bm{S_m}$~\eqref{detS} as a function of the real part of the frequency $M\omega_R$ for $M\mu= 0.15$. The function has only one minima corresponding to the quasibound-state frequency. 
% %
% \begin{figure}[htb]
% \begin{center}
% % \begin{tabular}{c}
% \epsfig{file=polar_dipole_BW_mu015.eps,width=7cm,angle=0,clip=true}
% % \end{tabular}
% \caption{Determinant of the matrix $\bm{S_m}$, $|\det\bm{S_m}|^2$ as a function of the real part of the frequency $M\omega_R$ for $M\mu= 0.15$. The minimum corresponds to the real part of the quasibound-state frequency, which shows that there is only one quasibound-state mode.\label{fig:polar_dipole_BW}}
% \end{center}
% \end{figure}
% %
%%%%%%%%%%%%%%%%%%%%%%%%%%%%%%%%%%
%%%%%%%%%%%%%%%%%%%%%%%%%%%%%%%%%%%%%%%%%%%%%%%%%%%
The real part of this special mode in region $M\mu\lesssim 0.4$ is very well fitted by
\be
\label{polar_di_Re}
\omega_R/\mu\approx 0.72(1-M\mu)\,.
\ee
For the imaginary part we find in the limit $M\mu\ll 1$,
\be\label{wIslope_po}
\omega_I/\mu\approx -(M\mu)^{3}\,.
\ee
That this mode is different is not completely unexpected since in the massless limit it becomes unphysical. This peculiar behavior seems to be the 
result of a nontrivial coupling between the states with spin projection $S=-1$ and $S=0$. Besides that, this mode has the largest binding energy ($\omega_R/\mu-1$) for all couplings $M\mu$, much higher than the ground states of the scalar, Dirac and vector fields (see Fig.7 of Ref.~\cite{Rosa:2011my}). However the decay rate is very large even for small couplings $M\mu$, corresponding to a very short lifetime for this state.

To summarize, the $l>0$ modes of Schwarzschild BHs in massive gravity theories are stable, with a rich and potentially interesting 
fluctuation spectrum, which could give rise to very long-lived clouds of tensor hair in the right circumstances.
We will show in Chapter~\ref{chapter:Kerr} that once rotation is included, this hair grows exponentially and extracts angular momentum away from the BH. Thus, while the monopole $l=0$ mode is unstable even in the static case, the $l>0$ modes suffer from a superradiant instability (see Part~\ref{part:super}) only above a certain threshold of the BH angular momentum.

%%%%%%%%%%%%%%%%%%%%%%%%%%%%%%%%%%%%
\section{Conclusions}
%%%%%%%%%%%%%%%%%%%%%%%%%%%%%%%%%%%%
The advent of new and powerful methods in BH perturbation theory and Numerical Relativity in the past few years 
allows one to finally tackle traditionally complex problems. Particularly important to beyond-the-Standard-Model 
physics are scenarios where ultralight bosonic degrees of freedom are present; simultaneously, massive degrees
of freedom turn out to be important outside particle physics, in particular
several extensions of GR encompassing massive mediators have been proposed.
Thus, the study of massive fluctuations around BHs is a timely topic.

Interesting nonlinear completions of the Fierz-Pauli theory have recently been put forward~\cite{deRham:2010ik,deRham:2010kj,Hassan:2011hr}.
While it is at this stage too early to claim a consistent theory of massive gravitons (these theories or at least certain sectors are either
pathological~\cite{2013arXiv1302.4367C,Deser:2012qx} or phenomenologically disfavored~\cite{Burrage:2012ja}), any nonlinear theory describing a massive spin-2 field --~including a massive graviton~-- will eventually reduce to Eqs.~\eqref{eqmotioncurved}--\eqref{constraint2} in the linearized regime.

Here we have explored the propagation of massive tensors in a Schwarzschild BH background as described by Eqs.~\eqref{eqmotioncurved}--\eqref{constraint2}, and shown that they lead to a generic spherically symmetric instability. These are strong, small-mass instabilities whose end-state is unknown.

Schwarzschild BHs also admit a very rich spectrum of long-lived stable states. Once the BH rotation is turned on, we will show in Chapter~\ref{chapter:Kerr} that these long-lived states can grow exponentially and extract angular momentum away from the BH.

Our work requires extensions and further analysis (in particular, the understanding of the time-development of the monopole instability requires nonlinear simulations), and should in fact be looked at as the first step in a broader program of understanding gravitational-wave emission in massive theories of gravity. 

A final word of caution should be made here. As pointed out in Chapter~\ref{sec:longintro}, due to the Vainshtein mechanism~\cite{Vainshtein:1972sx} present in non-linear theories of massive gravity, near some sources there is a radius below which perturbation theory cannot be trusted. However, for the BH solutions here presented the Vainshtein mechanism does not seem to be present, since those solutions are exactly the ones found in GR. Thus, we expect that the results we show are robust. In fact, as we will show in Chapter~\ref{chapter:BHhair}, the linear study of the spherically symmetric instability correctly predicts the existence of new solutions at the full nonlinear level.

%%%%%%%%%%%%%%%%%%%%%%%%%%%%%%%%%%%%
\chapter{Partially massless gravitons do not destroy general relativity black holes}\label{chapter:partial}
%%%%%%%%%%%%%%%%%%%%%%%%%%%%%%%%%%%%

%%%%%%%%%%%%%%%%%%%%%%%%%%%%%%%%%%%%
\section{Introduction}
%%%%%%%%%%%%%%%%%%%%%%%%%%%%%%%%%%%%

The last couple of years have witnessed a flurry of activity on theories with a propagating massive graviton.
With a very special status, ``partially massless (PM) theories'' have been considered for which the graviton mass is constrained to take a specific value dictated by the cosmological constant such that a new gauge symmetry emerges~\cite{Deser:1983mm,Deser:2001pe,Hassan:2012gz,Deser:2012qg,Deser:2013uy,deRham:2012kf,deRham:2013wv}.

As we showed in the previous Chapter, the simplest BH solution of massive bi(gravity), namely the bidiagonal Schwarzschild de Sitter BHs, are dynamically unstable. The instability is due to a propagating spherically symmetric degree of freedom and affects BHs with or without a cosmological constant. 
As already pointed out in the previous Chapter, and discussed in more detail below, such degree of freedom is absent in PM gravity~\cite{Higuchi:1986py,Deser:1983mm,Deser:2001pe}, so one might wonder whether Schwarzschild-de Sitter BHs are stable in such theories.
The purpose of this Chapter is to show that this is indeed the case.

%%%%%%%%%%%%%%%%%%%%%%%%%%%%%%%%%%%%
\section{Setup}
%%%%%%%%%%%%%%%%%%%%%%%%%%%%%%%%%%%%
Let us consider a massive spin-2 field on a curved spacetime. If the background is a vacuum solution of Einstein's equations with a cosmological constant $\Lambda$, then the field equations are given by the system of eqs.~\eqref{eqmotioncurved}--\eqref{constraint2}. This system propagates five degrees of freedom, corresponding to the healthy helicities of a massive spin-2 field.
The value 
\be
\mu^2=2\Lambda/3\label{PM_limit}
\ee
plays a special role in asymptotically de Sitter spacetimes and it is known as Higuchi limit~\cite{Higuchi:1986py,Deser:1983mm,Deser:2001pe}. When $\mu^2<2\Lambda/3$, the helicity-0 mode becomes itself a ghost, whereas when $\mu^2>2\Lambda/3$ all propagating degrees of freedom are physical. In the Higuchi limit the tracelessness of $h_{\mu\nu}$ is not enforced by the field equations~\eqref{eqmotioncurved}--\eqref{constraint2} and an extra gauge symmetry can be used to eliminate the helicity-0 mode. In this particular case, known as PM gravity~\cite{Deser:2012qg,Deser:2013uy}, the graviton propagates only four helicities.  Although advances in finding a consistent full nonlinear theory of PM gravity have been recently made in the context of bimetric theories~\cite{Hassan:2012gz,Hassan:2012rq,Hassan:2013pca,Akrami:2013km}, the attempt to find a nonlinear completion of PM gravity has been shown to suffer from some obstructions in the framework of massive gravity~\cite{deRham:2013wv,Deser:2013uy,Deser:2013gpa,Fasiello:2013woa,Joung:2014aba,Garcia-Saenz:2014cwa}. We note however that our linear analysis applies to \emph{any} nonlinear generalization of the Fierz-Pauli theory.

In the following we will study the field equations~\eqref{eqmotioncurved}--\eqref{constraint2} in a Schwarzschild-de Sitter background when the condition~\eqref{PM_limit} is satisfied.

%%%%%%%%%%%%%%%%%%%%%%%%%%%%%%%%%%%%%%%%%%%%%%%%%%%%%%%%%%%%%%%%%%
%\noindent{\bf{\em Field equations in PM gravity.}}
\subsection{Field equations in PM gravity}
%%%%%%%%%%%%%%%%%%%%%%%%%%%%%%%%%%%%%%%%%%%%%%%%%%%%%%%%%%%%%%%%%%
As we showed in the previous Chapter, Schwarzschild-de Sitter geometries are {\it generically} unstable against a 
monopole fluctuation~\cite{Brito:2013wya}. In the asymptotically flat case, the instability is equivalent to the Gregory-Laflamme instability~\cite{Gregory:1993vy} of a black string~\cite{Babichev:2013una}. Thus, GR BHs cannot describe static solutions whose linearized equations reduce to Eqs.~\eqref{eqmotioncurved}--\eqref{constraint2}. 

However, a notable exception to this outcome is given by PM theories, which are defined by the tuning~\eqref{PM_limit}.
In this case, the helicity-0 mode which is responsible for the instability can be gauged away and the theory propagates four degrees of freedom~\cite{Higuchi:1986py,Deser:1983mm,Deser:2001pe}.

The approach we developed in the previous Chapter can be easily extended to obtain a set of coupled master equations that fully characterize the linear stability properties of the background. Using the decomposition~\eqref{decom}, and using the background given by the metric~\eqref{eq:SdS}, we find that the axial sector is described by the following system:
\beq
&&\frac{d^2}{dr_*^2}Q+\left[\omega^2-f\left(\frac{\lambda+4}{r^2}-\frac{16M}{r^3}\right)\right]Q=S_Q\,,\label{oddf1_PM}\\
&&\frac{d^2}{dr_*^2}Z+\left[\omega^2-f\left(\frac{\lambda-2}{r^2}+\frac{2M}{r^3}\right)\right]Z=S_Z\,,\label{oddf2_PM}
\eeq
where $\lambda=l(l+1)$ and we have defined the tortoise coordinate $r_*$ via $dr/dr_*=f$. The functions $Q(r)\equiv f(r)h_1$ and $Z(r)\equiv h_2/r$ are combinations of the axial perturbations, whereas the source terms are given by
\begin{equation}
 S_Q=(\lambda-2)\frac{2f(r-3M)}{r^3}Z\,,\qquad S_Z= \frac{2}{r^2}f\,Q\,.
\end{equation}

The polar sector can be simplified by using an extra gauge symmetry arising in the Higuchi limit~\eqref{PM_limit}. In this case the field equations~\eqref{eqmotioncurved}--\eqref{constraint2} are invariant under~\cite{Higuchi:1986py}
%%%
\begin{equation}
 h_{\mu\nu}\to h_{\mu\nu}+\left(\bar\nabla_\mu\bar\nabla_\nu+\frac{\Lambda}{3}\bar g_{\mu\nu}\right)\xi
\end{equation}
%%%
where $\xi$ is a generic scalar gauge function of the spacetime coordinates. The symmetry above can be used to enforce $\eta_0\equiv0$ in the decomposition~\eqref{evenpart}.
In this gauge, the polar sector is fully described by a system of two coupled ordinary differential equations:
\beq
f^2\frac{d^2 \eta_1}{dr^2}+\hat\alpha_1 \frac{d \eta_1}{dr}+\hat\beta_1 \eta_1 &=& S_{\eta_1}\,,\\
f^2\frac{d^2 G}{dr^2}+\hat\alpha_2 \frac{d G}{dr}+\hat\beta_2 G &=& S_G\,,
\eeq
where the source terms are given by
\be
\label{source1}
S_{\eta_1}= (\lambda-2)\hat\sigma_1 \frac{d G}{dr}+(\lambda-2)\hat\rho_1 G\,,\qquad
 S_G = \hat\sigma_2 \frac{d \eta_1}{dr}+\hat\rho_2 \eta_1\,.
\ee
The coefficients $\hat\alpha_i,\,\hat\beta_i,\,\hat\sigma_i,\,\hat\rho_i$ are radial functions which also depend on $\omega$, $l$, $r_b$ and $r_c$.

The regular asymptotic solutions of the axial and polar systems are ingoing and outgoing waves, $\Phi\to e^{\mp i\omega r_*}$, at the BH horizon and at the cosmological horizon, respectively [$\Phi$ collectively denotes the master functions $Q$, $Z$, $\eta_1$ and $G$]. The complex eigenfrequencies $\omega=\omega_R+i\omega_I$, that simultaneously satisfy these boundary conditions are called QNMs~\cite{Berti:2009kk}.

%%%%%%%%%%%%%%%%%%%%%%%%%%%%%%%%%%%%%%%%%%%%%%%%%%%%%%%%%%%%%%%%%%%
\section{Results}
%%%%%%%%%%%%%%%%%%%%%%%%%%%%%%%%%%%%%%%%%%%%%%%%%%%%%%%%%%%%%%%%%%%
\subsection{The near-extremal Schwarzschild-de Sitter geometry}
%%%%%%%%%%%%%%%%%%%%%%%%%%%%%%%%%%%%%%%%%%%%%%%%%%%%%%%%%%%%%%%%%%%
The above equations are in general not analytically solvable.
Fortunately, the geometry is sufficiently rich that it admits a special limit where
one can indeed considerably simplify the equations and solve them analytically.
This regime is the near extremal Schwarzschild-de Sitter BH, defined as the spacetime for which the cosmological horizon $r_c$
is very close (in the $r$ coordinate) to the BH horizon $r_b$, i.e.
$\frac{r_c-r_b}{r_b}\ll 1$ (see~\eqref{eq:SdS}). As shown in Ref.~\cite{Cardoso:2003sw}, it is possible
to solve analytically a large class of Schr\"{o}dinger-type equations in this background
by adopting a perturbative approach in powers of $r_c-r_b$.
In the near-extremal limit, and recalling the definitions for the mass and surface gravity of the metric~\eqref{eq:SdS}, one can make the following approximations:~\cite{Cardoso:2003sw}
\be
r_0 \sim -2r_{b}^2\,\,;\qquad\Lambda\sim r_{b}^{-2}\,\,;\qquad
M \sim \frac{r_b}{3}\,\,;\qquad\kappa_b \sim \frac{r_c-r_b}{2r_{b}^2}\,.
\label{approximation1}
\ee
The key point of the approximation is to realize that in the near-extremal regime $r\sim r_b\sim r_c$,
as we are interested only in the region between the two horizons. Then, $r-r_0 \sim r_b -r_0 \sim 3r_0$
and thus
\begin{equation}
 r\sim\frac{r_c e^{2\kappa_b r_*}+r_b}{1+e^{2\kappa_b r_*}}\,,\qquad 
 f\sim\frac{(r_c-r_b)^2}{4r_{b}^2\cosh{(\kappa_b r_*)}^2}\,.
\end{equation}
Finally, the equations for massive gravitational perturbations
of near-extremal Schwarzschild de Sitter geometries reduce to 
\begin{equation}
\frac{d^2 \Phi}{d r_*^{2}} +\left\lbrack\omega^2-\frac{\kappa_{b}^2 U_0}{\cosh{(\kappa_b r_*)}^2}\right\rbrack\Phi=0 \,,
\label{waveequation}
\end{equation}
where $\Phi$ can be any of the metric variables. The potential $U_0$ is the same for the axial metric functions $Q,Z$
but it is different for the polar functions $\eta_1,G$. We find
\begin{equation}\label{U0}
U_0=\left\{ \begin{array}{ll}
            -4/3+\lambda\,,   & {\,\rm{axial}}\\
            \frac{27 \lambda ^3+36 \lambda  \left(9 r_b^2 \omega ^2-1\right)-16}{3 \left(9 \lambda ^2+12 \lambda +36 r_b^2 \omega ^2+4\right)}\,, & {\,\rm{polar}\,\rm{I}}	\\
	\frac{27 \lambda ^3-36 \lambda +72 r_b^2 \omega ^2-16}{3 \left(9 \lambda ^2+12 \lambda +36 r_b^2 \omega ^2+4\right)}\,,   &{\,\rm{polar}\,\rm{II}}	
\end{array}\right.
\end{equation}

The potential in (\ref{waveequation}) is the well known
P\"oshl-Teller potential~\cite{Poschl:1933zz}. The solutions of the corresponding Schrodinger-like equations
were studied and they are of the hypergeometric type
(for details see Section 4.1 in Ref.~\cite{Berti:2009kk}).
The eigenfrequencies are given by~\cite{Cardoso:2003sw,Berti:2009kk} 
\be
\frac{\omega}{\kappa_b}=  -\left(n+\frac{1}{2}\right)i+
\sqrt{U_0-\frac{1}{4}},\quad n=0,1,\dots\label{solution}
\ee
Using Eq.~\eqref{U0}, we obtain
\be
\frac{\omega}{\kappa_b}=  -\left(n+\frac{1}{2}\right)i+\sqrt{l(l+1)-\frac{19}{12}},\quad n=0,1,\dots\label{finalsclarelectr}
\ee
for \emph{both} axial and polar perturbations. Thus, we get the surprising result that in this regime all perturbations are isospectral~\cite{Chandra,Berti:2009kk}.
Because the polar potential $U_0$ is frequency-dependent, we also find a second, spurious, root at $\omega=\pm i\frac{2+3\lambda}{6r_b}$.
It is easy to check analytically that at these frequencies the wavefunction is not regular at one of the horizons, thus it does not belong to the spectrum.

%%%%%%%%%%%%%%%%%%%%%%%%%%%%%%%%%%%%%%%%%%%%%%%%%%%%%%%%%%%%%%%%%%
\subsection{Numerical results}
%%%%%%%%%%%%%%%%%%%%%%%%%%%%%%%%%%%%%%%%%%%%%%%%%%%%%%%%%%%%%%%%%%
%
\begin{figure}[htb]
\begin{center}
\begin{tabular}{cc}
\epsfig{file=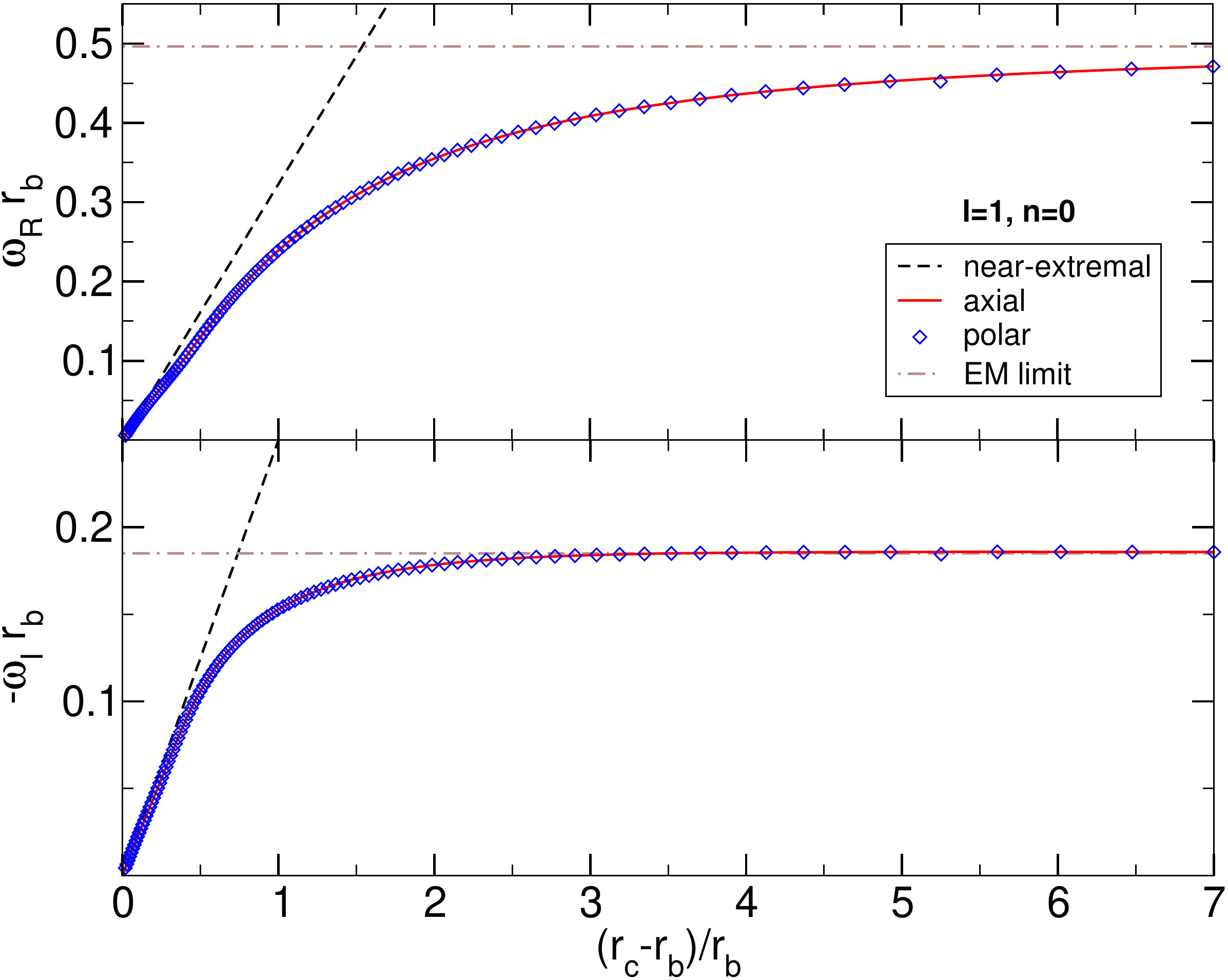,width=7.3cm,angle=0,clip=true}&
\epsfig{file=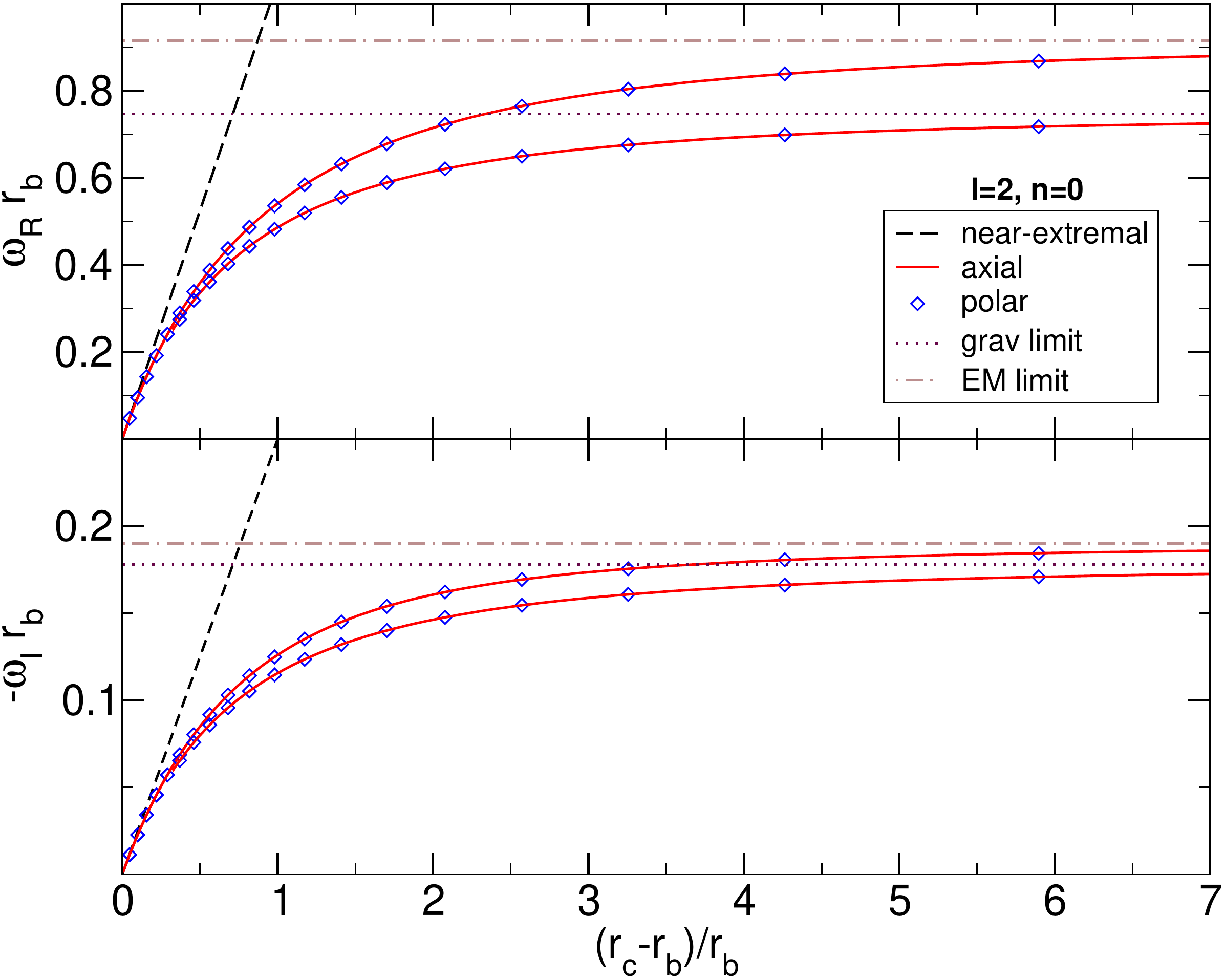,width=7.3cm,angle=0,clip=true}
\end{tabular}
\caption{Left panel: Real (top panel) and imaginary (bottom panel) part of the fundamental dipole mode of a Schwarzschild-de Sitter BH in PM gravity. The fundamental mode is the same for both the axial and polar sectors to within numerical accuracy. Similar results hold for the overtones. Numerical results are compared to the analytical expression~\eqref{finalsclarelectr}. The leftmost and rightmost parts of the $x-$axis are the extremal and the general-relativity, asymptotically flat limit ($\Lambda\propto\mu^2\to0$), respectively. In the $\Lambda\propto\mu^2\to0$ limit the $l=1$ modes approach the electromagnetic QNMs of a Schwarzschild BH~\cite{Berti:2009kk}. Right panel: same for $l=2$ modes. For $l>1$ there are two families of modes which, in the $\Lambda\propto\mu^2\to0$ limit approach the gravitational and the electromagnetic QNMs of a Schwarzschild BH, respectively. 
\label{fig:iso_l1}}
\end{center}
\end{figure}

Using two independent techniques (a matrix-valued direct integration and a matrix-valued continued-fraction method~\cite{Pani:2013pma}, see Appendix~\ref{app:modes}), we have numerically obtained the quasinormal spectrum for generic Schwarzschild-de Sitter geometries, looking explicitly for unstable modes, i.e., modes for which ${\rm{Im}}(\omega)>0$ and which therefore grow exponentially in time while being spatially bounded.
Our results are summarized in Fig.~\eqref{fig:iso_l1}, where we overplot the near-extremal analytical result (denoted by a black dashed line)
and the $\Lambda\to0$ limit (denoted by horizontal lines). Since $\Lambda\propto\mu^2$, this limit corresponds to the massless limit of PM gravity, i.e. to GR.

Except for $l=1$ modes, axial and polar perturbations are grouped in two different families which, in the $\Lambda\propto\mu^2\to0$ limit, reduce to gravitational and electromagnetic modes of a Schwarzschild BH, respectively. 
As predicted by our near-extremal analysis, the two families merge in the $r_c\to r_b$ limit.
For $l=1$ there is only one single family which reduces to the dipole electromagnetic modes of a Schwarzschild BH in the $\Lambda\propto\mu^2\to0$ limit.

An intriguing result, which would merit further study is the fact that axial and polar modes have exactly the \emph{same} quasinormal-mode spectrum
for any value of the cosmological constant, up to numerical accuracy. We were not able to produce an analytical proof of this. Isospectrality guarantees that the entire quasinormal spectrum can be obtained from the axial equations~\eqref{oddf1_PM} and~\eqref{oddf2_PM} only. 

To summarize, our numerical results are in excellent agreement with independent analytical/numerical analysis on two opposite regimes,
the general-relativity limit when the cosmological constant vanishes and the near-extremal limit when the two horizons coalesce.
We found no hints of instabilities in the full parameter space.

%%%%%%%%%%%%%%%%%%%%%%%%%%%%%%%%%%%%%%%%%%%%%%%%%%%%%%%%%%%%%%%%%%%%%%
\section{Discussion}
%%%%%%%%%%%%%%%%%%%%%%%%%%%%%%%%%%%%%%%%%%%%%%%%%%%%%%%%%%%%%%%%%%%%%%
In addition to the spectrum of stable modes presented above, we have searched for unstable, exponentially-growing modes and found none.
Thus, our analysis provides solid evidence for the linear stability of Schwarzschild-de Sitter BHs in PM gravity. This is in contrast with generic theories of massive spin-2 fields (including massive gravitons) in which  Schwarzschild-de Sitter BHs are unstable~\cite{Babichev:2013una,Brito:2013wya} (see Section~\ref{sec:monopoleinstability}).

From our results and from those of Refs.~\cite{Babichev:2013una,Brito:2013wya} the following interesting picture emerges. 
If a theory of massive gravity allows for the same bidiagonal BH solutions of GR, the latter are unstable against spherical perturbations. This is the case for Schwarzschild BHs in any consistent theory of a massive spin-2 field, including the recent nonlinear massive gravity~\cite{deRham:2010ik,deRham:2010kj} and bimetric theories~\cite{Hassan:2011zd}, with or without a cosmological constant. The end-state of the instability is an interesting open problem. 

We have shown here that a notable exception to this picture is represented by PM gravity, which is obtained by enforcing the constraint~\eqref{PM_limit}. The unstable monopole is absent in this theory and a complete analysis of nonspherical modes has revealed no instability. Remarkably, the spectrum of massive gravitational perturbations is isospectral in this theory, which is another piece of evidence for its special role within the family of massive gravities.

Our analysis is only valid at the linear level. It is still a matter of debate whether a nonlinear completion of PM gravity exist~\cite{Hassan:2012gz,Hassan:2012rq,Hassan:2013pca,Akrami:2013km,deRham:2013wv,Deser:2013uy,Deser:2013gpa,Fasiello:2013woa,Joung:2014aba,Garcia-Saenz:2014cwa}. Strong arguments indicate that any nonlinear completion would re-introduce the helicity-0 degree of freedom~\cite{deRham:2013wv,Deser:2013uy,Deser:2013gpa,Fasiello:2013woa,Joung:2014aba,Garcia-Saenz:2014cwa}. Whether this could lead to an instability at the nonlinear level is an interesting open question that we leave for future work.

%%%%%%%%%%%%%%%%%%%%%%%%%%%%%%%%%%%%%%
\chapter{Linear perturbations of nonbidiagonal black holes in massive (bi)gravity}\label{chapter:nonbi}
%%%%%%%%%%%%%%%%%%%%%%%%%%%%%%%%%%%%%%
%%%%%%%%%%%%%%%%%%%%%%%%%%%%%%%%%%%%%%
\section{Introduction}
%%%%%%%%%%%%%%%%%%%%%%%%%%%%%%%%%%%%%%

In Chapter~\ref{chapter:massive2} we showed that the bidiagonal Schwarzschild solution is generically unstable against radial perturbations. This instability is equivalent~\cite{Babichev:2013una} to the Gregory-Laflamme instability~\cite{Gregory:1993vy} of a five-dimensional black string.
Radial perturbations of nonbidiagonal solutions were considered in Ref.~\cite{Babichev:2014oua} showing that, unlike the bidiagonal case, these solutions are classically stable against radial perturbations. One open problem concerns the modal stability of nonbidiagonal solutions to nonradial perturbations.
In this Chapter we close this gap by considering generic gravitational perturbations of these solutions. Our main result is the proof that the QNM spectrum of these solutions is the same as that of a Schwarzschild BH in GR and, therefore, these solutions are classically mode stable\footnote{By ``modes'' we mean the quasinormal spectrum of perturbations, unlike the more generic perturbations which are also considered in this Chapter.  Similarly, by modal stability we mean that the quasinormal spectrum of perturbations does not contain unstable modes. Strictly speaking, the modal stability does not necessarily imply the full stability of a solution.} precisely as the Schwarzschild metric. 
Along the way we discuss various peculiar properties of the gravitational perturbations of these solutions.

%%%%%%%%%%%%%%%%%%%%%%%%%%%%%%%%%
\section{Nonbidiagonal spherically symmetric solutions in massive (bi)gravity}
%%%%%%%%%%%%%%%%%%%%%%%%%%%%%%%%%
We consider the massive (bi)gravity theory defined by the Lagrangian~\eqref{biaction}.
The two classes of static black-hole solutions in this theory, can be conveniently written in the bi-advanced Eddington-Finkelstein form~\cite{Babichev:2014oua}
\begin{eqnarray}\label{sol}
ds_g^2 & = & -\left(1-\frac{r_g}{r}\right)dv^2 +2dvdr+r^2 d\Omega^2,\label{metricg}\\
ds_f^2 & = & C^2\left[-\left(1-\frac{r_f}r \right)dv^2 +2dvdr+r^2 d\Omega^2\right], \label{metricf}
\end{eqnarray}
where $C$ is a constant conformal factor and $r_g$ and $r_f$ are the two (generically different) horizon radii of the two metrics. Using the field equations~\eqref{field_eqs1} and~\eqref{field_eqs2}, one finds that the only nondiagonal terms of $T^{\mu}_{\phantom{\mu}\nu}$ and $\mathcal{T}^\mu_{\phantom{\mu}\nu}$ read
\begin{equation}\label{Toff}
T^r_{\phantom{r}v} =-C^4\mathcal{T}^r_{\phantom{r}v}=\frac{C \left(\beta_1+2C\beta_2+C^2\beta_3\right) \left(r_g-r_f\right)}{2 r}\,.
\end{equation}
%%%
Clearly, these off-diagonal terms must vanish for the metrics~\eqref{metricg} and~\eqref{metricf} to be solutions of the vacuum field equations.
This implies either $r_g=r_f$, which is equivalent to the (bidiagonal) bi-Schwarzschild solution analyzed in~\cite{Babichev:2013una,Brito:2013wya} and Chapter~\ref{chapter:massive2}, or 
\begin{equation}
\label{relation} 
	\beta_1+2C\beta_2+C^2\beta_3 = 0.
\end{equation} 
%%%
The above condition fixes the value of the conformal factor $C$ for a given choice of the coupling constants $\beta_i$. In the rest of the Chapter we will focus on this case, which describes two metrics that cannot be simultaneously diagonalized. 
The case of a flat (Minkowski) non-dynamical metric $f_{\mu\nu}$ coupled to a Schwarzschild metric $g_{\mu\nu}$ falls within this class of solutions.  
The solution~(\ref{sol}) is not the most general analytic nonbidiagonal solution. As it has been shown in Ref.~\cite{Volkov:2012wp}  there is a family of 
nonbidiagonal solutions which contains a function satisfying a nonlinear partial differential equation (see also Ref.~\cite{Volkov:2014ooa}). 
Each regular solution of the partial differential equation gives a different solution for the metrics. 
We consider asymptotically-flat solutions, which implies a fine tuning of the coupling constants such that the two effective cosmological constants vanish. From~\eqref{cosmo_bigrav} this imposes
%%%%
\be
\beta_0 =-\left(3C\beta_1+3 C^2\beta_2+C^3\beta_3\right)\,,\qquad
\beta_4 =-\frac{\beta_1+3C\beta_2+3C^2\beta_3}{C^3}\,,
\ee
%%%
to balance the corresponding contributions coming from 
$T_{\mu\nu}$ and $\mathcal{T}_{\mu\nu} $. 

%%%%%%%%%%%%%%%%%%%%%%%%%%%%%%%%%
\section{Gravitational perturbations of the nonbidiagonal solution}
%%%%%%%%%%%%%%%%%%%%%%%%%%%%%%%%%

Let us now consider linear perturbations around the solutions \eqref{metricg} and \eqref{metricf} with the condition~\eqref{relation}, i.e. we focus on nonbidiagonal solutions.
We consider perturbations of the form:
\be
g_{\mu\nu}=\bar{g}_{\mu\nu}+h_{\mu\nu}^{(g)}\,,\qquad
f_{\mu\nu}=\bar{f}_{\mu\nu}+h_{\mu\nu}^{(f)}\,,
\ee
where the bar notation denotes, once again, background quantities and $h_{\mu\nu}$ are small perturbations of the background solutions. From the field equations~\eqref{field_eqs1} and~\eqref{field_eqs2}, the tensors $h_{\mu\nu}^{(g)}$ and
$h_{\mu\nu}^{(f)}$ satisfy the linearized equations 
\begin{equation}\label{perteqs}
	\delta G_{\mu\nu}+  
	\frac{M_v^4}{M_g^2}\delta T_{\mu\nu}=0\,,\qquad 
	\delta \mathcal{G}_{\mu\nu} + \frac{M_v^4}{M_f^2}\delta\mathcal{T}_{\mu\nu}=0\,,
\end{equation} 
%%%
where $G_{\mu\nu}(g)=\bar{G}_{\mu\nu}+\delta G_{\mu\nu}$, $\mathcal{G}_{\mu\nu}(f)=\bar{\mathcal{G}}_{\mu\nu}+\delta \mathcal{G}_{\mu\nu}$ and similarly for $\delta T_{\mu\nu}$ and $\delta\mathcal{T}_{\mu\nu}$.

As discussed in Chapter~\ref{chapter:massive2}, in a spherically symmetric background, the spin-2 perturbations $h_{\mu\nu}^{(g)}$ and
$h_{\mu\nu}^{(f)}$ can be decomposed in terms of axial and polar perturbations, as in Eq.~\eqref{decom}~\footnote{Note that since we are working in $(v,r)$-coordinates, we change $t\to v$ in this decomposition.}. 
Without loss of generality, we will also multiply the definition of  $h_{\mu\nu}^{(f)}$ by an overall $C^2$ factor. Spherical symmetry assures that the field equations do not depend on the azimuthal number $m$. In addition, perturbations with different parity and different harmonic index $l$ decouple from each other\footnote{To simplify the notation, we shall often omit the superscript ${}^{lm}$ in the perturbation functions.}. 

In the nonbidiagonal case~\eqref{relation}, by using this decomposition, it turns out that the mass terms in the perturbation equations~\eqref{perteqs} take the remarkably simple form
{\scriptsize
\begin{align}\label{nbd}
&\delta T^{\mu}_{\phantom{\mu}\nu}=
\frac{\mathcal{A} \left(r_g-r_f\right) }{4 r}\, e^{-i \omega v}\nonumber\\
&\begin{pmatrix}
 0 & 0 & 0 & 0 \\
 2K^{lm}_{(-)} Y_{lm} & 0 & -\left(h^{lm}_{1(-)}\frac{\partial_{\phi}Y_{lm}}{\sin\theta}+\eta^{lm}_{1(-)}\partial_{\theta}Y_{lm}\right) & h^{lm}_{1(-)}\sin\theta\partial_{\theta}Y_{lm}-\eta^{lm}_{1(-)}\partial_{\phi}Y_{lm} \\
 -\left(h^{lm}_{1(-)}\frac{\partial_{\phi}Y_{lm}}{\sin\theta}+\eta^{lm}_{1(-)}\partial_{\theta}Y_{lm}\right) & 0 & H^{lm}_{2(-)} Y_{lm} & 0 \\
 \frac{1}{r^2\sin\theta}\left(h^{lm}_{1(-)}\partial_{\theta}Y_{lm}-\eta^{lm}_{1(-)}\frac{\partial_{\phi}Y_{lm}}{\sin\theta}\right) & 0 & 0 & H^{lm}_{2(-)} Y_{lm}
\end{pmatrix}\,,
\end{align}
}
and $C^4\delta\mathcal{T}^{\mu}_{\phantom{\mu}\nu}=-\delta T^\mu_{\phantom{\mu}\nu}$, where $h_{\mu\nu}^{(-)} := h_{\mu\nu}^{(f)}- h_{\mu\nu}^{(g)}$ and $\mathcal{A} = 2C^2\left(\beta_2+C\beta_3\right)$.
%%%%%

By taking the divergence of Eq.~\eqref{perteqs} and using the Bianchi identities for the Einstein tensors, 
we obtain the constraint 
$ \nabla^\nu_{(f)}\delta \mathcal{T}_{\mu\nu}\propto \nabla^\nu_{(g)}\delta T_{\mu\nu} =0$, which, from Eq.~\eqref{nbd} in the nonbidiagonal case ($r_g\neq r_f$), yields the following relations:
\beq\label{divg}
&&H^{lm}_{2(-)}=0\,,\qquad
\left(r \eta^{lm}_{1(-)}\right)'=0\,,\nn\\
&&\left(r h^{lm}_{1(-)}\right)'=0\,,\qquad
\left(r K^{lm}_{(-)}\right)'+\frac{l(l+1)\eta^{lm}_{1(-)}}{2r}=0\,.
\eeq
The above equations can be immediately solved for
\beq
&&H^{lm}_{2(-)}=0\,,\qquad \eta^{lm}_{1(-)} = \frac{c_0}{r}\,, \label{constnbd1} \\
&& h^{lm}_{1(-)} = \frac{c_1}{r}\,,\qquad K^{lm}_{(-)}= \frac{c_2}{r}+\frac{l(l+1)c_0}{2r^2} \label{constnbd2} \,,
\eeq
where $c_0$, $c_1$, and $c_2$ are (generically complex) integration constants\footnote{We should note that, since we are working in the frequency-domain, $c_i$ are arbitrary functions of $\omega$ while in the time-domain they are arbitrary (real) functions of the advanced time $v$.}. The peculiar structure of $\delta T^{\mu}_{\phantom{\mu}\nu}$ is responsible for some highly nontrivial properties which are discussed in the section below.

%%%%%%%%%%%%%%%%%%%%%%%%%%%%%%%%%
\section{Equivalence of the QNMs to those of a Schwarzschild black hole in GR}\label{sec:QNMs}
%%%%%%%%%%%%%%%%%%%%%%%%%%%%%%%%%
In this section we show that the QNMs of the nonbidiagonal black-hole solution of massive gravity are the same as those of a Schwarzschild BH in GR.

The QNMs are the proper frequencies of vibration of a relativistic self-gravitating object, in analogy with the normal modes of oscillating stars in Newtonian gravity (cf. Refs.~\cite{Kokkotas:1999bd,Berti:2009kk,Konoplya:2011qq} for some reviews). Due to the emission of gravitational waves or to absorption by an event horizon, the QNMs are complex numbers whose real part defines the frequency of the perturbation, whereas the imaginary part defines the inverse of the damping time (or of the instability time scale in the case of unstable modes). It should be stressed that the QNMs do not form a complete set~\cite{Kokkotas:1999bd} so they do not describe the full response of the BH to external perturbations. In fact, we show below that QNMs of nonbidiagonal BHs in bimetric theories are exactly the same as in GR, while generic perturbations of bimetric BHs are different from those of GR BHs.

The QNMs can be computed as the eigenvalues of a boundary-value problem defined by Eq.~\eqref{perteqs} with suitable boundary conditions. For the case of static, asymptotically-flat BHs, one imposes that the perturbations behave as ingoing waves near the horizon, $\sim e^{-i (\omega t+k_- r_*)}$ and as outgoing waves near infinity, $\sim e^{i (k_+ r_*-\omega t)}$. Here, $r_*$ is the tortoise coordinate defined through $v=t+r_*$, where $t$ is a Schwarzschild-like time coordinate\footnote{For clarity, in this Section the metric perturbations $h_{\mu\nu}$ are written as functions of $t$. One can always do this by defining $v=t+r_*$.}. The constant $k_\pm$ (which we assume to be positive without loss of generality) is the momentum of the perturbations and it is related to the effective dispersion relation. For example, for an outgoing perturbation with effective mass $\mu$ propagating in Minkowski spacetime\footnote{For gravitational perturbations of GR Schwarzschild BHs $k_\pm=\pm\omega$, whereas for the static bidiagonal black-hole solutions of massive gravity $k_-=\omega$ and $k_+=\sqrt{\omega^2-\mu^2}$, as shown in Chapter~\ref{chapter:massive2}.}, $k_+=\sqrt{\omega^2-\mu^2}$.

Therefore, the QNMs of the bimetric system are defined by the following boundary conditions for the metrics $h^{(g)}_{\mu\nu}$ and $h^{(f)}_{\mu\nu}$,
%%%
\begin{eqnarray}
   \tilde{h}^{(g)}_{\mu\nu} \to A_{\mu\nu}^\pm e^{\pm i k_\pm r_{g*}}\,, \quad   \tilde{h}^{(f)}_{\mu\nu} \to B_{\mu\nu}^\pm e^{\pm i k_\pm r_{f*}} \,,\label{BCs}
\end{eqnarray}
%%%
where $A_{\mu\nu}^\pm$ and $B_{\mu\nu}^\pm$ are typically polynomials in $1/r$, the plus (minus) sign refers to the near-infinity (near-horizon) behavior, whereas the
tortoise coordinates are defined via $dr/dr_{g*}=\left(1-r_g/r\right)$ and $dr/dr_{f*}=\left(1-r_f/r\right)$.

Inspection of Eqs.~\eqref{constnbd1} and \eqref{constnbd2} together with the decomposition~\eqref{decom} immediately shows that the boundary conditions~\eqref{BCs} cannot be satisfied unless $c_i=0$ in Eqs.~\eqref{constnbd1} and \eqref{constnbd2}. For example, from Eqs.~\eqref{constnbd1}, \eqref{constnbd2}, \eqref{decom},~\eqref{oddpart} and~\eqref{evenpart}, we obtain
\begin{equation}
 \tilde h^{(f)}_{r\phi}-\tilde  h^{(g)}_{r\phi} = e^{-i\omega r_*} \left[\frac{c_0}{r}\partial_\phi Y_{lm}-\frac{c_1}{r}\sin\theta\partial_{\theta}Y_{lm}\right]\,,
\end{equation}
for the difference of the inverse-Fourier transformed quantities $\tilde h^{(f)}_{r\phi}$ and $\tilde h^{(g)}_{r\phi}$ (and similarly for other components). Therefore, it is clear that the difference $\tilde h^{(f)}_{r\phi}-\tilde  h^{(g)}_{r\phi}$ represents an ingoing wave of frequency $\omega$ in the whole space and the same property must hold independently for $\tilde h^{(f)}_{r\phi}$ and $\tilde h^{(g)}_{r\phi}$. Because $r_{g*}\sim r_{f*}\to-\infty$ near the corresponding horizon, the near-horizon boundary condition in Eq.~\eqref{BCs} is always satisfied with $k_-=\omega$. On the other hand, the near-infinity boundary condition,  $h^{(g)}_{\mu\nu}\sim h^{(f)}_{\mu\nu}\to e^{ik_+r}$, cannot be enforced\footnote{If we were using retarded Eddington-Finkelstein coordinates, the opposite situation would occur: the solution would describe an outgoing wave in the whole space, and the boundary conditions would be automatically satisfied at infinity but not at the event horizon. In both cases, the full set of boundary conditions~\eqref{BCs} cannot be enforced unless $c_i=0$.}. 

This simple observation implies that the boundary conditions for QNMs impose $c_0=c_1=c_2=0$ and, in turn, $\delta T^\mu_\nu=\delta {\cal T}^\mu_\nu=0$\footnote{In the special case ${\cal A}=0$, i.e., $\beta_2=-C \beta_3$, one always gets $\delta T^\mu_\nu=\delta {\cal T}^\mu_\nu=0$ and the perturbation equations reduce to the standard linearized Einstein's equations as noted in Ref.~\cite{Kobayashi:2015yda} (see also~\cite{Kodama:2013rea}  for the case with only one dynamical metric). This can be also related to an extra symmetry for spherically symmetric solutions in the case $\beta_2=-C \beta_3$~\cite{Volkov:2012wp,Babichev:2014fka}.}. Therefore, the eigenvalue problem reduces to the standard linearized Einstein's equations
%%%
\begin{equation}
	\delta G_{\mu\nu}=0\ , \ \ \ \delta \mathcal{G}_{\mu\nu} =0\ ,\label{perteqs0}
\end{equation} 
%%%
with the extra constraints coming from Eqs.~\eqref{constnbd1} and \eqref{constnbd2} with $c_0=c_1=c_2=0$ , namely
%%%
\beq
H^{lm}_{2(g)}=H^{lm}_{2(f)}\,,&\quad & \eta^{lm}_{1(g)} = \eta^{lm}_{1(f)} \,,\label{constnbd1b} \\
h^{lm}_{1(g)} = h^{lm}_{1(f)}\,,&\quad & K^{lm}_{(g)}=K^{lm}_{(f)} \,.\label{constnbd2b} 
\eeq
%%%

To complete our proof, we can use the freedom to choose a particular gauge. In this case it is convenient to choose a gauge such that $H^{lm}_{2(g)}=K^{lm}_{(g)}=\eta^{lm}_{1(g)}=h^{lm}_{1(g)}=0$. This can always be imposed by transforming~\cite{Zerilli:1971wd}
\begin{equation}
 h_{\mu\nu}^{(g)} \to h_{\mu\nu}^{(g)} -\nabla_{\mu}\xi_{\nu}-\nabla_{\nu}\xi_{\mu}\,,
\end{equation}
where  $\xi_\mu$ is the transformation four-vector. The latter can be decomposed into an axial vector component and into three polar vector components, which can be chosen to enforce the aforementioned relations $h^{lm}_{1(g)}=0$ and $H^{lm}_{2(g)}=K^{lm}_{(g)}=\eta^{lm}_{1(g)}=0$, respectively.
Since there is only one diffeomorphism invariance and two metrics, the components of the metric $f$ are not fixed  {\it a priori} by the above gauge choice.
%%%
However, Eqs.~\eqref{constnbd1b} and~\eqref{constnbd2b} imply $H^{lm}_{2(f)}=K^{lm}_{(f)}=\eta^{lm}_{1(f)}=h^{lm}_{1(f)}=0$. Therefore, Eq.~\eqref{perteqs0} reduces to two copies of the linearized Einstein equations in the gauge $H^{lm}_{2}=K^{lm}=\eta^{lm}_{1}=h^{lm}_{1}=0$. Note that this gauge is different from the standard Regge-Wheeler-Zerilli gauge, in which $G^{lm}_{2}=\eta^{lm}_{0}=\eta^{lm}_{1}=h^{lm}_{2}=0$~\cite{Regge:1957td,Zerilli:1971wd}. Nonetheless, the perturbation equations are precisely the same as in the case of GR.

Thus, we have just proved that the eigenvalue problem reduces to that of two Schwarzschild metrics with horizon radii $r_g$ and $r_f$ in GR. In particular, there will be no monopole and dipole modes, the QNMs exist only for $l\geq2$, and they correspond to 2 propagating degrees of freedom. 
%%%
As a by-product of this equivalence, the QNM spectrum does not contain any unstable mode and the nonbidiagonal black-hole solution of massive gravity is therefore mode stable for any gravitational perturbations (which can be decomposed into QNMs). 
Both properties (the absence of $l=0$ and $l=1$ modes and the modal stability) are in striking contrast to the case of bidiagonal solutions discussed in Chapter~\ref{chapter:massive2}, as we also show in the next section.

%%%%%%%%%%%%%%%%%%%%%%%%%%%%%%%%%
\section{Generic Gravitational Perturbations}
%%%%%%%%%%%%%%%%%%%%%%%%%%%%%%%%%
As shown in the previous section, the QNM spectrum of the nonbidiagonal BH in massive gravity coincides with that of a Schwarzschild BH in GR. This property is true for both the axial and polar sectors, which respectively reduce to a Regge-Wheeler and a Zerilli equation. Nonetheless, the full set of perturbations (and therefore the object's response to external sources) is generically different, both in the axial and in the polar sector. In this section we relax the boundary conditions at infinity to include ingoing (at infinity) perturbations, unlike the previous section where those perturbations were forbidden by boundary conditions corresponding to QNMs. Thus, our study will include more general perturbations which are useful to study the linear response of the BH to external perturbers. 
In the following we will consider the axial and polar sectors, and the cases $l=0$, $l=1$ and $l\geq2$, separately. 

%%%%%%%%%%%%%%%%%%%%%%%%%%%%%%%%%
\subsection{Polar sector}
%%%%%%%%%%%%%%%%%%%%%%%%%%%%%%%%%
Here we discuss the perturbation equations for the polar sector separately for $l=0$, $l=1$ and $l\geq2$.

%%%%%%%%%%%%%%%%%%%%%%%%%%%%%%%%%
\subsubsection{Polar monopole}\label{ssec:monopole}
%%%%%%%%%%%%%%%%%%%%%%%%%%%%%%%%%

Radial (i.e., $l=0$) perturbations were studied in Ref.~\cite{Babichev:2014oua}. In this case the perturbation functions $G^{lm}$, $\eta^{lm}_{0}$ and $\eta^{lm}_{1}$ are not defined because their corresponding angular part in Eq.~\eqref{evenpart} vanishes. For $\omega=0$, one gets $c_0=0$ and $h_{\mu\nu}^{(f)}=h_{\mu\nu}^{(g)}$ and there is one solution which corresponds to a trivial mass shift in both metrics $f_{\mu\nu}$ and $g_{\mu\nu}$. When $\omega\neq 0$, we find the same solution as in Ref.~\cite{Babichev:2014oua} when using the same gauge. For the sake of completeness, we here show the explicit form of this solution.

Unlike Ref.~\cite{Babichev:2014oua}, however, let us choose a gauge such that $H_{1(g)}=K_{(g)}=H_{2(g)}=0$. 
%%%%
From Eqs.~\eqref{constnbd1} and \eqref{constnbd2} we then have $H_{2(f)}=0$ and $K_{(f)}=c_2/r$. Finally, the field equations yield 
\begin{align}
&H_{1(f)}=i c_5\omega\,,\\
&H_{0(f)}=-i c_2\omega-\frac{c_2 r_f}{2 r^2}-2i\omega c_5\left(1-\frac{r_f}{r}\right)
+\frac{\mathcal{A}c_2 M_v^4(r_f-r_g)}{4M_f^2 C^2i\omega r}\,,\\
&H_{0(g)}=\frac{\mathcal{A}c_2 M_v^4(r_g-r_f)}{4M_g^2 i\omega r}\,,
\end{align}
where $c_5$ is an integration constant\footnote{Note that the result in \cite{Babichev:2014oua} is written in terms of $h^{\mu\nu}$, while here we work with $h_{\mu\nu}$, hence the apparent difference of the expressions.}. 
%%%
There are two free integration constants, $c_2$ and $c_5$, which are not fixed by the assumption of asymptotic flatness. This can be checked by 
calculating the curvature invariants. For example, the Kretschmann scalar $R_{abcd}R^{abcd}$ of the $l=0$ polar solution vanishes at large distances for any value of the integration constants. Moreover, in this gauge $c_5$ does not affect the $g_{\mu\nu}$ metric, and does not contribute either to the curvature of both metrics or to the energy-momentum tensors $\delta T_{\mu\nu}$ and $\delta\mathcal{T}_{\mu\nu}$. 

In other words, if one takes the $g_{\mu\nu}$ metric to be the physical one and couples it to matter, the constant $c_5$ would be completely decoupled and would not affect any observable physical quantity, at least to linear order (we discuss possible nonlinear effects in Sec.~\ref{sec:conclusion}).

On the contrary, the constant $c_2$ is physical. This constant cannot be gauged away from either of the metrics, contributes to $\delta T_{\mu\nu}$ and $\delta\mathcal{T}_{\mu\nu}$, and is therefore associated with observable quantities.

For any $c_2\neq 0$, due to the term $e^{-i\omega v}$ appearing in Eq.~\eqref{decom} (recall that here we are using $t\to v$), the solution above describes an ingoing wave which does not feel any effective potential and therefore does not change its propagation in the entire space.
This property is reminiscent of Minkowski spacetime, in which an ingoing wave is not backscattered due to the absence of a gravitational potential\footnote{The analogy with the Minkowski spacetime extends also to the computation of QNMs previously discussed. Minkowski spacetime does not possess proper modes of vibration due to the absence of an effective potential. However, one could imagine to add a test, perfectly-absorbing surface at some fixed location $r=r_0$, which would play the role of an event horizon. Similarly to what previously discussed, in this case one can impose purely absorbing boundary conditions at $r=r_0$ but it would be impossible to impose simultaneously the correct boundary conditions at infinity. Due to the absence of backscattering, Minkowski spacetime does not possess  QNMs even in the presence of a perfectly absorbing surface.}. 
This behavior is in contrast to the Schwarzschild case in GR, in which the radial mode is nondynamical. On the other hand, the radial perturbations of the bidiagonal metric are described by a Zerilli-like equation, given by Eq.~\eqref{evenl0}.
As already discussed, not only in this case is the perturbation dynamical, but it also leads to an instability.

%%%%%%%%%%%%%%%%%%%%%%%%%%%%%%%%%
\subsubsection{Polar dipole}
%%%%%%%%%%%%%%%%%%%%%%%%%%%%%%%%%
\label{ssec:dipole}
When $l=1$, the function $G^{lm}$ is not defined. Up to gauge freedom we can set $H_{2(g)}=\eta_{1(g)}=K_{(g)}=0$. By using the constraints~\eqref{constnbd1} and \eqref{constnbd2}, a straightforward calculation then leads to the following solution

\begin{align}
H_{0(f)}&= 2 r c_9-i \omega  c_2
-\frac{\left(2 c_9+\omega  \left(\omega  c_2+(2 i-4 r \omega ) c_6+6 i r c_9\right)\right) r_f}{2 r^2 \omega ^2}\nn\\
&-\frac{i {\cal A} M_v^4 \left(\omega  c_0+(3 r\omega-i ) c_2\right) }{12 C^2 r^2 \omega ^2 M_f^2}\left(r_f-r_g\right)\,,\\
H_{0(g)}&= \frac{\omega  (2 r \omega -i) c_7 r_g+c_8 \left(2 r^3 \omega
   ^2-(1+3 i r \omega ) r_g\right)}{r^2 \omega ^2}\nn\\
&+\frac{{\cal A} M_v^4 \left(c_2+i \omega  \left(c_0+3 r c_2\right)\right) }{12 r^2 \omega ^2 M_g^2}\left(r_f-r_g\right) \,,\\
H_{1(f)}&= c_6\,,\\
H_{1(g)}&= c_7\,,\\
\eta_{0(f)}&= \frac{i \omega  c_0}{2} +\frac{c_2}{2}+r \left(c_6+r c_9\right)+\frac{\left(i \omega  c_6+c_9\right) r_f}{r \omega ^2}
+\frac{{\cal A} M_v^4 \left(i \omega  c_0+c_2\right)}{12 C^2 r \omega ^2 M_f^2}\left(r_f-r_g\right)\,,\\
\eta_{0(g)}&= r \left(c_7+r c_8\right)+\frac{\left(i \omega  c_7+c_8\right) r_g}{r \omega ^2}
-\frac{{\cal A} M_v^4 \left(i \omega  c_0+c_2\right) }{12 r \omega ^2 M_g^2}\left(r_f-r_g\right)\,,
\end{align}
where $c_i$ are integration constants. 
The perturbations must be small in order to stay within the validity of the perturbation theory, i.e. $h^{(g)}_{\mu\nu}\ll g_{\mu\nu}$, and similar for perturbations of the second metric. This requirement leads to $c_6=c_7=c_8=c_9=0$. The only free constants are then $c_0$ and  $c_2$. Both these constants induce physical (observable) changes in the metric perturbations, unlike the monopole case, where only one constant is physical and the other one is a gauge constant. 
Nevertheless, similarly to the monopole case, this solution describes a purely ingoing wave which is not backscattered by the BH.

Also in this case the GR solution describes a gauge mode and is nondynamical, whereas the $l=1$ polar sector of the bidiagonal solution describes two propagating degrees of freedom governed by a pair of coupled equations, given by Eqs.~\eqref{polar_dipole1} and~\eqref{polar_dipole2}. Contrary to the nonbidiagonal case under consideration, the bidiagonal solution possesses $l=1$ polar QNMs which depend on the graviton mass.

%%%%%%%%%%%%%%%%%%%%%%%%%%%%%%%%%
\subsubsection{Polar perturbations with $l\geq 2$}
%%%%%%%%%%%%%%%%%%%%%%%%%%%%%%%%%
\label{ssec:quadropole}
The $l\geq2$ polar case is qualitatively similar to the $l\geq2$ axial case (considered below) although technically more involved. Also in this case we can adopt a gauge such that  $H^{lm}_{2(g)}=K^{lm}_{(g)}=\eta^{lm}_{1(g)}=0$ which, from Eqs.~\eqref{constnbd1} and \eqref{constnbd2}, implies  $H^{lm}_{2(f)}=0$, $K^{lm}_{(f)}=c_2/r+l(l+1)c_0/(2r^2)$ and $\eta^{lm}_{1(f)}=c_0/r$. After some algebra, the field equations can be solved for $H_{1(g)}^{lm}$, $H_{1(f)}^{lm}$, $G_{(g)}^{lm}$, $G_{(f)}^{lm}$, ${d\eta_{0(g)}^{lm}}/dr$ and ${d\eta_{0(f)}^{lm}}/dr$, whereas the functions $H_{0(g)}^{lm}$ and $H_{0(f)}^{lm}$ satisfy a set of two decoupled, second-order differential equations, namely
%%%%
%
\beq
{\cal D}_g [{\tilde \Phi}_g]-W_g {\tilde \Phi}_g &=& s_g\,,\label{polarg}\\
{\cal D}_f [{\tilde \Phi}_f]-W_f {\tilde \Phi}_f &=& s_f\,, \label{polarf}
\eeq
where ${\tilde \Phi}_g:=r^2 H_{0(g)}^{lm} /(r-r_g)$ and ${\tilde \Phi}_f:=r^2 H_{0(f)}^{lm} /(r-r_f)$ and we defined the differential operators
%%%
\begin{equation}
 {\cal D}_{g}=\frac{d^2}{dr_{g*}^2}-2i\omega \frac{d}{dr_{g*}} \,,\quad {\cal D}_{f}=\frac{d^2}{dr_{f*}^2}-2i\omega \frac{d}{dr_{f*}}\,.\\
\end{equation}
%%%%
In the above equations, the potentials read
\beq
W_g&=&\frac{l (l+1) \left(r-r_g\right)-2 i r \omega  \left(2 r-3 r_g\right)+r_g}{r^3}\,,\\
W_f&=&\frac{l (l+1) \left(r-r_f\right)-2 i r \omega  \left(2 r-3 r_f\right)+r_f}{r^3}\,,
\eeq
whereas the source terms are
%
%\begin{widetext}
 \begin{align}
s_g&= \frac{{\cal A} M_v^4 r \left[c_0 \lambda+2 c_2 r\right] \left(r_g-r_f\right)-4 B_1 M_g^2 \left(r_g+2 i r^2
   \omega \right)}{4 r^3 M_g^2}\nn\\
	&+\frac{4 c_4 r M_g^2 \left[r_g \left(\lambda+4 i r \omega \right)+2 i \left(\lambda-2\right) r^2
   \omega \right]}{4 r^3 M_g^2}\,,\\
   %%%%
s_f&= \frac{{\cal A} M_v^4 \left(c_0 \lambda+2 c_2 r\right) \left(r_f-r_g\right)}{4 C^2 r^2 M_f^2}
+\frac{2 r^2 \omega \left(-2 i B_2+2 i c_3 \left(\lambda-2\right) r+c_0 \lambda \omega +2 c_2 r \omega \right)}{2 r^3}\nn\\
&-\frac{r_f \left(2
   B_2-2 c_3 r \left[\lambda+4 i r \omega \right)+i c_0 \lambda \omega +c_2 (2+6 i r \omega )\right]}{2 r^3}\,,
\end{align}
%\end{widetext}
%%%%
where $\lambda:=l(l+1)$ and $B_1$ and $B_2$ are two further integration constants. Similar to the previous cases, the validity of the perturbation theory requires $c_2=c_3=c_4=0$, otherwise the functions $G_{(g)}^{lm}$, $G_{(f)}^{lm}$, $\eta_{0(g)}^{lm}$ and $\eta_{0(f)}^{lm}$ would grow linearly with $r$ at large distances. 
Note that Eqs.~\eqref{polarg} and \eqref{polarf} are decoupled from each other and, in the GR limit\footnote{The source terms vanish when ${\cal A}=0$ and when the integration constants $c_i$ and $B_i$ are set to zero. In the GR case, this choice can be done without loss of generality.}, they reduce to two copies of the same homogeneous differential equation. The latter is not in the standard Zerilli form~\cite{Zerilli:1970se} but, quite interestingly, is precisely the Bardeen-Press-Teukolsky equation for gravitational perturbations of the Schwarzschild metric in GR~\cite{Bardeen:1973xb,Teukolsky:1972my,Teukolsky:1973ha} (cf. Eq.~(5.2) in Ref.~\cite{Teukolsky:1973ha} when the black-hole spin is zero. Compare also Eq.~\eqref{teu_radial} in the Appendix with Eq.~\eqref{polarg_Teu} below). It is easy to check that this equation is isospectral to the Regge-Wheeler equation by performing a Chandrasekhar transformation~\cite{Chandra:1975xx} (see discussion in Sec.~\ref{app:Teu}). We have also checked this property numerically by transforming the homogeneous equations into a 4-term recurrence relation and by computing the modes through continued fractions~\cite{Berti:2009kk,Pani:2013pma}.

As for the $l\geq2$ axial case that we discuss below, the source terms $s_g$ and $s_f$ in Eqs.~\eqref{polarg} and \eqref{polarf} do not alter the QNM spectrum, in agreement with the generic argument presented in Sec.~\ref{sec:QNMs}. The situation is therefore rather different from that of the bidiagonal solution. In the latter case, the $l\geq2$ polar perturbations reduce to a set of three coupled ordinary-differential equations, Eqs.~\eqref{polar_eq1}~--~\eqref{polar_eq3}, which propagate three degrees of freedom and correspond to a quasinormal spectrum that depends on the graviton mass.
%%%%

%%%%%%%%%%%%%%%%%%%%%%%%%%%%%%%%%
\subsection{Axial sector}
%%%%%%%%%%%%%%%%%%%%%%%%%%%%%%%%%
The axial sector does not contain a monopole ($l=0$) and one is left with the axial dipole mode ($l=1$) and with the higher multipoles $l\geq2$, which we treat separately.

%%%%%%%%%%%%%%%%%%%%%%%%%%%%%%%%%
\subsubsection{Axial dipole mode}
%%%%%%%%%%%%%%%%%%%%%%%%%%%%%%%%%

When $l=1$, the angular functions $W_{lm}$ and $X_{lm}$ in Eq.~\eqref{oddpart} vanish (and therefore $h^{lm}_{2}$ is not defined), while $h_{1(f)}=c_1/r+h_{1(g)}$ from Eq.~\eqref{constnbd2}. The $(v,\theta)$ component of the field equations~\eqref{perteqs} yields
\beq
r^2 h''_{0(g)}&=&2 h_{0(g)}-i r \omega \left(r h'_{1(g)}+2 h_{1(g)}\right)\,, \label{eqs_axialdi1}\\
r^2 h''_{0(f)}&=&2 h_{0(f)}-i r \omega \left(r h'_{1(g)}+2 h_{1(g)}+c_1/r\right)\,.\label{eqs_axialdi2}
\eeq
The residual gauge freedom can be used to set one of the axial functions to zero. If we impose $h_{1(g)}=0$, from the constraints~\eqref{constnbd1} and \eqref{constnbd2} we obtain $h_{1(f)}=c_1/r$ and in such case Eqs.~\eqref{eqs_axialdi1} and \eqref{eqs_axialdi2} can be solved for
\begin{align}
&h_{0(g)}=r^2 c_3+\frac{i(r_g-r_f) c_1\mathcal{A}M_v^4}{12 M_g^2 \omega  r}\,,\\
&h_{0(f)}=r^2 c_4+\frac{i(r_f-r_g) c_1\mathcal{A}M_v^4}{12 M_f^2C^2\omega r} +\frac{i c_1\omega}{2}\,,\label{h0f}
\end{align}
where $c_3$ and $c_4$ are two further integration constants. 
The constants of integration $c_3$ and $c_4$ must be set to zero, otherwise the perturbative approach breaks down at large $r$.
On the other hand, $c_1\neq 0$ does not violate our (relaxed) assumptions on the metrics: indeed both metrics are asymptotically flat, as can be checked by computing curvature invariants. In particular, the Pontryagin density ${}^{*}\!RR:=\frac{1}{2}\epsilon^{abef}R_{abcd}R^{cd}{}_{ef}\sim c_1/r^7$. 
Note that $f_{\mu\nu}$ is asymptotically flat but not Minkowski in this case, due to the last term in Eq.~\eqref{h0f}. 

This solution is qualitatively similar to the $l=0$ polar case.
As in the $l=0$ polar case, the above solution represents a purely ingoing wave which does not feel any effective potential and, therefore, it is not backscattered by the geometry. 

Thus, the axial dipolar perturbation of the nonbidiagonal black-hole solution describes a dynamical (purely ingoing) wave. As such, this solution cannot be an eigenfunction of the boundary-value problem, in agreement with our previous analysis which showed that no dipolar QNMs exist for this solution. Nonetheless, this behavior is dramatically different from the case of GR --~in which the $l=1$ mode is pure gauge and therefore nondynamical~-- and also from the case of the bidiagonal solution. In the latter case, the dipolar axial sector is described by a single second-order Regge-Wheeler-like equation~\eqref{oddl1}. As we showed in Chapter~\ref{chapter:massive2}, in this case, the eigenvalue problem admits a novel set of QNMs.

%%%%%%%%%%%%%%%%%%%%%%%%%%%%%%%%%
\subsubsection{Axial perturbations with $l\geq 2$}
%%%%%%%%%%%%%%%%%%%%%%%%%%%%%%%%%

In this case, to simplify the equations, we define two new radial functions given by 
\beq
{\tilde Q}_g=r^3\left(\frac{h^{lm}_{0(g)}}{r^2}\right)'+i\omega r h^{lm}_{1(g)}\,, \label{Qg}\\
{\tilde Q}_f=r^3\left(\frac{h^{lm}_{0(f)}}{r^2}\right)'+i\omega r h^{lm}_{1(f)}\,.
\eeq
From the $(v,\theta)$ component of the field equations~\eqref{perteqs}, we can then obtain two algebraic equations for $h^{lm}_{2(g)}$ and $h^{lm}_{2(f)}$, which allow us to eliminate these functions from the other equations. From the $(r,\theta)$ components, we get two second-order differential equations for ${\tilde Q}_g$ and ${\tilde Q}_f$, namely 
\beq
{\cal D}_g [{\tilde Q}_g]-V_g {\tilde Q}_g &=&\frac{c_1 M_v^4 \mathcal{A}}{2 M_g^2 r^2}\left(r_g-r_f\right)\left(1-\frac{r_g}{r}\right),\label{axialg}\\
{\cal D}_f [{\tilde Q}_f]-V_f {\tilde Q}_f &=&\frac{c_1 M_v^4 \mathcal{A}}{2 M_f^2  r^2 C^2}\left(r_f-r_g\right)\left(1-\frac{r_f}{r}\right), \label{axialf}
\eeq
where the effective potentials read
\beq
V_g&=&\left(1-\frac{r_g}{r}\right)\left[\frac{l(l+1)}{r^2}-\frac{3r_g}{r^3}\right]\,, \label{Vg}\\
V_f&=&\left(1-\frac{r_f}{r}\right)\left[\frac{l(l+1)}{r^2}-\frac{3r_f}{r^3}\right]\,. \label{Vf}
\eeq
Note that the field equations allow us to compute only the master functions $\tilde Q_g$ and $\tilde Q_f$ and not the metric perturbations $h_{0(g)}$, $h_{1(g)}$ and $h_{0(f)}$, $h_{1(f)}$  separately. This is consistent with the existence of a residual gauge freedom. For example, the function $h_{1(g)}$ in Eq.~\eqref{Qg} can be set to zero through a gauge choice. In such a case, $h_{1(f)}=c_1/r$ from Eq.~\eqref{constnbd2}.

Note that, when $c_1=0$, Eqs.~\eqref{axialg} and \eqref{axialf} reduce to a pair of Regge-Wheeler equations~\cite{Regge:1957td}, and are thus identical to the case of GR, consistent with our general argument in Sec.~\ref{sec:QNMs}.
On the other hand, the terms proportional to $c_1$ act as a source of the Regge-Wheeler equation and cannot modify the proper frequencies of the system. This is discussed in more detail in Sec.~\ref{sec:time} below.

Also in this case it is interesting to compare the perturbations of the nonbidiagonal solutions with those of a Schwarzschild BH in GR and with those of the bidiagonal solution of massive gravity. In the former case, the perturbation describes a single propagating degree of freedom governed by Eq.~\eqref{axialg} with $c_1=0$. In the latter case, the $l\geq2$ axial sector is described by two propagating degrees of freedom, but they are governed by a \emph{coupled} system of equations, Eqs.~\eqref{axial_bi1} and~\eqref{axial_bi2}, which are also associated with a different set of quasinormal frequencies. Finally, the perturbation equations in the bidiagonal case depend on the graviton mass, similar to the $l=1$ case previously discussed, whereas the graviton mass in the nonbidiagonal case appears only in the source terms, but not in the effective potentials~\eqref{Vg} and \eqref{Vf} (the same property holds true in the $l\geq2$ polar case discussed above).

%%%%%%%%%%%%%%%%%%%%%%%%%%%%%%%%%%%%%%%%%%%%%%%%%%%%%%%%%%%%%%%%%%%%%%
\subsection{Isospectrality of QNMs}\label{app:Teu}
%%%%%%%%%%%%%%%%%%%%%%%%%%%%%%%%%%%%%%%%%%%%%%%%%%%%%%%%%%%%%%%%%%%%%%

One can easily prove that the $l\geq 2$ polar QNMs are isopectral to the $l \geq 2$ axial QNMs by showing that they are governed by the same equations.
Consider Eqs.~\eqref{polarg} and~\eqref{polarf}. By defining the radial functions ${\tilde Y}_g:=e^{-i\omega r_{*g}} H_{0(g)}^{lm} /(r-r_g)^2$ and ${\tilde Y}_f:=e^{-i\omega r_{*f}} H_{0(f)}^{lm} /(r-r_f)^2$ we find, after some algebra, the following equations:
%%%%
%
\begin{align}
{\cal L}_g [{\tilde Y}_g] &= e^{-i\omega r_{*g}}T_g\,,\label{polarg_Teu}\\
{\cal L}_f [{\tilde Y}_f] &= e^{-i\omega r_{*f}}T_f\,, \label{polarf_Teu}
\end{align}
where we defined the differential operators
%%%
\begin{align}
{\cal L}_{g}&=\left(r^2-r_g r\right)\frac{d^2}{dr^2}+6\left(r-\frac{r_g}{2}\right)\frac{d}{dr}
+\frac{r^4\omega^2-4ir^2\omega\left(r-\frac{r_g}{2}\right)}{\left(r^2-r_g r\right)}
+8ir\omega-l(l+1)+6\,,\nonumber\\
{\cal L}_{f}&=\left(r^2-r_f r\right)\frac{d^2}{dr^2}+6\left(r-\frac{r_f}{2}\right)\frac{d}{dr}
+\frac{r^4\omega^2-4ir^2\omega\left(r-\frac{r_f}{2}\right)}{\left(r^2-r_f r\right)}
+8ir\omega-l(l+1)+6\,,
\end{align}
%%%%
whereas the source terms are
%
%\begin{widetext}
 \beq
&&T_g=\frac{c_0{\cal A}\lambda M_v^4 (r_g-r_f)}{4 M_g^2 r (r-r_g)^2}+\frac{c_2 {\cal A} M_v^4 (r_g-r_f)}{2 M_g^2 (r-r_g)^2}
+c_4\frac{\lambda r_g+2 i r \omega  \left(\left(\lambda-2\right) r+2 r_g\right)}{r (r-r_g)^2}\nn\\
&&-B_1\frac{r_g+2 i r^2 \omega }{r^2 (r-r_g)^2}\,,\\
   %%%%
&&T_f= \frac{c_0\lambda \left({\cal A} M_v^4 r (r_f-r_g)+2 C^2 M_f^2 \omega  \left(2 r^2 \omega -i r_f\right)\right)}{4 C^2
   M_f^2 r^2 (r-r_f)^2}+c_3\frac{\lambda r_f+2 i r \omega  \left(\left(\lambda-2\right) r+2 r_f\right)}{r (r-r_f)^2}\nonumber\\
&&+c_2\frac{{\cal A}  M_v^4 r^2 (r_f-r_g)+2 C^2 M_f^2 \left(2 r^3 \omega ^2-r_f (1+3 i r \omega )\right)}{2 C^2 M_f^2 r^2
   (r-r_f)^2}
-B_2\frac{r_f+2 i r^2 \omega }{r^2 (r-r_f)^2}\,.
\eeq
%
%\end{widetext}
%%%%
In the GR limit ($T_{g,f}\to 0$), these two equations reduce to two copies of the Bardeen-Press-Teukolsky equation (in the form originally written by them) for gravitational perturbations of the Schwarzschild metric in GR~\cite{Bardeen:1973xb,Teukolsky:1972my,Teukolsky:1973ha}. By performing a Chandrasekhar transformation~\cite{Chandra:1975xx} of the form (as given in Ref.~\cite{Hughes:2000pf}):
\beq
r^2{\tilde Y}_g&=&{\cal D}^2_{-g}\left(r \tilde{X}_g\right)\label{chandra_g}\,,\\
r^2{\tilde Y}_f&=&{\cal D}^2_{-f}\left(r \tilde{X}_f\right)\label{chandra_f}\,,
\eeq
where ${\cal D}_{-g,f}\equiv d/dr-i r \omega/(r-r_{g,f})$, one finds that the functions $\tilde{X}_{g,f}$ satisfy the following Regge-Wheeler equations:
\beq
&&\frac{d^2 \tilde{X}_g}{dr_{g*}^2}+\left(\omega^2-V_g\right)  \tilde{X}_g=S^P_{g}\,,\label{polarXg}\\
&&\frac{d^2 \tilde{X}_f}{dr_{f*}^2}+\left(\omega^2-V_f\right)  \tilde{X}_f=S^P_{f}\,,\label{polarXf}
\eeq
with the effective potentials given in Eqs.~\eqref{Vg} and~\eqref{Vf}, whereas the source terms can be obtained by inserting the transformations~\eqref{chandra_g} and~\eqref{chandra_f} in Eqs.~\eqref{polarg_Teu} and~\eqref{polarf_Teu}, using~\eqref{polarXg} and~\eqref{polarXf} to eliminate $X_{g,f}$ and their derivatives, and then solving for $S^P_{g,f}$, which can be found analytically. Since their analytical expression is rather lengthy and their explicit form is not fundamental we do not show it here. Similar to the axial case, to the leading order the source terms decay as $S^P_{g,f}\sim 1/r^2$ when $r\to\infty$ [cf. Eq.~\eqref{source_evolution}]. 
Comparing this with the axial case, Eq.~\eqref{axialZg0}, one immediately sees that the only difference is in the source term, and thus under the same boundary conditions, the QNM spectrum of the polar and axial sector is the same (and coincides with that of a GR Schwarzschild BH).

%%%%%%%%%%%%%%%%%%%%%%%%%%%%%%%%%
\section{Time evolution} \label{sec:time}
%%%%%%%%%%%%%%%%%%%%%%%%%%%%%%%%%
In this section we consider the time evolution governed by the perturbation equation~\eqref{axialg} [or, equivalently, Eq.~\eqref{axialf}] in the time domain, in order to investigate the role of the source term appearing on the right-hand side of this equation. A similar analysis for the $l\geq2$ polar sector is more technically involved but it is qualitatively similar. (As shown in the previous Section, for the polar case the only difference is in the source term. One can show that although the sources are more complicated, their asymptotic behavior at the horizon and at infinity is similar to the axial case, and thus the waveforms are qualitatively similar.)

By introducing a new radial function $\tilde{Z}_g=e^{-i\omega r_{*g}}{\tilde Q}_g$, Eq.~\eqref{axialg} becomes
\be
\frac{d^2 \tilde{Z}_g}{dr_{g*}^2}+\left(\omega^2-V_g\right)  \tilde{Z}_g=\left(1-\frac{r_g}{r}\right)S_g\,,\label{axialZg0}\\
\ee
where 
\be\label{source_evolution}
S_g=e^{-i\omega r_{g*}}c_1\left(r_g-r_f\right)\frac{M_v^4 \mathcal{A}}{2 M_g^2 r^2}\,.
\ee

As previously discussed, the above source term appears in the perturbed nonbidiagonal solution and it would vanish in the case of GR. To investigate the impact of such a term on the waveform, we assume that the latter is produced by a driving force at $t=0$, which for simplicity we take to be a static Gaussian. In the frequency-domain this amounts to adding a source to the right-hand side of Eq.~\eqref{axialZg0}, namely
\be\label{source_evolution_2}
S_{\rm{Gaussian}}=A_0 e^{-\left(r_{*g}-r_0\right)^2/2\sigma^2}\,.
\ee

Thus, the full time-evolution equation reads
\be
\frac{d^2 \tilde{Z}_g}{dr_{g*}^2}+\left(\omega^2-V_g\right)  \tilde{Z}_g=\left(1-\frac{r_g}{r}\right)S\,,\label{axialZg}\\
\ee
where $S:=S_g+(1-r_g/r)^{-1}S_{\rm{Gaussian}}$.
To obtain the waveform $Z_g(t,r)$\footnote{Note that due to Eq.~\eqref{decom} (with $t\to v$) and the definition $\tilde{Z}_g=e^{-i\omega r_{*g}}{\tilde Q}_g$, $Z_g(t,r)$ is a function of $t:= v-r_{*g}$.} we use an inverse-Fourier transform,
\be\label{waveform_td}
Z_g(t,r)=\frac{1}{\sqrt{2\pi}}\int_{-\infty}^{+\infty} e^{-i\omega t}\tilde{Z}_g(\omega,r)d\omega\,,
\ee
where $\tilde{Z}_g(\omega,r)$ is computed using the Green's function technique outlined in Appendix~\ref{app:GF}. In principle, $c_1$ is an arbitrary function of $\omega$ which depends on the initial conditions of the perturbations $h_1$. For simplicity, here we consider the case where $c_1$ is a constant, which is sufficient for our argument. For this choice, in the time domain, the source~\eqref{source_evolution} is proportional to the Dirac delta function $\delta(v)$.

\begin{figure}[th]
\begin{center}
%\begin{tabular}{c}
\epsfig{file=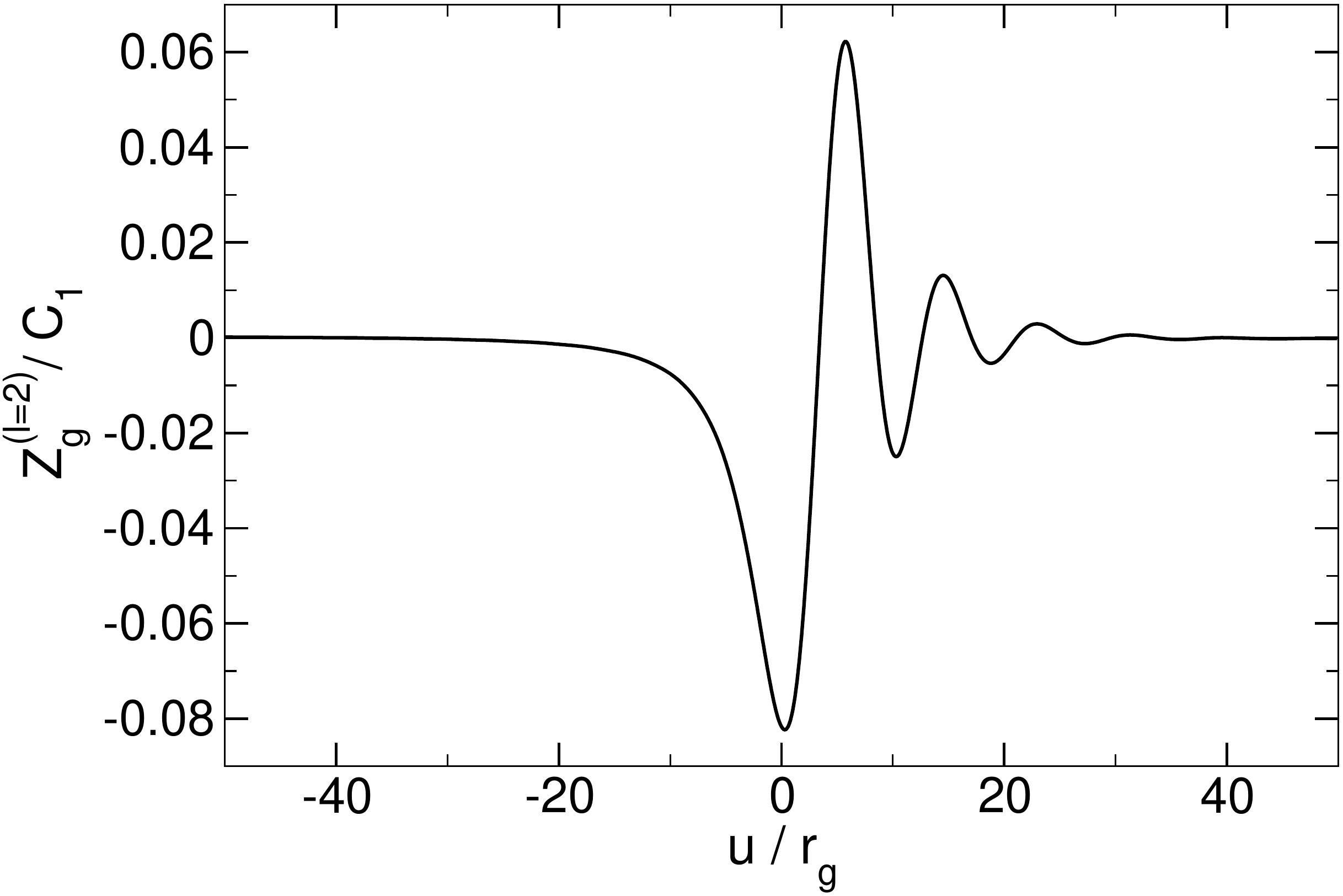,width=10cm,angle=0,clip=true}
%\end{tabular}
\end{center}
\caption{\label{fig:waveform} The waveform $Z_g(t,r)$ (in units of $C_1:=c_1\left(r_g-r_f\right)\frac{M_v^4 \mathcal{A}}{2 M_g^2}$) with $l=2$ as a function of $u:= t-r_{*g}$ (in units of $r_g$). This is the solution of Eq.~\eqref{axialZg0}, i.e. in the case where the external perturbation $S_{\rm{Gaussian}}$ is absent. It is easy to check that the ringdown signal is governed by the QNMs of a Schwarzschild BH in GR.
}
\end{figure}

Let us first consider the case in which no external source is present, i.e. we solve Eq.~\eqref{axialZg} with $A_0=0$ [or, equivalently, Eq.~\eqref{axialZg0}]. In this case the waveform is proportional to the combination $C_1:=\frac{c_1 M_v^4 \mathcal{A} }{2 M_g^2}\left(r_g-r_f\right)$. The waveform obtained with the Green's function method is shown in Fig.~\ref{fig:waveform}\footnote{In units where $G=c=1$, the constant $C_1:= c_1\left(r_g-r_f\right)M_v^4 \mathcal{A}/(2 M_g^2)$ is dimensionless.}. A straightforward Fourier analysis of the waveform shows that the ringdown signal~\cite{Berti:2009kk} is governed precisely by the QNMs of a Schwarzschild \textbf{BH} in GR. This is natural since Eq.~\eqref{axialZg0} is equivalent to the standard Regge-Wheeler equation in GR but with an external source term given by $S_g$. As in the case of a forced harmonic oscillator, the source can modify the waveform but not the proper modes of the system (for a similar analysis in a different modified theory of gravity, see Ref.~\cite{Molina:2010fb}), which are still described by the QNMs of the solution, i.e. by the same QNMs of a Schwarzschild BH in GR.

\begin{figure}[th]
\begin{center}
%\begin{tabular}{c}
\epsfig{file=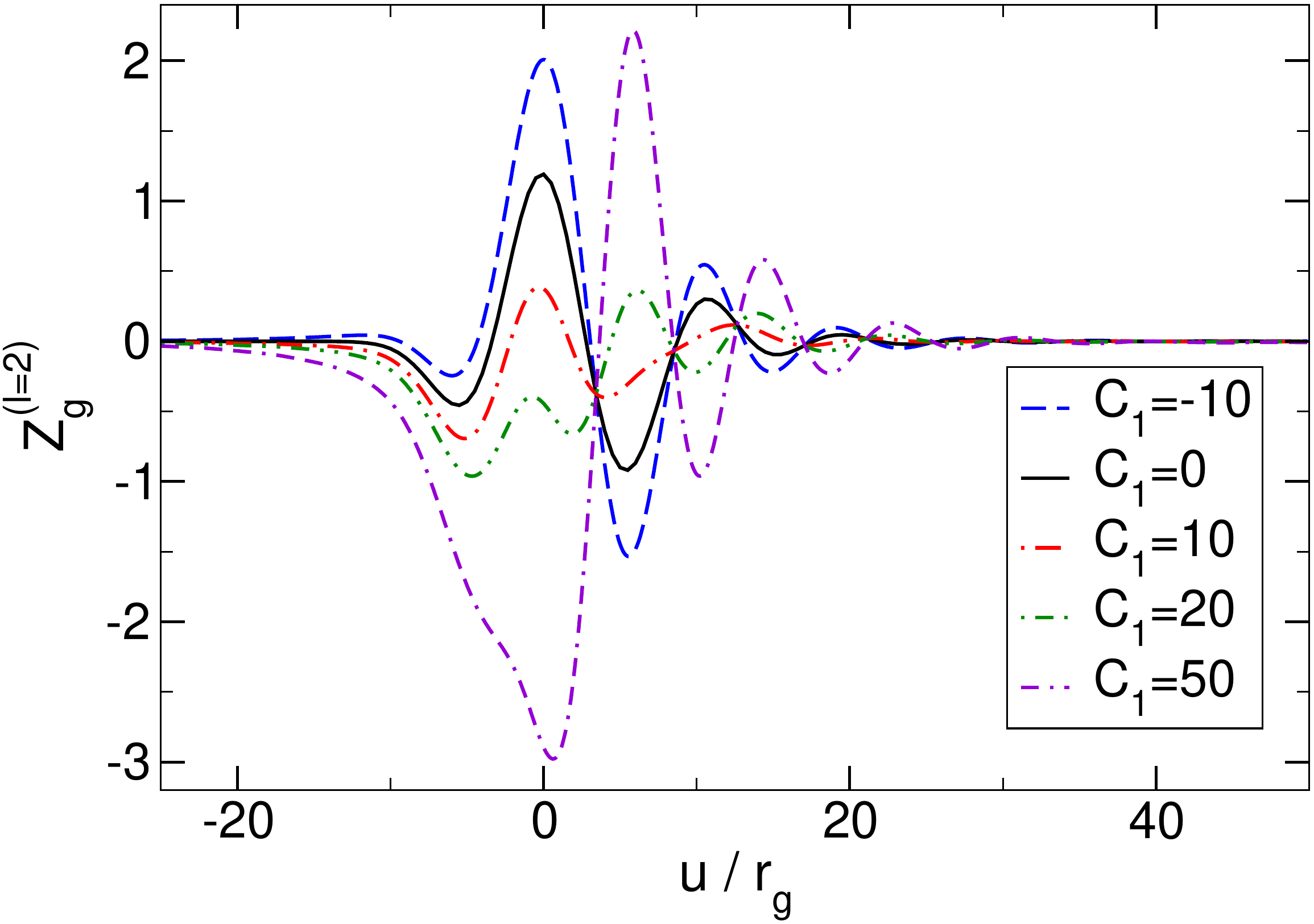,width=10cm,angle=0,clip=true}
%\end{tabular}
\end{center}
\caption{\label{fig:waveform_Gaussian} Waveform $Z_g(t,r)$ for $l=2$ as a function of $u:= t-r_{*g}$ and for different values of $C_1$. This is a solution of the full time-evolution equation~\eqref{axialZg} with an external Gaussian source~\eqref{source_evolution_2}, with  $A_0 r_g^2=0.4$, $r_0=5r_g$ and $\sigma=2.5 r_g$. From this waveform it is easy to check that, for any value of $C_1$, the ringdown signal is governed by the QNMs of a Schwarzschild BH in GR. }
\end{figure}

As is clear from the above discussion, adding an external source term like Eq.~\eqref{source_evolution_2} is simply equivalent to solving the standard Regge-Wheeler equation in GR but with an effective source term given by $S$ in Eq.~\eqref{axialZg}. The waveform obtained by solving Eq.~\eqref{axialZg} for different values of $C_1$ and for a representative external source term is shown in Fig.~\ref{fig:waveform_Gaussian}. Also in this case a straightforward frequency decomposition shows that, for any value of $C_1$, the ringdown waveform is governed by the QNMs of the Schwarzschild solution in GR, although the black-hole response to the external perturbation depends on $C_1$. This is in agreement with our proof given in Sec.~\ref{sec:QNMs}.

% 

%%%%%%%%%%%%%%%%%%%%%%%%%%%%%%%%%
\section{Conclusion and discussion}\label{sec:conclusion}
%%%%%%%%%%%%%%%%%%%%%%%%%%%%%%%%%
We derived the full set of linearized equations governing gravitational perturbations of the nonbidiagonal Schwarzschild solution in massive (bi)gravity. We showed that the quasinormal spectrum of these solutions coincides with that of a Schwarzschild BH in GR. This result is quite surprising and has some interesting consequences. In general, massive (bi)gravity propagates more degrees of freedom than GR (including massive modes), so one might naively expect that black-hole solutions possess more modes of vibration and that the latter would depend on the value of the graviton mass. This is indeed the case for bidiagonal solutions~\cite{Babichev:2013una,Brito:2013wya} (see Chapter~\ref{chapter:massive2}), but it is not the case for the nonbidiagonal solutions discussed here. 

Furthermore, as shown in Chapter~\ref{chapter:massive2}, the bidiagonal solution possesses an unstable radial mode, which is absent in the nonbidiagonal case~\cite{Babichev:2014oua}\footnote{As we will discuss in next Chapter, due to the instability of the bidiagonal solutions along with the existence of several other spherically symmetric solutions~\cite{Volkov:2012wp,Brito:2013xaa,Babichev:2015xha}, the outcome of gravitational collapse in massive gravity is still unclear (see also Ref.~\cite{Enander:2015kda} for arguments showing that gravitational collapse of stars might not lead to black-hole formation in these theories).}. Finally, as we showed in Chapter~\ref{chapter:massive2}, massive bosonic perturbations generically allow for quasi-bound, long-lived modes in the spectrum of spherically-symmetric BHs. We will show in Chapter~\ref{chapter:Kerr}, that such modes can turn (superradiantly) unstable when the BH rotates above a certain threshold~\cite{Brito:2015oca}. Remarkably, such long-lived modes are absent in the static nonbidiagonal solution. This suggests that, when spinning, this solution does not suffer from the superradiant instability (see Chapter~\ref{chapter:Kerr} for details). 
%If this conjecture is confirmed, the nonbidiagonal black-hole solution of massive gravity would be the first case of a spinning black-hole geometry which is mode-stable in a theory that propagates a massive bosonic field (see Chapter~\ref{chapter:Kerr} for details).

It is also natural to conjecture that the QNMs of the rotating BH found in Ref.~\cite{Babichev:2014tfa} are identical to those of a Kerr BH in GR, similar to the static case discussed in this Chapter.
 
The absence of extra QNMs, compared to GR, naturally raises the question about the number of propagating degrees of freedom for perturbations of nonbidiagonal BHs. In particular, one may worry about the disappearance of some degrees of freedom and, consequently, possible strong coupling.
Indeed, some modes are indeed absent, as compared to, e.g. bidiagonal BHs. 
At the same time, there are modes which do not feel any potential and therefore do not backscatter. 
The absence of backscattering implies that these modes do not satisfy the boundary conditions imposed for QNMs.
Nevertheless these ``free propagating'' modes depend on the initial conditions and their impact on the resulting metric cannot be removed by a gauge transformation.
We would like to stress here that these perturbations contain free functions, as ``normal'' propagating modes do, 
and initial conditions are required to impose them.
Indeed, each integration constant, e.g. $c_0$ and $c_2$  in Sec.~\ref{ssec:dipole}  are functions of $\omega$ and 
when converted to the time domain they yield free functions. 
Note that, on the contrary, in the special case $\beta_2=-C \beta_3$ studied in Refs.~\cite{Kodama:2013rea,Kobayashi:2015yda} the ``free propagating'' modes are absent, 
so the solution is certainly strongly coupled in this specific case.
In the general case, however, a separate study is required to find explicitly whether all the modes are truly dynamical, and hence to address the issue of possible strong coupling. 

It would also be interesting to go beyond the linear level, employed in this Chapter, and to consider nonlinear effects. 
This question is connected to the possible strong coupling issue. If some of the degrees of freedom happen not to propagate on the background of nonbidiagonal black-hole solutions (due to their peculiarity) one would naturally expect that at least some of them reappear at the nonlinear level. 
If this is indeed the case, then the nonbidiagonal solutions may be nonlinearly unstable. 
Nonlinear effects may change our discussion in Section~\ref{ssec:monopole}, where we argued that one of the two integration constants is a pure gauge, since it can be reabsorbed in the perturbations $f_{\mu\nu}$ and it does not give a contribution to the mass energy-momentum tensor. This constant might source physical perturbations of $g_{\mu\nu}$ at the nonlinear level, thus activating a physical degree of freedom. Nonlinear effects may also generate a potential for those modes which propagate from infinity down to the horizon without scattering. 

We should also mention that our study did not address the question of ghosts in the spectrum of perturbations. This issue may be addressed together with the question about the number of propagating degrees of freedom mentioned above, for example, by a Hamiltonian analysis. 

%%%%%%%%%%%%%%%%%%%%%%%%%%%%%%%%%%%%
\chapter{Black holes with massive graviton hair}\label{chapter:BHhair}
%%%%%%%%%%%%%%%%%%%%%%%%%%%%%%%%%%%%
%%%%%%%%%%%%%%%%%%%%%%%%%%%%%%%%%%%%
\section{Introduction}
%%%%%%%%%%%%%%%%%%%%%%%%%%%%%%%%%%%%
Schwarzschild BHs stand out among all possible solutions of GR as the only static,
asymptotically flat, regular solutions of vacuum Einstein equations. They are, in addition, stable solutions of the theory.
The Schwarzschild solution also solves many other field equations, such as some scalar-tensor theories,
$f(R)$ theories and Chern-Simons gravity (see e.g. Refs.~\cite{Psaltis:2007cw,Sotiriou:2011dz}). In fact, it is possible to show that the Schwarzschild solution is the only static, asymptotically flat, regular solution also in the vacuum of these theories\footnote{Note that generic scalar-tensor theories allow for the existence of regular static BH solutions~\cite{Babichev:2013cya,Sotiriou:2014pfa}. See also Ref.~\cite{Herdeiro:2015waa} for a recent review on the status of black-hole solutions with non-trivial scalar fields.}.

These uniqueness properties are in agreement with various ``no-hair'' proofs that Schwarzschild BHs cannot support minimally coupled
static regular scalar hair, nor fields mediating the weak or the strong interaction \cite{Bekenstein:1971hc,Bekenstein:1972ky}.

The case of spin-2 hair is much less clear. It was shown by Bekenstein that BHs cannot support massive spin-2 fields in theories with generic nonminimal couplings to curvature, at least
as long as the graviton mass is sufficiently large~\cite{Bekenstein:1972ky}. 
However, as proved by Aragone and Deser~\cite{Aragone:1971kh,Aragone:1979bm}, it is impossible to couple consistently a spin-2 field with a nonlinear gravitational theory. This result does not leave much room for BHs with spin-2 hair, unless the massive tensor field is itself the mediator of the gravitational interaction, i.e. in the case of massive theories of gravity~\cite{deRham:2010ik,deRham:2010kj,Hassan:2011hr}.

Even in the case of massive gravity, recent searches for nonlinear spherically symmetric solutions~\cite{Volkov:2012wp} seem to put a rest to the possibility of finding static, asymptotically flat BH solutions endowed with spin-2 hair.

On the other hand, the nonexistence of hairy BHs in massive gravities seems at odds with the findings, presented in Chapter~\ref{chapter:massive2}, that
bi-Schwarzschild BHs are unstable in generic theories with light massive spin-2 fields.
The instability is due to a propagating spherically symmetric degree of freedom and it is a long-wavelength instability. As shown in Fig.~\ref{fig:GL}, it only occurs for a nonvanishing mass coupling $\mu M_S\lesssim0.438$, with $\mu$ being the inverse of the Compton wavelength of the graviton and $M_S$ the mass of the background BH.
%%%%%

%%%%%%%%%%%%
Interestingly, for values of $M_S$ and $\mu$ that are phenomenologically relevant, the mass coupling $\mu M_S$ is always well within the instability region. Indeed, it is natural to consider the graviton mass of the order of the Hubble constant, $m_g=\hbar\mu\sim H\sim 10^{-33}{\mathrm{eV}}$, in order to account for an effective cosmological constant (see e.g. Ref.~\cite{Hinterbichler:2011tt}). This tiny value implies that a graviton with mass $m_g\sim H$ would trigger an instability for any Schwarzschild BH with mass smaller than $10^{22} M_\odot$! 
%%%%%%%%%%%
Even if the instability timescale $\tau$ can be extremely long ($\tau\sim1.43/\mu$ in the small-mass limit~\cite{Brito:2013wya}), as a matter of principle if Schwarzschild BHs are unstable in massive gravity, they must decay to {\it something} (or not even be formed in the first place) and, unless cosmic censorship is violated, the final state should be a spherically symmetric BH. 

This apparent conundrum prompts the following question: \emph{do spherically symmetric, asymptotically flat BH solutions surrounded by a graviton cloud exist in theories with a massive graviton?}
Here, we show that such solutions do indeed exist and were not found in the thorough analysis of Ref.~\cite{Volkov:2012wp} simply because they were not searched for explicitly.

%%%%%%%%%%%%%%%%%%%%%%%%%%%%%%%%%%%%
\section{Setup}
%%%%%%%%%%%%%%%%%%%%%%%%%%%%%%%%%%%%
Le us focus on the theory described by the action~\eqref{biaction}. The parameters $\beta_n$ are not all independent if flat space is to be a solution of the theory. They can be written in terms of two free parameters $\alpha_3$ and $\alpha_4$ defined as
\be
\beta_n=(-1)^{n+1}\left(\frac{1}{2}(3-n)(4-n)-(4-n)\alpha_3-\alpha_4\right)\,.
\ee
With this definition, and admitting asymptotic flatness, one obtains from~\eqref{cosmo_bigrav} and~\eqref{mass_g} that the graviton mass $\mu$ can be written in terms of the other parameters of the theory as
\be
\mu=\frac{M_v^2}{M_f} \sqrt{1+M_f^2/M_g^2}\,.
\ee

The Lagrangian~\eqref{biaction} gives rise to two sets of modified Einstein equations for $g_{\mu\nu}$ and $f_{\mu\nu}$, given by~\eqref{field_eqs1} and~\eqref{field_eqs2}. In addition it will be useful to also consider the conservation conditions given by Eqs.~\eqref{bianchi1}.

We consider static spherically symmetric solutions of Eqs.~\eqref{field_eqs1} and~\eqref{field_eqs2}. The most general ansatz for the metrics is given by\footnote{Note that, as discussed in Chapter~\ref{chapter:nonbi}, massive graviton theories also allow for spherically symmetric solutions whose metrics are not both diagonal in the same coordinates~\cite{Volkov:2012wp}. Since we are interested in the end state of the monopole instability found in Chapter~\ref{chapter:massive2} and discussed in Refs.~\cite{Babichev:2013una,Brito:2013wya}, we focus here on the ansatz~\eqref{ansatz_g}-\eqref{ansatz_f}.}
\beq
g_{\mu\nu}dx^{\mu}dx^{\nu}&=&-F^2\, dt^2 + B^{-2}\, dr^2 + R^2 d\Omega^2\,,\label{ansatz_g}\\
f_{\mu\nu}dx^{\mu}dx^{\nu}&=&-p^2\, dt^2 + b^{2}\, dr^2 + U^2 d\Omega^2\,,\label{ansatz_f}
\eeq
where $F\,,B\,,R\,,p\,,b$ and $U$ are radial functions. The gauge freedom allow us to reparametrize the radial coordinate $r$ such that $R(r)=r$. To simplify the equations we also introduce the radial function $Y(r)$ defined as $b=U'/Y$, where $'\equiv d/dr$. 

Inserting~\eqref{ansatz_g} and~\eqref{ansatz_f} into the equations of motion~\eqref{field_eqs1} and~\eqref{field_eqs2}, and using the conservation condition~\eqref{bianchi1}, we can reduce the problem to a system of three coupled first-order ordinary differential equations, which can be schematically written as (for a detailed derivation see~\cite{Volkov:2012wp})
\be
\left\{\begin{array}{l}
        B'=\mathcal{F}_1(r,B,Y,U,\mu,M_f,M_g,\alpha_3,\alpha_4)\\
	Y'=\mathcal{F}_2(r,B,Y,U,\mu,M_f,M_g,\alpha_3,\alpha_4)\\
	U'=\mathcal{F}_3(r,B,Y,U,\mu,M_f,M_g,\alpha_3,\alpha_4)
       \end{array}\right.\label{eqs_Volkov}\,.
\ee
The remaining two functions $F$ and $p$ can then be evaluated using
\beq
F^{-1}F'&=&\mathcal{F}_4(r,B,Y,U,\mu,M_f,M_g,\alpha_3,\alpha_4)\,,\label{eqs_Volkov_2a}\\
F^{-1}p&=&\mathcal{F}_5(r,B,Y,U,\mu,M_f,M_g,\alpha_3,\alpha_4)\,.\label{eqs_Volkov_2}
\eeq
The explicit form of the functions $\mathcal{F}_i$ is somewhat lengthy and not very instructive. The derivation of Eqs.~\eqref{eqs_Volkov}--\eqref{eqs_Volkov_2} and their final form is publicly available online in a {\scshape Mathematica} notebook~\cite{webpage}.

%%%%%%%%%%%%%%%%%%%%%%%%%%%%%%%%%%%%%%%%%%%%%%%
\subsection{Boundary conditions at the horizon}
%%%%%%%%%%%%%%%%%%%%%%%%%%%%%%%%%%%%%%%%%%%%%%%%
Since we are interested in BH solutions, we assume the existence of an event horizon at $r_H$, where $F(r_H)=B(r_H)=0$. From the discussion in~\cite{Deffayet:2011rh,Banados:2011hk} where it is shown that for the spacetime to be nonsingular the two metrics must share the same horizon, it follows that $Y$ and $p$ must also have a simple root at $r=r_H$. On the other hand, the function $U$ can have any finite value different from zero at the horizon. For numerical purposes we then assume a power-series expansion at the horizon,
\beq
B^2&=&\sum_{n\geq 1}a_n(r-r_H)^n, \quad Y^2=\sum_{n\geq 1}b_n(r-r_H)^n,\label{BCs1}\\
U&=&u_H\,r_H+\sum_{n\geq 1}c_n(r-r_H)^n\label{BCs3}\,.
\eeq
After inserting this into the system~\eqref{eqs_Volkov}, $a_n\,,b_n\,,c_n$ all can be expressed in terms of $u_H$ and $a_1$ only, where the constant $u_H$ is arbitrary while $a_1$ is given by the solution of a quadratic equation 
\be
\mathcal{A}a_1^2+\mathcal{B}a_1+\mathcal{C}=0\,,\label{eq:a1_sol}
\ee
where $\mathcal{A}\,,\mathcal{B}\,,\mathcal{C}$ are functions of $u_H,\,r_H,\,\mu,\,M_f,\,M_g$ and $\alpha_3,\,\alpha_4$. Since there are two solutions for this equation for each choice of the parameters, there exist two different branches of solutions for the metric functions. Moreover, reality of $a_1$ requires $\mathcal{B}^2>4\mathcal{A}\mathcal{C}$, and this condition restricts the parameter space.

Inserting~\eqref{BCs1}--\eqref{BCs3} into Eqs.~\eqref{eqs_Volkov_2a} and \eqref{eqs_Volkov_2}, we find
\beq
F^2&=&q^2(r-r_H)+q^2\sum_{n\geq 2}d_n(r-r_H)^n\,,\\
p^2&=&q^2\sum_{n\geq 1}e_n(r-r_H)^n\,,
\eeq
where $d_n$ and $e_n$ can be expressed in terms of $u_H$ and of the other parameters and $q$ is an integration constant, which can be set arbitrarily and is related to the time- scaling symmetry.

Equations~\eqref{eqs_Volkov} are invariant under the following transformations:   
\beq
B(r)&\to& B(\Omega r)\,,\quad  Y(r)\to Y(\Omega r)\,,\nn\\
U(r)&\to& \frac{1}{\Omega}U(\Omega r)\,,\quad \mu\to \frac{\mu}{\Omega}\,,
\eeq
where $\Omega$ is a constant. The parameter $u_H=U(r_H)/r_H$ remains invariant under the transformations above and the rescaling $r_H\to r_H/\Omega$. We use this rescaling to express all dimensionful quantities in terms of the mass of a Schwarzschild BH with horizon $r_H$, i.e. $M_S=r_H/2$.  We also consider without loss of generality $M_f=M_g$. 

%Another important quantity that can be used to check the validity of the solutions is the temperature of each horizon, which can be evaluated as~\cite{Volkov:2012wp} 
%
%\be
%T=T_g\equiv\frac{q\sqrt{a_1}}{4\pi}=T_f\equiv \frac{q\sqrt{b_1 e_1}}{4\pi c_1}\,.
%\ee
%
%These two temperatures can be shown to be the same for any value of the parameters, in agreement with the discussion of Ref.~\cite{Banados:2011hk}. To evaluate the temperature we fix the constant $q$ by requiring that $F(r)\to 1$ (or, equivalently, $p(r)\to 1$) as $r\to \infty$.

%%%%%%%%%%%%%%%%%%%%%%%%%%%%%%%%%%%%%%%%%%
\subsection{Asymptotically flat solutions}
%%%%%%%%%%%%%%%%%%%%%%%%%%%%%%%%%%%%%%%%%%
We require the solutions to be asymptotically flat such that as $r\to \infty$, we have $B=1+\delta B$, $Y=1+\delta Y$, $U=r+\delta U$, where the variations are taken to be small. Inserting this in the field equations~\eqref{eqs_Volkov}, we obtain to first order
\beq
\delta B&=&-\frac{C_1}{2r}+\frac{C_2(1+r\mu)}{2r}e^{-r\mu}\,,\label{inf1}\\
\delta Y&=&-\frac{C_1}{2r}-\frac{C_2(1+r\mu)}{2r}e^{-r\mu}\,,\label{inf2}\\
\delta U&=&\frac{C_2(1+r\mu+r^2\mu^2)}{\mu^2 r^2}e^{-r\mu}\,,\label{inf3}
\eeq
where $C_1$ and $C_2$ are integration constants.
Finally, we can find asymptotically flat solutions numerically using a shooting method. 
%%%%%%%%%%%%%%%%%%%%%%%%%%%%%%%%%%%%
\section{Results}
%%%%%%%%%%%%%%%%%%%%%%%%%%%%%%%%%%%%

%
\begin{figure}[htb]
\begin{center}
%\begin{tabular}{cc}
\epsfig{file=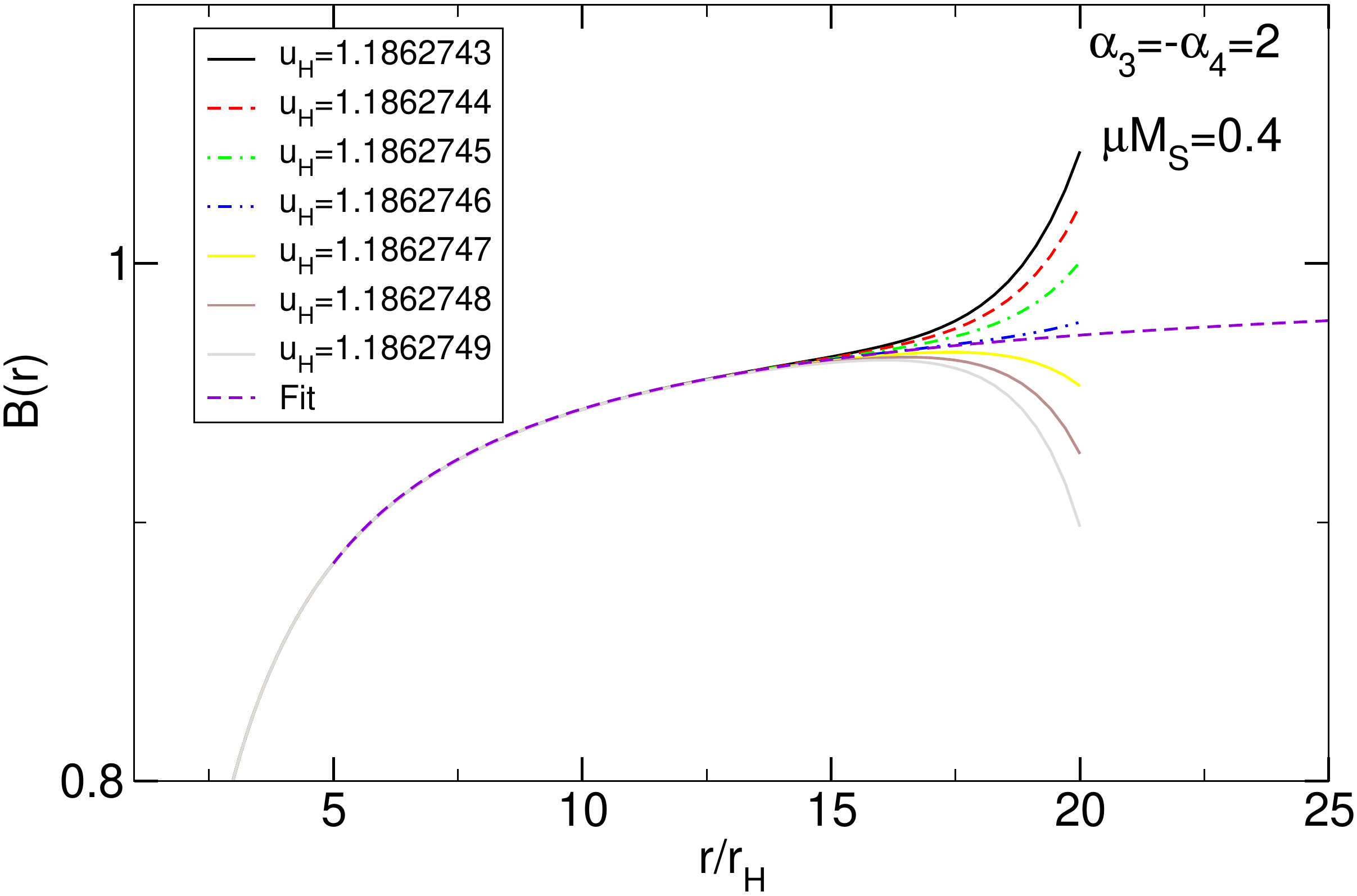,width=10cm,angle=0,clip=true}
%\epsfig{file=fit.eps,width=10cm,angle=0,clip=true}
%\end{tabular}
\caption{Metric function $B(r)$ for different values of $u_H$ close to an asymptotically flat solution, represented by the curve with the label \emph{Fit}. Any small deviation from the asymptotically flat solution, leads to a divergent behavior.\label{fig:fit}}
\end{center}
\end{figure}

\begin{figure}[htb]
\begin{center}
\begin{tabular}{cc}
\epsfig{file=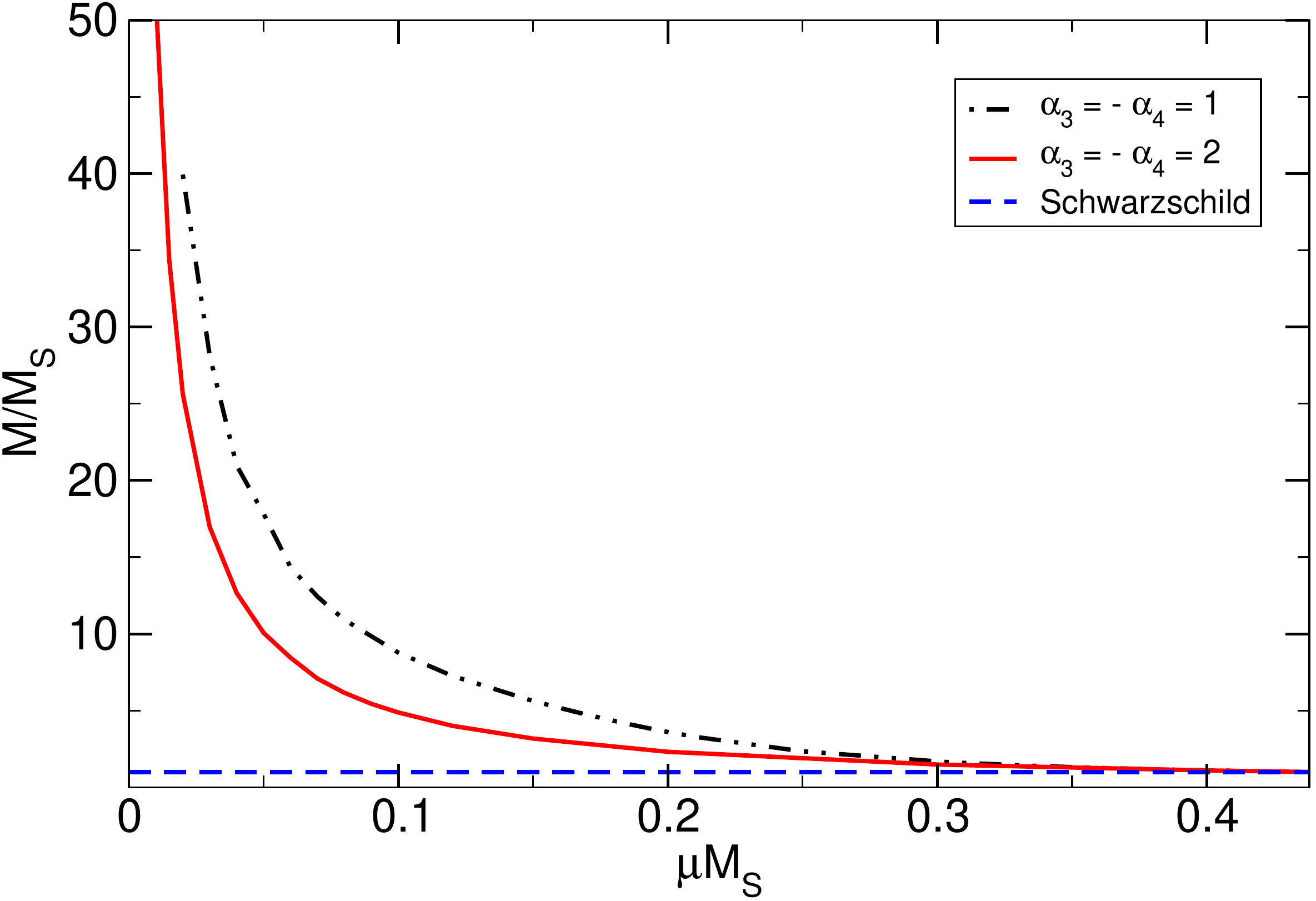,width=7.3cm,angle=0,clip=true}&
\epsfig{file=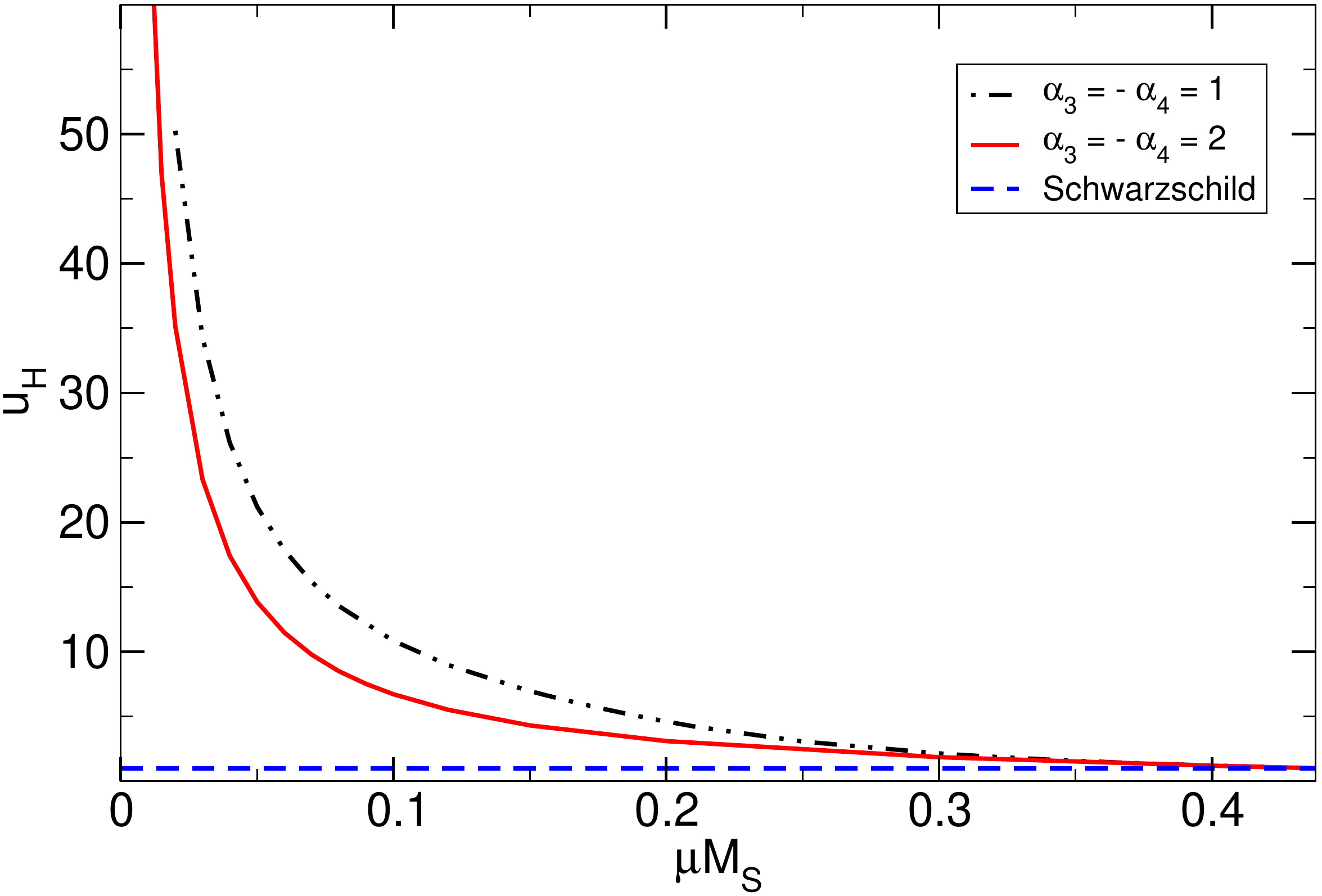,width=7.3cm,angle=0,clip=true}
\end{tabular}
\caption{Spacetime mass $M$ (left panel) and parameter $u_H$ (right panel) as functions of the graviton mass $\mu M_S$. Quantities are normalized by the mass of the corresponding Schwarzschild BH, $M_S$. For graviton masses close to the critical threshold $\mu M_S\sim 0.438$ the solutions merge smoothly with a Schwarzschild BH, as they should since the latter are marginally stable at this point (cf. Fig.~\ref{fig:GL}).  \label{fig:mass}}
\end{center}
\end{figure}
%%%%%%%%%%%%%%%%%%%%%%%%%%%%%%%%%%%%
For fixed values of $\mu,\,\alpha_3$ and $\alpha_4$, we integrate the system~\eqref{eqs_Volkov} starting from the horizon with the boundary conditions~\eqref{BCs1}--\eqref{BCs3}, towards large $r$ and find the values of the shooting parameter $u_H$ for which the solution matches the asymptotic behavior~\eqref{inf1}--\eqref{inf3}.
%%%
For each choice of $\mu,\,\alpha_3$ and $\alpha_4$, there are two branches of solutions, corresponding to the two roots of the quadratic equation~\eqref{eq:a1_sol}. In most cases only one of the branches will give an asymptotically flat solution. 

As expected, a trivial solution for any value of $\mu$, $\alpha_3$ and $\alpha_4$ is obtained when $u_H=1$, and it corresponds to the two metrics being equal and described by the Schwarzschild solution. However, we also find different solutions for which $u_H\neq 1$ and that correspond to regular, asymptotically flat BHs endowed with a nontrivial massive-graviton hair. We note that such solutions were not found in Ref.~\cite{Volkov:2012wp}, most likely because the free parameter $u_H$ was not adjusted in order to enforce asymptotic flatness.  In fact, as shown in Fig.~\ref{fig:fit}, in addition to the solution with asymptotic behavior~\eqref{inf1}--\eqref{inf3}, there is always another branch which diverges exponentially at spatial infinity. Thus, any small deviation from a regular solution leads to a singular behavior, making it numerically challenging to shoot for the correct solution.
 
The main results are summarized in Figs.~\ref{fig:mass}--\ref{fig:du}.
The first important result is that hairy solutions exist near the threshold $\mu M_S\lesssim 0.438$ for {\it any} value of $\alpha_3,\alpha_4$. This number precisely corresponds to the critical threshold for the Gregory-Laflamme instability~\cite{Gregory:1993vy} (cf. Fig.~\ref{fig:GL}). Solutions were expected to exist
close to this threshold and, in fact, this expectation has prompted our search at the nonlinear level.

We also find that for any value of $\alpha_3$ and $\alpha_4$, $M/M_S$ and $u_H$ are monotonically increasing functions of $(\mu M_S)^{-1}$ as shown in Fig.~\ref{fig:mass}. Here $M$ is the spacetime mass evaluated from the asymptotic behavior at infinity as $M=C_1/2$ (cf. Eqs.~\eqref{inf1}--\eqref{inf3}).

Above the threshold $\mu M_S\gtrsim 0.438$, the Schwarzschild solution is linearly stable. Consistent with the linear analysis, the only asymptotically flat solution that we were able to find in this region is the Schwarzschild solution, labeled by $u_H=1$ and $M=M_S$. 

%%%%%%%%%%%%%
\subsection{Parameter space}
%%%%%%%%%%%%%
The behavior at smaller $\mu M_S$ is more convoluted as it depends strongly on higher curvature terms: the nonlinear terms of the potential~\eqref{potential} become important and the solution differs substantially from the eigeinfunctions shown in Fig.~\ref{fig:GL}. Nevertheless, after an extensive analysis of the full two-dimensional parameter space $(\alpha_3\,,\alpha_4)$, we obtain the following classification: 
%
%\begin{enumerate}

(i) $\alpha_3\neq-\alpha_4 \lor \alpha_3=-\alpha_4\lesssim 1$ -- The solutions stop to exist below a cutoff $\mu_c M_S$, where the two branches of solutions near the horizon merge.

(ii) $1\lesssim\alpha_3=-\alpha_4\lesssim 2$ -- The solutions disappear only near $\mu M_S\sim 0.01$ and are singular at small $\mu M_S$, according to Refs.~\cite{Deffayet:2011rh,Banados:2011hk}, because some component of the metric $f_{\mu\nu}$ is vanishing where the metric $g_{\mu\nu}$ is regular (see Fig.~\ref{fig:du}).

%
%\item $3/2\lesssim\alpha_3=-\alpha_4\lesssim 2$ -- the solutions exist for any $\mu M_S$ but are singular at small $\mu M_S$ and there is no turning point ($T\to 0$ when $\mu M_S \to 0$, see right panel of Fig.~\ref{fig:temp});
(iii) $\alpha_3=-\alpha_4\gtrsim 2$ -- The solutions exist for arbitrarily small $\mu M_S$ and are nonsingular.
%
%\end{enumerate}
%%%%%%%%%%%%%%
%
\begin{figure}[htb]
\begin{center}
\begin{tabular}{c}
\epsfig{file=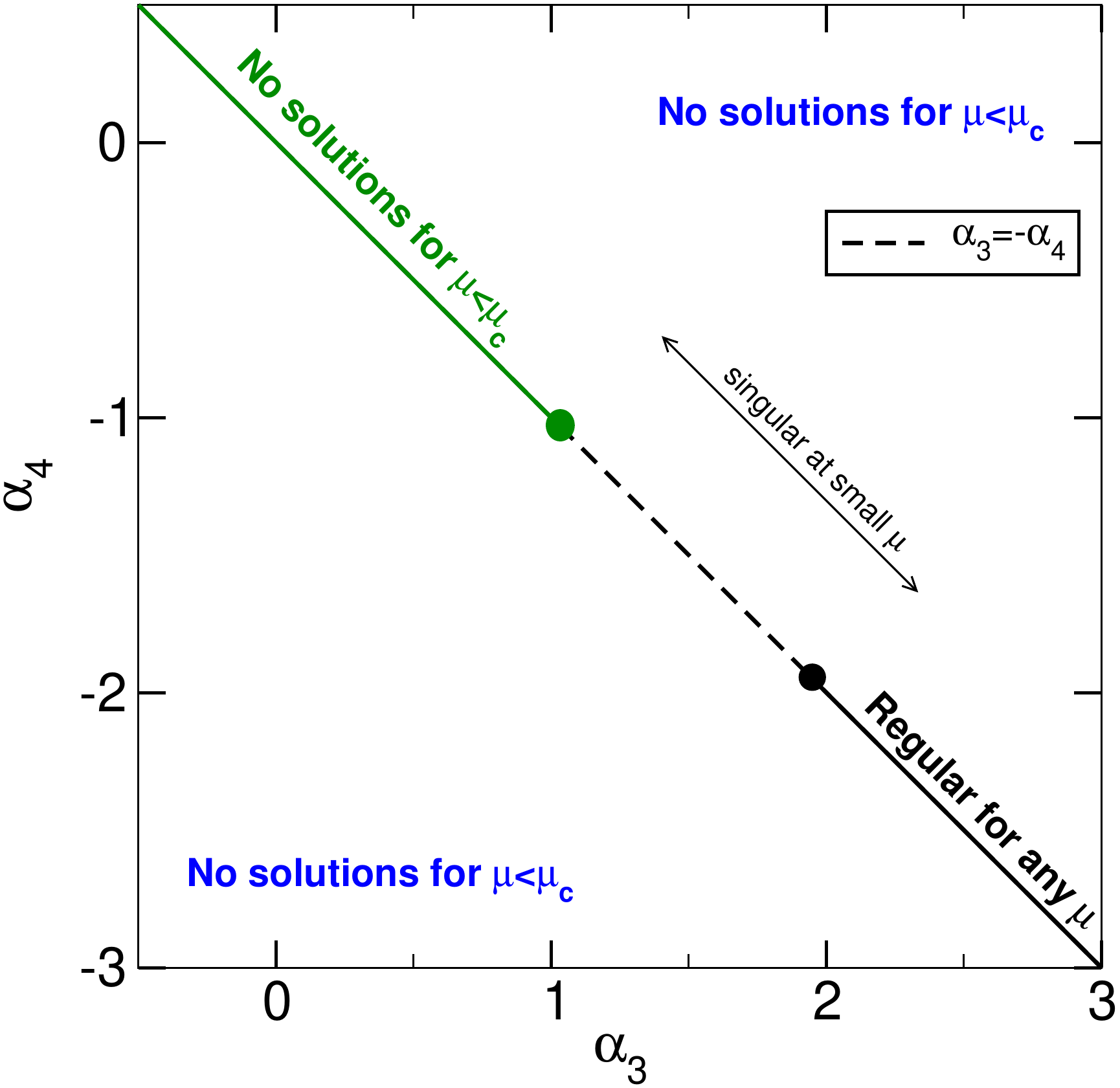,width=8cm,angle=0,clip=true}
\end{tabular}
\caption{Conjectured diagram of the parameter space for BHs with massive graviton hair in bimetric massive gravity. See main text for details.\label{fig:diagram}}
\end{center}
\end{figure}
This schematic classification of the parameter space is shown in Fig.~\ref{fig:diagram}.

It is important to emphasize that an analysis of the full parameter space is an extraordinary task. As such, it is extremely difficult to guarantee that the parameter space {\it is} divided as depicted in Fig.~\ref{fig:diagram}, as we cannot rule out certain choices of $(\alpha_3, \alpha_4)$ not belonging to the above classes. Also, the numerical integration becomes increasingly more challenging in the small-$\mu$ limit. We were able to obtain solutions for mass coupling as small as $\mu M_s\sim0.001$ and found no indication that, in the region $\alpha_3=-\alpha_4\gtrsim2$, such solutions cease to exist. However, our numerical procedure cannot guarantee that hairy BHs exist for arbitrarily small $\mu$.

\begin{figure}[htb]
\begin{center}
%\begin{tabular}{ccc}
%\epsfig{file=dU_vs_r.eps,width=6.9cm,angle=0,clip=true}&
%\epsfig{file=dU_vs_r_a3_1_a4_m1.eps,width=6.9cm,angle=0,clip=true}&
\epsfig{file=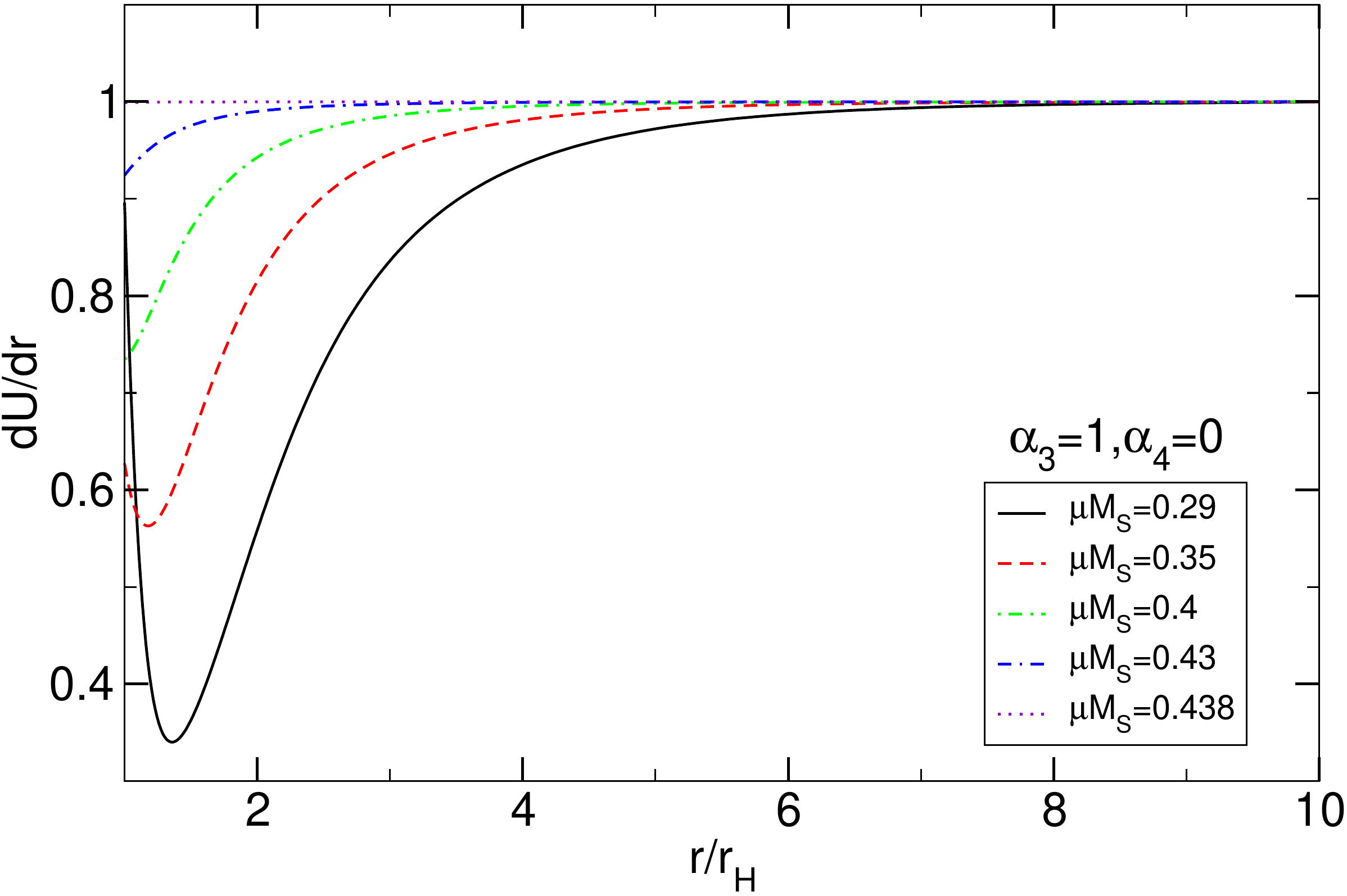,width=10cm,angle=0,clip=true}
%\end{tabular}
\caption{The function $U'(r)$ for different values of the mass  $\mu M_S$. The behavior is similar for any value $\alpha_3$ and $\alpha_4$ near the threshold $\mu M_S\sim 0.438$. Here we used $\alpha_3=1$ and $\alpha_4=0$.\label{fig:du1}}
\end{center}
\end{figure}

\begin{figure}[htb]
\begin{center}
\begin{tabular}{cc}
\epsfig{file=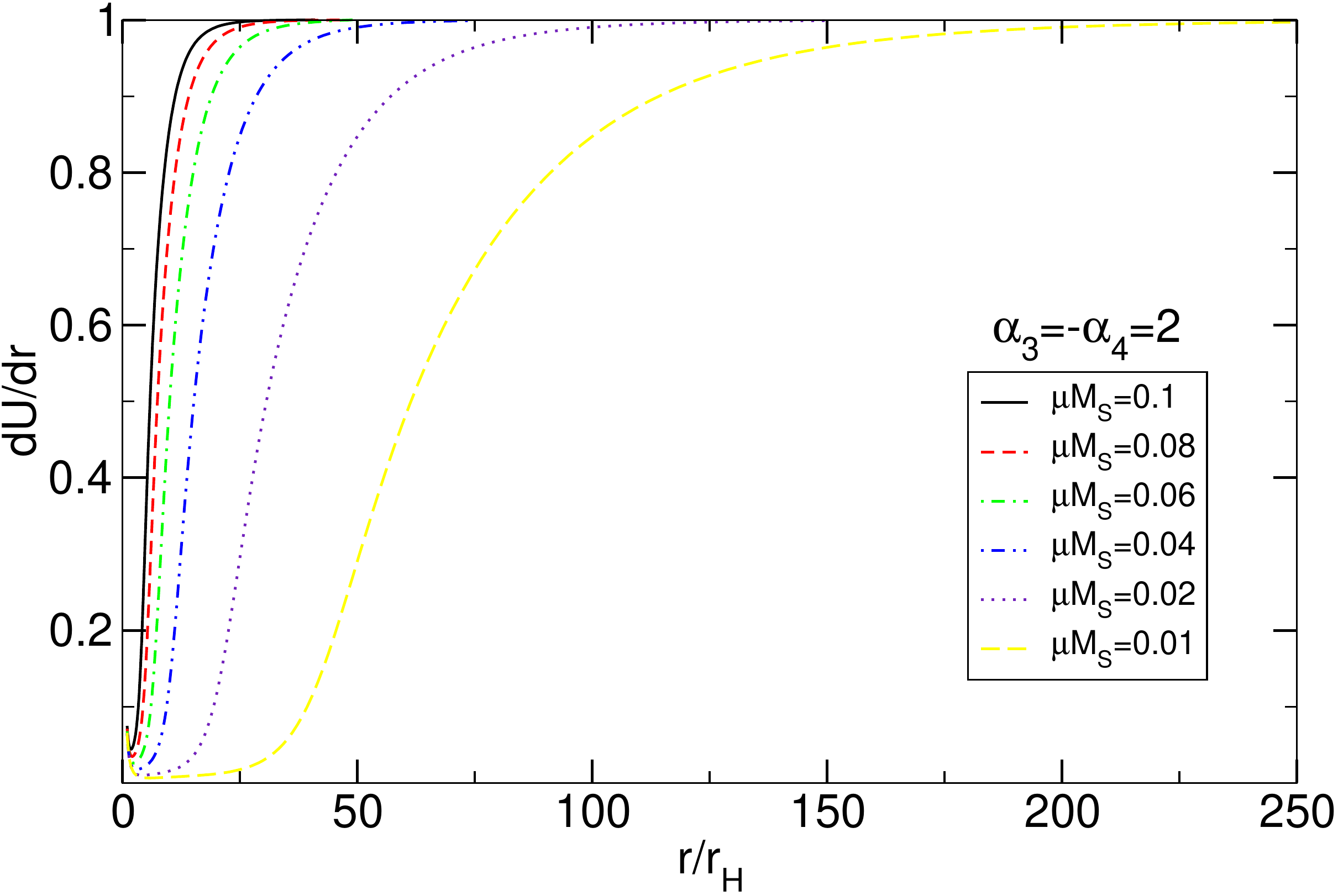,width=7.3cm,angle=0,clip=true}&
\epsfig{file=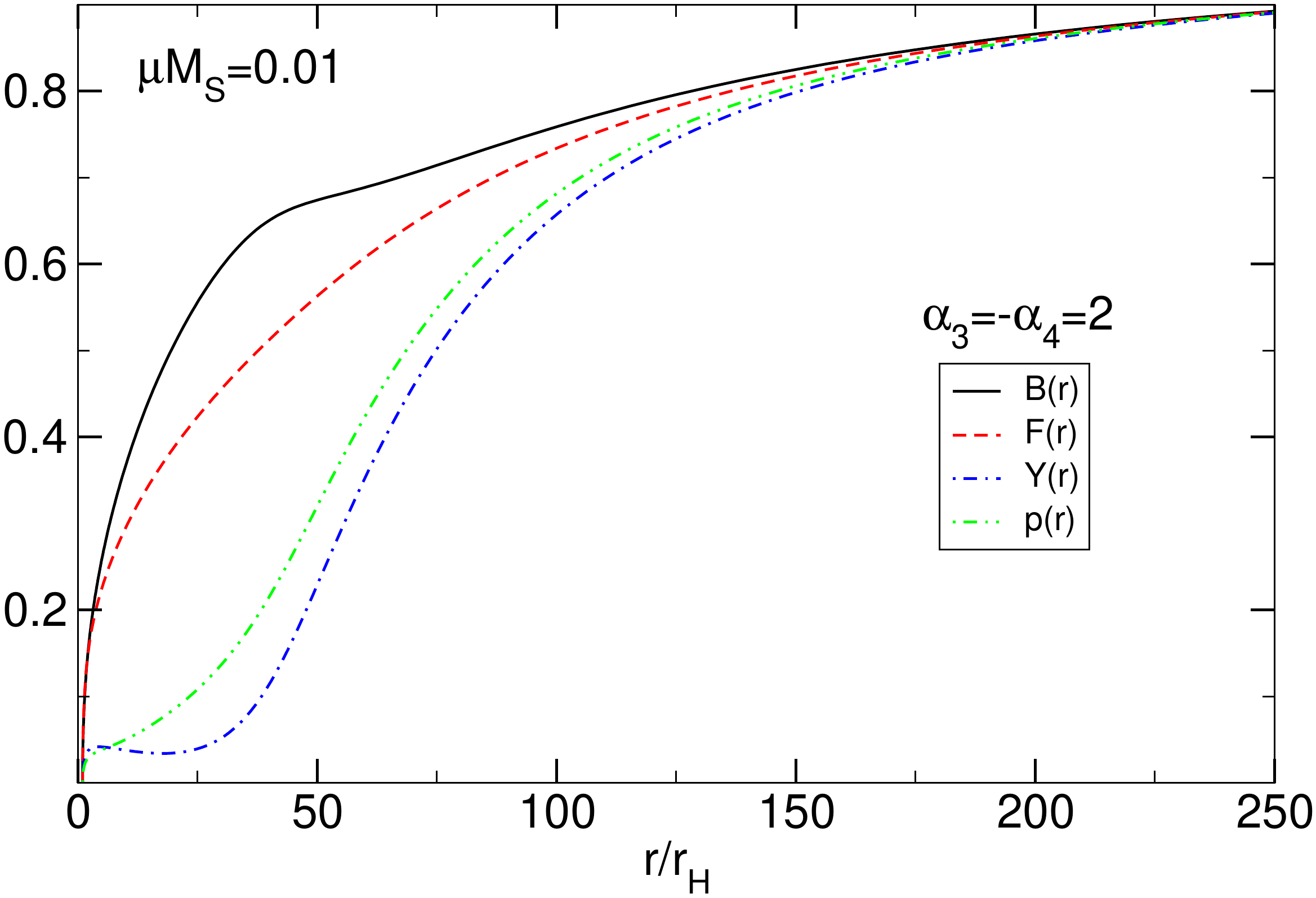,width=7.2cm,angle=0,clip=true}\\
\epsfig{file=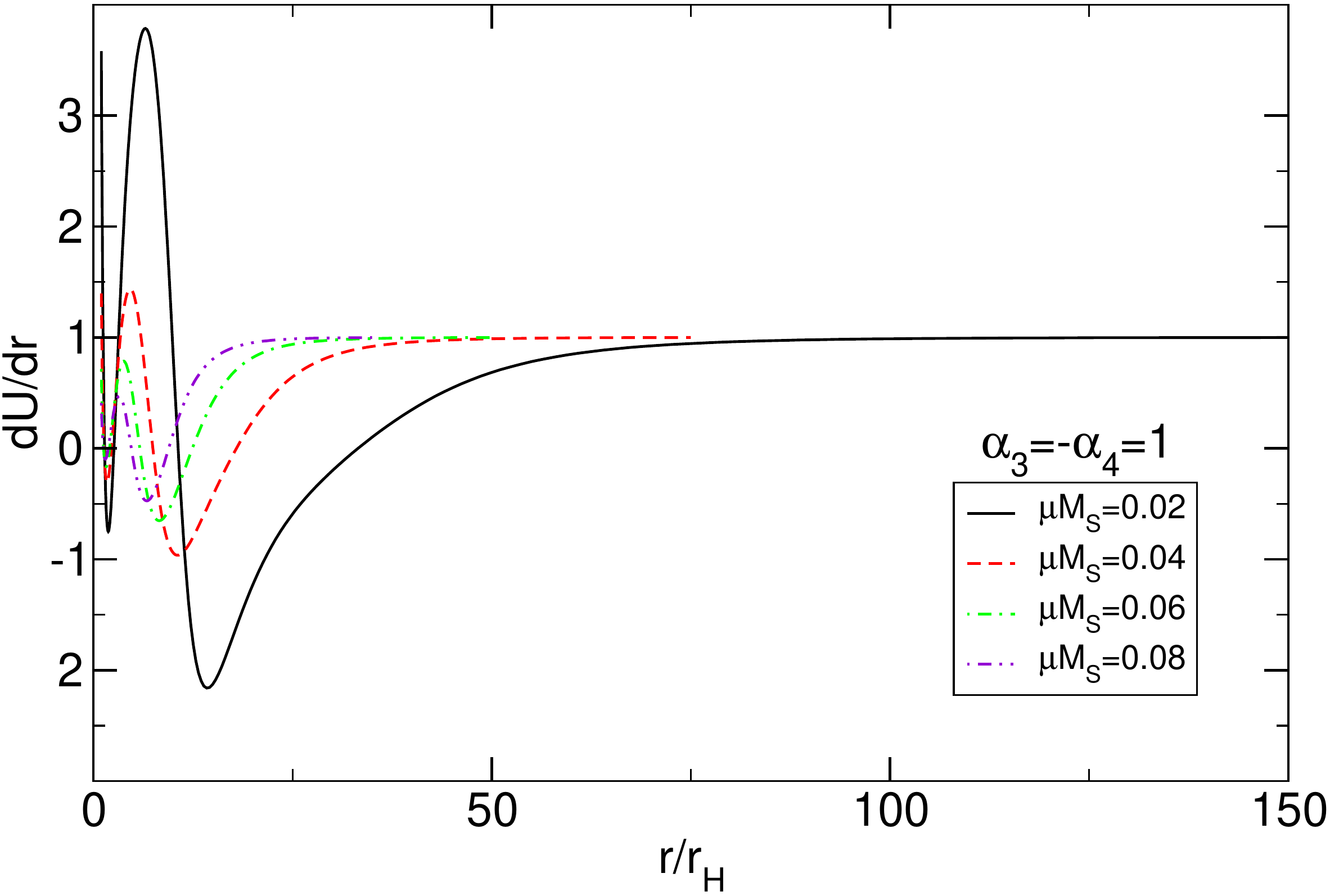,width=7.3cm,angle=0,clip=true}&
\epsfig{file=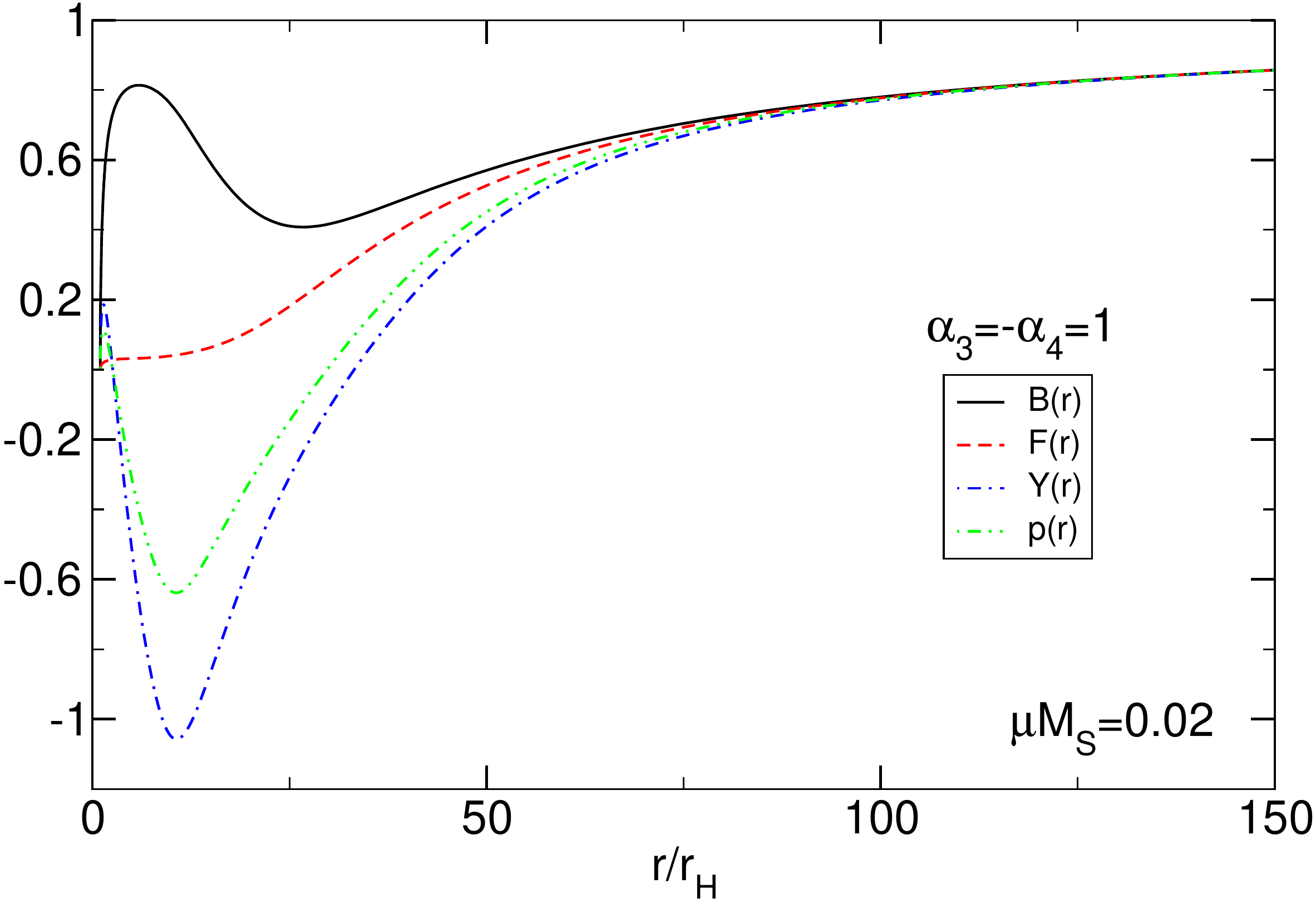,width=7.3cm,angle=0,clip=true}
\end{tabular}
\caption{Metric functions for small values of $\mu M_S$. The function $U'(r)$ can be very different depending on the specific values of the parameters. Top panels: $\alpha_3=2\,,\alpha_4=-2$. Bottom panels: $\alpha_3=1\,,\alpha_4=-1$.
\label{fig:du}}
\end{center}
\end{figure}

The change of behavior between different regions seems to be smooth, since near the boundaries the solutions do not change drastically. For example, in the vicinity of $\alpha_3=-\alpha_4=1$ the solutions behave all in the same way.
We compare the solutions for different choices of $\alpha_3$ and $\alpha_4$ in Figs.~\ref{fig:du1} and~\ref{fig:du}.


Nevertheless, the above classification seems very natural from the mathematical structure of the field equations. For instance,
the choice $\alpha_3=-\alpha_4$ corresponds to $\beta_3=0$, i.e. the higher-order term $V_3$ is absent in the potential~\eqref{potential}. Furthermore, in this case,
\beq
\beta_0&=&-6+3\alpha_3\,,\quad \beta_1=3-2\alpha_3\,,\\
\beta_2&=&-1+\alpha_3\,,\quad \beta_4=\alpha_4\,.
\eeq
Thus, the boundaries where the behavior of the solutions change qualitatively correspond to a change of sign of the parameters $\beta_n$. It is also not surprising to find that $\alpha_3=-\alpha_4=1,2$ are special points of the parameter space, because they correspond to the cases where $V_2$ and $V_0$ are, respectively, absent in the potential~\eqref{potential}.

Finally, the above picture does not hold in the limit where one of the metrics is taken to be a nondynamical Schwarzschild metric. In this case our numerical search suggests that, for any choice of $\alpha_3$ and $\alpha_4$, hairy BH solutions exist near the threshold but they do not exist for arbitrarily small $\mu M_S$. 
%This could be explained by the fact that in the decoupling limit of massive gravity ($\mu\to 0,\,M_g\to\infty$, keeping $(\mu^2M_g)^{1/3}$ fixed) the interactions of the helicity-0 coming from the potential~\eqref{potential} can be decomposed in Galileon-like terms~\cite{deRham:2010ik}, which cannot support nontrivial configurations around a spherically symmetric BH~\cite{Hui:2012qt}.

To summarize, although it is very challenging to infer the behavior of the solutions for all choices of the parameters $\alpha_n$, we have found convincing evidence that the term $V_3$ in the potential~\eqref{potential} plays an important role as it does not allow for hairy deformations of a Schwarzschild BH in the small graviton mass limit. This term is precisely the one that gives rise to a mixing between the helicity-0 and the helicity-2 components of the massive graviton in the decoupling limit~\cite{deRham:2010ik}.

%
%\begin{figure}[bht]
%\begin{center}
%\begin{tabular}{cc}
%\epsfig{file=func_vs_r.eps,width=7.3cm,angle=0,clip=true}&
%\epsfig{file=func_vs_r_a3_1_a4_m1.eps,width=7.3cm,angle=0,clip=true}
%\end{tabular}
%\caption{Metric functions for small masses $\mu M_S$. Left panel: $\alpha_3=2\,,\alpha_4=-2$. Right panel: $\alpha_3=1\,,\alpha_4=-1$.\label{fig:func}}
%\end{center}
%\end{figure}
%

%%%%%%%%%%%%%%%%%%%%%%%%%%%%%%%%%%%%%%%%%%%%%%%%%%%%%%%%%%%%%%%%%%%%%%
\section{Discussion}
%%%%%%%%%%%%%%%%%%%%%%%%%%%%%%%%%%%%%%%%%%%%%%%%%%%%%%%%%%%%%%%%%%%%%%
As far as we are aware, the nonlinear solutions we have found are the first example of graviton-hairy BH solutions in asymptotically flat spacetime.

It is a matter of debate if the theory we considered can in fact be a viable alternative to GR~(see e.g.~\cite{Deser:2012qx,Burrage:2012ja,Deser:2013gpa,Deser:2013eua} and also Sec.VI of Ref.~\cite{Gabadadze:2013ria} for a recent discussion on the status of massive gravity). Nevertheless, whatever the fate of the ghost-free massive and bimetric gravities, these solutions are interesting on their own as they provide the first example of an asymptotically flat graviton-hairy BH. Furthermore, we believe that several of the properties we have presented here are likely to be found in other hairy BH solutions of any putative nonlinear theory with a propagating massive spin-2 field.

These solutions are also natural candidates for the final state of the monopole instability discussed in Chapter~\ref{chapter:massive2}. The instability would presumably cause the Schwarzschild spacetime to evolve towards a hairy solution.
Depending on the parameters of the theory, however, different types of solutions exist in the highly nonlinear regime. This suggests that hairy, static, asymptotically flat BH solutions exist only in certain regions of the parameter space. This in turn makes nonlinear time evolutions
of Schwarzschild BHs highly desirable. It is of course possible that, in some regions of parameter space, Schwarzschild BHs
are not the preferred outcome of gravitational collapse or even that these theories do not allow for stable static BH solutions. These issues can only be addressed by performing nonlinear collapse simulations.
%%%%%%%%%%%%%%%%%%%%%%%%%%%%%%%%%%%%%%%%%%%%%%%%%%%%%%%%%%%%%%%%%%%%

%%%%%%%%%%%%%%%%%%%%%%%%%%%%%%%%%%%%%%%%%%%%%%%%%%%%%%%%%%%%%%%%%%%%
\part{Superradiant instabilities}\label{part:super}
%%%%%%%%%%%%%%%%%%%%%%%%%%%%%%%%%%%%%%%%
\chapter{Superradiance: an introduction}\label{chapter:BHsuperradiance}
%%%%%%%%%%%%%%%%%%%%%%%%%%%%%%%%%%%%%%%%
\emph{This part is based on Refs.~\cite{Brito:2013wya,Brito:2014nja,Brito:2014wla,Brito:2015oca}}.
%%%%%%%%%%%%%%%%%%%%%%%%%%%%%%%%%%%%%%%%%%%%%%%%%%%%%%%%%%
\section{Introduction}
%%%%%%%%%%%%%%%%%%%%%%%%%%%%%%%%%%%%%%%%%%%%%%%%%%%%%%%%
The origin of BH superradiance can be traced back to 1971, when Zel'dovich showed that scattering of radiation off rotating absorbing surfaces
results, under certain conditions, in waves with a larger amplitude~\cite{zeldovich1,zeldovich2}. This phenomenon is now widely known also as (rotational) superradiance and requires that the incident radiation, assumed monochromatic of frequency $\omega$, satisfies~\cite{zeldovich1,zeldovich2}
\be
\frac{\omega}{m} <\Omega\,,\label{eq:superradiance_condition}
\ee
with $m$ the azimuthal number with respect to the rotation axis and $\Omega$ the angular velocity of the body.
Rotational superradiance belongs to a wider class of classical problems displaying stimulated or spontaneous energy emission, such as the Vavilov-Cherenkov effect, the anomalous 
Doppler effect and other examples of ``superluminal motion''~\cite{Bekenstein:1998nt,Brito:2015oca}. When quantum effects were incorporated, it was argued that rotational superradiance would become a spontaneous process and that rotating bodies -- including BHs -- would slow down by spontaneous emission of photons satisfying~\eqref{eq:superradiance_condition}~\cite{zeldovich1}.
In parallel, similar conclusions were reached when analyzing BH superradiance from a thermodynamic viewpoint~\cite{Bekenstein:1973mi,Bekenstein:1998nt}.
From a historic perspective, the first studies of BH superradiance played a decisive role in the discovery of BH evaporation~\cite{Hawking:1974sw,Hawking:book}.

The possibility to extract energy from a spinning BH was first quantified by Roger Penrose~\cite{Penrose:1969} some years before the discovery of BH superradiance, and it is related to the fact that the energy of a particle within the ergoregion (see Section~\ref{Kerr_metric} for the definition of the ergoregion), as perceived by an observer at infinity, can be negative. In the process devised by Penrose, a particle decays into two pieces, one of which escapes to infinity, while the other particle, tuned to have negative energy, is absorbed at the horizon. This results in a net energy gain at the expense of the
rotational energy of the hole.
Soon after, Teukolsky and Press, performed the first quantitative study of BH superradiance~\cite{Teukolsky:1974yv}, showing that amplification occurs for generic monochromatic \emph{bosonic} waves satisfying the condition~\eqref{eq:superradiance_condition}. These first studies of BH superradiance also showed that this phenomenon is an exclusivity of bosonic waves, unlike fermions which cannot be superradiantly amplified~\cite{Unruh:1973,Chandra:1976,Iyer:1978}. 

Teukolsky and Press also predicted that confining superradiant modes would give rise to strong instabilities~\cite{Press:1972zz}. Several mechanisms are able to do so, such as enclosing a Kerr BH with a reflecting mirror, impose Anti-de Sitter (AdS) asymptotics, magnetic fields and even massive bosonic fields (see Ref.~\cite{Brito:2015oca} and references therein).
This superradiant instability is associated to the existence of new asymptotically flat, hairy BH solutions~\cite{Herdeiro:2014goa,Herdeiro:2015tia,Herdeiro:2016tmi} and to phase transitions between spinning or charged black objects in asymptotically AdS spacetime~\cite{Cardoso:2004hs,Dias:2011tj,Dias:2011at} or in higher dimensions~\cite{Shibata:2010wz}. Finally, superradiant instabilities are also fundamental in deciding the stability of BHs and the fate of the gravitational collapse in confining geometries~\cite{Brito:2015oca}.

In this second Part of the thesis we will discuss several systems which are prone to superradiant instabilities. In particular we will show that these instabilities have important applications to dark-matter searches and to physics beyond the Standard Model~\cite{Arvanitaki:2010sy,Pani:2012vp,Brito:2013wya}.
  
%%%%%%%%%%%%%%%%%%%%%%%%%%%%%%%%%%%%%%%%%%%%%%%%%%%%%%%%%%
\section{Superradiance in black hole physics}\label{sec:BHsuperradiance}
%%%%%%%%%%%%%%%%%%%%%%%%%%%%%%%%%%%%%%%%%%%%%%%%%%%%%%%%

The phenomenon of superradiance requires \emph{dissipation}~\cite{Brito:2015oca}. The latter can emerge in various forms, e.g. viscosity, friction, turbulence, radiative cooling, etc. All these forms of dissipation are associated with some medium or some matter field that provides the \emph{arena} for superradiance. It is thus truly remarkable that --~when spacetime is \emph{curved}~-- superradiance can also occur in vacuum, even at the classical level. 

Despite their simplicity, BHs are probably the most fascinating predictions of GR and enjoy some extremely nontrivial properties. The most important property (which also defines the very concept of BH) is the existence of an event horizon, a boundary in spacetime which separates two causally disconnected regions. Among the various properties of BH event horizons, the one that is most relevant for the present discussion is that BHs behave in many respects as a viscous one-way membrane in flat spacetime. This is the so-called BH membrane paradigm~\cite{MembraneParadigm}. Thus, the existence of an event horizon provides vacuum with an intrinsic dissipative mechanism, which is naturally prone to superradiance. As we shall see, the very existence of event horizons allows to extract energy from the vacuum, basically in any relativistic theory of gravity.

%%%%%%%%%%%%%%%%%%%%%%%%%%%%%%%%%%%%%%%%%%%%%%%%%%%%%%%%%%%%%%%%%%%%%%%%%%%%%%%%%%%%%%%%%%%%%%%%%%
\subsection{Spinning, neutral BHs}\label{Kerr_metric}
%%%%%%%%%%%%%%%%%%%%%%%%%%%%%%%%%%%%%%%%%%%%%%%%%%%%%%%%%%%%%%%%%%%%%%%%%%%%%%%%%%%%%%%%%%%%%%%%%%

To set the stage for later use, and to introduce some useful quantities, consider Einstein's equations in vacuum without a cosmological constant. The uniqueness theorems guarantee that the only regular,
asymptotically flat solution to the background equations is given by the Kerr family of spinning BHs. In the following Chapters we will often consider test fields on top of this solution. In standard Boyer-Lindquist coordinates this geometry reads (for details on the Kerr spacetime see, e.g.,~\cite{Wiltshire:2009zza})
\begin{align}
\label{Kerr}
&ds_{\rm{Kerr}}^2=-\left(1-\frac{2Mr}{\Sigma}\right)dt^2+\frac{\Sigma}{\Delta}dr^2-\frac{4Mr}{\Sigma}a\sin^2\theta d\phi dt \nn\\
&+\Sigma d\theta^2+\left[(r^2+a^2)\sin^2\theta+\frac{2M r}{\Sigma}a^2\sin^4\theta \right]d\phi^2\,,
\end{align}
where $\Sigma=r^2+a^2\cos^2\theta$, $\Delta=(r-r_+)(r-r_-)$, $r_\pm=M\pm \sqrt{M^2-a^2}$.
This metric describes the gravitational field of a spinning BH with mass $M$ and angular
momentum $J=a M$. The roots of $\Delta$ determine the event horizon, located at $r_+=M+\sqrt{M^2-a^2}$, and a Cauchy horizon at $r_-=M-\sqrt{M^2-a^2}$. The static surface $g_{tt}=0$ defines the ergosphere given by $r_{\rm{ergo}}=M+\sqrt{M^2-a^2\cos^2\theta}$. Requiring the presence of an event horizon in this spacetime, Kerr BHs have a maximum possible spin given by $a=M$.

A fundamental parameter of a spinning BH is the angular velocity of its event horizon, which for the Kerr solution is given by
\begin{equation}
\Omega_{\rm{H}}=\frac{a}{r_+^2+a^2}\,.  \label{Omega}
\end{equation}

The physical interpretation of this quantity can be understood in the following way: consider an observer with timelike four-velocity which falls into the BH with zero angular momentum. This observer is known as the ZAMO (Zero Angular Momentum Observer). These observers have an angular velocity, as measured at infinity, given by
\be
\Omega \equiv \frac{\dot{\phi}}{\dot{t}}=-\frac{g_{t\phi}}{g_{\phi\phi}}=\frac{2Mar}{r^4+r^2a^2+2a^2Mr}\,.
\ee
At infinity $\Omega=0$ consistent with the fact that these are zero angular momentum observers. However, $\Omega\neq 0$ at any finite distance and at the horizon one finds 
\be
\Omega_{\rm{H}}^{\rm{ZAMO}}=\frac{a}{r_+^2+a^2}\,.
\ee
Thus, observers are frame-dragged and forced to co-rotate with the geometry.

The Kerr geometry is also endowed with a surface outside the horizon where $g_{tt}=0$ called the ergosurface, located at
\be
r_{\rm{ergo}}=M+\sqrt{M^2-a^2\cos^2\theta}\,.\label{ergoKerr}
\ee
In particular, it is defined by $r=2M$ at the equator and $r=r_+$ at the poles. The region between the event horizon and the ergosurface is the ergoregion.
%The ergosurface is an infinite-redshift surface, in the sense that any light ray emitted from the ergosurface will be infinitely redshifted when observed at infinity.
The ergosurface is the static limit, as no static observer is allowed inside the ergoregion.
Indeed, the Killing vector $\xi^{\mu}=(1,0,0,0)$ becomes spacelike in the ergoregion $\xi^{\mu}\xi^{\mu}g_{\mu\nu}=g_{tt}>0$.
We define a static observer as an observer (i.e., a timelike curve) with tangent vector proportional to
$\xi^{\mu}$. The coordinates $(r,\theta,\phi)$ are constant along this wordline.
Such an observer cannot exist inside the ergoregion, because $\xi^{\mu}$ is spacelike there. In other words, inside the ergoregion an observer
cannot stay still, but is forced to rotate with the BH.

%%%%%%%%%%%%%%%%%%%%%%%%%%%%%%%%%%%%%%%%%%%%%%%%%%%%%%%%%%%%%%%
\subsection{Black-hole superradiance in a nutshell}\label{sec:ABC}
%%%%%%%%%%%%%%%%%%%%%%%%%%%%%%%%%%%%%%%%%%%%%%%%%%%%%%%%%%%%%%%

Let us consider a model that captures the basic ingredients of superradiant scattering in curved spacetime.
We assume that the spacetime is stationary and axisymmetric. Consider a test bosonic field in this background. At linear order in the field's amplitude, one can generically write down a single master variable $\Psi$ which obeys a Schroedinger-type equation of the form
\begin{equation}
  \frac{d^2 \Psi}{dr_*^2}+V_{\rm{eff}} \Psi=0\,,\label{wave} 
\end{equation}
where the potential $V_{\rm{eff}}(r)$ is model dependent and encodes the curvature of the background and the properties of the test field. The coordinate $r_*$ maps the region $r\in[r_+,\infty[$ to the entire real axis. Given the symmetries of the background, we consider a scattering experiment of a monochromatic wave with frequency $\omega$ and azimuthal and time dependence $e^{-i\omega t+im\phi}$. Assuming $V_{\rm{eff}}$ is constant at the boundaries, Eq.\eqref{wave} has the following asymptotic behavior 
\begin{equation}
 \Psi \sim\left\{
\begin{array}{ll}
{\cal T}e^{-i k_H r_*}+{\cal O}e^{i k_H r_*} & {\rm{as}}\ r\rightarrow r_+ \,, \\
{\cal R}e^{i k_\infty r_*}+ {\cal I}e^{-i k_\infty r_*}& {\rm{as}}\ r\rightarrow \infty\,.
\end{array}
\right. \label{bound2}
\end{equation}
where $r_+$ is the horizon radius in some chosen coordinates, $k_H^2=V_{\rm{eff}}(r\to r_+)$ and $k_\infty^2=V_{\rm{eff}}(r\to \infty)$.
These boundary conditions correspond to an incident wave of amplitude ${\cal I}$ from spatial infinity giving rise to a reflected wave of amplitude ${\cal R}$ and a transmitted wave of amplitude ${\cal T}$ at the horizon. 
The ${\cal O}$ term describes a putative outgoing flux across the surface at $r=r_+$. Although the presence of a BH horizon would imply ${\cal O}\equiv 0$, let us keep this term for the sake of the argument, and in order to allow for a nonvanishing outgoing flux in absence of an event horizon.

Let us assume that the potential is real\footnote{This condition does not hold for electromagnetic and gravitational perturbations of a Kerr BH, whereas it holds for scalar perturbations of spinning and charged BHs. See Appendix~\ref{app:Teu_eqs}. When such condition does not hold, a more sophisticated analysis is needed, but similar arguments can be made.}. Thus, there exists another solution $\bar\Psi$ to Eq.~(\ref{wave}) which satisfies the complex conjugate boundary conditions. 
The solutions $\Psi$ and $\bar\Psi$ are linearly independent and standard theory of ODEs tells us that their Wronskian is independent of $r_*$. Thus, the Wronskian evaluated near the horizon, $W= -2i k_H\left(|{\cal T}|^2-|{\cal O}|^2\right)$, must equal the one evaluated at infinity, $W=2i k_\infty(|{\cal R}|^2-|{\cal I}|^2)$, so that 
%%%%%
\begin{equation}
 |{\cal R}|^2=|{\cal I}|^2-\frac{k_H}{k_\infty}\left(|{\cal T}|^2-|{\cal O}|^2\right)\,,\label{reflectivity}
\end{equation}
%%%
independently from the details of the potential in the wave equation.

In the case of a one-way membrane boundary conditions at the horizon, i.e. ${\cal O}=0$, one gets $|{\cal R}|^2<|{\cal I}|^2$ when $k_H/k_\infty>0$, as is to be expected for scattering off perfect absorbers. However, for $k_H/k_\infty<0$, the wave is superradiantly amplified, $|{\cal R}|^2>|{\cal I}|^2$~\cite{Teukolsky:1974yv}. In the case of a massless scalar field around a Kerr BH, one finds that $k_H=\omega-m\Omega_{\rm{H}}$ and $k_{\infty}=\omega$ (see Appendix~\ref{app:Teu_eqs}), showing that $k_H<0$ when the condition~\eqref{eq:superradiance_condition} is met. 

Again, we stress how {\it dissipation} is a crucial ingredient for superradiance: without ingoing boundary conditions at the horizon, no superradiant scattering can occur~\cite{zeldovich1,zeldovich2,Bekenstein:1973mi,Richartz:2009mi,Cardoso:2012zn}. In absence of a horizon (for example in the case of rotating perfect-fluid stars if no dissipation is included~\cite{Richartz:2013unq,Cardoso:2015zqa}), regularity boundary conditions must be imposed at the center of the object. By applying the same argument as above, the Wronskian at the center vanishes, which implies $|{\cal R}|^2=|{\cal I}|^2$, i.e. no superradiance. If the rotating object does not possess a horizon, superradiance can only come from some other dissipation mechanism, like friction due the atmosphere or viscosity, which anyway require a precise knowledge of the microphysics governing the interior of the object.
%%%
Equivalently, we can argue that $|{\cal O}|^2$ and $|{\cal T}|^2$ are respectively proportional to the outgoing and transmitted energy flux across the surface at $r_+$. In absence of dissipation, energy conservation implies that the outgoing flux will equal the transmitted one, i.e. $|{\cal O}|^2=|{\cal T}|^2$ and Eq.~(\ref{reflectivity}) would again prevent superradiance, $|{\cal R}|^2=|{\cal I}|^2$.

Superradiant scattering seems to imply that energy is being extracted from the background which -- at linearized
order where superradiance is observed -- is kept fixed. When backreaction effects are included, energy is indeed extracted from the BH. For rotating BHs, both the mass and angular momentum of the background BH decrease~\cite{East:2013mfa,Brito:2015oca}.

The amount of energy extracted through superradiance strongly depends on the spin of the field. The maximum amplification factors are about $0.4\%$, $4.4\%$ and $138\%$ for scattering of massless scalar, electromagnetic and gravitational waves, respectively. The maximum amplification occurs close to the maximum spin of a Kerr BH and very close to the superradiant threshold, $\omega\sim m\Omega_{\rm{H}}$.

%%%%%%%%%%%%%%%%%%%%%%%%%%%%%%%%%%%%%%%%%%%%%%%%%%%%%%%%%%%%%%%%%%%%%%%%%%%%%%%%%%%%%%%%%%%%%%%%%%%%%%%%
\section{Black holes \& superradiant instabilities}\label{sec:bombs}
%%%%%%%%%%%%%%%%%%%%%%%%%%%%%%%%%%%%%%%%%%%%%%%%%%%%%%%%%%%%%%%%%%%%%%
As already mentioned, superradiant amplification lends itself to extraction of energy from BHs, and can also
be looked at as the chief cause of a number of important instabilities in BH spacetimes.
Some of these instabilities lead to hairy BH solutions~\cite{Herdeiro:2014goa,Herdeiro:2015tia,Herdeiro:2016tmi,Sanchis-Gual:2015lje,Bosch:2016vcp}, whereas others extract rotational energy from the BH, spinning it down~\cite{Brito:2014wla} (this will be discussed in more detail in Chapter~\ref{chapter:detectors}).

%%%%%%%%%%%%%%%%%%%%%%%%%%%%%%%%%%%%%%%%%%%%%%%%%%%%%%%%%%%%%%%%%%%%%%%%%%%%%%%%%%%%%%%%%%%%%%%%%%%%%%%%
\subsection{Spinning black holes in confining geometries are unstable}\label{sec:model}
%%%%%%%%%%%%%%%%%%%%%%%%%%%%%%%%%%%%%%%%%%%%%%%%%%%%%%%%%%%%%%%%%%%%%%%%%%%%%%%%%%%%%%%%%%%%%%%%%%%%%%%%
%
\begin{figure}[ht]
\begin{center}
\epsfig{file=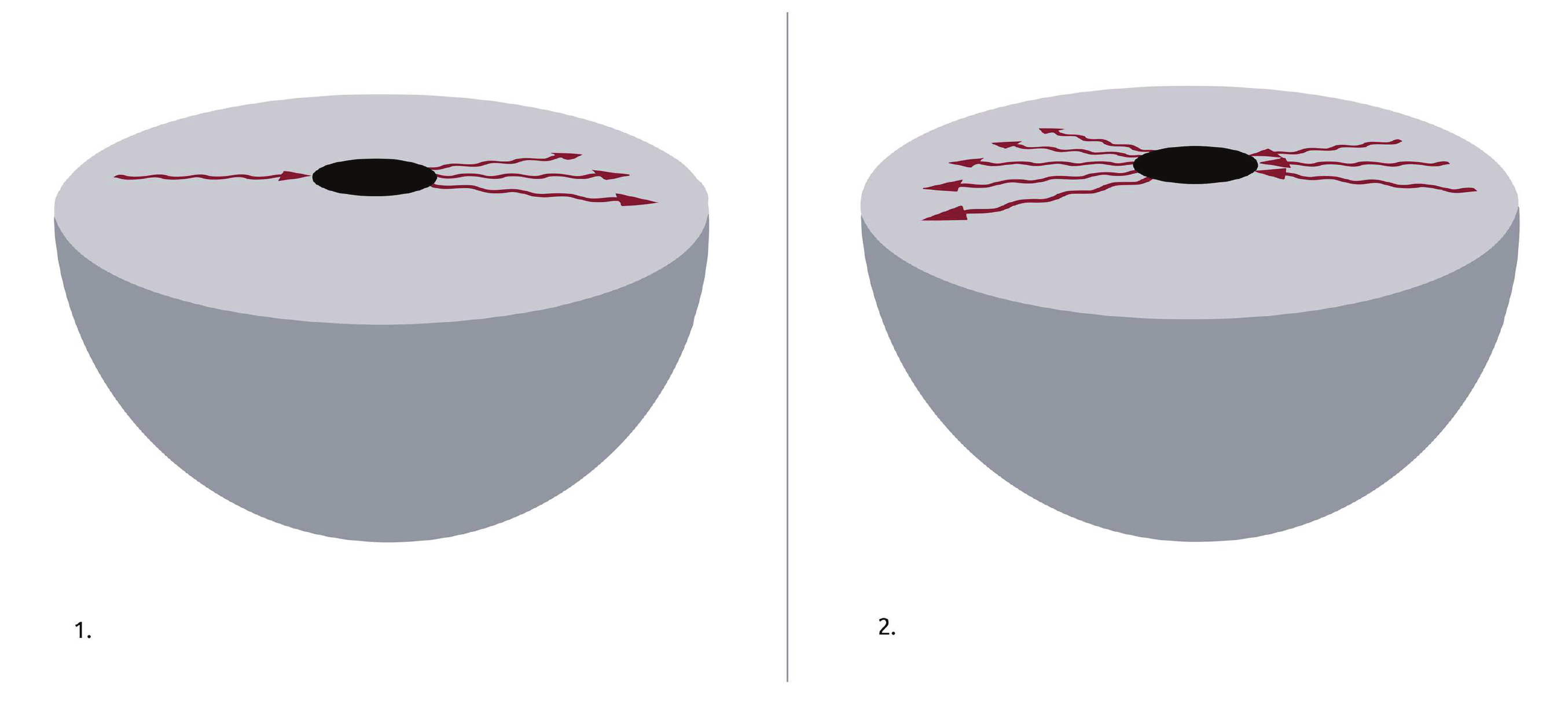,width=0.6\textwidth,angle=0}
\caption{Scheme of a confined rotating BH, and how an initially small fluctuation-- the single red arrow -- grows by successive reflections at the confining wall and amplifications by superradiance in the ergoregion.
\label{fig:bhbomb}}
\end{center}
\end{figure}
It was recognized early on that confinement will generically turn superradiant amplification into an instability mechanism. The idea is very simple and is depicted in Fig.~\ref{fig:bhbomb}: superradiance amplifies any incoming pulse, and the amplification process occurs near the ergoregion. If the pulse is now confined (say, by a perfectly reflecting mirror at some distance) it is ``forced'' to interact -- and be amplified -- numerous times, giving rise to an exponentially increasing amplitude, i.e. to an instability. 

The details of the confinement are irrelevant and a simple picture in terms of a small perfect absorber immersed in a confining box can predict a number of features. A confining box supports stationary, \emph{normal} modes. Once a small BH is placed inside, one expects that the normal modes will become quasinormal and acquire a small imaginary part, describing absorption -- or amplification -- at the horizon of the small BH. Thus, it seems that one can separate the two scales -- BH and box size -- and describe quantitatively the system in this way~\cite{Brito:2014nja}.

Normal modes supported by a box have a wavelength comparable to the box size $r_0$, in other words a frequency $\omega_R\sim 1/r_0$. For small BHs, $M/r_0\ll 1$, we then have $M\omega\ll 1$, i.e., we are in the low-frequency limit. In this limit, the equation for wave propagation can be solved via matched asymptotics~\cite{Cardoso:2005mh}. 
Let $\mathcal{A}$ denote the absorption probability at the horizon of a rotating BH (which can be computed analytically in the small frequency regime~\cite{Cardoso:2005mh,1973ZhETF..64...48S,Staro1,1973ZhETF..65....3S,Staro2}).
By definition, a wave with initial amplitude $A_0$ is scattered with amplitude $A=A_0\left(1-|\mathcal{A}|^2\right)$ after one interaction with the BH.
In the superradiant regime $|\mathcal{A}|^2<0$.
Consider now a wave trapped inside the box and undergoing a large number of reflections. After a time $t$ the wave interacted $N=t/r_0$ times with the BH, and its amplitude changed to
$A=A_0\left(1-|\mathcal{A}|^2\right)^N\sim A_0\left(1-N|\mathcal{A}|^2\right)$. We then get
\be
A(t)=A_0\left(1-t|\mathcal{A}|^2/r_0\right)\,. \label{ampl0}
\ee
The net effect of this small absorption at the event horizon is to add a small imaginary part to the frequency, $\omega=\omega_R+i\omega_I$ (with $|\omega_I|\ll\omega_R$). 
In this limit, $A(t)\sim A_0 e^{-|\omega_I| t}\sim A_0(1-|\omega_I| t)$. Thus we immediately get that
\be
\omega_I=|\mathcal{A}|^2/r_0\,. \label{ampl}
\ee

For example, for a non-rotating BH~\cite{Cardoso:2005mh}
\beq
\left|\mathcal{A}\right|^2&=& 4\pi \left(\frac{M\omega_R}{2}\right)^{2+2l} \frac{\Gamma^2[1+l+s]\Gamma^2[1+l-s]}{\Gamma^2[1+2l]\Gamma^2[l+3/2]}\label{crosssection}\\
&\sim&\left({M}/{r_0}\right)^{2l+2}\ll1\,
\eeq
where $s=0,1,2$ for scalar, electromagnetic and gravitational fields, respectively. Comparing with Eq.~\eqref{ampl}, we obtain
\be\label{absor_wIm}
M\omega_I\sim-(M/r_0)^{2l+3}\,. 
\ee
% % 

When the BH is rotating, rotation can be taken into account by multiplying the previous result by the superradiant factor $1-m\Omega/\omega$. In fact, low-frequency waves co-rotating with the BH are amplified by superradiance. Starobinsky has shown that, at least for moderate spin, the result in Eq.~\eqref{crosssection} still holds with the substitution~\cite{1973ZhETF..64...48S,Staro1,1973ZhETF..65....3S,Staro2}
\be
\omega^{2l+2}\to (\omega-m\Omega_{\rm{H}})\omega^{2l+1}\,,\label{absor_wIm_rot}
\ee
where we recall that $\Omega_{\rm{H}}$ is the horizon angular velocity. 

In other words, this intuitive picture immediately predicts that confined rotating BHs are {\it generically}
unstable and estimates the growth rate. The dependence of the growth rate on the confining radius $r_0$ is estimated to be independent on the spin of the field, and this behavior is observed in a variety of systems. The details need, of course, a careful consideration of the corresponding perturbation equations; nevertheless, as we will show, such conclusions hold for several different scenarios~\cite{Cardoso:2004nk,Cardoso:2004hs,Cardoso:2013pza,Cardoso:2005vk,Brito:2014nja}.

%%%%%%%%%%%%%%%%%%%%%%%%%%%%%%%%%%%%%%%%%%%%%%%%%%%%%%%%%%%%%%%%%%%%%%%%%%%%%%%%%%%%%%%%%%%%%%%%%%%%%%%%
\subsection{General formalism}
%%%%%%%%%%%%%%%%%%%%%%%%%%%%%%%%%%%%%%%%%%%%%%%%%%%%%%%%%%%%%%%%%%%%%%%%%%%%%%%%%%%%%%%%%%%%%%%%%%%%%%%%

At linearized level, BH superradiant instabilities are associated with perturbations of a fixed BH background which grow exponentially in time. Because the background is typically stationary, a Fourier-domain analysis proves to be very convenient. In a stationary and axisymmetric background, a given perturbation $\Psi(t,r,\theta,\phi)$ can be Fourier transformed as 
\begin{equation}
 \Psi(t,r,\theta,\phi)=\frac{1}{2\pi}\sum_m\int d\omega \tilde\Psi_m(\omega,r,\theta) e^{-i\omega t} e^{im\phi}\,, \label{Fourier}
\end{equation}
%%%
and the perturbation function $\tilde\Psi_m$ will satisfy a set of PDEs in the variables $r$ and $\theta$. For the special case of a Kerr BH and for most types of fields, such PDEs can be miraculously separated using spheroidal harmonics~\cite{Teukolsky:1972my,Teukolsky:1973ha} (see Appendix~\ref{app:Teu_eqs}), whereas in more generic settings, other methods, that we will discuss in Chapters~\ref{chapter:magnetic} and~\ref{chapter:Kerr}, have to be used~\cite{Pani:2013pma}.

The system of equations for $\tilde\Psi_m$ together with suitable boundary conditions at the BH horizon and at spatial infinity define an eigenvalue problem for the frequency $\omega$. Due to the boundary conditions at the BH horizon and at spatial infinity, the eigenfrequencies (or quasinormal modes) are generically complex, $\omega=\omega_R+i\omega_I$~\cite{Berti:2009kk}.

In the rest of this Part we will discuss various superradiant instabilities obtained by solving the corresponding perturbation equations in the frequency domain and finding the complex eigenspectrum. Through Eq.~\eqref{Fourier}, an instability corresponds to an eigenfrequency with $\omega_I>0$ and the instability time scale is $\tau\equiv 1/\omega_I$. In the case of superradiant modes this always occurs when the real part of the frequency satisfies the superradiant condition, e.g. $\omega_R<m\Omega_{\rm{H}}$ for a spinning BH.

%%%%%%%%%%%%%%%%%%%%%%%%%%%%%%%%%%%%%%%%%%%%%%%%%%%%%%%%%%
\section{Outline of Part II}\label{sec:outline2}
%%%%%%%%%%%%%%%%%%%%%%%%%%%%%%%%%%%%%%%%%%%%%%%%%%%%%%%%

In the second Part of this thesis we will study several superradiantly unstable systems.
We start by exploring the original black-hole bomb scenario in Chapter~\ref{sec:mirror}, where we study electromagnetic perturbations of a Kerr BH enclosed by a spherical mirror. We will show that this system is unstable, and the results are in full agreement with the model of Section~\ref{sec:model}. 

In Chapter~\ref{chapter:magnetic} we study another system where confinement is manifest. We consider massless scalar perturbations of a solution of Einstein-Maxwell's equations, describing spinning Kerr--Newman magnetized BHs. We show that these solutions are unstable due to the superradiant instability, in analogy with the black-hole bomb case.
At infinity this solution resembles a solution of the Einstein--Maxwell equations describing a uniform magnetic field held together by its own gravitational pull. This solution, which was found by Melvin~\cite{Melvin:1963qx,Melvin:1965zza} and further studied by Thorne~\cite{PhysRev.139.B244}, is known in the literature as the Melvin spacetime. Like the AdS spacetime, the Melvin solution admits normal modes, because the asymptotic boundary of the Melvin solution acts as a confining box for perturbations. Once a BH is added to the spacetime, absorption or amplification at the horizon is possible, in analogy with what happens for a small BH immersed in AdS~\cite{Horowitz:1999jd,Cardoso:2001bb}. 
In particular, we show that rotating magnetized BHs are unstable against the superradiant instability, in full agreement with the arguments of Section~\ref{sec:model}.

Chapter~\ref{chapter:Kerr} is devoted to the study of massive spin-2 perturbations of the Kerr metric. In general, the radial and angular part of the perturbation equations on a spinning geometry are difficult --~if possible at all~-- to separate within the standard Teukolsky approach~\cite{Berti:2009kk,Pani:2013pma}. The same obstacle is encountered for massive spin-1 (Proca) perturbations of a Kerr BH. To tackle this problem we have extended the slow-rotating technique of Refs.~\cite{Pani:2012bp,Pani:2012vp} to the case of massive spin-2 perturbations (see also~\cite{Pani:2013ija} for the case of gravito-electromagnetic perturbations of Kerr-Newman BHs).
By expanding the perturbation equations to first order in the BH spin, we find strong evidence for the existence of unstable modes in the spectrum. This instability is different from the one affecting Schwarzschild BHs (discussed in Chapter~\ref{chapter:massive2}) and it is associated to nonspherical modes which becomes unstable above a certain BH spin. The instability can be four orders of magnitude stronger than in the Proca case and up to seven orders stronger than in the massive scalar case. These results provide strong indications that massive spin-2 fields trigger the strongest superradiant instability in vacuum BH solutions.
Furthermore, the unstable, spherically-symmetric mode active for Schwarzschild BHs presented in Chapter~\ref{chapter:massive2} is unaffected by rotation, at first order. Thus, we present {\it two} mechanisms by which Kerr BHs are rendered unstable in massive theories of gravity.

Finally, in Chapter~\ref{chapter:detectors} we study the development of the superradiant instability using an adiabatic approximation. We study the impact of both gravitational-wave emission and gas accretion on the evolution of the instability. This analysis shows that: (i) gravitational-wave emission does not have a significant effect
on the evolution of the BH, (ii) accretion plays an important role and (iii) although the mass of the bosonic cloud developed through superradiance can be a
sizeable fraction of the black-hole mass, its energy-density is very low and backreaction is negligible. Thus, massive BHs are well described by the Kerr geometry even
if they develop bosonic clouds through superradiance. Using Monte Carlo methods and very conservative assumptions, we provide strong support to the validity of the
linearized analysis.

%%%%%%%%%%%%%%%%%%%%%%%%%%%%%%%%%%%%%%%%%%%%%%%%%%%%%%%%%%%%%%%%%%%%%%%%%%%%%%
\chapter{Black holes enclosed in a mirror: Electromagnetic black-hole bomb}\label{sec:mirror}
%%%%%%%%%%%%%%%%%%%%%%%%%%%%%%%%%%%%%%%%%%%%%%%%%%%%%%%%%%%%%%%%%%%%%%%%%%%%%%

%%%%%%%%%%%%%%%%%%%%%%%%%%%%%%%%%%%%%%%%%%%%%%%%%%%%%%%%%%%%%%%%%%%%%%%%%%%%%%
\section{Introduction}
%%%%%%%%%%%%%%%%%%%%%%%%%%%%%%%%%%%%%%%%%%%%%%%%%%%%%%%%%%%%%%%%%%%%%%%%%%%%%%

One of the first conceptual experiments related to BH superradiance concerns a spinning BH surrounded by a perfectly reflecting mirror~\cite{zeldovich2,Press:1972zz,Cardoso:2004nk}.
As discussed in the previous Chapter, confinement turns this system unstable against superradiant modes\footnote{
Any initial fluctuation grows exponentially, as we argued previously, leading to an ever increasing field density and pressure inside the mirror. The exponentially increasing pressure eventually disrupts the confining mirror, leading to an ``explosion,'' and to this system being termed a \emph{black-hole bomb}~\cite{Press:1972zz}.}. A perfectly reflecting wall is an artificial way of confining fluctuations, but is a useful guide to the other more realistic and complex systems that we will consider in the next Chapters.

For scalar fields, the relevant equation~\eqref{wave} can be solved imposing suitable in-going or regularity boundary conditions at the horizon (namely ${\cal O}=0$ in the boundary conditions~\eqref{bound2}) and a no-flux condition at the mirror boundary $r=r_m$ in Boyer-Lindquist coordinates. The latter can be realized in two different ways: either with Dirichlet $\Psi(r_m)=0$ (see Ref.~\cite{Cardoso:2004nk} for a full analysis of this case) or Neumann $\Psi'(r_m)=0$ conditions for the corresponding master wavefunction. The more realistic situation of electromagnetic waves trapped by a conducting spherical surface is slightly more involved and will be explained in this Chapter.

%%%%%%%%%%%%%%%%%%%%%%%%%%%%%%%%%%%%%%%%%%%%%%%%%%%%%%%%%%%%%%%%%%%%%%%%%%%%
\section{Electromagnetic fluctuations around a rotating black hole enclosed in a mirror\label{sec:EM_BCs}}
%%%%%%%%%%%%%%%%%%%%%%%%%%%%%%%%%%%%%%%%%%%%%%%%%%%%%%%%%%%%%%%%%%%%%%%%%%%%

%%%%%%%%%%%%%%%%%%%%%%%%%%%%%%%%%%%%%%%%%%%%%%%%%%%%%%%%%%%%%%%%%%%%%%%%%%%%
\subsection{Static black hole enclosed in a mirror}
%%%%%%%%%%%%%%%%%%%%%%%%%%%%%%%%%%%%%%%%%%%%%%%%%%%%%%%%%%%%%%%%%%%%%%%%%%%%

Consider first the evolution of a Maxwell field in a Schwarzschild background with metric given by
\begin{equation}
ds^{2}= -f(r) dt^{2}+ \frac{dr^{2}}{f(r)}+r^{2}(d\theta^{2}+\sin^2\theta d\phi^{2})\,,
\label{sch_lineelement}
\end{equation}
where, $f(r)=1-2M/r$ and $M$ is the BH mass. The perturbations are governed by Maxwell's equations:
\begin{equation}
{F^{\mu\nu}}_{;\nu}=0\,, \quad F_{\mu\nu}=A_{\nu,\mu}-A_{\mu,\nu}\,,
\label{maxwell}
\end{equation}
where a comma stands for ordinary derivative and a semi-colon
for covariant derivative. Since the background is spherically symmetric,
we can expand $A_{\mu}$ in 4-dimensional vector spherical harmonics (see~\cite{Ruffini}):

{\small
\be
A_{\mu}(t,r,\theta,\phi)=\sum_{l,m}\left[\left(
 \begin{array}{c} 0 \\ 0 \\
 a^{lm}(t,r)\bar{\bm{S}}_{lm}\end{array}\right)
+\left(\begin{array}{c}f^{lm}(t,r)Y_{lm}\\h^{lm}(t,r)Y_{lm} \\
 k^{lm}(t,r) \bar{\bm{Y}}_{lm}\end{array}\right)\right]\,,
\label{expansion}
\ee
}
with the vector spherical spherical harmonics given by,
\be
\bar{\bm{Y}}^\intercal_{lm}=\left(\partial_\theta Y_{lm}, \partial_\phi Y_{lm}\right)\,,\quad
\bar{\bm{S}}^\intercal_{lm}=\left(\frac{1}{\sin\theta}\partial_\phi Y_{lm}, -\sin\theta\partial_\theta Y_{lm}\right)\,,
\ee
and where $Y_{lm}$ are the usual scalar spherical harmonics, $m$ is the azimuthal number and $l$ the angular quantum number. The first term in the right-hand side has parity $(-1)^{l+1}$,
and the second term has parity $(-1)^{l}$. We shall call the former the axial modes and the latter the polar modes. 

Upon defining
\be 
\Upsilon^{lm}=\frac{r^2}{l(l+1)}\left(\partial_t h^{lm}-\partial_r f^{lm}\right)\,,\label{upsilon}
\ee
and inserting~\eqref{expansion} into Maxwell's equations~\eqref{maxwell}, and after some algebra, we get the following system of equations
\beq
& & \frac{\partial^{2} a^{lm}(t,r)}{\partial r_*^{2}} + \left\lbrack -\frac{\partial^{2}}{\partial
t^{2}}-V(r)\right\rbrack a^{lm}(t,r)=0 \,,\\
& &\frac{\partial^{2}\Upsilon^{lm}(t,r)}{\partial r_*^{2}} + \left\lbrack -\frac{\partial^{2}}{\partial
t^{2}}-V(r)\right\rbrack \Upsilon^{lm}(t,r)=0 \,,\\
& & V=f\frac{l(l+1)}{r^2}\,.\label{potentialmaxwell}
\eeq
If we assume a time dependence $a^{lm}\,,\Upsilon^{lm}\propto e^{-i\omega t}$, the equation for
electromagnetic perturbations of the Schwarzschild geometry takes the form
\be
\frac{\partial^{2}\Psi}{\partial r_*^{2}} + \left\lbrack \omega^2-V\right\rbrack \Psi=0
\,,\label{wavemaxwell}
\ee
where the tortoise coordinate is defined through $dr/dr^*=f(r)$, $\Psi=a^{lm}$ for axial modes and $\Psi=\Upsilon$ for polar modes. The
potential $V$ appearing in equation~\eqref{wavemaxwell} is given by Eq~\eqref{potentialmaxwell}.

Let us now assume we have a spherical conductor at $r=r_m$. The conditions to be satisfied are then that the electric/magnetic field as seen by an observer at rest with respect to the conductor has no tangential/parallel components, $E_{\theta}\propto F_{\theta \,t}=0,\,E_{\phi}\propto F_{\phi \,t}=0,\,B_{r}\propto F_{\phi \,\theta}=0$.
This translates into
\be
\partial_t a^{lm}(t,r_m)=0\,,\quad f^{lm}(t,r_m)-\partial_t k^{lm}(t,r_m)=0\,.
\ee
Using Maxwell's equations~\eqref{maxwell}, we get the relation
\be
f^{lm}(t,r_m)-\partial_t k^{lm}(t,r_m)=\frac{f}{l(l+1)}\partial_r\left(r^2\partial_r f^{lm}-r^2\partial_t h^{lm}\right)\,.
\ee
Finally, using Eq.~\eqref{upsilon} we get
\be
\partial_r\Upsilon=0\,.
\ee
In other words, the boundary conditions at the surface $r=r_m$ are $\Psi=0$ and $\partial_r\Psi=0$ for axial and polar perturbations respectively. This can be used to easily compute the electromagnetic modes inside a resonant cavity in flat space. Taking $M=0$ in Eq.~\eqref{wavemaxwell} we find the exact solution
\be
\Psi=\sqrt{r}\left[C_1 J_{l+1/2}(r \omega)+C_2 Y_{l+1/2}(r \omega)\right]\,,
\ee
where $C_{i}$ are constants and $J_n(r\omega)$ and $Y_n(r\omega)$ are Bessel functions of the first and second kind, respectively. Imposing regularity at the origin $r=0$ implies $C_2=0$. The Dirichlet boundary condition $\Psi=0$ at $r=r_m$, which can easily be shown to correspond to the \emph{transverse electric} modes (modes with $E_r=0$)~\cite{Jackson}, then gives
\be\label{TE_modes}
\omega_{\mathrm{TE}}=\frac{j_{l+1/2,n}}{r_m}\,,
\ee
where $j_{l+1/2,n}$ are the zeros of the Bessel function $J_{l+1/2}$ and $n$ is a non-negative integer. On the other hand the eigenfrequencies for the Neumann boundary condition $\partial_r\Psi=0$, which corresponds to the \emph{transverse magnetic} modes (modes with $B_r=0$)~\cite{Jackson}, can be computed solving
\be
\left\{\partial_r\left[\sqrt{r}J_{l+1/2}(r\omega)\right]\right\}_{r=r_m}=\frac{(l+1)J_{l+1/2}(r_m\omega)-r_m\omega J_{l+3/2}(r_m\omega)}{\sqrt{r_m}}=0\,.
\ee
Defining $\tilde{j}_{l+1/2,n}$ as being the zeroes of $\partial_r\left[\sqrt{r_m}J_{l+1/2}(r_m\omega)\right]$ we find
\be\label{TM_modes}
\omega_{\mathrm{TM}}=\frac{\tilde{j}_{l+1/2,n}}{r_m}\,.
\ee
The eigenfrequencies for $l=1$ and $n=0$ are shown in Fig.~\ref{Fig:BQNM} where we see that when $r_m\gg M$, the real part of the quasinormal frequencies of a BH enclosed in a mirror asymptotically reduces to the flat space result.  

One can write down a relation between the Regge-Wheeler function $\Psi$~\cite{Chrzanowski:1975wv,Ori:2002uv,Hughes:2000pf} and the Teukolsky radial function $R$ (cf. Eq.~\eqref{teu_eigen}) given by
\begin{eqnarray}
\frac{\Psi}{r(r^2-2Mr)^{s/2}}&=&\left(r\sqrt{\Delta}\right)^{|s|}\mathcal{D}_-^{|s|}\left(r^{-|s|}R\right)\,,\, s<0,\nn\\
\frac{\Psi}{r(r^2-2Mr)^{s/2}}&=&\left(\frac{r}{\sqrt{\Delta}}\right)^{s}\mathcal{D}_+^{s}\left[\left(\frac{r^2-2Mr}{r}\right)^{s}R\right]\,,\, s>0,\nn\\\label{hughes}
\end{eqnarray}
where $\mathcal{D}_{\pm}=d/dr\pm i \omega/f$. Using these relations and Teukolsky's radial equation~\eqref{teu_radial}, one finds that the Dirichlet and the Neumann boundary conditions for $\Psi$, correspond to the Robin boundary conditions for the radial function $R$ given respectively by
\beq
\partial_r R_{-1}&=&\frac{r-2M+i r^2\omega}{r(r-2M)}R_{-1}\,,\label{cond1}\\
\partial_r R_{-1}&=&\frac{r\omega[2M+r(-1-i r \omega)]-i l(l+1) (2M-r)}{(2M-r)r^2\omega}R_{-1}\,.\label{cond2}
\eeq

After having understood the nonrotating case, below we turn to the rotating case. The main difficulty lies in describing the electromagnetic physical quantities in terms of the Newman-Penrose quantities. We will show that doing so, will allow us to generalize the conditions~\eqref{cond1} and~\eqref{cond2}. 

%%%%%%%%%%%%%%%%%%%%%%%%%%%%%%%%%%%%%%%%%%%%%%%%%%%%%%%%%%%%%%%%%%%%%%%%%%%%
\subsection{Electromagnetic black-hole bomb}
%%%%%%%%%%%%%%%%%%%%%%%%%%%%%%%%%%%%%%%%%%%%%%%%%%%%%%%%%%%%%%%%%%%%%%%%%%%%

In the Newman-Penrose formalism, the electromagnetic field is characterized by three complex scalars from which one can obtain the electric and magnetic field (see Appendix~\ref{app:Teu_eqs}). The details of this procedure are not important for us here so we refer the reader to Ref.~\cite{King:1977}. In the frame of a ZAMO observer (cf. subsection~\ref{Kerr_metric}), the relevant electric and magnetic field components read~\cite{King:1977}
\beq
E_{(\theta)}&=&\left[\frac{\Delta^{1/2}(r^2+a^2)}{\sqrt{2}\rho^* A^{1/2}(r^2+a^2\cos^2\theta)}\left(\frac{\phi_0}{2}-\frac{\phi_2}{\rho^2\Delta}\right)+{\mathrm{c.c.}}\right]
-\frac{2 a \Delta^{1/2}}{A^{1/2}}\sin\theta\, {\mathrm{Im}}(\phi_1)\,,\nn\\
E_{(\phi)}&=&\left[-i\Delta^{1/2}\rho\left(\frac{\phi_0}{2\sqrt{2}}+\frac{\phi_2}{\sqrt{2}\rho^2\Delta}\right)+{\mathrm{c.c.}}\right]\,,\nn\\
B_{(r)}&=&\left[\frac{a\sin\theta}{\sqrt{2}\rho A^{1/2}}\left(\phi_2-\Delta\rho^2\frac{\phi_0}{2}\right)+{\mathrm{c.c.}}\right]
+2\frac{r^2+a^2}{A^{1/2}}{\mathrm{Im}}(\phi_1)\,,
\eeq
where $\rho=-(r-i a\cos\theta)^{-1}$, $A=(r^2+a^2)^2-a^2\Delta\sin^2\theta$ and $\Delta=r^2-2Mr+a^2$.

If we assume a conducting spherical surface surrounding the BH at $r=r_m$, then Maxwell's equations require that
$E_{(\theta)}=E_{(\phi)}=B_{(r)}=0$ at $r=r_m$ and we are left with the boundary conditions at the conductor:
\be
\rho\Phi_0=\frac{\rho^*\Phi_2^*}{\Delta}\,,\quad
\rho^*\Phi_0^*=\frac{\rho\Phi_2}{\Delta}\,,\quad
{\mathrm{Im}}(\phi_1)=0\,,
\ee
where we defined $\Phi_0=\phi_0$ and $\Phi_2=2\rho^{-2}\phi_2$ . This also implies that:
\be\label{bc_NP}
|\Phi_0|^2=\frac{|\Phi_2|^2}{\Delta^2}\,.
\ee

To solve this equation we use the decomposition
\beq\label{decom1}
\Phi_0&=&\sum_{lm}\int d\omega\, e^{-i\omega t+im\phi}R_{s\,l\,m\,\omega}S_{s\,l\,m\,\omega}(\theta)\,,\nn\\
\Phi_2&=&\sum_{lm}\int d\omega\, e^{-i\omega t+im\phi}R_{-s\,l\,m\,\omega}S_{-s\,l\,m\,\omega}(\theta)\,,
\eeq
where the radial and the angular function, $R$ and $S$, satisfy Teukolsky's Eqs.\eqref{teu_radial} and~\eqref{spheroidal} of Appendix~\ref{app:Teu_eqs}, respectively.
The function $R_{s=1}$ can be written as a linear combination of $R_{s=-1}$ and its derivative through the Starobinski-Teukolsky identities~\cite{Teukolsky:1974yv,1973ZhETF..65....3S,Staro2}
\be\label{ST_iden}
\mathcal{D}_0\mathcal{D}_0{}_{-1}R=B R_{1}\,,
\ee
where $B=\sqrt{Q^2+4ma\omega-4a^2\omega^2}$, $Q=A_{-1lm}+a^2\omega^2-2am\omega$, with $A_{-1lm}$ defined in Eq.~\eqref{spheroidal}, and the linear operator is given by
\be
\mathcal{D}_0=\partial_r-i\frac{K}{\Delta}\,.
%\mathcal{L}_n=\partial_{\theta}+m\csc\theta-a\omega\sin\theta+n\cot\theta.
\ee
%
%Furthermore, from Teukolsky's equations one can derive the following identities~\cite{Chrzanowski:1975wv}
%
%\be\label{iden}
%R^*_{s\,l\,m\,\omega}= (-1)^{m+s} R_{s\,l\,-m\,-\omega}\,,\quad
%
%S^*_{s\,l\,m\,\omega}= (-1)^{m+s} S_{-s\,l\,-m\,-\omega}\,.
%\ee
%

Finally, using~\eqref{decom1},~\eqref{ST_iden} and integrating eq.~\eqref{bc_NP} over the sphere\footnote{By integrating Eq.~\eqref{bc_NP} over the sphere, we are actually only requiring it to be satisfied on average over all angles. This turns out to have a clear physical meaning. In fact, the same boundary conditions could have been obtained by requiring a vanishing radial energy flux at $r=r_m$, as one can easily check by comparing~\eqref{bc_NP} with Eq.~(9) in Ref.~\cite{Wang:2015goa}.} we find the following conditions for the two polarizations:
\beq\label{BC_bomb}
&&\partial_r R_{-1}=\frac{-i\Delta\left[\pm B+A_{-1lm}+\omega  \left(a^2 \omega -2 a m+2 i
   r\right)\right]}{2 \Delta\left(a^2 \omega -a m+r^2 \omega \right)}R_{-1}\nonumber\\
&&		+\frac{\left(a^2 \omega -a m+r^2 \omega \right) \left(2 i a^2 \omega -2 i a m+2 M+2 i r^2 \omega+ \partial_r\Delta-2  r\right)}{2 \Delta\left(a^2 \omega -a m+r^2 \omega \right)}R_{-1}\,.
\eeq

Note that the perturbations can be written in terms of two Newman-Penrose scalars, $\phi_2$ and $\phi_0$, which are two linearly dependent complex functions. This explains the existence of two different boundary conditions, as would have been expected given the two degrees of freedom of electromagnetic fields. For $a=0$ we recover the condition~\eqref{cond1} when using the minus sign, while for the plus sign we recover the condition~\eqref{cond2}; accordingly, we label these modes as axial and polar modes, respectively.

\begin{figure*}[hbt]
\begin{center}
\begin{tabular}{cc}
\epsfig{file=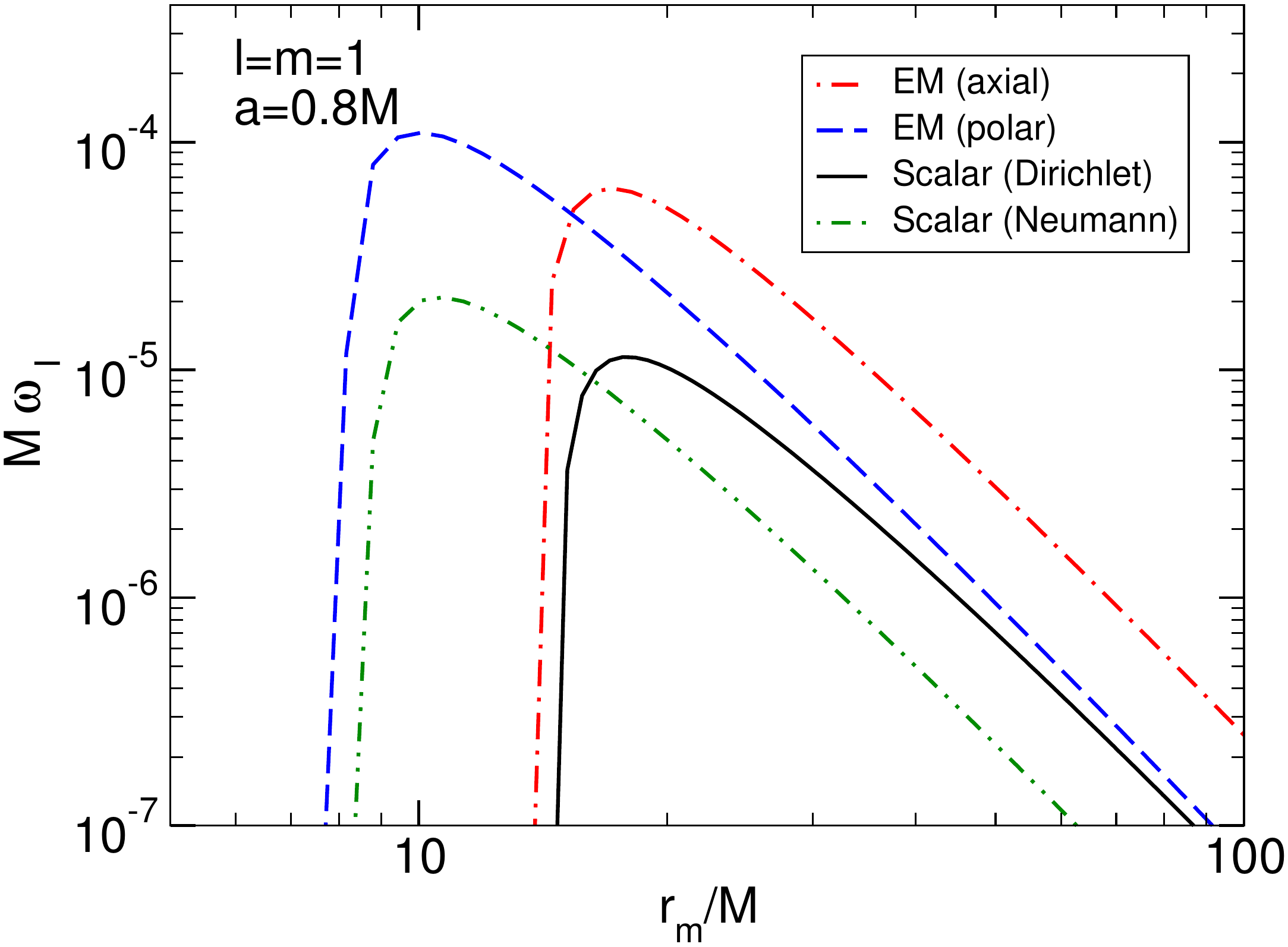,width=0.46\textwidth,angle=0,clip=true}&
\epsfig{file=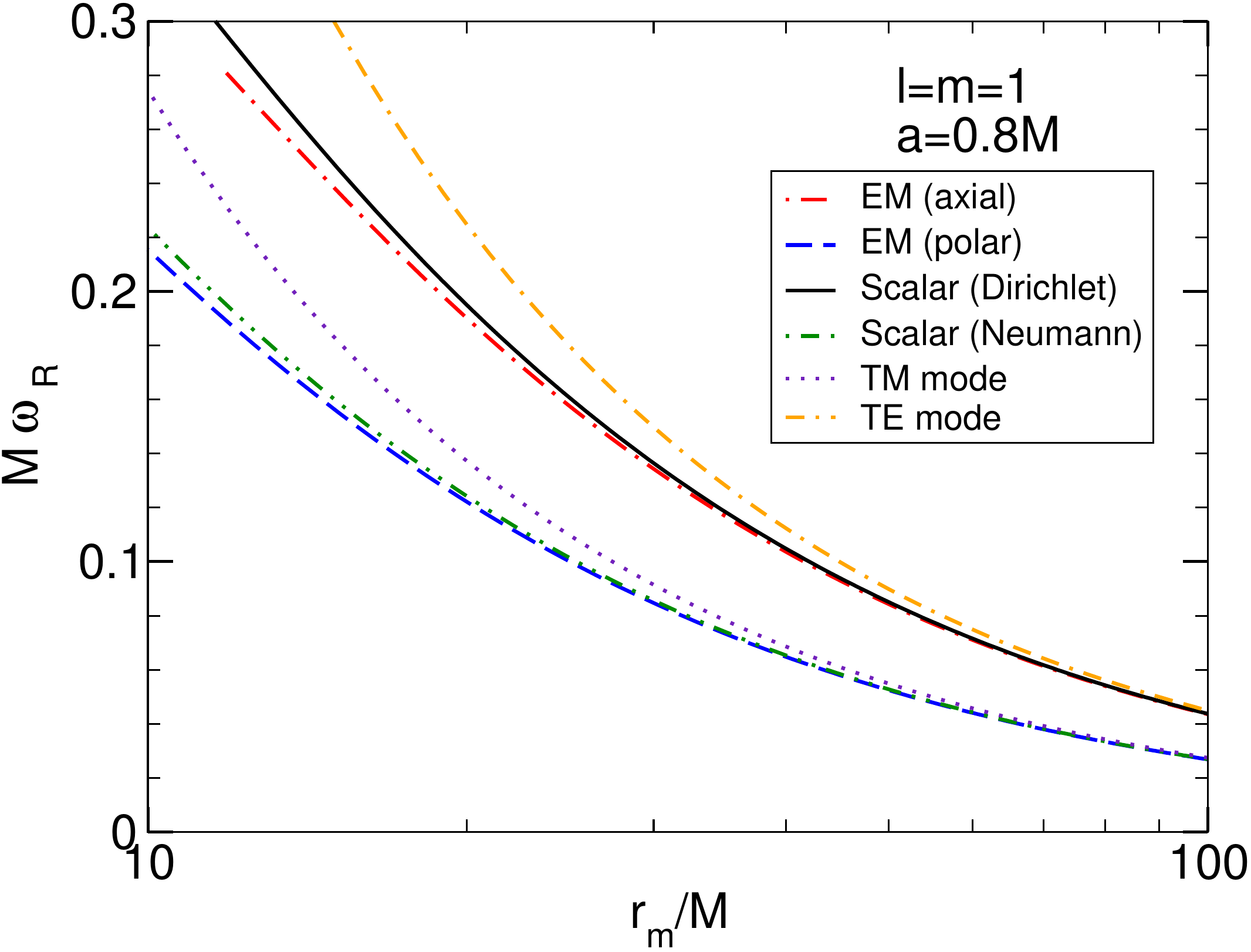,width=0.46\textwidth,angle=0,clip=true}
\end{tabular}
\caption{Fundamental ($n=0$) QNM frequency for scalar and electromagnetic perturbations of a confined Kerr BH as a function of the mirror's location $r_m$, for $l = m = 1$ and $a=0.8M$. For $r_m$ larger than a critical value the modes are unstable. We show the two different polarizations for the electromagnetic BH bomb compared to the modes of a scalar field for Dirichlet and Neumann boundary conditions at the boundary. For comparison we also show the flat space transverse electric (TE) and transverse magnetic (TM) modes inside a resonant cavity, as computed in Eqs.~\eqref{TE_modes} and~\eqref{TM_modes}.\label{Fig:BQNM}}
\end{center}
\end{figure*}

The boundary conditions described above are only satisfied for a discrete number of QNM eigenfrequencies $\omega$. Our results for the characteristic frequencies are shown in
Fig.~\ref{Fig:BQNM} for $l=m=1$ and $a=0.8M$. As the generic argument presented in Section~\ref{sec:model} anticipated, confined BHs develop an instability, i.e.
some of the characteristic frequencies satisfy $\omega_I>0$\footnote{We recall that the time-dependence of the field is $\sim e^{-i\omega t}$, and a positive imaginary component
of the frequency signals an instability.}. Figure~\ref{Fig:BQNM} (left panel) shows that the time scale dependence on $r_m$ is the same for electromagnetic and scalar fluctuations,
as predicted in Section~\ref{sec:model}. Note that the electromagnetic growth rates $1/\omega_I$ are about one order of magnitude smaller than those of scalar fields. This is consistent with the fact that the maximum superradiant amplification factor for vector fields is approximately one order of magnitude larger than for scalars.

As also anticipated with the heuristic argument of Section~\ref{sec:model}, the instability time scale grows with $r_m^{2l+2}$
and the oscillation frequency $\omega_R$ is inversely proportional to the mirror position and reduces to the flat space result when $r_m\gg M$. Thus, for very small $r_m$ the superradiant condition $\omega<m\Omega_{\rm{H}}$ is violated and the superradiant instability is quenched.
In the limit of very large cavity radius $r_m/M$ our results reduce to the TE and TM modes of a spherical cavity in flat space~\cite{Jackson}.

%

%%%%%%%%%%%%%%%%%%%%%%
\section{Conclusions}
%%%%%%%%%%%%%%%%%%%%%
The BH bomb scenario discussed in this chapter can serve as a model to describe astrophysical BHs surrounded by plasmas or accretion disks. Ionized matter is a good low-frequency electromagnetic waves reflector and can thus play the role of the mirror (this was first realized by Teukolsky~\cite{teukolskythesis}. See also Ref.~\cite{Pani:2013hpa}). A very important question which still needs clarification concerns the effectiveness of the instability in these realistic situations. The matter surrounding the BH comes under the form of thin or thick accretion disks and not as spherically shaped mirrors. Confining the field along some angular direction means forbidding low angular eigenvalue modes, implying that only higher-angular eigenvalue modes (with longer time scales, cf. Eq.~\eqref{absor_wIm_rot}) are unstable~\cite{Putten,Aguirre:1999zn}.

Although the geometrical constraint imposed by accretion disks does not completely quench the instability, it can be argued that absorption effects at the mirror could~\cite{Aguirre:1999zn}. Consider an optimistic setup for which the electromagnetic wave is amplified by $\sim 1\%$ each time that it interacts with the BH~\cite{Teukolsky:1974yv}. A positive net gain only ensues if the wall has a reflection coefficient of $99\%$ or higher. 
On the other hand, this argument assumes that the mirror itself does not amplify the waves. But if it is rotating, it may also contribute to further amplification (an interesting example of amplification induced by a rotating cylinder is discussed in Ref.~\cite{Bekenstein:1998nt}).
Clearly, further and more realistic studies need to be made before any conclusion is reached about the effectiveness of ``BH bomb'' mechanisms in astrophysical settings.

Finally, we note that the Robin boundary conditions~\eqref{BC_bomb} that we found, are analogous to the ones found in Refs.\cite{Wang:2015fgp,Wang:2015goa} for electromagnetic perturbations in AdS, where a no-flux boundary condition was imposed, and to the ones in Ref.~\cite{Dias:2013sdc,Cardoso:2013pza} for gravitational perturbations in AdS. In both cases there are two families of boundary conditions, reminiscent of the two polarizations of electromagnetic and gravitational waves. These results also confirm the close similarity between small BHs in asymptotically AdS spacetimes and the BH-bomb scenario.

%%%%%%%%%%%%%%%%%%%%%%%%%%%%%%%%%%%%%%%%%%%%%%%%%%%%%%%%%%%%%%%%%%%%
\chapter{Superradiant instability of black holes immersed in a magnetic field}\label{chapter:magnetic}
%%%%%%%%%%%%%%%%%%%%%%%%%%%%%%%%%%%%%%%%%%%%%%%%%%%%%%%%%%%%%%%
\section{Introduction}
%%%%%%%%%%%%%%%%%%%%%%%%%%%%%%%%%%%%%%%%%%%%%%%%%%%%%%%%%%%%%%%
The existence of strong magnetic fields around astrophysical BHs is believed to be at the origin of some of the most energetic events of our Universe, such as the emission of relativistic jets. The Blandford-Znajek process is widely believed to be one leading mechanism at the origin of these phenomena~\cite{Blandford:1977ds}. This process allows the extraction of energy from a spinning BH due to the presence of a magnetic field supported by the material accreted by the BH.

A full understanding of the interactions between the accretion disk, the surrounding magnetic field and the BH is a complex problem and requires the use of sophisticated general-relativistic magnetohydrodynamic simulations (cf. e.g. Refs.~\cite{2012MNRAS.423.3083M,2013MNRAS.436.3741P}), nonetheless a qualitative picture can be drawn by studying stationary magnetized BH solutions in general relativity. For example, the approximate solution found by Wald~\cite{Wald:1974np}, which describes a Kerr BH immersed in a test uniform magnetic field aligned with the BH spin axis, has served as a model to understand the interaction of BHs with magnetic fields. Several remarkable phenomena --~such as charge induction~\cite{Wald:1974np} and a Meissner-like effect~\cite{PhysRevD.12.3037}~-- can be understood by studying this simple solution~\cite{Penna:2014aza}. 

In addition to the perturbative solution found by Wald, a class of \emph{exact} solutions of the Einstein--Maxwell equations, describing BHs immersed in a uniform magnetic field, was discovered by Ernst, who developed a powerful method to construct them starting from vacuum solutions of Einstein's equations~\cite{Ernst:1976BHM}. Although the Ernst spacetimes are not asymptotically flat, they can also be used to model the properties of BHs immersed in strong magnetic fields in a simple way. 

Even though the Wald and Ernst solutions were discovered 40 years ago, the dynamics of linear perturbations in these backgrounds is still largely unexplored.
One of the main motivations to study perturbations of magnetized rotating BHs is the possibility, first proposed by Galt'sov and Petukhov~\cite{Galtsov:1978ag}, that the magnetic field can trigger superradiant instabilities~\cite{Press:1972zz,Cardoso:2013krh}. 
As already mentioned, superradiant instabilities need essentially two ingredients to occur: (i) a monochromatic bosonic wave with low-frequency $\omega$ satisfying the superradiance condition~\eqref{eq:superradiance_condition}; and (ii) a mechanism to trap superradiant modes near the BH. The first condition allows the extraction of rotational energy from the BH, spinning it down, while the second condition is necessary to ``keep the extraction going,'' thereby triggering the instability.

Several confining mechanisms to trap the modes have been investigated, starting from an artificial mirror around the BH (the so-called ``BH bomb''~\cite{Press:1972zz,Cardoso:2004nk,Cardoso:2013krh} discussed in Chapter~\ref{sec:mirror}), to more natural ones like massive bosonic fields~\cite{Damour:1976kh,Zouros:1979iw,Detweiler:1980uk,Hod:2012px,Herdeiro:2014goa}, where the mass term plays the role of the mirror (discussed in Chapter~\ref{chapter:Kerr}), or the asymptotically AdS spacetime, where the AdS boundary confines the perturbations inside the bulk~\cite{Cardoso:2004hs,Cardoso:2006wa,Dias:2013sdc,Cardoso:2013pza}.

Magnetic fields can confine the radiation in a similar way. Working in a $B r\ll 1$ expansion (with $B$ being the magnetic field strength and $r$ the radial coordinate, both in geometric units), Refs.~\cite{Galtsov:1978ag,Konoplya:2007yy,Konoplya:2008hj} showed that a scalar field propagating on the Ernst background is equivalent to a massive scalar perturbation propagating on a Schwarzschild or Kerr metric with an effective mass $\mu_{\mathrm{eff}}=B m$, with $m$ the azimuthal number. As such, the magnetic field triggers the same superradiant instability associated to massive fields. However, such approximation becomes inaccurate at distances comparable to or larger than $\sim 1/B$. As we show, this profoundly affects the dynamics of the perturbations, because the spectrum is defined by physically-motivated boundary conditions imposed at large distances $r\gg 1/B$.

In this Chapter we take a step further to understand how strong magnetic fields affect BH spacetimes. We study scalar perturbations of the Ernst solutions with no approximation for the first time. We show that magnetized BHs can indeed support superradiant unstable modes and that this instability can be orders of magnitude stronger than the one estimated using the approximation of Refs.~\cite{Galtsov:1978ag,Konoplya:2007yy,Konoplya:2008hj} in terms of an effective mass $\mu_{\mathrm{eff}}=Bm$. In the exact case, the perturbation equations do not seem to be separable and this prevents the use of most methods to compute quasinormal modes (QNMs) (see~\cite{Kokkotas:1999bd,Nollert:1999ji,Berti:2009kk} for reviews). We circumvent this problem using powerful techniques developed in the past few years (see e.g.~\cite{Pani:2013pma}), which allow us to solve the full linearized dynamics for any value of $B$.

%%%%%%%%%%%%%%%%%%%%%%%%%%%%%%%%%%%%%%%%%%%%%%%%%%%%%%%%%%%%%%%%%%%%%%%%%%%%%%%%%
\section{The Melvin spacetime and its normal oscillation modes}\label{sec:Melvin}
%%%%%%%%%%%%%%%%%%%%%%%%%%%%%%%%%%%%%%%%%%%%%%%%%%%%%%%%%%%%%%%%%%%%%%%%%%%%%%%%%
We start by studying the geodesic motion and the normal modes of the Melvin spacetime, which are instructive to understand the QNMs of a BH immersed in a magnetic field. In cylindrical coordinates the Melvin metric is given by~\cite{Melvin:1963qx}
\be
\label{Melvin}
ds^2=\Lambda_M^2\left(-dt^2+d\rho^2+dz^2\right)+\frac{\rho^2}{\Lambda_M^2}d\phi^2\,,
\ee
where $\Lambda_M=1+B^2\rho^2/4$. This solution describes a uniform magnetic field aligned along the $z$-axis. 

Let us start with a brief geodesic analysis of the metric~\eqref{Melvin}. Staticity and axial symmetry of the metric imply the existence of a conserved energy $E$ and angular momentum parameter $L$, defined as
\begin{equation}
\Lambda_M^2\dot{t}=E\,, \qquad \frac{\rho^2}{\Lambda_M^2}\dot{\phi}=L\,,
\end{equation}
where a dot stands for derivative with respect to an affine parameter. Null particles then obey the equation
\be
\dot{\rho}^2=V_\rho\equiv \frac{E^2}{\Lambda_M^4}-\frac{L^2}{\rho^2}\,,
\ee
or simply
\be
\left(\frac{d\rho}{dt}\right)^2=1-\frac{\Lambda_M^4\left(L/E\right)^2}{\rho^2}\,.
\ee
Circular ($V_\rho=dV_\rho/d\rho=0$) geodesics for massless particles are only possible for $\rho^2=4/(3B^2)$ and correspond to an angular frequency 
\be
\Omega\equiv \frac{d\phi}{dt}=\frac{16\sqrt{3}\,B}{9\sqrt{4}}\sim 1.5396 B \label{eq:nullgeodesics}\,.
\ee

In the geometric-optics regime, normal modes with $m\gg 1$ in the Melvin spacetime are expected~\cite{Cardoso:2008bp} to reduce to the geodesic result described by Eq.~\eqref{eq:nullgeodesics}, i.e,
\be
\omega_{\mathrm{normal}}=m\Omega=\frac{16\sqrt{3}\,mB}{9\sqrt{4}}\sim 1.5396\,mB\,.\label{eq:nullgeodesics2}
\ee

Let us now find the normal modes of a probe scalar field propagating in the Melvin metric~\eqref{Melvin}. The Klein-Gordon equation for a massless field has the form
\be\label{KG}
\Box\Phi\equiv\frac{1}{\sqrt{-g}}\left(g^{\mu\nu}\sqrt{-g}\Phi_{;\mu}\right)_{;\nu}=0\,.
\ee
By making the following ansatz for the scalar field
\be
\Phi(t,\rho,z,\phi)=\frac{Q(\rho)}{\sqrt{\rho}} e^{i k z}e^{i m \phi} e^{-i \omega t}\,,
\ee
the Klein-Gordon equation \eqref{KG} reads
\be\label{melvinKG}
Q''(y)+\left[{\tilde{\omega}}^2-\frac{m^2 \left(y^2+4\right)^4-64}{256 y^2}\right]Q(y)=0\,,
\ee
where ${\tilde{\omega}}^2=(\omega^2-k^2)/B^2$ and $y=B\rho$. Note that with these redefinitions the magnetic field $B$ scales out of the problem. Furthermore, the eigenvalue problem is invariant under $m\to -m$, $\omega\to -\omega$, $k\to -k$ and $Q(y)\to Q^*(y)$ so we consider only modes with $m>0$.

After imposing appropriate boundary conditions, Eq.~\eqref{melvinKG} defines a boundary value problem that admits normal modes. Near the origin the solution behaves as
\begin{equation}\label{or_melvin}
 Q(y)\sim A_1 y^{m+1/2}+A_2 y^{-m+1/2}\,,
\end{equation}
%%%
and regularity at the origin imposes $A_2=0$. The asymptotic behavior at infinity is given by
\begin{equation}
 Q(y)\sim y^{-3/2}\left[C e^{y^4m/64}+ D e^{-y^4m/64}\right]\,, \label{inf_melvin}
\end{equation}
and the only acceptable physical solution corresponds to $C=0$. 

To find the normal frequencies of this spacetime we integrate numerically Eq.~\eqref{melvinKG} starting from the boundary condition~\eqref{or_melvin} and imposing $C=0$ in the asymptotic solution~\eqref{inf_melvin}. This selects a discrete spectrum of frequencies which are summarized in Tables~\ref{tab:spectrum_melvin1} and~\ref{tab:spectrum_melvin2}.
\begin{table}[t]
% \scriptsize
\centering \caption{Scalar normal modes of the Melvin spacetime for $m=1$ and different overtone number $n$.} 
\vskip 12pt
\begin{tabular}{@{}ccccccccccccccccccccc@{}}
\hline \hline
\multicolumn{7}{c}{${\tilde{\omega}}$}\\ 
\hline
$n=0$      & $n=1$   & $n=2$    & $n=3$   &$n=4$    & $n=5$  & $n=6$\\
\hline 
2.04862    & 2.91334 & 3.68457  & 4.39629	&5.06541  &5.70187 & 6.31212\\
\hline \hline
\end{tabular}
\label{tab:spectrum_melvin1}
\end{table}

\begin{table}[t]
% \scriptsize
\centering \caption{Fundamental ($n=0$) scalar normal modes of the Melvin spacetime for different azimuthal number $m$.} 
\vskip 12pt
\begin{tabular}{@{}ccccccccccccccccccccc@{}}
\hline \hline
\multicolumn{6}{c}{${\tilde{\omega}}$}\\ 
\hline
$m=1$      & $m=2$   & $m=3$    & $m=4$   &$m=5$    & $m=6$  \\
\hline 
2.04862    & 3.59874 &5.14195   &	6.68336 &8.22404  &	9.76436\\
\hline \hline
\end{tabular}
\label{tab:spectrum_melvin2}
\end{table}
The most important points to retain from these results are that (i) Melvin spacetimes are (marginally) stable and are described by a set of {\it normal} modes; (ii)
for large $m$ our results are well consistent with the expansion ${\tilde{\omega}}=1.5396m+ 0.5301-0.02113/m$, in excellent agreement with the geodesic analysis in Eq.~\eqref{eq:nullgeodesics2}.

At this point it is important to stress that these modes only exist due to the behavior of the metric at $\rho\to\infty$, which is not asymptotically flat. Indeed, considering $y\equiv B\rho\ll 1$ and neglecting terms at $\mathcal{O}(y^2)$, we find that Eq.~\eqref{melvinKG}  describes a scalar field propagating in Minkowski spacetime with effective mass $\mu_{\mathrm {eff}}=mB$. A free massive field in flat space does not form stationary bound states and, therefore, no modes are predicted for the Melvin spacetime within this approximation. These modes solely exist due to the boundary condition~\eqref{inf_melvin} imposed by the magnetic field at large distances, $B\rho\gg1$. The situation is analogous to what happens in AdS spacetime. Normal modes exist in pure AdS space due to the timelike boundary at spatial infinity, which allows null rays to reach the boundary in a finite time and be reflected back. In this case the AdS radius selects the frequencies of these modes. In the same way, in the Melvin spacetime perturbations are confined by the magnetic field which behaves like an infinite ``wall'' at a radius $r_0\sim 1/B$. This allows for the existence of a discrete set of normal modes. As we discuss in the next sections, such modes would be missed by a perturbative analysis similar to what was done in Refs.~\cite{Galtsov:1978ag,Konoplya:2008hj,Konoplya:2007yy}, where QNMs of a BH immersed in the Melvin universe have been computed perturbatively to order $B^2$.

%%%%%%%%%%%%%%%%%%%%%%%%%%%%%%%%%%%%%%%%%%%%%%%%%%%%%%%%%%%%%%%%%%%%%%%%%%
\section{The linear stability of the Ernst spacetime}\label{sec:Ernst}
%%%%%%%%%%%%%%%%%%%%%%%%%%%%%%%%%%%%%%%%%%%%%%%%%%%%%%%%%%%%%%%%%%%%%%%%%%
%%%%%%%%%%%%%%%%%%%%%%%%%%%%%%%%%%%%%%%%%%%%%%%%%%%%%%%%%%%%%%%%%%%
\subsection{The Ernst background spacetime}
%%%%%%%%%%%%%%%%%%%%%%%%%%%%%%%%%%%%%%%%%%%%%%%%%%%%%%%%%%%%%%%%%%%
In 1976 Ernst found a class of exact BH solutions of the Einstein--Maxwell equations immersed in the Melvin spacetime~\cite{Ernst:1976BHM}. The simplest of these solutions corresponds to a magnetized Schwarzschild BH, also known as the Ernst metric, which is given by
\be\label{Ernst}
ds^2=\Lambda_M^2\left(-f(r)dt^2+\frac{dr^2}{f(r)}+r^2d\theta^2\right)+\frac{r^2\sin\theta^2}{\Lambda_M^2}d\phi^2\,,
\ee
where $f(r)=1-\frac{2M}{r}$. In the limit $M\to 0$ this metric reduces to the Melvin solution~\eqref{Melvin}, with $\rho=r \sin\theta$ and $z=r \cos\theta$, while in the limit $B\to 0$ it reduces to the standard Schwarzschild solution. Due to the presence of the magnetic field this spacetime is not asymptotically flat, but instead approaches the Melvin metric as $r/M\to\infty$. The vector potential giving rise to the homogeneous magnetic field reads
\be
A_{\mu}dx^{\mu}=-\frac{B r^2 \sin^2\theta}{2\Lambda_M}d\phi\,.
\ee

The event horizon is located at $r_H=2M$ and its area is given by $A_H=4\pi r_H^2$, as in the Schwarzschild BH, but due to the $\theta$--dependence of the $g_{\phi\phi}$ component, the horizon takes the form of a cigar-shaped object~\cite{Wild:1980zz}. However the magnetic field only starts to distort significantly the spacetime at distances of the order of $B^{-1}$.

%%%%%%%%%%%%%%%%%%%%%%%%%%%%%%%%%%%%%%%%%%%%%%%%%%%%%%%%%%%%%%%%%%%
\subsection{Linearized analysis}
%%%%%%%%%%%%%%%%%%%%%%%%%%%%%%%%%%%%%%%%%%%%%%%%%%%%%%%%%%%%%%%%%%%
In a Melvin background the scalar field equation can be separated using cylindrical coordinates. However, due to the presence of the BH, cylindrical symmetry is lost in the Ernst metric, making the separation of the radial and angular part apparently impossible. Nevertheless we can use the method discussed in~\cite{Dolan:2012yt} to separate the equation at the expense of introducing couplings between different modes (see also Ref.~\cite{Pani:2013pma} for a review).

We begin by splitting the angular and radial dependence of the field as
\be
\Phi(t,r,\theta,\phi)=\sum_{jm}\frac{Q_j(r,t)}{r}Y_{jm}(\theta,\phi)\,,
\ee
where $Y_{jm}(\theta,\phi)$ denotes the usual spherical harmonics. 
Because the background is axisymmetric, the eigenfunctions are degenerate in the azimuthal number $m$.
Inserting the ansatz above in the Klein-Gordon equation~\eqref{KG} and considering the background \eqref{Ernst}, we find
\be\label{KG_ernst_1}
\sum_{jm} Y_{jm}(\theta,\phi)\left[\frac{d^2Q_j}{dr^2_*}-\frac{d^2Q_j}{dt^2}-V_{\mathrm{eff}}(r,\theta) Q_j\right]=0\,,
\ee
where $r_*$ is the tortoise coordinate, defined via $dr/dr_*=f$, and
\begin{align}\label{potential_Ernst}
&V_{\mathrm{eff}}(r,\theta)=f(r)\left\{\frac{j(j+1)}{r^2}+\frac{2M}{r^3}
+\frac{B^2 m^2}{256}\left[\left(B^2 r^2+8\right)
\left(B^4 r^4+8 B^2 r^2+32\right) \right.\right.\nn\\
&\left.\left.- B^2 r^2 \left(3 B^4 r^4+32 B^2 r^2+96\right) \cos ^2{\theta}
+B^4 r^4 \left(3 B^2 r^2+16\right) \cos ^4\theta  -B^6 r^6 \cos ^6\theta \right]\right\}\,.
\end{align}

Since the spacetime is not spherically symmetric, the angular and radial parts of the Klein-Gordon equation cannot be separated using a basis of spherical harmonics. Nonetheless, the problem can be reduced to a $1+1$--problem using the fact that terms with $\cos^n\theta$ lead to couplings between different multipoles~\cite{Dolan:2012yt}. To show this, we first multiply Eq.~\eqref{KG_ernst_1} by $Y_{lm}^*(\theta,\phi)$ and integrate over the sphere. Then, making use of the fact that the Clebsch-Gordan coefficients,
\be
c^{(n)}_{jlm}\equiv\left\langle l m\left|\cos^n\right|j m\right\rangle\,,
\ee
are zero unless $j=l$ or $j=l-n,....,l+n$,
we finally arrive at the following equation
\begin{align} \label{KG_ernst}
&\frac{d^2Q_l(r,t)}{dr^2_*}-\frac{d^2Q_l(r,t)}{dt^2}-\sum_{i=-3}^{3}V_{l+2i} Q_{l+2i}(r,t)=0\,,
\end{align}
where the radial potentials read
%
% \begin{widetext}
%
\beq
V_{l}&=&f\left\{\frac{l(l+1)}{r^2}+\frac{2M}{r^3}
+\frac{B^2 m^2}{256}\left[\left(B^2 r^2+8\right) \left(B^4 r^4+8 B^2 r^2+32\right)\right.\right.\nn\\
&&-B^2 r^2\left(B^4 r^4 c^{(6)}_{ll}-B^2 r^2 \left(3 B^2 r^2+16\right) c^{(4)}_{ll}\right.\nn\\
&&\left.\left.\left.+\left(3 B^4 r^4+32 B^2 r^2+96\right) c^{(2)}_{ll}\right)\right]\right\}\,, \\
V_{l\pm 2}&=&f\left[-\frac{B^4 m^2}{256}  r^2 \left(B^4 r^4 c^{(6)}_{l\pm 2l}-B^2 r^2 \left(3 B^2 r^2+16\right) c^{(4)}_{l\pm 2l}\right.\right.\nn\\
&&\left.\left.+\left(3 B^4 r^4+32 B^2 r^2+96\right)   c^{(2)}_{l\pm 2l}\right)\right]\,,\\
V_{l\pm 4}&=&f\left[\frac{B^6 m^2}{256}  r^4 \left(\left(3 B^2 r^2+16\right) c^{(4)}_{l\pm 4l}-B^2 r^2 c^{(6)}_{l\pm 4l}\right)\right]\,,\\
V_{l\pm 6}&=&-f\left[\frac{B^8 m^2}{256}  r^6 c^{(6)}_{l\pm 6l}\right]\,,
\eeq
%
% \end{widetext}
%
where for ease of notation we have suppressed the index $m$ of the Clebsch-Gordan coefficients, but it is understood that the latter depend also on $m$.

This system of equations admits long-lived modes. To find them we can either evolve the system in time (as discussed in Sec.~\ref{sec:time2} below) or compute them in the frequency domain. In the frequency domain we consider the following time dependence for the field:
\be
Q_j(r,t)=Q_j(r)e^{-i\omega t}\,.
\ee
Imposing regularity boundary conditions at the horizon and at infinity, Eq.~\eqref{KG_ernst} defines an eigenvalue problem for the complex frequency $\omega=\omega_R+i\omega_I$. The eigenfrequencies are also termed the QNM frequencies and form a discrete spectrum~\cite{Kokkotas:1999bd,Nollert:1999ji,Berti:2009kk}, which generically depends on $m$, $B$ and on the overtone number $n$. Since the presence of the magnetic field breaks the spherical symmetry of the Schwarzschild background, the harmonic index $l$ is not a conserved ``quantum number'' and, for a given $m$, Eq.~\eqref{KG_ernst} effectively describes an \emph{infinite} system of equations where all the eigenfunctions $Q_j$ ($j=0,1,2,...m$) are coupled together.

It is straightforward to show that at the horizon the system decouples and regularity requires purely ingoing waves,
\be\label{hor_ernst}
Q_l(r)\sim e^{-i\omega r_*}\,,\quad r\to r_H\,.
\ee

The behavior at infinity is more intricate since different multipoles are coupled. However this is a difficulty introduced by the spherical coordinates. Expanding the potential~\eqref{potential_Ernst} at infinity and defining $\rho=r\sin\theta$, we easily see that the asymptotic solution reduces to~\eqref{inf_melvin}. We can then use standard methods for systems of coupled equations (see e.g Refs.~\cite{Rosa:2011my,Pani:2012bp,Brito:2013wya} and the review~\cite{Pani:2013pma}) to find the QNM frequencies. Since the full system~\eqref{Ernst} contains an infinite number of equations, in practice we must truncate the series at some given $L$, i.e. we assume $Q_j\equiv0$ when $j>L$. Convergence is then checked by increasing the truncation order. The results shown have converged to the number of digits displayed and have been obtained with two different methods, a ``direct integration'' and a ``Breit-Wigner'' approach~\cite{Pani:2013pma}.

%%%%%%%%%%%%%%%%%%%%%%%%%%%%%%%%%%%%%%%%%%%%%%%%%%%%%%%%%%%%%%%%%%%
\subsection{Results}
%%%%%%%%%%%%%%%%%%%%%%%%%%%%%%%%%%%%%%%%%%%%%%%%%%%%%%%%%%%%%%%%%%%
%
\begin{table}[t]
%%%%%%%%%%%%%%%%%
% \scriptsize
\centering \caption{Fundamental ($n=0$) QNMs of the Ernst BH solution computed in the frequency domain
for $l=m=1$ and different values of $B$.}
\vskip 12pt
%%%%%%%%%%%%%%%%%%%%%%
\begin{tabular}{@{}ccccccccccccccccccccc@{}}
\hline \hline
$BM$    & $M\omega_R$ & $-M\omega_I$\\
\hline
$0.025$	&	$0.0510$   & $1.2 \cdot 10^{-8}$\\
$0.050$	&	$0.1002$   & $6.7 \cdot 10^{-7}$\\
$0.075$	&	$0.1473$   & $9.1 \cdot 10^{-6}$\\
$0.100$	&	$0.1919$   & $7.0 \cdot 10^{-5}$\\
$0.125$	&	$0.2337$   & $3.7 \cdot 10^{-4}$\\
$0.150$	&	$0.2721$   & $1.4 \cdot 10^{-3}$\\
\hline
\hline
\end{tabular}
\label{tab:Ernst_spectrum1} 
\end{table}

We have performed a detailed numerical analysis of the scalar eigenfrequencies of the Ernst BH as functions of $B$, $m$ and overtone number $n$. Some results are shown in Tables~\ref{tab:Ernst_spectrum1} and~\ref{tab:Ernst_spectrum2}. Even though the background metric is not spherically symmetric, a notion of harmonic index $l$ is still meaningful. In the following we define a mode with given $(l,m)$ as the one corresponding to a set $Q_j$ (with $j=0,1,2,...m$) for which the eigenfunction $Q_l$ is the one with largest relative amplitude. Although this practical definition becomes ambiguous for large values of $B$, we find that such hierarchy in $l$ holds in a large region of the parameter space. For the same reason, the multipolar series in Eq.~\eqref{Ernst} converges rather fast in $L$, even for moderately large values of $B$.

For $BM\ll 1$, the real part behaves approximately as
\be\label{wreal}
\omega_R\sim \left[0.75n+1.2m+0.25l+0.7\right]B\,,
\ee
whereas we infer for the imaginary part a dependence of the form
\be\label{wim}
M\omega_I\sim -\gamma(BM)^{2l+3}\,,
\ee
where $\gamma$ is a numerical coefficient that depends on $n$, $l$ and $m$. For $l=m=1$, we find $\gamma\approx 2.2$ for $n=0$ and $\gamma\approx 9.3$ for $n=1$, respectively.
\begin{table}[t]
%%
% \scriptsize
\centering \caption{Quasinormal modes of the Ernst spacetime computed in the frequency domain for $l=m=1$, $B M=0.1$ and different overtone number $n$.}
\vskip 12pt
\begin{tabular}{@{}ccccccccccccccccccccc@{}}
\hline \hline
$n$ & $M\omega_R$ & $-M\omega_I$\\
\hline
$0$	&	$0.1919$   & $7.0 \cdot 10^{-5}$\\
$1$	&	$0.2674$  & $7.5 \cdot 10^{-4}$\\
$2$	&	$0.3213$  & $2.9 \cdot 10^{-3}$\\
$3$	&	$0.3656$  & $4.5 \cdot 10^{-3}$\\
\hline
\hline
\end{tabular}
\label{tab:Ernst_spectrum2} 
\end{table}
%%%%%%%%%%%%%%%%%%%%%%%%%%%%%%%%%%%

%%%%%%%%%%%%%%%%%%%%%%%%%%%%%%%%%%%%%%%%%%%%%%%%%%%%%%%%%%%%%%%%%%%%%%
\subsection{Relation with previous results in the literature}
%%%%%%%%%%%%%%%%%%%%%%%%%%%%%%%%%%%%%%%%%%%%%%%%%%%%%%%%%%%%%%%%%%%%%%
The behavior~\eqref{wreal} and~\eqref{wim} is different from the results of Refs.\cite{Galtsov:1978ag,Konoplya:2007yy,Konoplya:2008hj}. The approximation employed in these works changes the asymptotic behavior at infinity in such a way that the only role of the external magnetic field is to introduce an effective mass, $\mu_{\mathrm{eff}}=Bm$, for the scalar field. Consequently, the QNM spectrum was found to be equivalent to that of massive scalar perturbations of a Schwarzschild BH. Massive fields admit quasibound state modes with a hydrogenic spectrum~\cite{Detweiler:1980uk,Dolan:2007mj} 
\begin{equation}
 \omega_R^{\mathrm{mass}}\sim\mu_{\mathrm{eff}}\,,\qquad M\omega^{\mathrm{mass}}_I\sim -(M\mu_{\mathrm{eff}})^{4l+6}\,. \label{wmass}
\end{equation}
%%%
While the real part is consistent with the exact behavior~\eqref{wreal}, the scaling of the imaginary part with $B$ is different from Eq.~\eqref{wim}. 

Indeed, solving the full system~\eqref{KG_ernst}, we find that the QNM spectrum has the same qualitative behavior as the modes of a small BH in AdS~\cite{Cardoso:2004hs} or of a BH inside a mirror~\cite{Cardoso:2004nk}.  This is not surprising since the asymptotic behavior plays a crucial role in defining the eigenfrequencies. In fact these frequencies are supported by an effective ``wall'' created by the magnetic field at $r_0\sim 1/B$. In the AdS and in the mirror cases the mirror is given by the AdS radius and by the mirror radius, respectively. We can therefore understand the modes of the Ernst BH as being a small correction to the modes of the Melvin spacetime. The BH event horizon behaves like a perfect absorber~\cite{Thorne:1986iy} and its role is mostly to change one of the boundaries, leading to the slow decay of the field~\footnote{Strictly speaking, since the regular behavior of Eq.~\eqref{inf_melvin} is a damped exponential, such modes could be dubbed as \emph{quasibound} states, in analogy to the case of massive fields~\cite{Dolan:2007mj,Pani:2012vp,Brito:2013wya}. However, in the case of the Ernst solution these are the \emph{only} eigenfrequencies that solve the exact problem and the distinction with the QNMs is irrelevant.}. This is consistent with qualitative picture given in the subsection~\ref{sec:model} [cf. Eq.~\eqref{absor_wIm}]. 

The behavior predicted by the model in subsection~\ref{sec:model} is different from the case of massive perturbations, because the latter can only confine low-frequency modes with $\omega_R\lesssim \mu$ (where $\mu$ is the mass of the field), whereas an ``effective box'' confines radiation of any frequency. From a mathematical viewpoint, this property requires different boundary conditions for the perturbations at infinity.

Therefore, our analysis gives an explicit example of a very natural fact: the eigenvalue spectrum of a given metric is highly sensitive to the asymptotic behavior of the spacetime. Any approximation that changes the boundary conditions might drastically affect the spectrum. Indeed, for the same multipole number $l$ the decay rate of a massive field [cf. Eq.~\eqref{wmass}] can be orders of magnitude smaller than the exact results given in Eq.~\eqref{wim}. Due to this difference, in Sec.~\ref{sec:Ernst_rot} we shall see that, when rotation is turned on, not only the exact modes of the Ernst BH become unstable due to superradiance, but also that the instability time scale can be orders of magnitude \emph{shorter} than that associated to a massive field~\cite{Cardoso:2005vk,Dolan:2007mj} --~and, consequently, the instability is stronger than that discussed in Ref.~\cite{Konoplya:2008hj}.

%%%%%%%%%%%%%%%%%%%%%%%%%%%%%%%%%%%%%%%%%%%%%%%%%%%%%%%%%%%%%%%%%%%%%%
\subsection{Time-domain analysis}\label{sec:time2}
\begin{figure*}[htb]
\begin{center}
\begin{tabular}{cc}
\epsfig{file=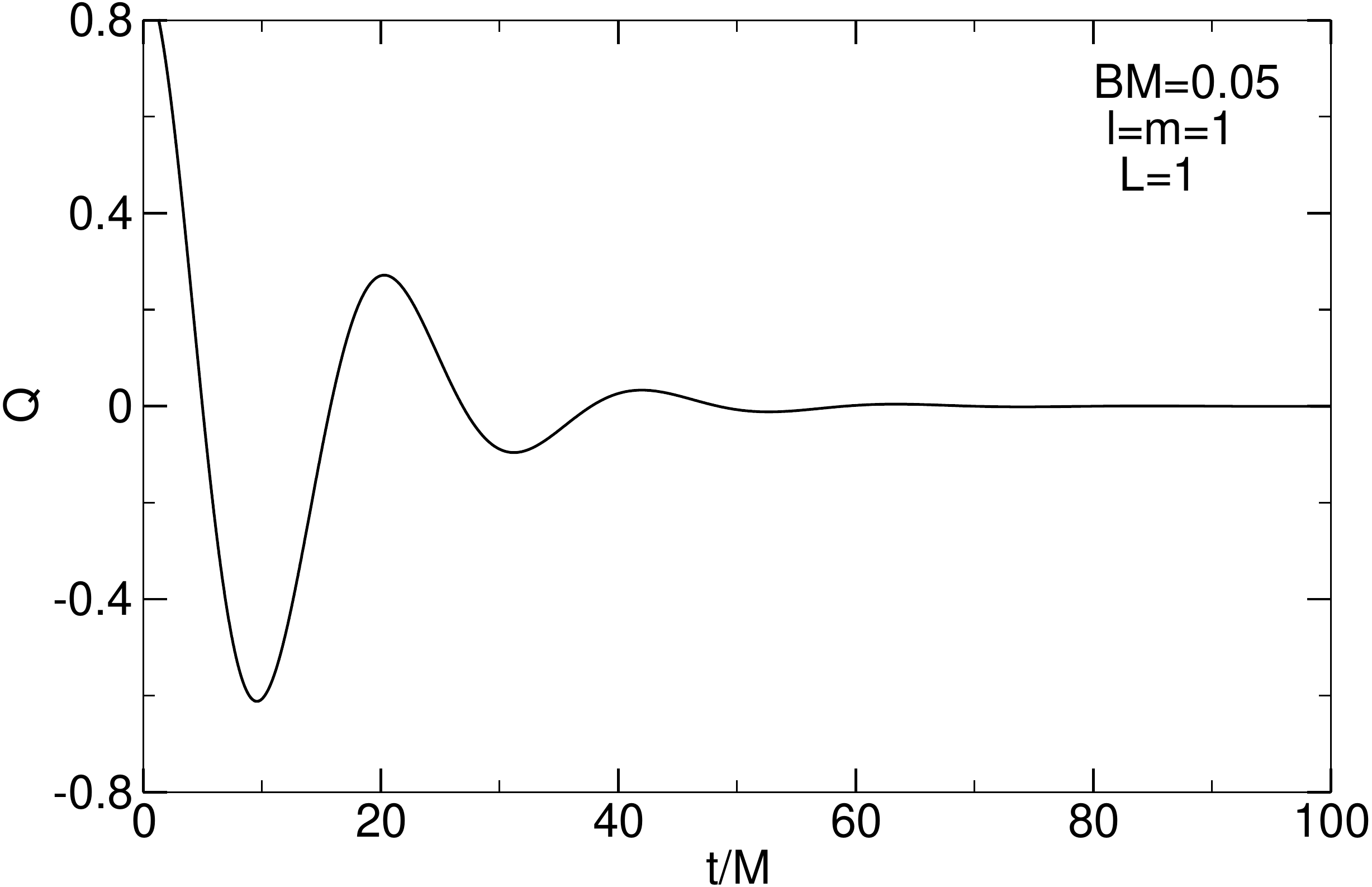,width=7.3cm,angle=0,clip=true}&
\epsfig{file=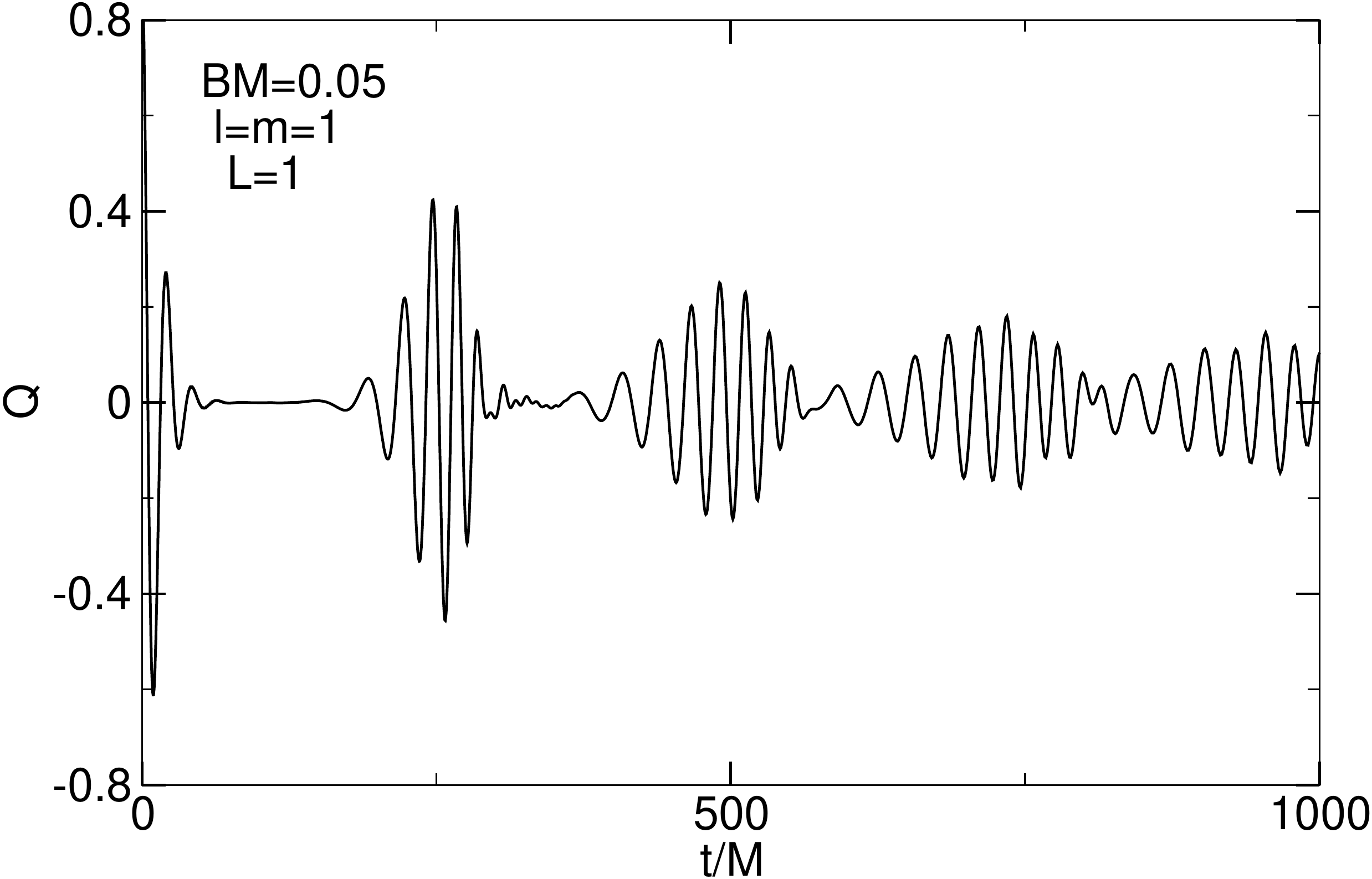,width=7.3cm,angle=0,clip=true}\\
\epsfig{file=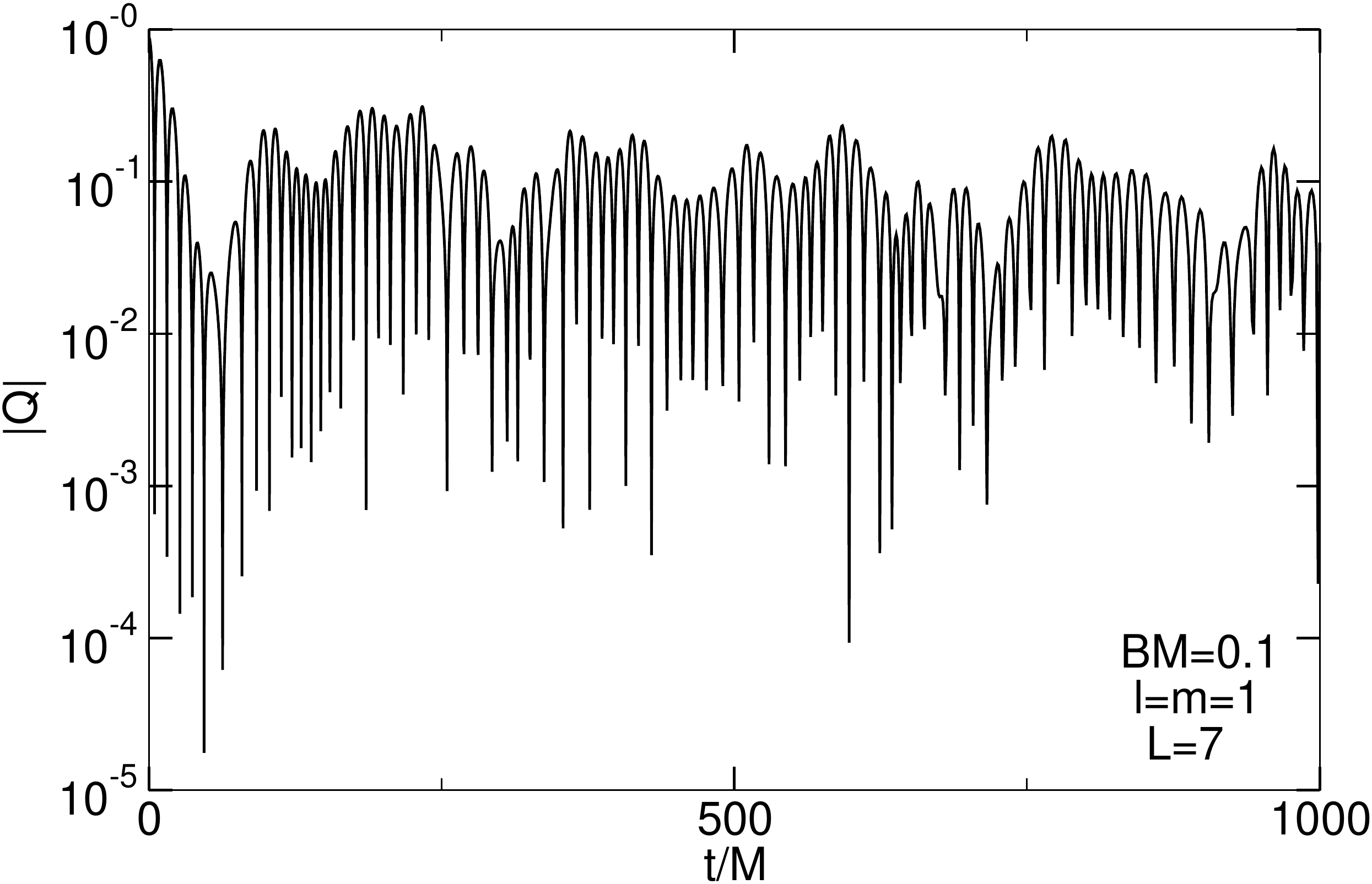,width=7.3cm,angle=0,clip=true}&
\epsfig{file=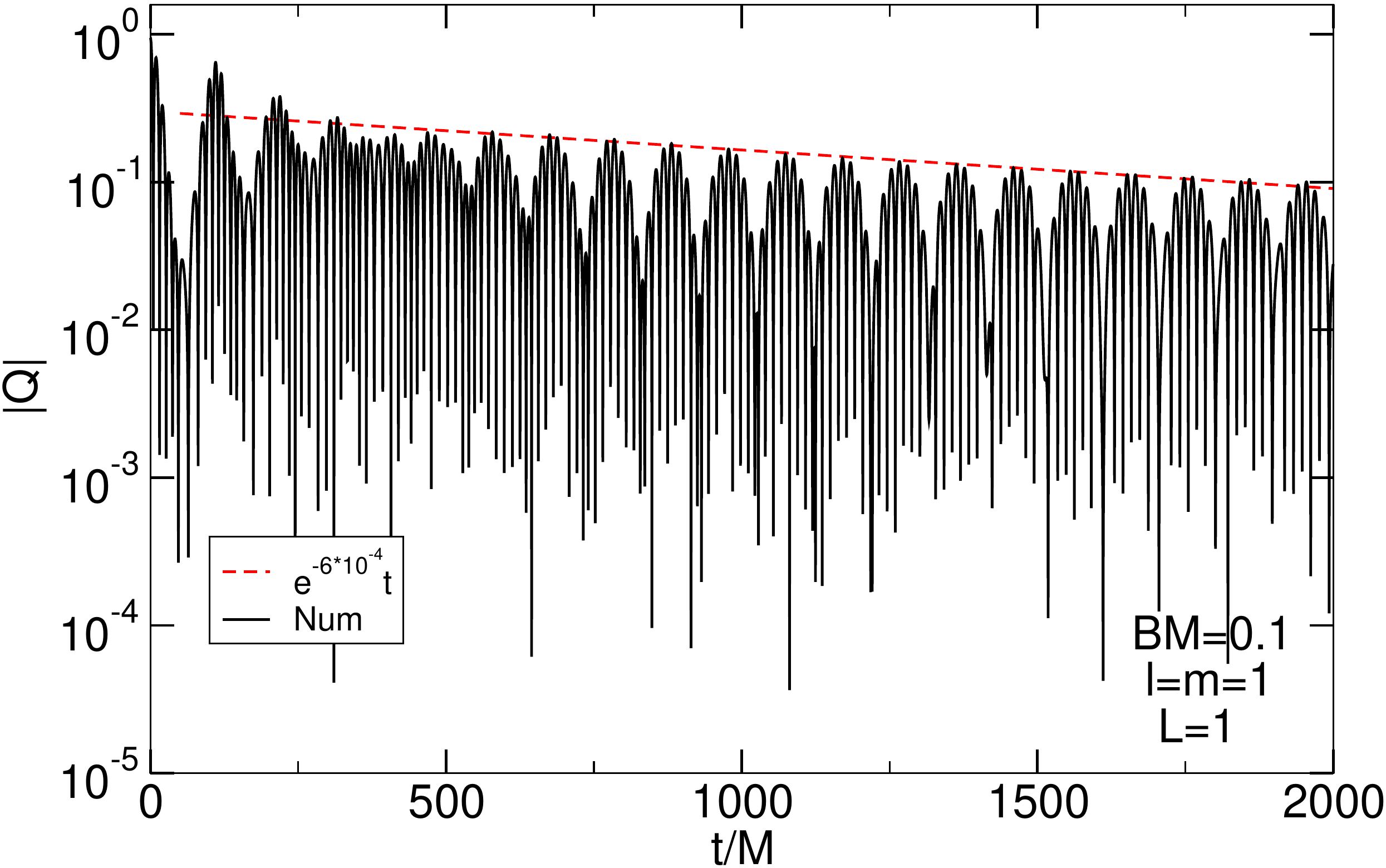,width=7.3cm,angle=0,clip=true}
\end{tabular}
\caption{Waveforms for a small Gaussian packet $Q_j(0,r)=\delta_{j1}\exp\left[(r_*-r_{c})^2/(2\sigma^2)\right]$ (with $\sigma=6M$ and $r_{c}=6M$ propagating on a Ernst spacetime for different values of $BM$. Top left: BH ringdown at early times for $BM=0.05$ and $l=m=1$. Top right: After a time $t\sim 1/B$, the ``Melvin-like'' modes are excited. Bottom panels: Waveform for $BM=0.1$ and $l=m=1$, truncating the series at $L=7$ (left panel) and $L=1$ (right panel). In the bottom right panel we also show the fit to the decay rate of the scalar field, showing good agreement with the frequency domain analysis.\label{Fig:waveform}}
\end{center}
\end{figure*}

%%%%%%%%%%%%%%%%%%%%%%%%%%%%%%%%%%%%%%%%%%%%%%%%%%%%%%%%%%%%%%%%%%%%%%
For completeness, in this section we discuss the results of a time-domain analysis of the system~\eqref{KG_ernst}.
Some examples of waveforms obtained in the time domain are shown in Fig.~\ref{Fig:waveform}. We consider an initial Gaussian wave packet $Q_j(0,r)=\delta_{j1}\exp\left[(r_*-r_{c})^2/(2\sigma^2)\right]$ with $\sigma=6M$ and $r_{c}=6M$, whose time evolution is governed by the system~\eqref{KG_ernst}. The discretization of spatial derivatives is performed using a 2nd order finite difference scheme and integration in time is done with a 4th order accurate Runge-Kutta method.

At early times and for small values of $B$, the waveform is dominated by some ringdown modes~\cite{Berti:2009kk} (top left panel of Fig.~\ref{Fig:waveform}). These ringdown frequencies are \emph{not} given by the modes previously computed, being in fact very similar to the Schwarzschild QNM frequencies~\cite{Berti:2009kk}.
{\it These} numbers are in very good agreement with what was found in Ref.~\cite{Konoplya:2007yy} and are indeed consistent with the fact that for $B M\ll 1$ the ringdown frequencies are almost unaffected by the long-range modification of the potential due to the magnetic field.  

However, we stress that such frequencies do not show up in the frequency domain analysis because they do not satisfy the asymptotic boundary conditions~\eqref{inf_melvin}. Indeed, after a time of order $t\sim 1/B$, the wave is reflected back by the effective wall at $r_0\sim 1/B$ with a smaller amplitude due to the absorption by the event horizon (top right panel of Fig.~\ref{Fig:waveform}). These reflections give rise to the QNM spectrum previously computed in the frequency domain and are related to the Melvin normal modes. Due to the couplings between different multipoles, various frequencies dominate the waveform, making it difficult to extract accurate information about these modes in the time domain.

The fact that the ringdown frequencies at intermediate times are different from the QNMs of the spacetime is another example of an interesting phenomena discussed in detail in Ref.~\cite{Barausse:2014tra} in the context of ``dirty'' BHs.

%
%%%%%%%%%%%%%%%%%%%%%%%%%%%%%%%%%%%%%%%%%%%%%%%%%%%%%%%%%%%%%%%%%%%%%%%%%%%%%%%%%%%%%%%%%%%%%%%%
\section{Superradiant instability of the magnetized Kerr-Newman solution}\label{sec:Ernst_rot}
%%%%%%%%%%%%%%%%%%%%%%%%%%%%%%%%%%%%%%%%%%%%%%%%%%%%%%%%%%%%%%%%%%%%%%%%%%%%%%%%%%%%%%%%%%%%%%%%
After having understood the QNMs of a nonrotating magnetized BH, we now turn our attention to the spinning case. A magnetized rotating BH is a complex object. For example, it was shown by Wald~\cite{Wald:1974np} that when a spinning neutral BH is immersed in a magnetic field it is energetically favorable for it to acquire a charge given by $q=-2\tilde{a} M^2 B$, where $q$ and $\tilde{a}$ correspond to the charge and rotation parameters of the unmagnetized Kerr-Newman solution. This result was established neglecting backreaction effects of the magnetic field onto the BH spacetime. Nonetheless, the result was quickly generalized, when Ernst and Wild found the first exact solution of a magnetized Kerr BH~\cite{Ernst:1976:KBH}. This solution was latter shown to suffer from conical singularities at the poles by Hiscock~\cite{Hiscock:1981np}, but he realized that this singularity could be removed by redefining the azimuthal angle $\phi$ (see Appendix~\ref{app:KN}).

Studying perturbations of the full magnetized Kerr-Newman solution (see e.g. Ref.~\cite{Gibbons:2013yq}) is a formidable task. However the problem becomes tractable if we consider an expansion in the rotation parameter $\tilde{a}$. In the following we will consider a slowly-rotating magnetized BH with Wald's charge, to second order in the spin. Note that the slow-rotation approximation of the linear perturbations is fully consistent to second or higher order in $\tilde{a}$, as discussed in Ref.~\cite{Pani:2012vp}.

The Klein-Gordon equation~\eqref{KG} on the slowly-rotating Kerr--Newman background is discussed in Appendix~\ref{app:KN}. The final result reads:
%%%%%%%%%%%%%%
\beq
&&\sum_l Y_{lm}(\theta,\phi)\left\{\frac{d^2Q_l(r)}{dr^2_*}+\left[\omega^2-V_{\mathrm{eff}}-\tilde{a}m\omega\,W\right]Q_l(r)\right.\nn\\
&&\left.+\tilde{a}^2\left[ \sum_{i=0}^4\mathcal{V}_{2i}(r)\cos^{2i}\theta\right] Q_l(r)\right\}=0\,, \label{KG_KN}
\eeq
%%%
where $V_{\mathrm{eff}}=V_{\mathrm{eff}}(r,\theta)$ is given by Eq.~\eqref{potential_Ernst}, the first-order function
%%%
\be
 W(r,\theta)=\frac{4M^2}{r^3}+\frac{8 M^2 B^2 }{r}-\frac{M^2 B^4}{4}(5r+22M)
-\frac{M^2 B^4}{4}\cos^2\theta(2+ \cos^2\theta)(r-2M)\,,
\ee
and the second-order radial coefficients $\mathcal{V}_i(r)$ are given in Appendix~\ref{app:KN}.
To separate this equation we use the same technique discussed in Sec.~\ref{sec:Ernst}, leading to an infinite set of radial equations with couplings between different multipoles up to $l\pm 8$.
%%%%%%%%%%%%%%

Defining a tortoise coordinate $dr/dr_*=F$ (where $F$ is a metric variable defined in Appendix~\ref{app:KN}), the purely ingoing wave condition at the horizon reads
\be\label{bchorizon}
Q_l\sim e^{-ik_{H} r_*}\,,\quad r\to r_+\,,
\ee
where $k_H=\omega-m\Omega_{\rm H}$, $r_+$ is the event horizon radius to second order in $\tilde{a}$ [cf. Eq.~\eqref{rp}], and 
%
%\be
$\Omega_{\rm H}=-\lim_{r\to r_+} g^{(0)}_{t\phi}/g^{(0)}_{{\phi\phi}}$,
%\ee
%
is the angular velocity at the horizon of locally nonrotating observers, with $g_{\mu\nu}^{(0)}$ being the background metric.

\begin{figure}[hbt]
\begin{center}
%\begin{tabular}{c}
\epsfig{file=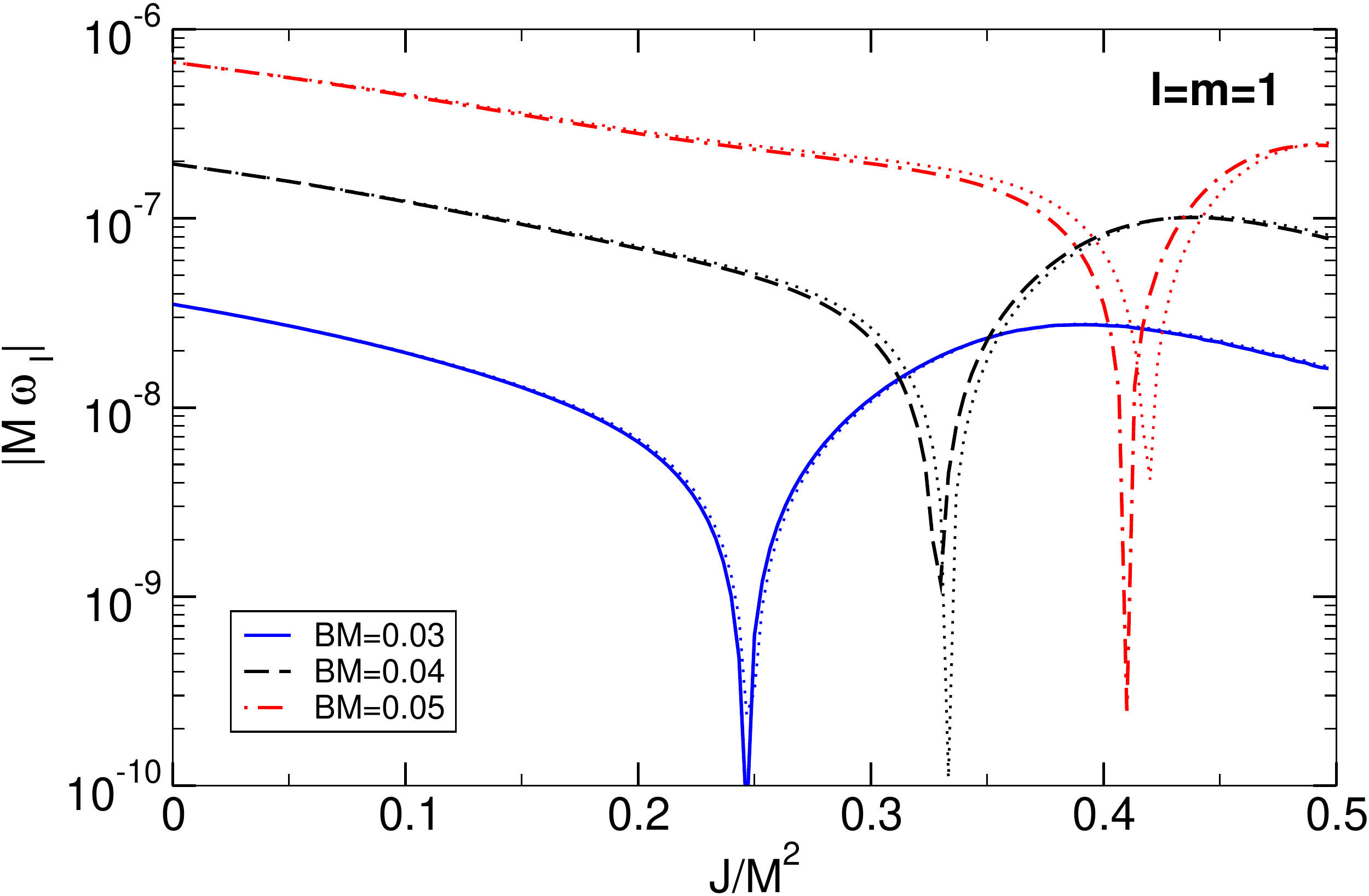,width=10cm,angle=0,clip=true}
%\end{tabular}
\caption{Imaginary part of the fundamental modes of a Kerr--Newman--Ernst BH with Wald's charge $q=-2JB$, computed at second order in the rotation and truncating the series at $L=9$, as a function of the BH rotation rate $\tilde{a}=J/M^2$, for $l=m=1$, and different values of the magnetic field. The dotted thinner lines correspond to a magnetized BH without charge. The only effect of the charge is to change the superradiance threshold.\label{Fig:super}}
\end{center}
\end{figure}

We have integrated the eigenvalue problem defined by Eq.~\eqref{KG_KN} numerically. A representative result is shown in Fig.~\ref{Fig:super} where we plot the imaginary part of the fundamental eigenvalue as a function of the BH spin $\tilde{a}\equiv J/M^2$ and for different values of $B$. As discussed in Appendix~\ref{app:KN}, the charge $q$ affects the superradiance threshold. Accordingly, the imaginary part crosses the axis when the superradiant conditions~\eqref{superwald} or~\eqref{supernocharge} are met, for a BH with $q=-2\tilde{a} M^2 B$ or $q=0$, respectively. Although not shown, the real part of the modes depends only mildly on the spin and it is well approximated by Eq.~\eqref{wreal}.

The results shown in Fig.~\ref{Fig:super} are obtained truncating the multipolar series at $L=9$, which guarantees convergence in the entire region of the parameter space under consideration.

In the limit $BM\ll 1$, one can estimate the instability time scale by considering modes~\eqref{wreal} and~\eqref{wim} in the nonrotating case and extrapolating the results to higher values of the spin. The same kind of extrapolation has been done in Refs.~\cite{Pani:2012bp,Brito:2013wya}, where it was found to be sufficiently accurate, for example it captures the onset of the instability and the order of magnitude of the time scale. 
This argument is further supported by the simple model we discussed in Sec.~\ref{sec:model}, and predicts that the imaginary part of the modes scales as
\be
\omega_I M\sim \gamma\left(\tilde{a}m-\frac{2\omega_R r_+}{1+8B^2M^2-16B^4M^4}\right)\left(BM\right)^{2(l+1)}\,.\label{wI_rot}
\ee
Because the time dependence of the perturbation is $\sim e^{i\omega_R t+\omega_I t}$, when the condition~\eqref{superwald} is satisfied $\omega_I>0$ and the perturbation grows exponentially in time.
In other words, as predicted in Sec.~\ref{sec:model}, rotating BHs in Melvin spacetimes are unstable, with an instability time scale given by $1/\omega_I$.

The estimate~\eqref{wI_rot} is in agreement with our numerical results to ${\cal O}(\tilde{a}^2)$. Although our analysis is perturbative in the spin, the results at order $\tilde{a}^2$ are found to be in remarkably good agreement with the exact ones for other systems~\cite{Pani:2012vp,Witek:2012tr}, suggesting that Eq.~\eqref{wI_rot} might be valid beyond its nominal regime of validity. In the next section we take Eq.~\eqref{wI_rot} as an order-of-magnitude estimate to discuss the astrophysical relevance of the superradiant instability triggered by an external magnetic field.

Finally we note that, unlike the case of a massive field, the fundamental mode ($n=0$) does not necessarily have the smallest instability time scale. In fact, the nonrotating results suggest that higher $n$ have larger imaginary parts (see Table~\ref{tab:Ernst_spectrum2}), which translates to a stronger instability in the spinning case. Nonetheless, due to the superradiant condition~\eqref{superwald} and the scaling of $\omega_R$ with $n$ given by Eq.~\eqref{wreal}, only the modes with small $n$ will be superradiant.

%%%%%%%%%%%%%%%%%%%%%%%%%%%%%%%%%%%%%%%%%%%%%%%%%%%%%%%%%%%%%%%%%%%%%%%%%%%%%%%%%%%%%%%%%%%%%%%%
\section{Astrophysical implications of the superradiant instability triggered by magnetic fields}
%%%%%%%%%%%%%%%%%%%%%%%%%%%%%%%%%%%%%%%%%%%%%%%%%%%%%%%%%%%%%%%%%%%%%%%%%%%%%%%%%%%%%%%%%%%%%%%%%  
To measure the strength of a magnetic field in an astrophysical context, we can define the characteristic magnetic field $B_M=1/M$ associated to a spacetime curvature of the same order of the horizon curvature. Restoring physical units, we obtain 
\be\label{magnetic}
B_M\sim 2.4\times 10^{19} \left(\frac{M_{\odot}}{M}\right) {\mathrm{Gauss}}\,.
\ee
The strongest magnetic fields around compact objects observed in the Universe are of the order of $10^{13}$--$10^{15} {\mathrm{Gauss}}$~\cite{McGill}. In natural units this corresponds to $B/B_M\sim 10^{-6}$--$10^{-4}$. However, $B_M$ is generically much larger than the typical magnetic field believed to be produced by accretion disks surrounding massive BHs. For supermassive BHs with $M\sim 10^9 M_\odot$ a magnetic field $B\sim 10^4 {\mathrm{Gauss}}\sim 10^{-6} B_M$ seems to be required to explain the observed luminosity of some active galactic nuclei, assuming a specific model for the interaction between the BH and the accretion disk~\cite{2010arXiv1002.4948P}. Likewise, the typical values of the magnetic field strength near stellar-mass BHs is estimated to be $B\sim 10^8 {\mathrm{Gauss}}\sim 10^{-10} B_M$. In other words, the magnetic field near massive BHs typically satisfy $B\ll B_M$. This justifies the small-$B$ estimates given in the previous sections but, on the other hand, it also implies that the superradiant instability time scale would typically be very long.
The purpose of this section is to quantify these statements and to investigate the (superradiant) instability triggered by uniform magnetic fields for astrophysical BHs.

In an astrophysical context our results should be taken with care. The Ernst metric is not asymptotically flat, since it describes a BH immersed in a magnetic field which is supported by some form of ``matter'' at infinity. In a realistic situation, the magnetic field is supported by an accretion disk. The Ernst metric therefore may be a relatively good approximation to the geometry of an astrophysical BH only up to a cutoff distance associated with the matter distribution. In other words, the characteristic length scale $r_0\sim 1/B$ should be smaller than the characteristic distance $r_M$ of the matter distribution around the BH. 
Considering that the accretion disk is concentrated near the innermost stable circular orbit, this would imply that our results can be trusted only when $r_0\lesssim r_M\sim 6M$, i.e. for $BM\gtrsim 0.1$. As we discussed above, this is a very large value for typical massive BHs. On the other hand, the Ernst metric is more accurate to describe configurations in which the disk extends much beyond the gravitational radius, as is the case in various models. In this case, however, the magnetic field will not be uniform and the matter profile has to be taken into account.  

Nevertheless, and since we wish to make a point of principle, we will use the results obtained in the previous sections for a Kerr BH immersed in a uniform magnetic field to predict interesting astrophysical implications.

\begin{figure}[t]
\begin{center}
%\begin{tabular}{c}
\epsfig{file=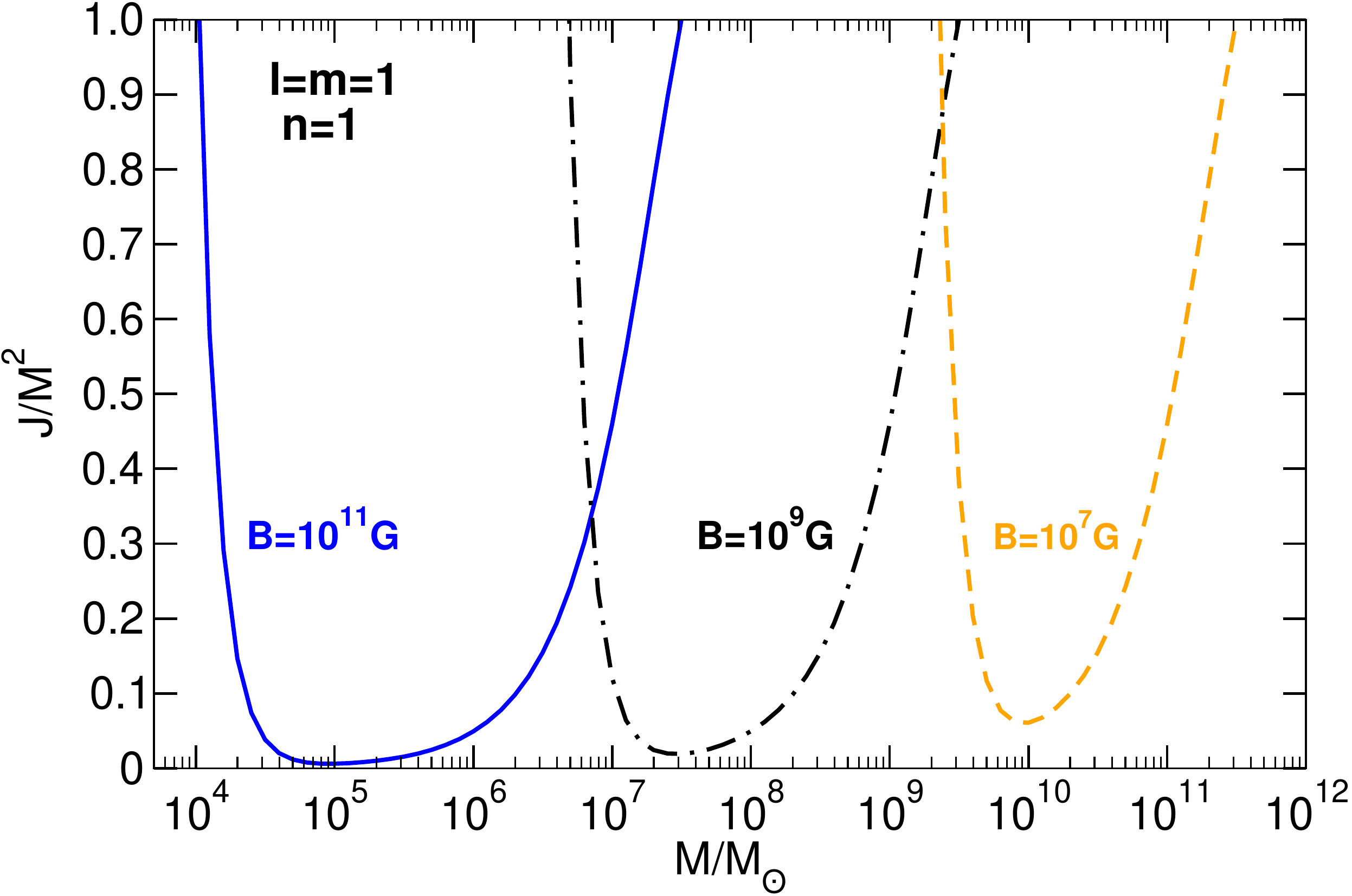,width=10cm,angle=0,clip=true}
%\end{tabular}
\caption{Contour plots in the BH Regge plane~\cite{Arvanitaki:2010sy} corresponding to an instability time scale shorter than $\tau_{\mathrm{Salpeter}}\sim 4.5\times 10^7{\mathrm{yr}}$ for different values of the magnetic field strength $B$
for modes with $l=m=n=1$. BHs lying above each of these curves would be unstable on an observable time scale. The threshold lines are obtained using Eq.~\eqref{wI_rot} in the range $10^{-4}\lesssim BM\lesssim 0.2$.\label{Fig:Regge}}
\end{center}
\end{figure}
As a result of the superradiant instability, the energy density of the radiation in the region $r\lesssim1/B$ would grow in time at the expense of the BH angular momentum. Therefore, the most likely end state of the instability is a spinning BH with dimensionless spin parameter slightly below the superradiant threshold\footnote{Note that in the case of the full Ernst metric, since radiation cannot escape, the end state is most likely similar to the one in AdS, a rotating BH in equilibrium with the outside radiation~\cite{Cardoso:2006wa}.}. 
This implies an upper limit on the spin of magnetized BHs which depends on the magnetic field, but it is certainly lower than the Kerr bound $\tilde{a}<1$. 
However, this argument remains valid only if the instability extracts the BH angular momentum at higher rate than any possible spin-up effect. For supermassive BHs, the most efficient mechanism to increase the BH spin is prolonged accretion. Therefore, to produce observable effects, the superradiance instability time scale should be shorter than the typical accretion time scale. For accretion at the Eddington rate, the typical time scale is the Salpeter time, $\tau_{\mathrm{Salpeter}}\sim 4.5\times 10^7{\mathrm{yr}}$.

This type of argument, together with supermassive BH spin measurements (cf. e.g. Refs.~\cite{Brenneman:2011wz,2013Natur.494..449R}), was used to impose stringent constraints on the allowed mass range of axionic~\cite{Arvanitaki:2010sy}, massive vector~\cite{Pani:2012bp,Pani:2012vp} and massive tensor~\cite{Brito:2013wya} fields, as we will discuss in more detail in the next two Chapters. Likewise, one could use spin measurements of supermassive BHs to impose constraints on the allowed range of the magnetic field strength. 
In Fig.~\ref{Fig:Regge} we show the spin--mass diagram (so-called BH Regge plane~\cite{Arvanitaki:2010sy}) with contour curves corresponding to an instability time scale $1/\omega_I$ of the order of the Salpeter time. For a given magnetic field $B$, BHs lying above the corresponding threshold curve would be unstable on an observable time scale. 

Spin measurements of supermassive BHs would allow us to locate data points on the Regge plane, thus excluding a whole range of possible magnetic fields. 
Since the contours extend almost up to $J/M^2\sim 0$, one interesting consequence of our results is that essentially \emph{any} observation of a spinning supermassive BH (even with spin as low as $J/M^2\sim 0.1$) would provide some constraint on $B$. However, these observations can possibly exclude only very large values of $B$. For example a putative observation of a supermassive BH with $M\sim 10^9 M_\odot$ and $J/M^2\gtrsim 0.5$ can potentially exclude the range $10^7 {\mathrm{Gauss}}\lesssim B\lesssim 10^9 {\mathrm{Gauss}}$.

We conclude this section with a note of caution. The threshold lines shown in Fig.~\ref{Fig:Regge} were obtained using Eq.~\eqref{wim} in the range $10^{-4}\lesssim BM\lesssim 0.2$, but the validity range of Eq.~\eqref{wim} might be smaller. Indeed, a different behavior is expected for large magnetic fields, $BM\gg1$. In the opposite regime, using the magnetized Ernst solution with $BM\sim 10^{-4}$ to approximate a realistic configuration requires the source of the magnetic field to extend at least up to $r_M\sim 10^4M\sim 0.5[M/(10^9 M_\odot)]{\mathrm{pc}}$. While we expect that our simplistic analysis can provide the correct order of magnitude for the instability, a more refined study would be needed to assess its validity in the full range of $B$.

%%%%%%%%%%%%%%%%%%%%%%%%%%%%%%%%%%%%%%%%%%%%%%%%%%%%%%%%%
\section{Conclusions}\label{sec:conclusions}
%%%%%%%%%%%%%%%%%%%%%%%%%%%%%%%%%%%%%%%%%%%%%%%%%%%%%%%%%%
The main purpose of this Chapter was to show how strong magnetic fields near spinning BHs can trigger superradiant instabilities and to start exploring the possible implications of such effect. 

To understand this issue, we have computed the normal modes of scalar perturbations of the Melvin spacetime and the QNMs of BHs immersed in a uniform magnetic field. We showed that the magnetic field can confine perturbations leading to long-lived modes, which can trigger superradiant instabilities when the BH spins above a certain threshold. The instability time scale can be orders of magnitude shorter than that associated to the same kind of instabilities triggered by massive fields. In fact, a BH immersed in a uniform magnetic field is very similar to the original BH bomb proposal~\cite{Press:1972zz} and to the case of small BHs in AdS.

In this work we considered only scalar perturbations.
Due to the presence of the magnetic field, gravitational and electromagnetic perturbations of magnetized BHs are coupled, and even a linear stability analysis is rather involved. Nevertheless, in analogy with the AdS case~\cite{Cardoso:2004hs,Cardoso:2006wa,Dias:2013sdc,Cardoso:2013pza}, we expect the instability of gravito-electromagnetic perturbations of a magnetized Kerr--Newman BH to follow the same scaling as scalar perturbations~\eqref{wI_rot}. This expectation is also supported by the model presented in Sec.~\ref{sec:model}. Since gravitational and electromagnetic perturbations extract energy from the BH more efficiently than a scalar field~\cite{Teukolsky:1974yv} we expect them to trigger a slightly stronger instability. Such problem could be tackled extending the results of Refs.~\cite{Pani:2013ija,Pani:2013wsa}, where the gravito-magnetic modes of a Kerr--Newman BH in vacuum were computed to first order in the spin.

In an astrophysical context, our results should be used with care. The Ernst spacetime is not asymptotically flat and, in a realistic situation, it must be matched with a Minkowski spacetime at large distance. This will add some amount of dissipation which is forbidden in the exact Ernst solution. The validity region of the Ernst metric depends on the extension of the source of the magnetic field. Besides that, in realistic situations the presence of an accretion disk can strongly affect the dynamics of electromagnetic perturbations, for example by quenching growing modes or introducing a cutoff plasma frequency for superradiant photons~\cite{Pani:2013hpa}. 

Nevertheless, we hope our work motivates further studies on the subject. To fully understand the magnitude and end state of the instability, general relativistic magnetohydrodynamic simulations (cf. e.g. Refs.~\cite{2012MNRAS.423.3083M,2013MNRAS.436.3741P}) are necessary. Another related subject that deserves further study are the possible effects of this instability on the Blandford-Znajek process. It would also be interesting to understand if the Meissner effect that affects magnetic fields around highly spinning BHs~\cite{PhysRevD.12.3037,Bicak:1980du,KarasJMP}, and that is still a matter of debate~\cite{Penna:2014aza}, can change the picture in the near-extremal limit.  

Finally, it is possible that a similar superradiant mechanism is at work in rotating stars immersed in strong magnetic fields:
strong fields provide the confinement necessary to grow the superradiant modes, and a putative dissipation at the star
would provide superradiance~\cite{Cardoso:2012zn,Cardoso:2015zqa}. The time scale for energy dissipation in neutron stars is governed by shear viscosity and estimated to be of the order of~\cite{1987ApJ...314..234C}
\be
\tau_{\eta}\sim 10^{9}\left(\frac{10^{14} \,{\mathrm{g}}\,{\mathrm{cm}}^{-3}}{\rho}\right)^{5/4}\left(\frac{T}{10^9\,{\mathrm{K}}}\right)^2\left(\frac{R}{10^6\,{\mathrm{cm}}}\right) \,{\mathrm{sec}}\,.
\ee
where $\rho$, $T$ and $R$ are the central density of the neutron star, the temperature and the radius, respectively.
By comparison, the time scale for energy dissipation in BHs scales like the light crossing time and is over 14 orders of magnitude smaller
for a stellar-mass BH. Thus, the instability is expected to be of extremely long time scale. Nevertheless, imprints of the (confined) perturbations should appear as new modes of vibration.

%%%%%%%%%%%%%%%%%%%%%%%%%%%%%%%%%%%%%%%%%%%%%%%%%%%%%%%%%%%%%%%%%%%%

%%%%%%%%%%%%%%%%%%%%%%%%%%%%%%%%%%%%%%%%%%%%%%%%%%%%%%%%%%%%%%%%%%%%
\chapter{Massive spin-2 perturbations of slowly rotating Kerr black holes}\label{chapter:Kerr}
%%%%%%%%%%%%%%%%%%%%%%%%%%%%%%%%%%%%%%%%%%%%%%%%%%%%%%%%%%%%%%%%%%%%%%%%%%%%%%%%%%%%
\section{Introduction}\label{sec:kerr}
%%%%%%%%%%%%%%%%%%%%%%%%%%%%%%%%%%%%%%%%%%%%%%%%%%%%%%%%%%%%%%%%%%%%%%%%%%%%%%%%%%%%

As we discussed in the previous Chapters, the interaction of bosonic fields with spinning BHs gives rise to interesting phenomena, related to 
BH superradiance~\cite{zeldo1,zeldo2,Press:1972zz,Cardoso:2012zn}. Due to the dissipative nature of the BH horizon and to the existence of negative-energy states in the ergoregion of a spinning BH, low-frequency $\omega$ monochromatic bosonic waves scattered off rotating BHs are amplified whenever the superadiant condition~\eqref{eq:superradiance_condition} is met.

In this Chapter we will be particularly interested in superradiance-triggered BH instabilities which are sustained by massive fields.
Ultralight bosons have received widespread attention recently as they are found in several extensions of the Standard Model,
for instance in the string axiverse scenario~\cite{Arvanitaki:2009fg,Arvanitaki:2010sy} where a plethora of massive pseudo-scalar fields called axions covers each decade of mass range down to the Hubble scale and fields with $10^{-22} {\rm eV}<m_S<10^{-10}{\rm eV}$ are of particular interest for BH physics~\cite{Kodama:2011zc}. In parallel, massive hidden $U(1)$ vector fields also arise in extensions of the standard model~\cite{Goodsell:2009xc,Jaeckel:2010ni,Camara:2011jg,Goldhaber:2008xy}, highlighting the importance of understanding the physics of such fields around BHs.

Superradiant instabilities were studied extensively for scalar fields both in the frequency- and in the 
time-domain~\cite{Detweiler:1980uk,Cardoso:2004nk,Cardoso:2005vk,Dolan:2007mj,Witek:2012tr,Dolan:2012yt,Cardoso:2013fwa}.
The non-separability of the field equations for a massive vector field in a Kerr background has hampered its study for decades (see for instance
Ref.~\cite{Rosa:2011my} for some references on the nonrotating case). Very recently however, progress has been made. 
In the frequency domain slow-rotating expansions were used to prove that massive vectors {\it are}
superradiantly unstable~\cite{Pani:2012bp,Pani:2012vp}, and these results were confirmed using evolutions of wavepackets around Kerr BHs ~\cite{Witek:2012tr}. It was shown that the massive vector field instability can be orders of magnitude stronger than the massive scalar field. 

The instability is regulated by two parameters, the BH spin $a/M$ and the dimensionless parameter $M\mu$ (in units $G=c=1$), where $M$ is the
BH mass and $m_g=\mu\hbar$ is the bosonic field mass. For ultralight scalar fields around massive BHs, the instability timescale can be of the order of seconds for solar-mass BHs and of the order of hundreds years for a supermassive BH with $M\sim 10^9 M_\odot$~\cite{Detweiler:1980uk,Arvanitaki:2009fg,Arvanitaki:2010sy}, typically much shorter than the evolution timescale of astrophysical objects. The instability timescale for spin-1 massive fields can be up to three orders of magnitude shorter~\cite{Pani:2012bp,Pani:2012vp,Witek:2012tr}. 
To summarize, this mechanism can be very efficient for extraction of angular momentum away from the BH.
As a consequence, observations of massive spinning BHs can effectively be used to impose bounds on ultralight boson masses~\cite{Pani:2012bp} (see Section~\ref{sec:bounds_RP} in the next Chapter). In this Chapter we show that the same kind of arguments can be used to impose a conservative bound on the graviton mass of $m_g=\hbar\mu\lesssim 5\times 10^{-23} {\mathrm{eV}}$. This is one order of magnitude better that the bound imposed by the gravitational-wave observation GW150914 by Advanced LIGO~\cite{Abbott:2016blz}, but several orders of magnitude lower than the current best (highly model-dependent) bound, $m_g\lesssim 10^{-32} {\mathrm{eV}}$~\cite{PDG}.

In Ref.~\cite{Pani:2012bp} a method to study generic perturbations of slowly rotating BHs was developed. Here we will extend this method to massive spin-2 perturbations of slowly rotating Kerr BHs. As shown in Chapter~\ref{chapter:massive2}, massive spin-2 fields can be consistently described within the framework of theories of massive gravity and bigravity theories. These theories admit the Kerr geometry as a solution (or more strictly speaking, two copies of the Kerr solution that solve Eqs.~\eqref{eqs_pro} when $\Lambda_g=\Lambda_f=0$). Around this background, massive spin-2 perturbations are described by the system of eqs.~\eqref{eqmotioncurved}--~\eqref{constraint2}. 

In this Chapter we will study this system of equations in a Kerr background to first order in $\tilde{a}\equiv a/M\equiv J/M^2$, with $M$ and $J$ being the mass and angular momentum of the Kerr geometry, as defined in~\eqref{Kerr}. Our analysis can be generalized to higher order in the BH angular momentum, but a first-order approximation can be shown to be already quite accurate even for moderately large spins~\cite{Pani:2012bp}. 

The technique that we will use consists in a decomposition of the perturbation equations in tensor spherical harmonics and in an expansion in the parameter $\tilde{a}$. The method was originally developed to study the gravitational perturbations of slowly-rotating stars~\cite{Kojima:1992ie,1993ApJ...414..247K,1993PThPh..90..977K} and it has been recently applied to BH spacetimes~\cite{Pani:2012vp,Pani:2012bp,Pani:2013ija}. As a result of using a basis of spherical harmonics in a nonspherical background, the perturbation equations display parity-mixing and coupling among perturbations with different harmonic indices. However, as we will show, and as discussed in Ref.~\cite{Pani:2012vp}, to first order in $\tilde{a}$ the eigenvalue spectrum is described by two decoupled sets, one for the axial and one for the polar perturbations, and all harmonic indices decouple.

%%%%%%%%%%%%%%%%%%%%%%%%%%%%%%%%%%%%%%%%%%%%%%%%%%%%%%%%%%%%%%%%%%%%%%%%%%%%%%%%%%%%%%%%%%%%%%%%%%%%%
\section{Linearized field equations for a spin-2 field on a slowly rotating Kerr BH}\label{app:kerr}
%%%%%%%%%%%%%%%%%%%%%%%%%%%%%%%%%%%%%%%%%%%%%%%%%%%%%%%%%%%%%%%%%%%%%%%%%%%%%%%%%%%%%%%%%%%%%%%%%%%%%

We will follow Kojima~\cite{Kojima:1992ie} to write down the field equations for a spin-2 field in a slowly rotating BH. Since this background is still ``almost'' spherically symmetric we can use the decomposition \eqref{decom} and insert it in the linearized field equations~\eqref{eqmotioncurved}--~\eqref{constraint2}, in the background given by the metric~\eqref{Kerr} expanded at first order in $\tilde{a}\equiv a/M$. We can then separate the equations in three different groups.

From the $(tt)$, $(tr)$, $(rr)$, the sum of $(\theta\theta)$ and $(\phi\phi)$ components of Eq.~\eqref{eqmotioncurved}, the $t$ and $r$ components of the transverse condition~\eqref{constraint1}, and the traceless condition~\eqref{constraint2}, we have 
\be
\label{eqscalar}
\left(A_{lm}^{(I)}+\tilde{A}_{lm}^{(I)}\cos\theta\right)Y^{lm}+B_{lm}^{(I)}\sin\theta \partial_{\theta} Y^{lm}
+C_{lm}^{(I)}\partial_{\phi}Y^{lm}=0\quad (I=0,\,\dots\,,6)\,,
\ee
where a sum over ($l,m$) is implicit, the functions $A_{lm}^{(I)}$ and $C_{lm}^{(I)}$ are some linear combinations of the polar functions $H_0$, $H_1$,$H_2$, $\eta_0$, $\eta_1$, $K$ and $G$. On the other hand $\tilde{A}_{lm}^{(I)}$ and $B_{lm}^{(I)}$ are some linear combinations of the axial functions $h_0$, $h_1$, $h_2$. 

From the $(t\theta)$, $(t\phi)$, $(r\theta)$, $(r\phi)$ components of Eq.~\eqref{eqmotioncurved}, and the $\theta$, $\phi$ components of Eq.~\eqref{constraint1}, we have 
\begin{align}
\label{eqvector1}
&\left(\alpha_{lm}^{(J)}+\tilde{\alpha}_{lm}^{(J)}\cos\theta\right)\partial_{\theta}Y^{lm}\nn\\
&-\left(\beta_{lm}^{(J)}+\tilde{\beta}_{lm}^{(J)}\cos\theta\right)\left(\partial_{\phi}Y^{lm}/\sin\theta\right)+\eta_{lm}^{(J)}(\sin\theta Y^{lm})\nn\\
&+\xi_{lm}^{(J)}X^{lm}+\chi_{lm}^{(J)}(\sin\theta W^{lm})=0 \quad (J=0,1,2)\,,
\end{align}
and
\begin{align}
\label{eqvector2}
&\left(\beta_{lm}^{(J)}+\tilde{\beta}_{lm}^{(J)}\cos\theta\right)\partial_{\theta}Y^{lm}\nn\\
&+\left(\alpha_{lm}^{(J)}+\tilde{\alpha}_{lm}^{(J)}\cos\theta\right)\left(\partial_{\phi}Y^{lm}/\sin\theta\right)+\zeta_{lm}^{(J)}(\sin\theta Y^{lm})\nn\\
&+\chi_{lm}^{(J)}X^{lm}-\xi_{lm}^{(J)}(\sin\theta W^{lm})=0\quad (J=0,1,2)\,,
\end{align}
where the functions $\alpha_{lm}^{(J)}$, $\tilde{\beta}_{lm}^{(J)}$, $\zeta_{lm}^{(J)}$ and $\xi_{lm}^{(J)}$ are some linear combination of the polar functions, while  $\beta_{lm}^{(J)}$, $\tilde{\alpha}_{lm}^{(J)}$, $\eta_{lm}^{(J)}$ and $\chi_{lm}^{(J)}$ belong to the axial sector.

From the $(\theta\phi)$ and the subtraction of $(\theta\theta)$ and $(\phi\phi)$ components of~\eqref{eqmotioncurved}, we have
\begin{align}
\label{eqtensor1}
&f_{lm}\partial_{\theta}Y^{lm}+g_{lm}\left(\partial_{\phi}Y^{lm}/\sin\theta\right)\nn\\
&+\left(s_{lm}+\hat{s}_{lm}\partial_{\phi}\right)\left(X^{lm}/\sin^2 \theta \right)
+\left(t_{lm}+\hat{t}_{lm}\partial_{\phi}\right)\left(W^{lm}/\sin\theta\right)=0\,,
\end{align}
and
\begin{align}
\label{eqtensor2}
&g_{lm}\partial_{\theta}Y^{lm}-f_{lm}\left(\partial_{\phi}Y^{lm}/\sin\theta\right)\nn\\
&-\left(t_{lm}+\hat{t}_{lm}\partial_{\phi}\right)\left(X^{lm}/\sin^2 \theta\right)
+\left(s_{lm}+\hat{s}_{lm}\partial_{\phi}\right)\left(W^{lm}/\sin\theta\right)=0\,,
\end{align}
where $f_{lm}$, $s_{lm}$ and $\hat{s}_{lm}$ are some linear combinations of polar functions and $g_{lm}$, $t_{lm}$ and $\hat{t}_{lm}$ from the axial functions.

It is easy to see that at zeroth-order in the rotation the perturbation equations reduce to
\be
A_{lm}^{(I)}=\alpha_{lm}^{(J)}=s_{lm}=0\,,\quad (I=0,\,\dots\,,6,\,J=0,1,2)\,,
\ee
for the polar sector and to
\be
\beta_{lm}^{(J)}=t_{lm}=0\,,\quad (\,J=0,1,2)\,,
\ee
for the axial sector, respectively. These equations correspond to the ones obtained for the Schwarzschild case.

To separate the angular variables we use the identities
\begin{align}
\cos\theta Y^{lm}&= Q_{l+1\,m} Y^{l+1\,m}+Q_{lm} Y^{l-1\,m}\,,\\
\sin\theta\partial_\theta Y^{lm}&= Q_{l+1\,m}\,l\, Y^{l+1\,m}-Q_{lm}(l+1) Y^{l-1\,m}\,,
\end{align}
with
\be
Q_{lm}=\sqrt{\frac{l^2-m^2}{4l^2-1}}\,,
\ee
and the orthogonality properties of scalar, vector and tensor harmonics. 
The separation of the angular dependence of Einstein's equations for a
slowly-rotating star was performed in Ref.~\cite{Kojima:1992ie}. Since the above equations are formally the same as those considered in Ref.~\cite{Kojima:1992ie}, they can be separated in exactly the same way (see Ref.~\cite{Pani:2013pma} for a review).
Below we omit the index $m$, because in an axisymmetric background it is possible to decouple the perturbation equations so that all quantities have the same value of $m$. 

From Eq.~\eqref{eqscalar} we have~\cite{Kojima:1992ie,Pani:2013pma}
\be
\label{descalar}
A_{l}^{(I)}+i m C_{l}^{(I)}+Q_{l}\left(\tilde{A}_{l-1}^{(I)}+(l-1)B_{l-1}^{(I)}\right)
+Q_{l+1}\left(\tilde{A}_{l+1}^{(I)}-(l+2)B_{l+1}^{(I)}\right)=0\,.
\ee

Defining $\lambda=l(l+1)$, equations~\eqref{eqvector1} and \eqref{eqvector2} give 
\begin{align}
\label{devector1}
&\lambda\alpha_{l}^{(J)}+i m\left[(l-1)(l+2)\xi_{l}^{(J)}-\tilde{\beta}_{l}^{(J)}-\zeta_{l}^{(J)}\right]\nn\\
&+Q_l (l+1)\left[(l-2)(l-1)\chi_{l-1}^{(J)}+(l-1)\tilde{\alpha}_{l-1}^{(J)}-\eta_{l-1}^{(J)}\right]\nn\\
&-Q_{l+1}\,l\left[(l+2)(l+3)\chi_{l+1}^{(J)}-(l+2)\tilde{\alpha}_{l+1}^{(J)}-\eta_{l+1}^{(J)}\right]=0\,,
\end{align}
and
\begin{align}
\label{devector2}
&\lambda\beta_{l}^{(J)}+i m\left[(l-1)(l+2)\chi_{l}^{(J)}+\tilde{\alpha}_{l}^{(J)}+\eta_{l}^{(J)}\right]\nn\\
&-Q_l (l+1)\left[(l-2)(l-1)\xi_{l-1}^{(J)}-(l-1)\tilde{\beta}_{l-1}^{(J)}+\zeta_{l-1}^{(J)}\right]\nn\\
&+Q_{l+1}\,l\left[(l+2)(l+3)\xi_{l+1}^{(J)}+(l+2)\tilde{\beta}_{l+1}^{(J)}+\zeta_{l+1}^{(J)}\right]=0\,.
\end{align}
Finally, Eqs.~\eqref{eqtensor1} and \eqref{eqtensor2} yield 
\be
\label{detensor1}
\lambda \left(s_{l}+i m\hat{s}_{l}\right)-i m f_{l}-Q_l (l+1) g_{l-1}+Q_{l+1}l\,g_{l+1}=0\,,
\ee
\be
\label{detensor2}
\lambda \left(t_{l}+i m\hat{t}_{l}\right)+i m g_{l}-Q_l (l+1) f_{l-1}+Q_{l+1}l\,f_{l+1}=0\,.
\ee

Because the background is nonspherically symmetric, the radial equations above display mixing between perturbations with opposite parity and different harmonic index. To first order, perturbations with given parity and harmonic index $l$ are coupled to perturbations with opposite parity and indices $l\pm1$. However, as discussed in Ref.~\cite{Pani:2012vp}, these couplings do not contribute to the eigenvalue spectrum to first order in $\tilde{a}$. 

%%%%%%%%%%%%%%%%%%%%%%%%%%%%%%%%%%%%%%%%%%%
\subsection{Axial equations at first order}
%%%%%%%%%%%%%%%%%%%%%%%%%%%%%%%%%%%%%%%%%%%

Neglecting the coupling to the opposite parity with harmonic indices $l\pm 1$, using Eqs.~\eqref{devector2} and~\eqref{detensor2}, and by defining:
\begin{eqnarray}
 h_1(r)&=& \frac{Q(r)}{f(r)}\left(1-\frac{\tilde{a}m M^2 \left(\lambda+2\right) }{\lambda r^3 \omega }\right)\,,\\
 h_2(r)&=& Z(r)r\left(1-\frac{\tilde{a}m M^2 \left(\lambda-2\right) }{\lambda r^3 \omega }\right)\,,
\end{eqnarray}
%%%
we obtain that a fully consistent solution at first order is such that $Z$ and $Q$ satisfy the following equations:
%%%
\begin{eqnarray}\label{axial1}
 \frac{d^2Q}{dr_*^2}+V_Q Q(r)&=&S_Q Z(r)\,,\\
\label{axial2}
 \frac{d^2Z}{dr_*^2}+V_Z Z(r)&=&S_Z Q(r)
\end{eqnarray}
%%%
with
\begin{align}
&V_Q=\omega^2-\frac{4 \tilde{a}m M^2\omega }{r^3}-
f\left[\frac{\lambda+4}{r^2}-\frac{16 M}{r^3}+\mu ^2+\tilde{a}m M^2\frac{6 (4 r-9 M)(\lambda+2)}{\lambda r^6 \omega }\right]\,,\\
& V_Z=\omega^2-\frac{4 \tilde{a}m M^2\omega }{r^3}-
f\left[\frac{\lambda-2}{r^2}+\frac{2M}{r^3}+\mu ^2+\tilde{a}m M^2\frac{6 (\lambda -2)  (r-3M)}{\lambda  r^6 \omega }\right]\,,\\
&S_Q= 2(\lambda-2)f\left[\frac{r-3M}{r^3}
-\tilde{a}m M^2\frac{\left(6 M (4+\lambda )-r \left(10+3 \lambda +3 r^2 \omega ^2\right)\right)}{\lambda r^6 \omega }\right]\,,\\
&S_Z=2f\left[\frac{1}{r^2}+\tilde{a}m M^2\frac{\left(-10+3 \lambda +3 r^2 \mu ^2\right)}{\lambda  r^5 \omega }\right]\,.
\end{align}
%%%
These equations reduce to Eqs.~\eqref{axial_bi1} and~\eqref{axial_bi2} in the nonrotating limit.
In the dipole case $l=1$, $\lambda=2$, the function $Z$ vanishes and we are left with a single decoupled equation:
\begin{equation}\label{axialdi_kerr}
 \frac{d^2Q}{dr_*^2}+V_Q Q(r)=0\,.
\end{equation}

%%%%%%%%%%%%%%%%%%%%%%%%%%%%%%%%%%%%%%%%%%%
\subsection{Polar equations at first order}
%%%%%%%%%%%%%%%%%%%%%%%%%%%%%%%%%%%%%%%%%%%
The polar equations can be obtained from Eqs.~\eqref{descalar},~\eqref{devector1} and~\eqref{detensor1}.
In line with the non-rotating case, for the polar sector we obtain at first order in $\tilde{a}$ three coupled equations for $K$, $\eta_1$ and $G$, which generalize Eqs.~\eqref{polar_eq1},~\eqref{polar_eq2}, and~\eqref{polar_eq3}, but in this case the coefficients $\hat\alpha_i,\,\hat\beta_i,\,\hat\gamma_i,\,\hat\delta_i,\,\hat\sigma_i,\,\hat\rho_i$ are also functions of $m\tilde{a}$. Due to the length of the equations we do not show them explicitly here but we made them available online in {\scshape Mathematica} notebooks~\cite{webpage}.
%%%

%%%%%%%%%%%%%%%%%%%%%%%%%%%%%%%%%%%%%%%%%%%%
\section{Superradiance and quasibound states}
%%%%%%%%%%%%%%%%%%%%%%%%%%%%%%%%%%%%%%%%
Interesting phenomena, such as BH superradiance, are already manifest at first order in the BH angular momentum.

As for the Schwarzschild case, at the horizon we must impose purely ingoing waves,
\be
\label{BC_hor_Kerr}
\Phi_j(r)\sim e^{-ik_H r_*}\,,
\ee
as $r_*\to -\infty$, where
\be
k_H=\omega-m\Omega_{\rm H}=\omega-\frac{m\tilde{a}}{4M}+\mathcal{O}(\tilde{a}^3)\,. 
\ee
Here $\Phi_j$ describes generically the perturbation functions, and the horizon angular velocity $\Omega_{\rm H}=a/(2Mr_+)$ was expanded to first-order in rotation. When $k_H<0$ an observer at infinity will see waves emerging from the BH~\cite{Teukolsky:1973ha}. This corresponds to the superradiant condition $\omega<m\Omega_{\rm H}$~\cite{Teukolsky:1974yv}, which at first-order in the rotation amounts to
\be
\tilde{a}>\frac{4M\omega_R}{m}\,,
\ee
where $\omega_R$ is the real part of the mode frequency, $\omega=\omega_R+i\omega_I$. All the polar and axial equations can be brought to a form such that the near-horizon solution is given by Eq.~\eqref{BC_hor_Kerr}. We thus expect that superradiance will also occur for massive spin-2 fields even at first-order in the rotation. At infinity, $r\to \infty$, the asymptotic behavior is the same than for a Schwarzschild BH (see Eq.~\eqref{BC_inf_2} and corresponding discussion). 

As already discussed, superradiant scattering leads to instabilities of bosonic massive fields~\cite{Detweiler:1980uk,Dolan:2007mj,Pani:2012bp,Witek:2012tr,Dolan:2012yt}. This instability was explicitly shown for scalars and vectors, and here we will show that it is also present for tensor fields.
We recall that with our convention, unstable modes correspond to $\omega_I>0$. These superradiant instabilities occur only for waves trapped in the vicinity of the BH, i.e., quasibound states, so we focus on these states in the next section (corresponding to $C_j=0$ in Eq.~\eqref{BC_inf_2}). 

The continued fraction method (see Appendix~\ref{app:modes}) can be used to determine the quasibound state frequencies of the axial equations by imposing an appropriate \emph{ansatz} which in this case is given by
\be
\label{ansatz2}
\Phi_j(\omega,r)=f(r)^{-2ik_H}r^{\nu}e^{-qr}\sum_n{a^{(j)}_n}f(r)^n\,,
\ee
where $\nu=-q+\omega^2/q$. To compute the quasinormal mode frequencies we use $q=-\sqrt{\mu^2-\omega^2}$ and for the quasibound state frequencies $q=\sqrt{\mu^2-\omega^2}$.
Inserting Eq.~\eqref{ansatz2} into Eq.\eqref{axialdi_kerr} leads to a six-term recurrence relation which can be reduced to a three-term recurrence relation by successive Gaussian elimination steps~\cite{Leaver:1990zz,Onozawa:1995vu}. For $l\geq 2$ we find a six-term matrix-valued recurrence relation which can also be brought to a three-term recurrence relation using a matrix-valued Gaussian elimination. The explicit form of the coefficients is not shown here for brevity but it is available online~\cite{webpage}.

Although the continued-fraction method works very well for quasibound states, the multiple matrix inversion of almost singular matrices (since some matrices are proportional to $\tilde{a}$) makes it very difficult to compute the very small imaginary part of the axial quasibound states. We therefore use the direct integration method for both the polar and axial quasibound states which gives more accurate results in this case, and use the continued-fraction method to check the robustness of our results.

%%%%%%%%%%%%%%%%%%%%%%%%%
\begin{figure*}[htb]
\begin{center}
% \begin{tabular}{c}
\epsfig{file=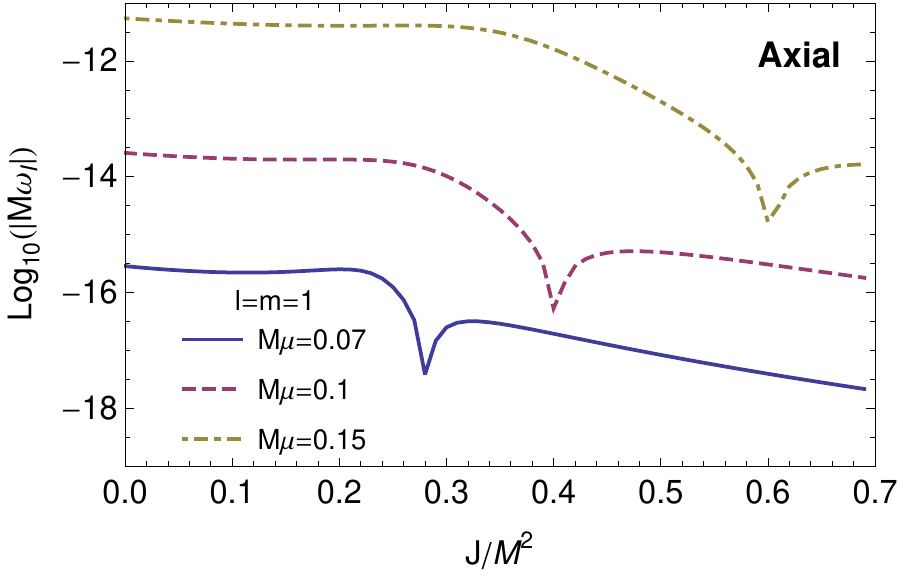,width=7.3cm,angle=0,clip=true}
\epsfig{file=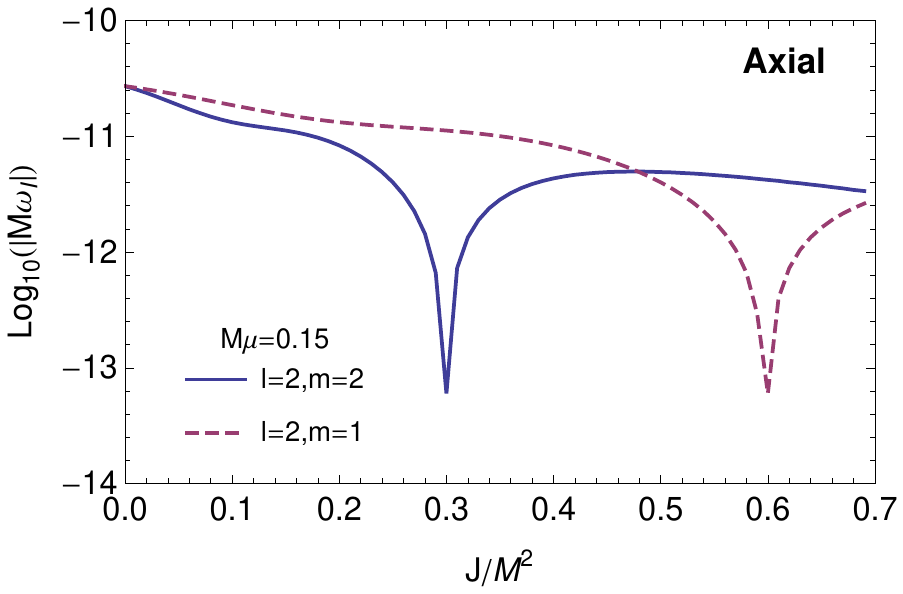,width=7.3cm,angle=0,clip=true}\\
\epsfig{file=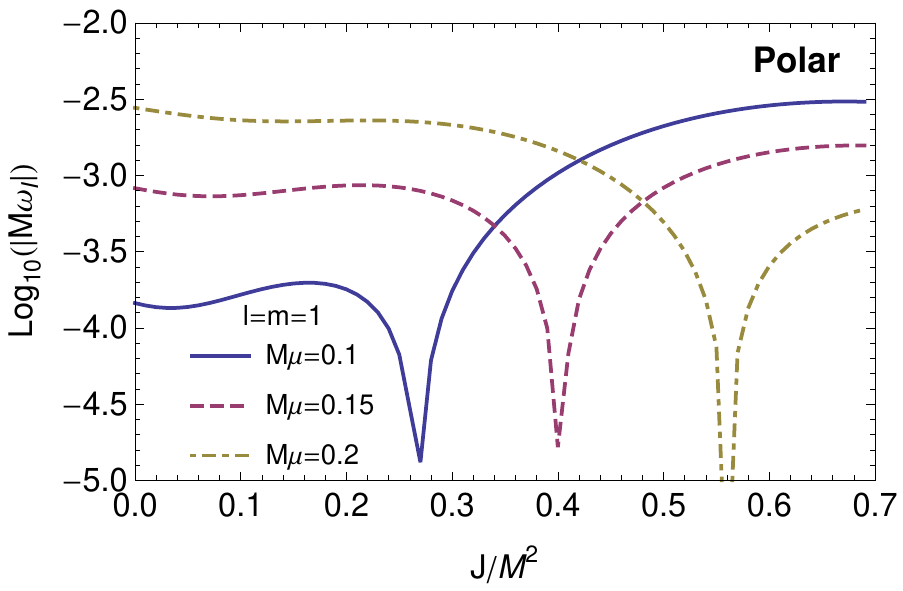,width=7.3cm,angle=0,clip=true}
\epsfig{file=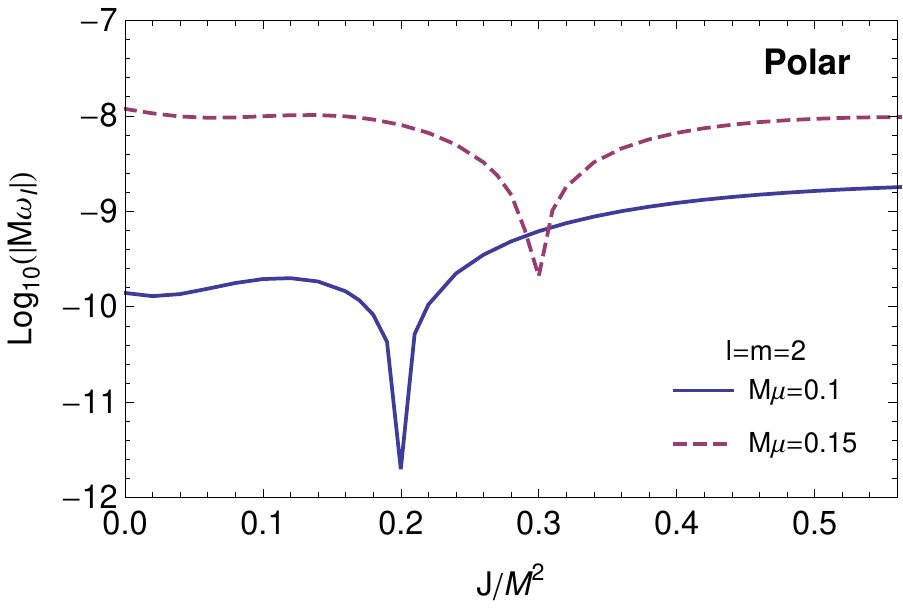,width=7.3cm,angle=0,clip=true}
% \end{tabular}
\caption{Absolute value of the imaginary part of the axial and polar quasibound modes as a function of the BH rotation rate $\tilde{a}$ for different values of $l$ and $m$ and different values of the mass coupling $\mu M$, computed at first order. Left top panel: axial dipole for $l=m=1$. Right top panel: axial mode $S=-1$ for a mass coupling $M\mu=0.15$ and different values of $m$.
Left bottom panel: polar dipole mode for $l=m=1$. Right bottom panel: polar mode $l=m=2$, $S=-2$. 
For any mode with $m\geq 0$, the imaginary part crosses the axis and become unstable when the superradiance condition is met.\label{fig:axial}}
\end{center}
\end{figure*}
%%%%%%%%%%%%%%%%%%%%%%%%%%
%%%%%%%%%%%%%%%%%%%%%%%%%%%%%%%
\section{Results}
%%%%%%%%%%%%%%%%%%%%%%%%%%%%%%%
%
\begin{figure}[htb]
\begin{center}
% \begin{tabular}{c}
\epsfig{file=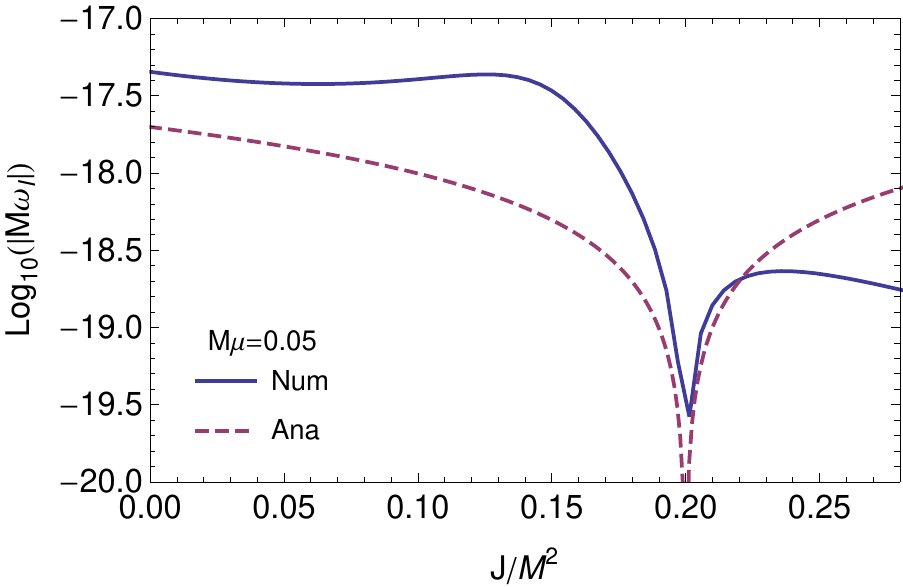,width=10cm,angle=0,clip=true}
% \end{tabular}
\caption{Comparison between the numerical and analytical results for the axial mode $l=m=1$, $n=0$ as a function of the BH rotation rate $\tilde{a}$ for a mass coupling of $M\mu=0.05$. The solid line shows the numerical data and the dashed shows the analytical formula.\label{fig:ana_vs_num}}
\end{center}
\end{figure}
In the top panels of Fig.~\ref{fig:axial} we show the absolute value of the imaginary part as a function of the rotation parameter for the axial modes $l=1$, $S=1$ and $l=2$, $S=-1$ (cf. subsection~\ref{subsec:quasibound}). Although a second-order approximation would be needed to describe the superradiant regime in a self-consistent way~\cite{Pani:2012vp}, the first-order approximation predicts very well the onset of the instability and should give the correct order of magnitude of the instability timescale. For axial modes the instability is very weak: even in the most favorable cases the instability is almost five orders of magnitude weaker than that associated to axial Proca modes~\cite{Pani:2012bp,Pani:2012vp}. This also  makes it difficult to track numerically the axial spin-2 modes with sufficient precision. For small masses the real part of the frequency is roughly independent on the spin.  
This is supported by analytical results for the axial dipole mode, which can be evaluated analytically in the small-mass limit at first order in $\tilde{a}$ (cf. Appendix~\ref{app:ana}). The analytical formula for the imaginary part of the fundamental mode reads
\be
M\omega_I\approx \frac{40}{19683}(\tilde{a}-2r_+\mu)(M\mu) ^{11}\,.
\ee
In Fig.~\ref{fig:ana_vs_num} we compare the analytical formula with the numerical results for the fundamental overtone and mass coupling $M\mu=0.05$. Although the imaginary part is tiny, the agreement is good in the $\mu\to0$ limit. Near the superradiant regime the agreement is only qualitative, as expected since the analytical formula is only valid for $\tilde{a}m/(M\mu)\lesssim l$.

The bottom panels of Figure~\ref{fig:axial} show the imaginary part as a function of the BH angular momentum for the polar dipole $l=1$ and the polar mode $l=2$, $S=-2$ (cf. subsection~\ref{subsec:quasibound}). In this case the imaginary part of the mode is larger, and these modes are easier to evaluate numerically. The instability for the mode $l=2$, $S=-2$ is roughly two orders of magnitude weaker than the strongest instability of a Proca field~\cite{Pani:2012bp} (cf. Eq.~\eqref{wIslope_2} below). Once more the polar dipole mode is the most interesting case as it has the largest imaginary part, corresponding to an extremely short instability timescale. This agrees with the analysis in the nonrotating case of subsection~\ref{subsec:quasibound}, where we found that the behavior of this mode is different from the rest of the spectrum.

\begin{figure}[htb]
\begin{center}
% \begin{tabular}{c}
\epsfig{file=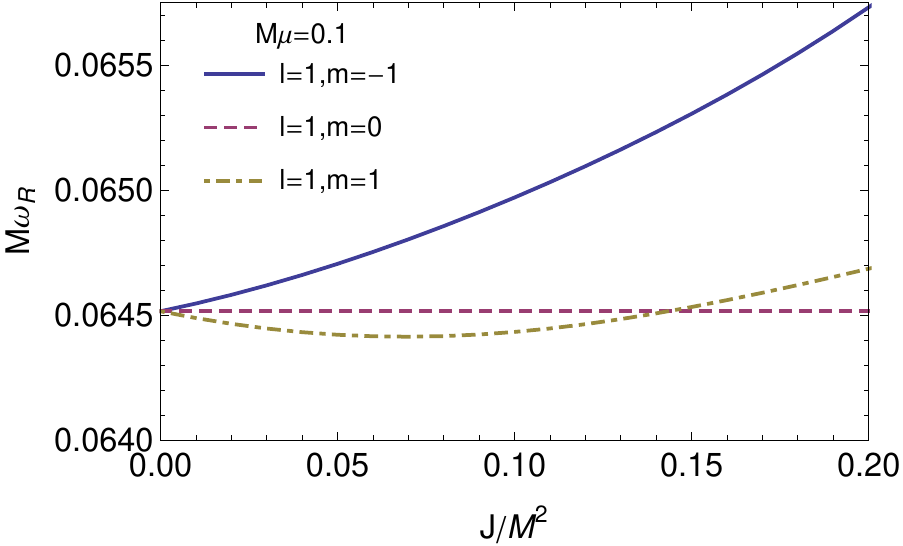,width=10cm,angle=0,clip=true}
% \end{tabular}
\caption{Real part of the polar dipole quasibound mode as a function of the BH rotation rate $\tilde{a}$ for different values of the azimuthal number $m$ and mass coupling $\mu M=0.1$, computed at first order.\label{fig:polar_Re_dipole}}
\end{center}
\end{figure}

As shown in the bottom panels of Fig.~\ref{fig:axial}, the polar dipole mode displays a peculiar behavior in the superradiant regime, where the power-law dependence is inverted, i.e., the instability is stronger for the lowest mass coupling $M\mu$. This suggests that extrapolating the first-order results to the superradiant case is probably less accurate for this mode.
This is confirmed by the behavior of the real-part of the frequency as a function of the spin, as shown in Fig.~\ref{fig:polar_Re_dipole}. At first-order the eigenfrequencies can be expanded as
\be
\omega_R=\omega_0+\tilde{a}m\omega_1+\mathcal{O}(\tilde{a}^2)\,,
\ee
where $\omega_0$ is the eigenfrequency in the nonrotating space-time and $\omega_1$ is the first-order correction which is an even function of $m$~\cite{Pani:2012vp}. Hence at first-order we would expect that the curves for $l=m$ and $l=-m$ are symmetric when reflected around the $m=0$ curve. For the polar dipole this only happens for very small masses. Note also that, contrarily to the rest of the spectrum, the real part of the polar dipole mode acquires a nonnegligible dependence on $\tilde{a}$, even in the small $\mu$ limit. In fact the analytical results for the axial dipole suggest that the first-order approximation is only valid for $\tilde{a}m/(M\omega_R)\lesssim l$. Since in this case $M\omega_R$ is much smaller that $M\mu$, the extrapolation to the superradiant regime is less accurate in the polar dipole case. Nonetheless, using the exact results in the nonrotating case (cf. subsection~\ref{subsec:quasibound}) and a linear extrapolation of the first-order corrections, we estimate the following scaling for the imaginary part of the polar dipole mode:
\be
M\omega_I\sim \gamma_{{\mathrm{polar}}}(\tilde{a}m-2r_+\omega_R)(M\mu)^{3}\,,~\label{wIpoldip}
\ee
where $\gamma_{{\mathrm{polar}}}\sim{\cal O}(1)$ and $\omega_R$ is the zeroth order real frequency given by Eq.~\eqref{polar_di_Re}. This behavior becomes less accurate deep inside the superradiant regime. Although such extrapolation is extremely rough, a similar estimate has been done in the scalar and in the Proca case and it turned out to be very accurate~\cite{Pani:2012bp}. In the scalar case a fit similar to Eq.~\eqref{wIpoldip} agrees with exact results (obtained solving the Klein-Gordon equation on an exact Kerr metric~\cite{Dolan:2007mj}) within a few percents; and, in the Proca case, it reproduces the results of exact numerical simulations (again in the quasiextremal, $\tilde{a}\sim0.99$ case) within a factor two~\cite{Witek:2012tr}. 

In the case at hand, even if Eq.~\eqref{wIpoldip} eventually turns out to be accurate only at the order-of-magnitude level, this would anyway mean that spin-2 fields can trigger the strongest superradiant instability among other bosonic perturbations.
The instability timescale is four orders of magnitude shorter than the shortest timescale for Proca unstable modes~\cite{Pani:2012bp} (cf. Eq.~\eqref{wIslope_2} below). A second-order analysis would be important to confirm this result, but it will also be very challenging. A most promising extension is to perform a full numerical analysis (along the lines of Ref.~\cite{Witek:2012tr}) in the case of massive spin-2 fields around highly spinning Kerr BHs.
% %
% \begin{figure*}[htb]
% \begin{center}
% % \begin{tabular}{c}
% \epsfig{file=polar_dipole_Im_1st.eps,width=7cm,angle=0,clip=true}
% \epsfig{file=polar_Im_1st.eps,width=7cm,angle=0,clip=true}
% % \end{tabular}
% \caption{Absolute value of the imaginary part of the polar quasibound modes as a function of the BH rotation rate $\tilde{a}$ for different values of $l$ and $m$ and different values of the mass coupling $\mu M$, computed at first order. In the left panel we show the polar dipole mode for $l=m=1$. In the right panel we show the polar mode $l=m=2$, $S=-2$. For $m\geq 0$, the modes cross the axis and become unstable when the superradiance condition is met.\label{fig:polar}}
% \end{center}
% \end{figure*}
% %
%
%

%%%%%%%%%%%%%%%%%%%%%%%%%%%%%%%%%
\subsection{Small note on perturbations of nonbidiagonal Kerr in massive (bi)gravity}
%%%%%%%%%%%%%%%%%%%%%%%%%%%%%%%%%

The results of Chapter~\ref{chapter:nonbi} show that, unlike in the bidiagonal case, gravitational perturbations of nonbidiagonal static black holes in massive gravity do not allow for (quasi)-bound states. This is due to the fact that: (i) the perturbations with $l=0,1$ do not feel any effective potential and (ii) perturbations with $l\geq2$ propagate exactly in the effective potential of a Schwarzschild black hole in GR; in particular, such effective potential does not depend on the graviton mass. 

As shown here, one of the consequences of bosonic quasi-bound states in the spectrum is the existence of a superradiant instability.  
A generalization of the nonbidiagonal solution~\eqref{sol} describing a rotating black hole was found in Ref.~\cite{Babichev:2014tfa} (a further generalization describing the Kerr-(anti-)de Sitter black holes was presented in Ref.~\cite{Ayon-Beato:2015qtt}). Due to the absence of quasi-bound states in the static case, for this family of solutions our results strongly suggest that no superradiant instability exists, at least in the slowly-rotating regime.
%

%%%%%%%%%%%%%%%%%%%%%%%%%%%%%%%%%%%%%%%%%%%%%%%%%%%%%%%%%%%%%%%%%%%%%%%%%%%%%%%%%%%%%%%%%%%%%%%%%%%%%%%%%%%%%%%%%%%%%
\section{A unified picture of superradiant instabilities of massive bosonic fields} \label{sec:massive_unified}
%%%%%%%%%%%%%%%%%%%%%%%%%%%%%%%%%%%%%%%%%%%%%%%%%%%%%%%%%%%%%%%%%%%%%%%%%%%%%%%%%%%%%%%%%%%%%%%%%%%%%%%%%%%%%%%%%%%%%
The results presented in this Chapter, jointly with previous works on spin-0~\cite{Cardoso:2005vk,Dolan:2007mj,Witek:2012tr} and spin-1 fields~\cite{Pani:2012vp,Pani:2012bp} suggest the following unified picture describing the superradiant instability of massive bosonic fields around a spinning BH. For any bosonic field propagating on a spinning BH, there exists a set of quasibound states whose frequency satisfies the superradiance condition $\omega_R<m\Omega_{\rm H}$. These modes are localized at a distance from the BH which is governed by the Compton wavelength $1/\mu$ and decay exponentially at large distances. In the small gravitational coupling limit, $M\mu\ll 1$ (where $\mu$ denotes the mass of the field), the spectrum of these modes resembles that of the hydrogen atom: 
\be
\omega_R/\mu \sim 1-\frac{(M\mu)^2}{2(j+1+n)^2}\,, \label{hydrogenic_2}
\ee
where $j=l+S$ is the total angular momentum of the state with spin projections $S=-s,-s+1,\ldots,s-1,s$, $s$ being the spin of the field. For a given $l$ and $n$, the total angular momentum $j$ satisfies the quantum mechanical rules for addition of angular momenta, $|l-s|\leq j\leq l+s$, and the spectrum is highly degenerate.

In the nonspinning case, the decay rate of these modes is well described by
\begin{equation}
 \omega_I/\mu\propto -(M\mu)^{\eta} \qquad \eta=4l+2S+5\,. \label{wIslope_2}
\end{equation}
%%%
In the spinning case, the imaginary part of the modes in the small $M\mu$ limit is described by the equation above with an extra factor $\left(2 r_+\mu-m a/M\right)$, which changes the sign of the imaginary part in the superradiant regime. Indeed, when $\omega_R<m\Omega_{\rm H}$ the imaginary part becomes positive and $\omega_I$ corresponds to the growth rate of the field ($\tau\equiv\omega_I^{-1}$ being the instability time scale). 

According to Eq.~\eqref{wIslope_2}, the shortest instability time scale occurs for $l=1$ and $S=-1$. The only exception to the scaling~\eqref{hydrogenic_2} and \eqref{wIslope_2} is given by the dipole polar mode of a spin-2 field, whose frequency is given by Eq.~\eqref{polar_di_Re} and the scaling of the imaginary part is similar to Eq.~\eqref{wIslope_2} but with $\eta=3$, as given by Eq.~\eqref{wIslope_po}.

Despite the recent progress in understanding these instabilities, so far only the massive spin-0 case has been studied in the full parameter space~\cite{Cardoso:2005vk,Dolan:2007mj,Witek:2012tr} and further work is needed to reach the same level of understanding for higher-spin fields. Massive spin-1 instabilities are known in detail to second order in the BH spin~\cite{Pani:2012vp,Pani:2012bp}. Beyond the slow-rotation approximation, the only work dealing with Proca instability of highly-spinning Kerr BHs is of numerical nature~\cite{Witek:2012tr}. The case of massive spin-2 fields is even less explored~\cite{Brito:2013wya}. The results here presented, obtained at first-order in the spin, are the first and only results available at the moment. We believe the progress made in recent years and the wide theoretical and phenomenological interest in light bosons should serve as an extra motivation to explore these instabilities further.

%%%%%%%%%%%%%%%%%%%%%%%%%%%%%%%%%%%%%%%%%%%%%%%%%%%%%%%%%%%%%%%%%%%%%%%%%%%%%%
\section{Discussion}
%%%%%%%%%%%%%%%%%%%%%%%%%%%%%%%%%%%%%%%%%%%%%%%%%%%%%%%%%%%%%%%%%%%%%%%%%%%%%%	

As shown in Chapter~\ref{chapter:massive2}, massive spin-2 fields propagating in a Schwarzschild BH background admit a very rich spectrum of long-lived stable states. Once rotation is turned on, these long-lived states can grow exponentially and extract angular momentum away from the BH. Thus, Kerr BHs are also unstable against a second mechanism: superradiance.
We showed that the instability is triggered when the superradiant condition is met, thus providing one further and strong piece of solid evidence that superradiant instabilities occur for any bosonic massive field. 

The polar gravitational sector is particularly interesting, as it displays the shortest instability timescale among other bosonic fields.
Our results are formally only valid in the small BH rotation limit, but previous second-order calculations for massive vector
fields suggest that a first-order analysis provides reasonably accurate results even beyond its regime of validity. The most crucial point in this regard
is the functional dependence of the instability timescale for the supposedly more unstable polar dipole mode, which we estimate to be:
\be
\tau_{{\mathrm{tensor}}}=\omega_I^{-1}\sim \frac{M(M\mu)^{-3}}{\gamma_{{\mathrm{polar}}}(\tilde{a}-2r_+\omega_R)}\,.
\ee
This timescale is four orders of magnitude shorter than the corresponding Proca field instability \cite{Pani:2012bp,Pani:2012vp}. 

In the next Chapter we will show that BH superradiant instabilities together with supermassive BH spin measurements
can be used to impose stringent constraints on
the allowed mass range of massive fields~\cite{Pani:2012bp,Pani:2012vp}. The observation of spinning BHs 
implies that the instability timescale is larger than typical competing spin-up effects.
For supermassive BHs a conservative estimate of these timescales is given by the Salpeter
timescale for accretion at the Eddington rate, $\tau_{\rm{Salpeter}}\sim 4.5\times 10^7$ years.
We find that the current best bound comes from Fairall 9 \cite{Schmoll:2009gq}, for which the polar instability implies a conservative
bound $m_g=\hbar\mu\lesssim 5\times 10^{-23} {\mathrm{eV}}$.
Unlike bounds for hypothetical massive photons, which may interact strongly with matter, the previous bound should not be strongly affected by the presence of
accretion disks around BHs, as the coupling of gravitons and other spin-2 fields to matter is very feeble.

This work requires extensions and further analysis, in particular the understanding of the time-development of the superradiant instability requires nonlinear
simulations. In a simpler context, we will partially address this issue in the next Chapter.
%%%%%%%%%%%%%%%%%%%%%%%%%%%%%%%%%%%%%%%%%%%%%%%%%%%%%%%%%%%%%%%%%%%%

%%%%%%%%%%%%%%%%%%%%%%%%%%%%%%%%%%%%%%%%%%%%%%%%%%%%%%%%%%%%%%%%%%%%
\chapter{Astrophysical Black Holes as Particle Detectors}\label{chapter:detectors}
%%%%%%%%%%%%%%%%%%%%%%%%%%%%%%%%%%%%%%%%%%%%%%%%%%%%%%%%%%%%%%%%%%%%%%%%%%%%%%%%
\section{Introduction}
%%%%%%%%%%%%%%%%%%%%%%%%%%%%%%%%%%%%%%%%%%%%%%%%%%%%%%%%%%%%%%%%%%%%%%%%%%%%%%%%
One of the most solid hypothesis of Einstein's GR is that \emph{BHs have no hair}~\cite{wheeler} and that all isolated, vacuum BHs in the Universe are described by the two-parameter Kerr family. Observing any deviation from this Kerr hypothesis --~a goal within the reach of upcoming and current gravitational-wave~\cite{Doeleman:2008qh,LIGO,VIRGO,KAGRA,ET,ELISA} and electromagnetic~\cite{Lu:2014zja,GRAVITY} facilities~-- would inevitably imply novel physics beyond either GR or the Standard Model of particle physics.

It has been recently pointed out that stationary spinning BHs can develop ``hair'' in the presence of massive bosonic fields~\cite{Hod:2012px,Herdeiro:2014goa,Herdeiro:2015tia,Herdeiro:2016tmi}. These new BH configurations exist at the threshold of the superradiant instability of the Kerr BH against massive bosonic fields~\cite{Damour:1976kh,Detweiler:1980uk,Zouros:1979iw,Cardoso:2004nk,Pani:2012bp,Cardoso:2013krh} and they can be interpreted as the nonlinear extension of linear bound states of frequency $\omega=m\Omega_{\rm{H}}$, where $m$ is the azimuthal wave number and $\Omega_{\rm{H}}$ is the angular velocity of the BH horizon. Such configurations require a complex field, with time and azimuthal dependence $\sim e^{im(\phi-\Omega_{\rm{H}} t)}$ otherwise a net scalar flux at the horizon and gravitational-wave flux at infinity would prevent the geometry from being stationary.
Formation scenarios for such configurations based on collapse or Jeans-like instability arguments are, notwithstanding, hard to devise.

\begin{figure}[ht]
\begin{center}
\begin{tabular}{c}
\epsfig{file=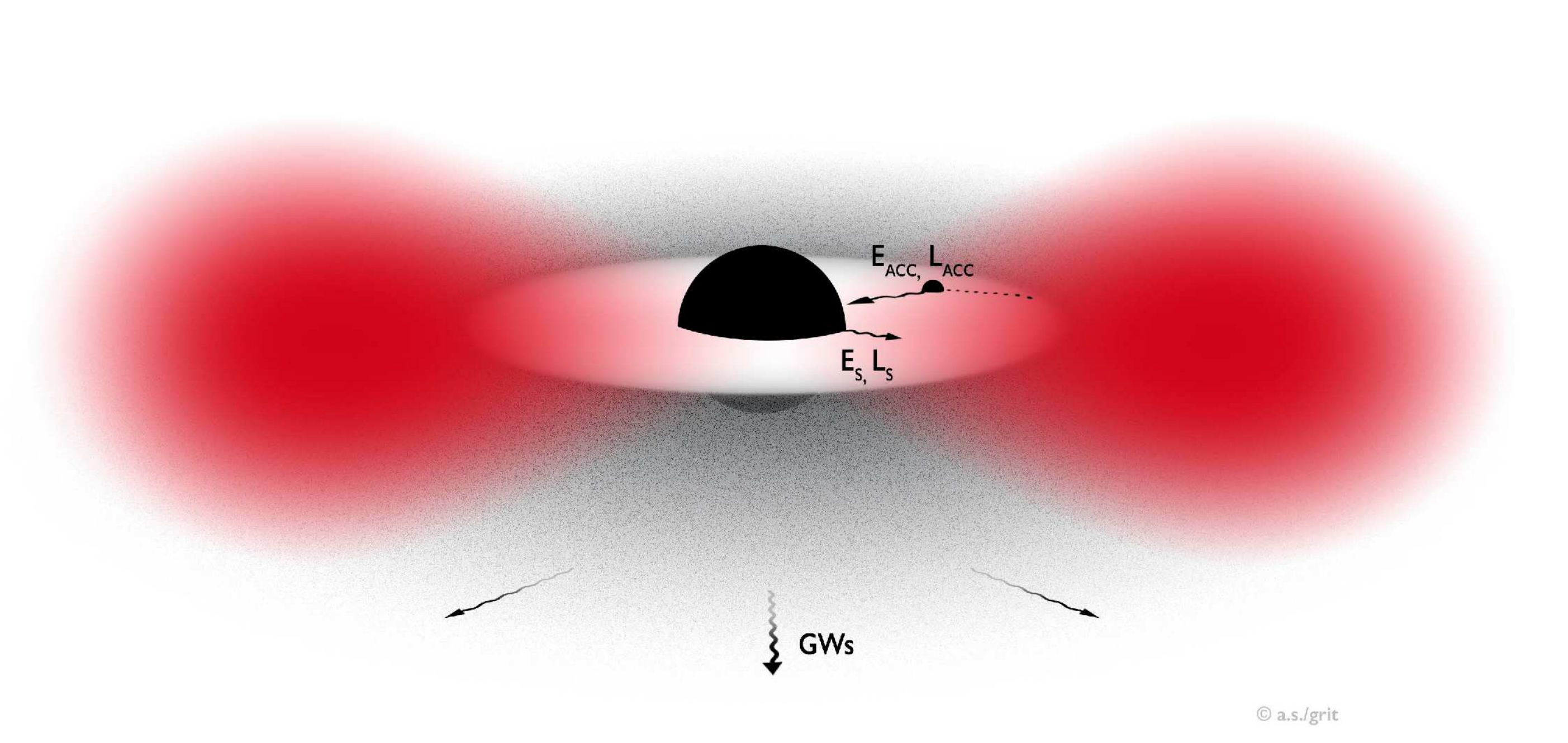,width=0.9\textwidth,angle=0,clip=true}
\end{tabular}
\end{center}
\caption{\label{fig:draw}
Pictorial description of a bosonic cloud around a spinning BH in a realistic astrophysical environment. The BH loses energy $E_S$ and angular momentum $L_S$ through superradiant extraction of scalar waves and emission of gravitational waves, while accreting gas from the disk, which transports energy $E_{\rm{ACC}}$ and angular momentum $L_{\rm{ACC}}$. Notice that accreting material is basically in free fall 
after it reaches the innermost stable circular orbit. The cloud is localized at a distance $\sim 1/M\mu^2>2M$.}
\end{figure}
However, quantum or classical fluctuations of \emph{any} massive bosonic field trigger a superradiant instability of the Kerr metric, whose time scale $\tau$ can be extremely short. For a BH with mass $M$, the shortest instability time scale is $\tau\sim \left(\frac{M}{10^6 M_\odot}\right){\rm{yr}}$ for a ultralight scalar field~\cite{Cardoso:2005vk,Dolan:2007mj,Pani:2012vp,Witek:2012tr,Brito:2013wya}, and shorter for vector~\cite{Pani:2012vp,Pani:2012bp,Witek:2012tr} and tensor fields~\cite{Brito:2013wya} for which superradiance is more efficient.
Little is known about the nonlinear development of the instability but it is expected that, within such short time, a nonspherical bosonic cloud would grow near the BH extracting energy and angular momentum, until superradiance stops and the cloud is slowly re-absorbed by the BH and dissipated through gravitational-wave emission~\cite{Arvanitaki:2010sy, Witek:2012tr,Okawa:2014nda,Cardoso:2013krh}. Although (at least for a real, stationary scalar field) the no-hair theorems~\cite{HawkingBook,Heusler:1995qj,Sotiriou:2011dz,Graham:2014ina} guarantee that the final state of the instability has to be a Kerr BH with lower spin and no hair, it is important to understand the time scales involved in this process, because a scalar cloud surviving for cosmological times would be practically indistinguishable from a full-fledged BH hair and would have various important consequences.

A further motivation to explore realistic evolutions of the instability derives from the surprising connections between strong-field gravity and particle physics.
In recent years superradiant instabilities of astrophysical BHs have been used --~together with precision measurements of BH mass and spin (see e.g.~\cite{Brenneman:2011wz})~-- to constrain stringy axions and ultralight scalars~\cite{Arvanitaki:2010sy,Kodama:2011zc} (these constraints being complementary to those coming from cosmological observations~\cite{Bozek:2014uqa,Hlozek:2014lca}), to derive bounds on light vector fields~\cite{Pani:2012vp} and on the mass of the graviton~\cite{Brito:2013wya} (cf. Chapter~\ref{chapter:Kerr}), as well as to put intrinsic bounds on magnetic fields near BHs~\cite{Brito:2014nja} (cf. Chapter~\ref{chapter:magnetic}) and on the fraction of primordial BHs in dark matter~\cite{Pani:2013hpa}. However, all these predictions are based on a linearized analysis, neglecting backreaction and other competitive effects --~such as gravitational-wave emission and gas accretion~-- which can have an impact on the development of the process (see Fig.~\ref{fig:draw} for a pictorial view of the system under consideration). In this Chapter we take the first step to understand the evolution of the superradiant instability of a Kerr BH and to identify the relevant time scales for this problem.

Our main conclusion is that the linearized analysis used so far to impose constraints on particle masses is \emph{accurate} for both real and complex fields. In fact, a linearized analysis
of the instability might remain accurate in the entire regime of initial conditions, even when the mass of the scalar ``cloud'' that forms is of the order of the BH mass. The reason
is that the scalar field is typically distributed over a very large volume, implying very small densities and consequent small backreaction effects. For this reason, the BH geometry is very well described by the Kerr metric, even in the presence of massive bosonic clouds.

%%%%%%%%%%%%%%%%%%%%%%%%%%%%%%%%%%%%%%%%%%%%%%%%%%%%%%%%%%%%%%%%%%%%%%%%%%%%%%%%
\section{Bosonic clouds around BHs: a quasi-adiabatic approximation \label{sec:formalism}}
%%%%%%%%%%%%%%%%%%%%%%%%%%%%%%%%%%%%%%%%%%%%%%%%%%%%%%%%%%%%%%%%%%%%%%%%%%%%%%%%
For concreteness --~and also because it is where most of the work on BH superradiance is framed~-- we focus on the action for a minimally coupled massive scalar field, which can be either real or complex (we use Planck units):
\be
S=\int d^4x \sqrt{-g} \left( \frac{R}{16\pi}-\frac{1}{2}g^{\mu\nu}\Phi^{\ast}_{,\mu}\Phi^{}_{,\nu} - \frac{\mu^2}{2}\Phi^{\ast}\Phi\right)\,,
\ee
although the qualitative aspects of our analysis are valid also for other massive bosonic fields\footnote{Here we neglect possible scalar self-interactions beyond the mass term. Nonlinearities can give rise to interesting effects, for example a condensate of axion-like particles governed by a sine-Gordon potential $V(\Phi)=f_a^2 \mu^2[1-\cos(\Phi/f_a)]$ would collapse and produce a ``bosenova'' explosion when $M_S\approx 1600 (f_a/M_P)^2 M$~\cite{Yoshino:2012kn,Kodama:2011zc}, where $f_a$ is a model-dependent decay constant and $M_P$ is the Planck mass.}.
The resulting field equations are $\nabla_{\mu}\nabla^{\mu}\Phi =\mu^2\Phi$ and $G^{\mu\nu}=8\pi T^{\mu\nu}$ with
\begin{equation}
T^{\mu \nu}=\Phi^{\ast,(\mu}\Phi^{,\nu)}-\frac{1}{2}g^{\mu\nu}\left( \Phi^{\ast}_{,\alpha}\Phi^{,\alpha}+{{\mu}^2}\Phi^{\ast}\Phi\right)\,. \label{Tmunu}
\end{equation}
A full nonlinear evolution of this system in the case of a spinning BH was recently performed~\cite{Okawa:2014nda}; following the development of the instability is extremely challenging because of the time scales involved: 
$\tau_{\rm{BH}}\sim M$ is the light-crossing time, $\tau_S\sim 1/\mu$ is the typical oscillation period of the scalar cloud and $\tau \sim M/(M\mu)^9$ is the instability time scale in the small-$M\mu$ limit. In the most favorable case for the instability, $\tau\sim 10^6\tau_S$ is the minimum evolution time scale required for the superradiant effects to become noticeable\footnote{The minimum instability time scale corresponds to $M\mu\sim0.42$ (see e.g.~\cite{Dolan:2007mj}). Although this value is beyond the analytical, small-$M\mu$ approximation, the numerical result is in good agreement with an extrapolation of the analytical formula~\cite{Pani:2012bp}.}. Thus, current nonlinear evolutions (which typically last at most $\sim 10^3 \tau_S$~\cite{Okawa:2014nda}) have not yet probed
the development of the instability, nor the impact of gravitational-wave emission.

However, in such configuration the system is suitable for a quasi-adiabatic approximation: over the dynamical time scale of the BH the scalar field can be considered almost stationary and its backreaction on the geometry can be neglected as long as the scalar energy is small compared to the BH mass.  
Therefore, we consider a perturbative expansion in powers of the scalar field and check consistency a posteriori.

%%%%%%%%%%%%%%%%%%%%%%%%%%%%%%%%
\subsection{Linearized analysis}
%%%%%%%%%%%%%%%%%%%%%%%%%%%%%%%%
At leading order, the geometry is described by the Kerr spacetime and the scalar evolves in this fixed background. In the Teukolsky formalism~\cite{Teukolsky:1972my,Teukolsky:1973ha} (see Appendix~\ref{app:Teu_eqs}), the Klein-Gordon equation can be separated by use of spin-0 spheroidal wavefunctions~\cite{Berti:2005gp},
\be
\Phi=\int d\omega e^{-i\omega t+im\phi}{_0}S_{lm}(\theta)R(r)\,, \nn
\ee
and is equivalent to the following differential equations,
\beq
{\cal D}_\theta[{_0}S]
&+&\left[a^2(\omega^2-\mu^2)\cos^2\theta
-\frac{m^2}{\sin^2\theta}+\lambda\right]{_0}S=0\,,\nn\\
{\cal D}_r[R]&+&\left[\omega^2(r^2+a^2)^2-4aMrm\omega+a^2m^2-\Delta(\mu^2r^2+a^2\omega^2+\lambda)\right]R=0\,.\nn
\eeq
where ${\cal D}_\theta=(\sin\theta)^{-1}\partial_\theta\left(\sin\theta\partial_\theta\right)$, ${\cal D}_r=\Delta\partial_r\left(\Delta\partial_r\right)$, $\Delta=(r-r_+)(r-r_-)$, $r_\pm=M\pm \sqrt{M^2-a^2}$ and $a$ is related to the BH angular momentum $J=a M$.
A numerical solution to the above coupled system is straightforward~\cite{Berti:2009kk,Cardoso:2005vk}. For small mass couplings $M\mu$, it can be shown that the corresponding eigenvalue problem admits a hydrogenic-like solution~\cite{Detweiler:1980uk,Pani:2012vp,Pani:2012bp} with $\lambda\sim l(l+1)$ and [cf. eqs.~\eqref{hydrogenic_2} and~\eqref{wIslope_2}]
\be\label{omega}
\omega\sim \mu-\frac{\mu}{2}\left(\frac{M\mu}{l+n+1}\right)^2+\frac{i}{\gamma_l}\left(\frac{am}{M}-2\mu r_+\right)(M\mu)^{4l+5}\,,
\ee
where $n=0,1,2...$ and $\gamma_1=48$ for the dominant unstable $l=1$ mode. Note that the eigenfrequencies are complex, $\omega=\omega_R+i\omega_I$, unless the superradiant condition is saturated, at 
\begin{equation}
 a=a_{\rm{crit}}\approx \frac{2\mu M r_+}{m}\,.\label{acrit}
\end{equation}
%%%%

In the small-$\mu$ limit the eigenfunctions read~\cite{Detweiler:1980uk,Yoshino:2013ofa}
\be
R(\mu,a,M,r)=A_{ln} g(\tilde{r})\,,
\ee
where $g(\tilde{r})$ is an universal function of the dimensionless quantity $\tilde{r}=2r M {\mu}^2/(l+n+1)$ and can be written in terms of Laguerre polynomials
%%%
\begin{equation}
 g(\tilde r)=\tilde{r}^l e^{-\tilde{r}/2}L_{n}^{2l+1}(\tilde{r})\,.\label{univg}
\end{equation}
%%%
We have verified that this is a good description of the numerical eigenfunctions for moderately large $\mu M\lesssim0.2$ even at large BH spin. Notice that the eigenfunction peaks at $r_{\rm{cloud}}\sim \frac{(l+n+1)^2}{(M\mu)^2}M$~\cite{Arvanitaki:2010sy} (see also~\cite{Benone:2014ssa}) and thus extends way beyond the horizon, where rotation effects can be neglected. 
For definiteness, and because it is the single most unstable mode, we focus for now on $l=m=1$ and $n=0$. In this case $g(\tilde r)=\tilde{r} e^{-\tilde{r}/2}$.

As we noted, there are clearly two scales in the problem. One is dictated by the oscillation time $\tau_S=1/\omega_R\sim1/\mu$, the other by the instability growth time scale, $\tau=1/\omega_I\gg\tau_S$.
As such, we will consider these scales to be well separated, and will assume that the cloud is stationary and described by
\be
\Phi=A_0g(\tilde{r})\cos\left(\phi-\omega_Rt\right)\sin\theta\,, \label{scalar}
\ee
where $A_0\equiv A_{10}$. In Eq.~\ref{scalar} we assumed a {\it real} scalar field, because this is the configuration that maximizes gravitational-wave emission:
complex scalars in a nearly stationary regime will exhibit no time-dependent stress-energy tensor, and therefore do not emit gravitational waves in this approximation (this case is briefly discussed in Sec.~\ref{sec:hairy} below).
As we will show, even real scalars give rise to very small gravitational-wave energy fluxes.

For convenience, by using Eq.~\ref{Tmunu}, the amplitude $A_0$ can be expressed in terms of the mass $M_S$ of the scalar cloud,
\be
M_S=\int r^2\sin\theta\rho=\frac{2\pi A_0^2}{3} \left(2{\cal I}_0+\frac{2{\cal I}_2}{M^2{\mu}^2}+{\cal I}'_2\right)\,,
\ee
where we defined the dimensionless integrals 
\begin{eqnarray}
 {\cal I}_n&=& \int_0^\infty d\tilde{r} \tilde{r}^n g(\tilde{r})^2\,,\quad
 {\cal I}'_n= \int_0^\infty d\tilde{r} \tilde{r}^n g'(\tilde{r})^2\,,
\end{eqnarray}
and the energy density $\rho\equiv -T_0^0$ reads
\begin{eqnarray}
 \rho &=& \frac{A_0^2}{2r^2}\left\{{\mu}^4 M^2 r^2 \sin ^2(\theta ) g'(\tilde{r})^2 \cos ^2(\phi -\omega_R t)\right.\nn\\
 &&\left.+g(\tilde{r})^2 \left[\left(\cos ^2(\theta )+{\mu}^2 r^2 \sin ^2(\theta )\right) \cos ^2(\phi -\omega_R t)\right.\right.\nn\\
 &&\left.\left.+\left[1+r^2 \omega_R^2 \sin ^2(\theta )\right] \sin ^2(\phi -\omega_R t)\right]\right\}\,, \label{rho}
\end{eqnarray}
where a prime denotes derivative with respect to the argument. 
In the small ${\mu}M$ limit one obtains
\begin{equation}\label{amplitude}
A_0^2=\frac{3}{4\pi {\cal I}_2}\left(\frac{M_S}{M}\right) ({\mu} M)^4\,. 
\end{equation}
In deriving the formulas above we have assumed that spacetime is flat. This approximation is accurate as long as the cloud is localized far away from the BH, i.e. when $\mu M\ll1$ (cf. Ref.~\cite{Yoshino:2013ofa} where a similar approximation is discussed). When $\mu M\ll 1$, the relation~\ref{amplitude} is valid also in the full Kerr case.
%%%%%%%%%%%%%%%%%%%%%%%%%%%%%
\subsection{Gravitational-wave emission}
%%%%%%%%%%%%%%%%%%%%%%%%%%%%%

A nonspherical monochromatic cloud as in Eq.~\ref{scalar} will emit gravitational waves with frequency $ 2\pi/\lambda_c\sim 2\omega_R\sim 2\mu$, the wavelength $\lambda_c$ being in general \emph{smaller} than the size of the source, $r_{\rm{cloud}}$. Thus, even though the cloud is nonrelativistic, the quadrupole formula does not apply because the emission is incoherent~\cite{Arvanitaki:2010sy,Yoshino:2013ofa}. However, due to the separation of scales between the size of the cloud and the BH size for $\mu M\ll 1$, the gravitational-wave emission can be analyzed taking the source to lie in a nonrotating (or even flat~\cite{Yoshino:2013ofa}) background.

In the fully relativistic regime, the gravitational radiation generated is best described by the Teukolsky formalism
for the gravitational perturbations. In the Teukolsky formalism the perturbation equations can be reduced to a second-order differential
equation for the Newman-Penrose scalar $\Psi_4$ (see Appendix~\ref{app:Teu_eqs}). We can decompose $\Psi_4$ as
\be
\Psi_4(t,r,\Omega)=\sum_{lm} r^{-4}\int^\infty_{-\infty}d\omega\sum_{lm}~_{-2}R_{lm}(r)~_{-2}Y_{lm}(\Omega)e^{-i\omega t}\,, \label{psi4expansion}
\ee
where ${_s}Y_{lm}(\theta,\phi)$ are the spin-$s$ weighted spherical harmonics~\cite{Berti:2005gp}. The radial function $~_{-2}R(r)$ satisfies the inhomogeneous equation
\beq\label{teu_eq}
&&r^2f ~_{-2}R''-2(r-M)~_{-2}R'+\left[f^{-1}\left(\omega^2r^2-4i\omega (r-3M)\right)\right.\nn\\
&&\left.-(l-1)(l+2)\right]~_{-2}R=-T_{lm\omega}\,,
\eeq
where $f=1-2M/r$. The source term $T_{lm\omega}$ is related to the scalar field stress-energy tensor $T_{\mu\nu}$ through the tetrad projections, $T_{\mu\nu}n^{\mu}n^{\nu}\equiv T_{nn}$, $T_{\mu\nu}n^{\mu}\bar{m}^{\nu}\equiv T_{n\bar{m}}$ and $T_{\mu\nu}\bar{m}^{\mu}\bar{m}^{\nu}\equiv T_{\bar{m}\bar{m}}$, where
\beq
n^{\mu}&=&\frac{1}{2}\left (1,-f,0,0 \right )\,,\\
\bar{m}^{\mu}&=&\frac{1}{\sqrt{2}\,r}\left (0,0,1,-\frac{i}{\sin\theta}
\right )\,. \eeq 
We define
\beq 
_{S}T&\equiv& \frac{1}{2\pi} \int\, d\Omega\, dt \, {\cal T}_{S}~_{S}\bar{Y}_{lm}e^{i\omega t}\,,
\eeq
where ${\cal T}_S=T_{nn}$, $T_{n\bar{m}}$ and $T_{\bar{m}\bar{m}}$ for $S=0,-1,-2$, respectively. The source is then given by~\cite{Poisson:1993vp}
\begin{eqnarray}
\frac{T_{lm\omega}}{2\pi}&=&2\left[(l-1)l(l+1)(l+2)\right]^{1/2}r^4~_{0}T\nn\\
&+&2\left[2(l-1)(l+2)\right]^{1/2}r^2 f \mathcal{L}\left(r^3 f^{-1}~_{-1}T\right)\nn\\
&+&r f\mathcal{L}\left[r^4 f^{-1}\mathcal{L}\left(r~_{-2}T\right)\right]\,,
\end{eqnarray}
where $\mathcal{L}\equiv f \partial_r+i\omega$. Using Eqs.~\ref{Tmunu} and~\ref{scalar} we find that,
as expected, for the scalar configuration~\ref{scalar} the only modes that contribute are $l=|m|=2$ with frequencies $\omega=\pm 2\omega_R$.

Once the source term is known, $\Psi_4$ can be computed using a Green's function approach. To construct the Green function we need two linearly independent solutions of the homogeneous equation. A physically motivated choice is to consider the solution $~_{-2}R^{\infty}$ which describes outgoing waves at infinity and $~_{-2}R^H$ which describes ingoing waves at the event horizon. By making use of the fact that the Wronskian $W=\frac{\partial_r R^{\infty} R^{H}-\partial_r R^H R^{\infty} }{r^2 f}=2i \omega B_{\mathrm{in}}$ is constant by virtue of the field equations, the correct solution of the inhomogeneous problem at infinity reads (see Appendix~\eqref{app:GF})
%%%
\begin{equation}
~_{-2}R(r\to \infty)\sim \frac{~_{-2}R^{\infty}}{2i\omega B_{\mathrm{in}}}\int_{2M}^{\infty} dr \frac{~_{-2}R^H T_{lm\omega}}{r^4 f^2}\,,
\end{equation}
%%%
where $~_{-2}R^{\infty}(r\to \infty)\sim r^3 e^{i\omega r}$ and $~_{-2}R^H(r\to \infty)\sim B_{\mathrm{out}} r^3 e^{i\omega r}+B_{\mathrm{in}} e^{-i\omega r}/r$. From the asymptotic solution of Eq.~\ref{teu_eq}, we find
\be
B_{\mathrm{in}}=-\frac{C_1}{8\omega^2}(l-1)l(l+1)(l+2)e^{i(l+1)\pi/2}\,,
\ee
where $C_1$ is a constant. The solution $~_{-2}R^H$ can be found more easily by solving the Regge-Wheeler equation (see e.g.~\cite{Poisson:1993vp}) for small frequencies and using the fact that $~_{-2}R^H=r^2f\mathcal{L}\left(f^{-1}\mathcal{L}r R^{\rm{RW}}\right)$, where $R^{\rm{RW}}$ is the Regge-Wheeler function that, at small frequencies, reads~\cite{Poisson:1993vp}
\be 
R^{\rm{RW}}\sim C_1\omega r j_l(\omega r)\,,
\ee
where $j_l$ denote the spherical Bessel functions of the first kind.
Finally, the luminosity can be computed from 
\be
\dot{E}_{\mathrm{GW}}=\int d\Omega\frac{r^2}{4\pi\omega^2}|\Psi_4|^2=\int d\omega\frac{|Z|^2}{2\pi\omega^2|W|^2}\,,
\ee
where $Z\equiv \int dr \frac{~_{-2}R^H T_{22\omega}}{r^4 f^2}$ and we used the fact that the modes with $m=\pm 2$ give the same contribution to the luminosity. 
The final result reads
\be
\dot{E}_{\mathrm{GW}}=\frac{2}{45} \pi ^2 \left(484+9 \pi ^2\right) M^6 A_0^4 \mu^6\,, \label{dEdtF2}
\ee
which, by using Eq.~\ref{amplitude} and $\mathcal{I}_2\sim 24$ and in the small-$\mu$ limit reduces to
\be
\dot{E}_{\rm{GW}}=\frac{484+9 \pi ^2}{23040}\left(\frac{M_S^2}{M^2}\right)(M\mu)^{14}\,. \label{dEdtF}
\ee
The different prefactor relative to that derived in Ref.~\cite{Yoshino:2013ofa} is due to the fact that we considered a Schwarzschild background instead of a flat metric. Indeed, our result~\ref{dEdtF} is in better agreement with the numerical results.
Note that such flux is an \emph{upper bound} relative to the exact numerical results which are valid for any $\mu$ and any BH spin~\cite{Yoshino:2013ofa}. In the following we will use Eq.~\ref{dEdtF} as a very conservative assumption, since the gravitational-wave flux is generically smaller.

A similar computation for the angular momentum dissipated in gravitational waves gives
\be
\dot{J}_{\rm{GW}}=\frac{1}{\omega_R} \dot{E}_{\rm{GW}}\,, \label{dJdtF}
\ee
in agreement with the general result for a monochromatic wave of the form~\ref{scalar}.

%%%%%%%%%%%%%%%%%%%%%%%%
\subsection{Accretion}
%%%%%%%%%%%%%%%%%%%%%%%%
Astrophysical BHs are not in isolation but surrounded by matter fields in the form of gas and plasma. On the one hand, addition of mass and angular momentum to the BH via accretion competes with superradiant extraction. On the other hand, a slowly-rotating BH which does not satisfy the superradiance condition might be spun up by accretion and might become superradiantly unstable precisely \emph{because} of angular momentum accretion. Likewise, for a light BH whose coupling parameter $\mu M$ is small, superradiance might be initially negligible but it can become important as the mass of the BH grows through gas accretion. It is therefore crucial to include accretion in the treatment of BH superradiance, as we do here for the first time\footnote{We consider only gas accretion, which is a more efficient spin-up mechanism than, for example, accretion by tidal disruption of binary stars~\cite{2012ApJ...749L..42B}. This is a conservative assumption to study the evolution of superradiant instabilities, because spin-up by accretion competes with superradiant extraction of the angular momentum. We also neglect other processes that can affect the evolution of the mass and spin, like BH mergers~\cite{Barausse:2012fy}.}.

We make the most conservative assumption by using a model in which mass accretion occurs at a fraction of the Eddington rate (see e.g.~\cite{Barausse:2014tra}):
\begin{equation}
 \dot M_{\rm{ACC}} \equiv f_{\rm{Edd}} \dot M_{\rm{Edd}}\sim 0.02 f_{\rm{Edd}} \frac{M(t)}{10^6 M_\odot} M_\odot {\rm{yr}}^{-1}\,,\label{dotMaccr}
\end{equation}
%%%
where we have assumed an average value of the radiative efficiency $\eta\approx0.1$, as required by Soltan-type arguments, i.e. a comparison between the luminosity of active galactic nuclei and the mass function of BHs~\cite{LyndenBell:1969yx,Soltan:1982vf}. The Eddington ratio for mass accretion, $f_{\rm{Edd}}$, depends on the details of the accretion disk surrounding the BH and it is at most of the order unity for quasars and active galactic nuclei, whereas it is typically much smaller for quiescent galactic nuclei (e.g. $f_{\rm{Edd}}\sim 10^{-9}$ for SgrA$^{*}$). If we assume that mass growth occurs via accretion through Eq.~\ref{dotMaccr}, the
BH mass grows exponentially with $e$-folding time given by a fraction $1/f_{\rm{Edd}}$ of the Salpeter time scale, $\tau_{\rm{Salpeter}}=\frac{\sigma_T}{4\pi m_p}\sim4.5\times 10^7$~yr, where $\sigma_T$ is the Thompson cross section and $m_p$ is the proton mass. Therefore, the minimum time scale for the BH spin to grow via gas accretion is roughly $\tau_{\rm{ACC}}\sim \tau_{\rm{Salpeter}}/f_{\rm{Edd}}\gg \tau_{\rm{BH}}$ and also in this case the adiabatic approximation is well justified.

Regarding the evolution of the BH angular momentum through accretion, we make the conservative assumption that the disk lies on the equatorial plane and extends down to the innermost stable circular orbit (ISCO). 
If not, angular momentum increase via accretion is suppressed and superradiance becomes (even) more dominant.
Ignoring radiation effects, the evolution equation for the spin reads~\cite{Bardeen:1970zz}
\begin{equation}
\dot J_{\rm{ACC}} \equiv \frac{L(M,J)}{E(M,J)} \dot M_{\rm{ACC}}\,,\label{dotJaccr}
\end{equation}
%%%
where $L(M,J)=2M/(3\sqrt{3})\left(1+2 \sqrt{3 r_{\rm{ISCO}}/{M}-2}\right)$ and $E(M,J)=\sqrt{1-2M/3r_{\rm{ISCO}}}$ are the angular momentum and energy per unit mass, respectively, of the ISCO of the Kerr metric, located at $r_{\rm{ISCO}}=r_{\rm{ISCO}}(M,J)$ in Boyer-Lindquist coordinates. 

In the absence of superradiance the BH would reach extremality in finite time, whereas radiation effects set an upper bound of $a/M\sim 0.998$~\cite{Thorne:1974ve}. To mimic this upper bound in a simplistic way, we introduced a smooth cutoff in the accretion rate for the angular momentum. This cutoff merely prevents the BH to reach extremality and does not play any role in the evolution discussed in the next section.

\begin{figure*}[ht]
\begin{center}
\begin{tabular}{cc}
\epsfig{file=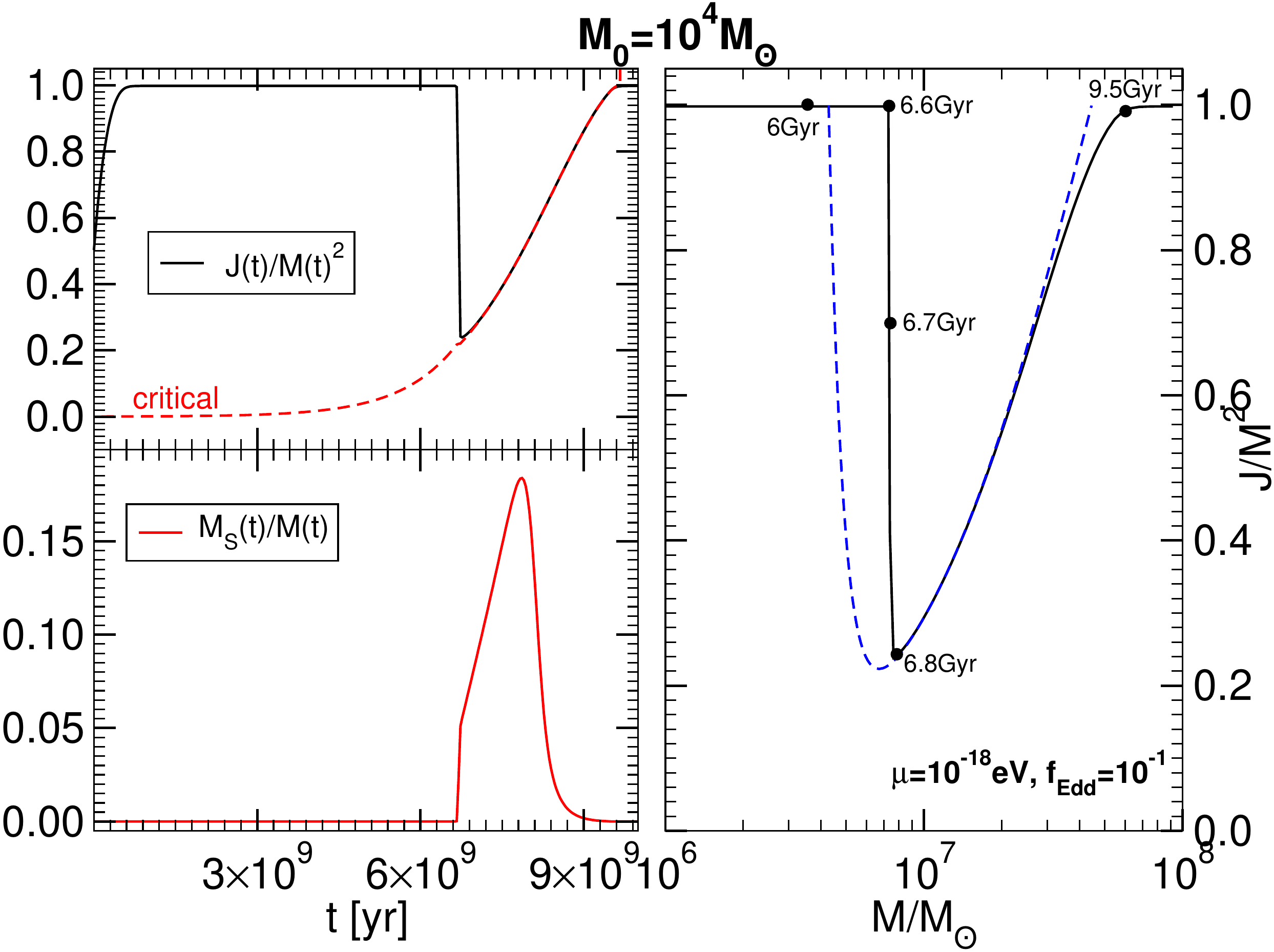,width=0.48\textwidth,angle=0,clip=true}&
\epsfig{file=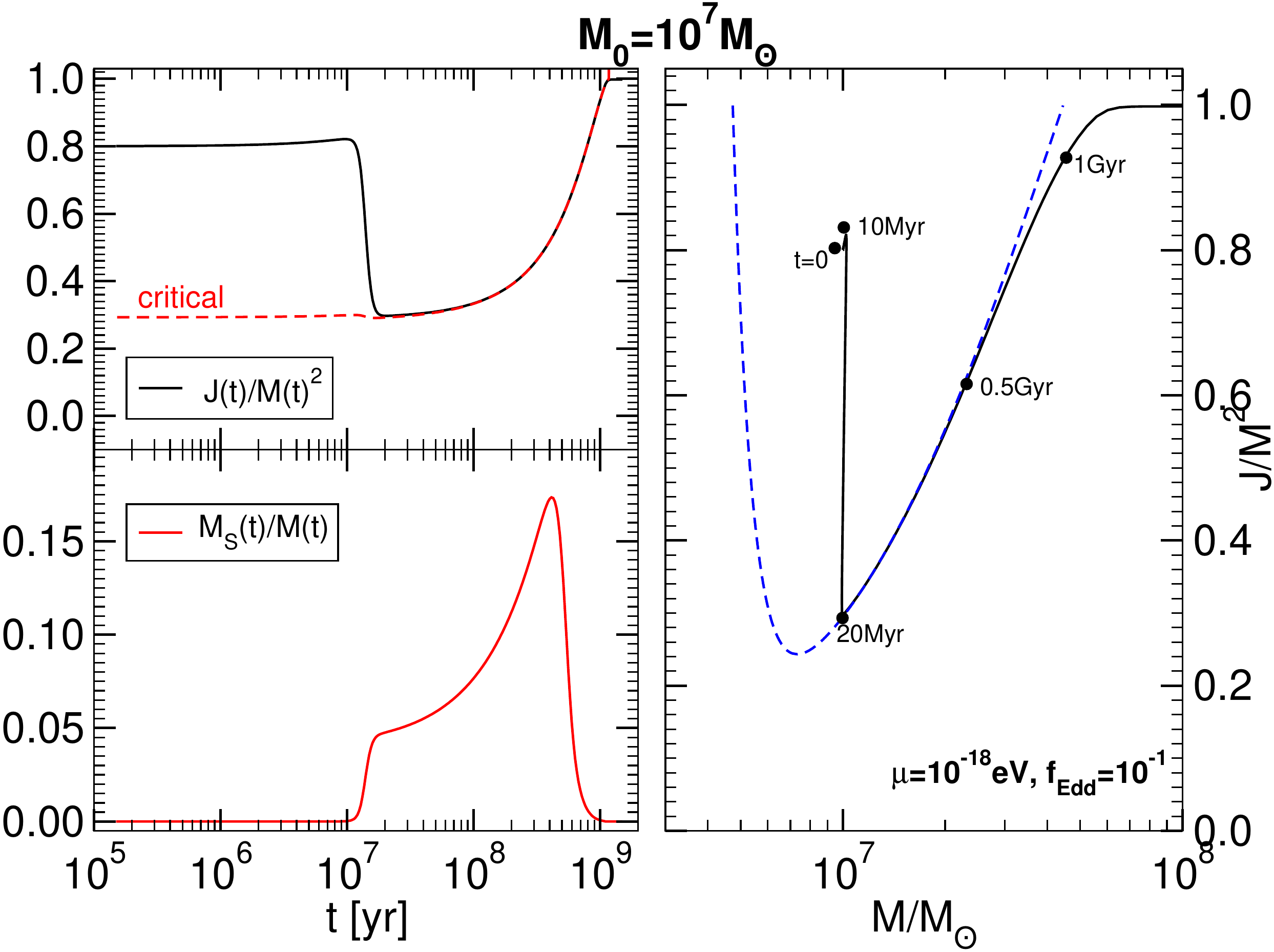,width=0.48\textwidth,angle=0,clip=true}
%%%%%
\end{tabular}
\end{center}
\caption{\label{fig:evolution}
Evolution of the BH mass and spin and of the scalar cloud due to superradiance, accretion of gas and emission of gravitational waves. The two sets of plots show two different cases. In Case I (left set) the initial BH mass $M_0=10^4 M_\odot$ and the initial BH spin $J_0/M_0^2=0.5$. The BH enters the instability region at about $t\sim 6{\rm{Gyr}}$, when its mass $M\sim10^7 M_\odot$ and its spin is quasi-extremal. The set of plots on the right shows Case II, in which $M_0=10^7M_\odot$ and $J_0/M_0^2=0.8$, and the evolution starts already in the instability region for this scalar mass $\mu=10^{-18}{\rm{eV}}$. For both cases, the left top panels show the dimensionless angular momentum $J/M^2$ and the critical superradiant threshold $a_{\rm{crit}}/M$ (cf. Eq.~\ref{acrit}); the left bottom panels show the mass of the scalar cloud $M_S/M$ (note the logarithmic scale in the x-axis for Case II); and the right panels show the trajectory of the BH in the Regge plane~\cite{Arvanitaki:2010sy} during the evolution. The dashed blue line denotes the depleted region as estimated by the linearized analysis, i.e. it marks the threshold at which $\tau\sim\tau_{\rm{ACC}}$. 
}
\end{figure*}
%

%%%%%%%%%%%%%%%%%%%%%%%%%%%%%%%%%%%%%%%%
\section{Evolution of the cloud}\label{sec:evolution}
%%%%%%%%%%%%%%%%%%%%%%%%%%%%%%%%%%%%%%%%
We are now in a position to discuss the evolution of the scalar cloud within the quasiadiabatic approximation. 
The scalar energy flux that is extracted from the horizon through superradiance is
\be
\dot{E}_S=2M_S\omega_I\,,
\ee
where $M\omega_I=\frac{1}{48}({a/M-2{\mu} r_+})(M{\mu})^9$
%%%
%%%
for the $l=m=1$ fundamental mode and clearly $\dot{E}_S>0$ only in the superradiant regime.  

Two further contributions come from the emission of gravitational waves through fluxes~\ref{dEdtF} and~\ref{dJdtF} and from gas accretion through the accretion rates~\ref{dotMaccr} and \ref{dotJaccr}. Energy and angular momentum conservation require that
%%%%
\begin{eqnarray}
 \dot{M}+\dot{M}_S&=&-\dot{E}_{\rm{GW}}+\dot M_{\rm{ACC}}\,,\\
 \dot{J}+\dot{J}_S&=&-\frac{1}{{\mu}}\dot{E}_{\rm{GW}}+\dot J_{\rm{ACC}}\,,
\end{eqnarray}
%%%%
where we have used $\dot{J}_{\rm{GW}}=\dot{E}_{\rm{GW}}/\omega_R\sim\dot{E}_{\rm{GW}}/\mu$, we have neglected the subdominant contributions of the mass of the disk and of those gravitational waves that are absorbed at the horizon, and we have approximated the local mass and angular momentum by their ADM counterparts. The latter approximation is valid as long as backreaction effects are small, as we discuss below.
The evolution of the system is governed by the two equations above supplemented by 
\beq
\dot{M}&=& -\dot{E}_S+\dot M_{\rm{ACC}} \,,\\
\dot{J}&=& -\frac{1}{{\mu}}\dot{E}_S +\dot J_{\rm{ACC}}\,,
\eeq
which describe the superradiant extraction of energy and angular momentum and the competitive effects of gas accretion at the BH horizon. These equations assume that the scalar cloud is not directly (or only very weakly)  coupled to the disk.

Representative results for the evolution of the system are presented in Fig.~\ref{fig:evolution} where we consider the scalar-field mass $\mu=10^{-18}{\rm{eV}}$ and mass accretion near the Eddington rate, $f_{\rm{Edd}}=0.1$. We consider two cases: (I) the left set of plots corresponds to a BH with initial mass $M_0=10^4 M_\odot$ and initial spin $J_0/M_0^2=0.5$, whereas (II) the right set of plots corresponds to $M_0=10^7 M_\odot$ and $J_0/M_0^2=0.8$.

In Case I, superradiance is initially negligible because $\mu M_0\sim 10^{-4}$ and superradiant extraction is suppressed. Thus, the system evolves mostly through gas accretion, reaching extremality ($J/M^2\sim0.998$) within the time scale $\tau_{\rm{ACC}}\sim 10\tau_{\rm{Salpeter}}$. At about $t\sim 6{\rm{Gyr}}$, the BH mass is sufficiently large that the superradiant coupling $\mu M$ becomes important. This corresponds to the BH entering the region delimited by a dashed blue curve in the Regge plane~\cite{Arvanitaki:2010sy} shown in Fig.~\ref{fig:evolution} for Case I. At this stage superradiance becomes effective very quickly: a scalar cloud grows exponentially near the BH (left bottom panel), while
mass and angular momentum are extracted from the BH (left top panel). This abrupt phase lasts until the BH spin reaches the critical value $a_{\rm{crit}}/M$ and superradiance halts. Because the initial growth is exponential, the evolution does not depend on the initial mass and initial spin of the scalar cloud as long as the latter are small enough, so that in principle also a quantum fluctuation would grow to a sizeable fraction of the BH mass in finite time. 

Before the formation of the scalar condensate, the evolution is the same regardless of gravitational-wave emission and the only role of accretion is to bring the BH into the instability window.  After the scalar growth, the presence of gravitational-wave dissipation and accretion produces two effects: (i) the scalar condensate loses energy through the emission of gravitational waves, as shown in the left bottom panel of Fig.~\ref{fig:evolution}; (ii) gas accretion returns to increase the BH mass and spin.

However, because accretion restarts in a region in which the superradiance coupling $\mu M$ is nonnegligible, the ``Regge trajectory'' $J(t)/M(t)^2\sim a_{\rm{crit}}/M$ (cf. Eq.~\ref{acrit}) is an attractor for the evolution and the BH ``stays on track'' as its mass and angular momentum grow. For Case I, this happens between $t\sim 6.8{\rm{Gyr}}$ and $t\sim9.5{\rm {Gyr}}$, i.e. until the spin reaches the critical value $J/M^2\sim 0.998$ and angular momentum accretion saturates.

A similar discussion holds true also for Case II, presented in the right set of plots in Fig.~\ref{fig:evolution}. In this case, the BH starts already in the instability regime, its spin grows only very little before superradiance becomes dominant, and the BH angular momentum is extracted in about $10{\rm{Myr}}$. After superradiant extraction, the BH evolution tracks the critical value $a_{\rm{crit}}/M$ while the BH accretes over a time scale of $1{\rm{Gyr}}$. 

%%%%%%%%%%%%%%%%%%%%%%%%%%%%%%%%%%
\subsection{The role of accretion}
%%%%%%%%%%%%%%%%%%%%%%%%%%%%%%%%%%

While gravitational-wave emission is always too weak to affect the evolution of the BH mass and spin (nonetheless being responsible for the decay of the scalar condensate as shown in Fig.~\ref{fig:evolution}), accretion 
plays a more important role. From Fig.~\ref{fig:evolution}, it is clear that accretion produces two effects. First, for BHs which initially are not massive enough to be in the superradiant instability region, accretion brings them to the instability window by feeding them mass as in Case I. Furthermore, when $J/M^2\to a_{\rm{crit}}/M$ the superradiant instability is exhausted, so that accretion is the only relevant process and the BH inevitably spins up again. This accretion phase occurs in a very peculiar way, with the dimensionless angular momentum following the trajectory $J/M^2\sim a_{\rm{crit}}/M$ over very long time scales.

Therefore, a very solid prediction of BH superradiance is that supermassive BHs would move on the Regge plane following the bottom-right part of the superradiance threshold curve. The details of this process depend on the initial BH mass and spin, on the scalar mass $\mu$ and on the accretion rate.

Thus, in order to verify the theoretical bounds on the existence of light bosons~\cite{Arvanitaki:2010sy,Kodama:2011zc,Pani:2012vp,Brito:2013wya}, a relevant problem concerns the \emph{final} BH state at the time of observation. In other words, given the observation of an old BH and the measurement of its mass and spin, would these measurements be compatible with the evolution depicted in Fig.~\ref{fig:evolution}? 

\begin{figure*}[ht]
\begin{center}
\begin{tabular}{c}
\epsfig{file=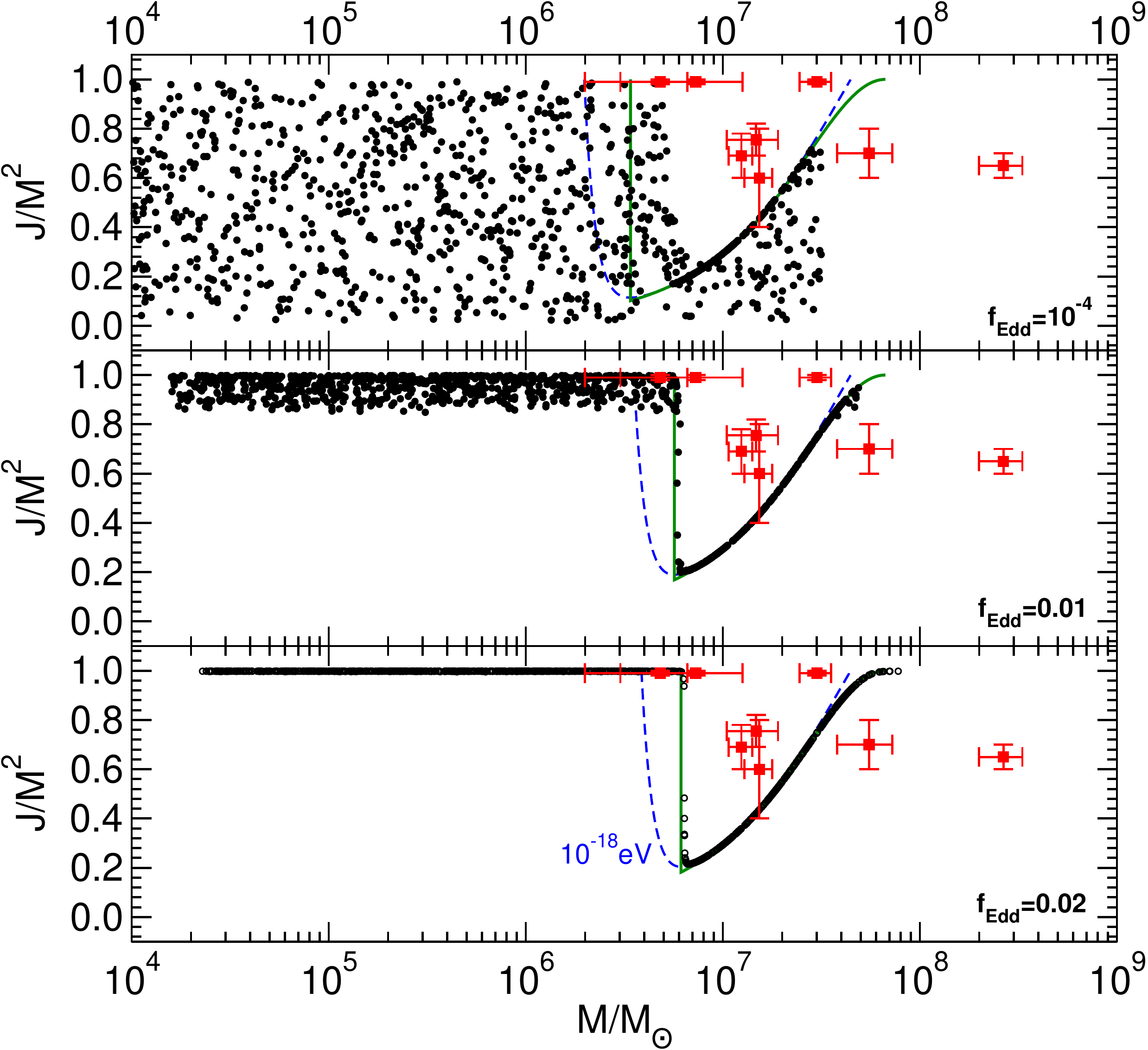,width=0.6\textwidth,angle=0,clip=true}
\end{tabular}
\end{center}
\caption{\label{fig:ReggeMC}
The final BH mass and spin in the Regge plane for initial data consisting of $N=10^3$ BHs with initial mass and spin randomly distributed between $\log_{10}M_0\in[4,7.5]$ and $J_0/M_0^2\in[0.001,0.99]$. The BH parameters are then extracted at $t=t_F$, where $t_F$ is distributed on a Gaussian centered at $\bar t_{F}\sim 2\times 10^9{\rm{yr}}$ with width $\sigma=0.1\bar t_{F}$. We considered $\mu=10^{-18}{\rm{eV}}$. The dashed blue line is the prediction of the linearized analysis obtained by comparing the superradiant instability time scale with the accretion time scale, $\tau\approx\tau_{\rm{Salpeter}}/f_{\rm{Edd}}$, whereas the solid green line denotes the region defined through Eq.~\ref{region}. Old BHs do not populate the region above the green threshold curve. The experimental points with error bars refer to the supermassive BHs listed in~\cite{Brenneman:2011wz}.
}
\end{figure*}

To assess this question, we have used Monte Carlo methods.
In Fig.~\ref{fig:ReggeMC} we show the final BH mass and spin in the Regge plane for $N=10^3$ evolutions for a scalar field mass $\mu=10^{-18}{\rm{eV}}$. These were obtained with random distributions of the initial BH mass between $\log_{10}M_0\in[4,7.5]$ and $J_0/M_0^2\in[0.001,0.99]$ extracted at $t=t_F$, where $t_F$ is distributed on a Gaussian centered at $\bar t_{F}\sim 2\times 10^9{\rm{yr}}$ with width $\sigma=0.1\bar t_{F}$. We consider three different accretion rates and, in each panel, we superimpose the bounds derived from the linearized analysis, i.e. the threshold line when the instability time scale equals the accretion time scale, $\tau\sim \tau_{\rm{ACC}}$. As a comparison, in the same plot we include the experimental points for the measured mass and spin of some supermassive BHs listed in Ref.~\cite{Brenneman:2011wz}.

Various comments are in order. First, it is clear that the higher the accretion rate the better the agreement with the linearized analysis. This seemingly counter-intuitive result can be understood by the fact that higher rates of accretion make it more likely to find BHs that have undergone a superradiant instability phase over our observational time scales. In fact, for high accretion rates it is very likely to find supermassive BHs precisely on the ``Regge trajectory''~\cite{Arvanitaki:2010sy} given by $J/M^2\sim a_{\rm{crit}}/M$ (cf. Eq.~\ref{acrit}).

Furthermore, for any value of the accretion rate, we always observe a depleted region (a ``hole'') in the Regge plane~\cite{Arvanitaki:2010sy}, which is not populated by old BHs. While the details of the simulations might depend on the distribution of initial mass and spin, the qualitative result is very solid and is a generic feature of the evolution. For the representative value $\mu=10^{-18}{\rm{eV}}$ adopted here, the depleted region is incompatible with observations~\cite{Brenneman:2011wz}. Similar results would apply for different values\footnote{Note that, through Eq.~\ref{dotMaccr}, the mass accretion rate only depends on the combination $f_{\rm{Edd}} M$, so that a BH with mass $M=10^6 M_\odot$ and $f_{\rm{Edd}}\sim10^{-3}$ would have the same accretion rate of a smaller BH with $M=10^{4} M_\odot$ accreting at rate $f_{\rm{Edd}}\sim10^{-1}$. Because this is the only relevant scale for a fixed value of $\mu M$, in our model the evolution of a BH with different mass can be obtained from Fig.~\ref{fig:evolution} by rescaling $f_{\rm{Edd}}$ and $\mu$.} of $\mu$ in a BH mass range such that $\mu M\lesssim1$. Therefore, as discussed in Refs.~\cite{Arvanitaki:2010sy,Pani:2012vp,Brito:2013wya}, observations of massive BHs with various masses can be used to rule out various ranges of the boson-field mass $\mu$.

Finally, Fig.~\ref{fig:ReggeMC} suggests that when accretion and gravitational-wave emission are properly taken into account, the holes in the Regge plane are smaller than what naively predicted by the relation $\tau\approx\tau_{\rm{ACC}}$, i.e. by the dashed blue curve in Fig.~\ref{fig:ReggeMC}. Indeed, we find that a better approximation for the depleted region is 
%%%
\begin{equation}
\left\{ \frac{J}{M^2}\gtrsim \frac{a_{\rm{crit}}}{M}\sim 4\mu M \right\} \quad  \cup \quad \left\{M \gtrsim \left({\frac{96}{\mu ^{10} \tau_{\rm{ACC}}}}\right)^{1/9} \right\}\,,\label{region}
\end{equation}
%%%%
whose boundaries are shown in Fig.~\ref{fig:ReggeMC} by a solid green line. These boundaries correspond to the threshold value $a_{\rm{crit}}$ (cf. Eq.~\ref{acrit}) for superradiance and to a BH mass which minimizes the spin for which $\tau\approx \tau_{\rm{ACC}}$, for a given $\mu$~\cite{Pani:2012bp}. As shown in Fig.~\ref{fig:ReggeMC}, the probability that a BH populates this region is strongly suppressed as the accretion rate increases.

%

%%%%%%%%%%%%%%%%%%%%%%%%%%%%%%%%%%%%%%%%%%%%%
\subsection{Estimating backreaction effects}
%%%%%%%%%%%%%%%%%%%%%%%%%%%%%%%%%%%%%%%%%%%%%

Our analysis neglects the gravitational effects of the scalar cloud and of a putative accretion disk on the BH geometry. The latter assumption is well justified because the disk density profile is roughly (see e.g.~\cite{Barausse:2014tra}):
%%%
\begin{equation}
 \frac{\rho_{\rm{disk}}}{\mbox{ kg}/\mbox{m}^3}\approx \left\{\begin{array}{l}
                           3.4\times 10^{-6} \left(\frac{10^6 M_\odot}{M}\right) \frac{f_{\rm{Edd}}}{\tilde{r}^{3/2}} \\
169 \frac{f_{\rm{Edd}}^{11/20}}{\tilde{r}^{15/8}} \left(1-\sqrt{\frac{\tilde{r}_{\rm{in}}}{\tilde{r}}}\right)^{11/20} \left(\frac{10^6 M_\odot}{M}\right)^{7/10} 
                          \end{array}\right.\nn \,,
\end{equation}
%%%
for geometrically-thick disks and for thin disks, respectively, where $\tilde{r}=r/M$ and $\tilde{r}_{\rm{in}}\sim 6$ is the radius of the inner edge of the disk in gravitational radii. These densities are negligible relative to the typical energy-density of the BH, $1/M^2\sim 10^8  (10^6 M_\odot/M)^2 {\rm{kg/m}}^3$, so that the deformation of the geometry due to the presence of the disk is unimportant.
%%%

\begin{figure}[ht]
\begin{center}
\begin{tabular}{c}
\epsfig{file=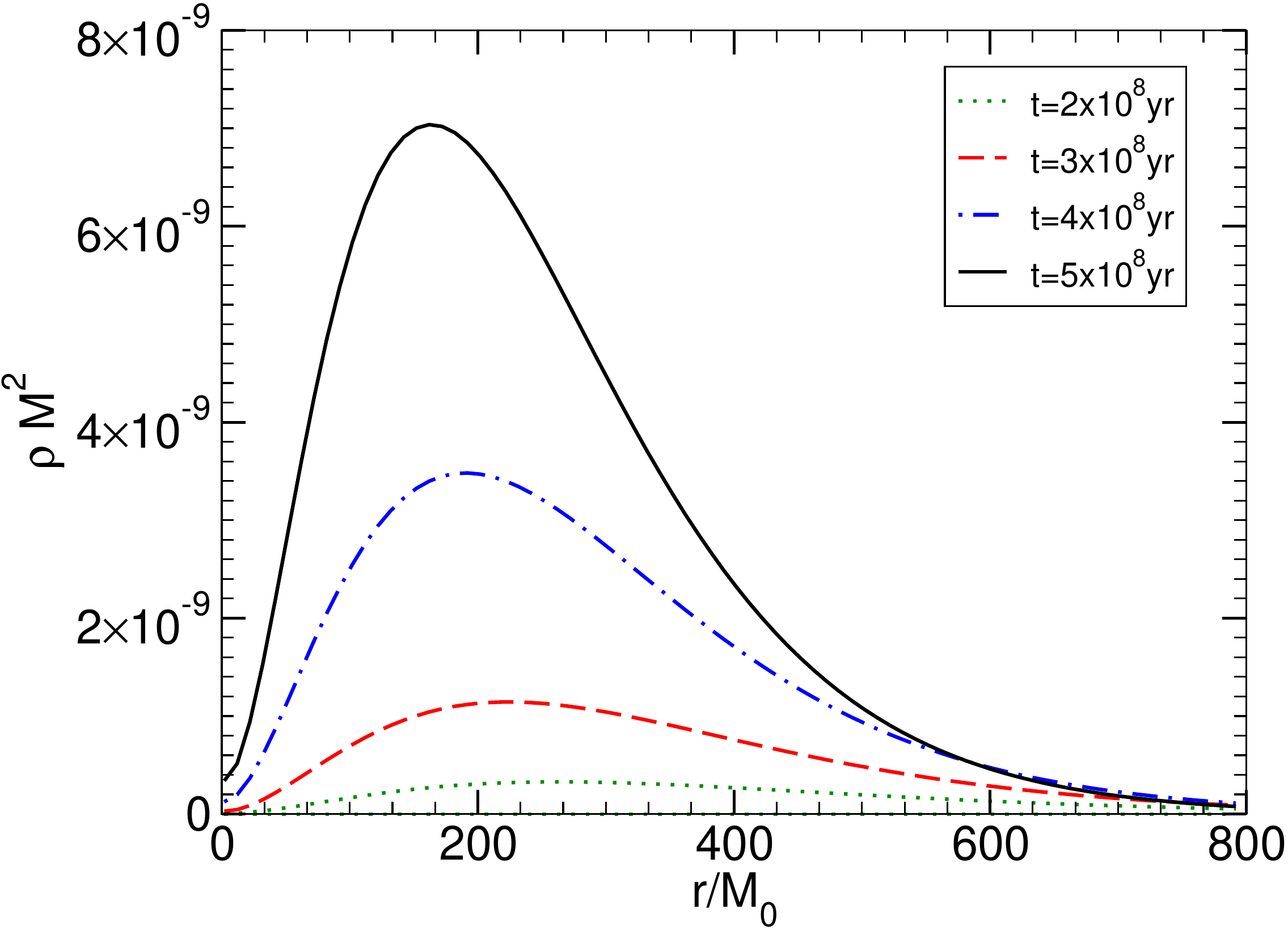,width=0.6\textwidth,angle=0,clip=true}
\end{tabular}
\end{center}
\caption{\label{fig:density}
The energy-density profile of the scalar cloud on the equatorial plane and at azimuthal angle $\phi=0$ in units of the BH density, $1/M^2\sim 10^6 {\rm{kg/m}}^3$, at different time snapshots for the evolution shown in the right panel of Fig.~\ref{fig:evolution}.
}
\end{figure}

On the other hand, from the evolution of Fig.~\ref{fig:evolution} it is clear that the scalar cloud attains a sizeable fraction of the total BH mass, so that backreaction effects might be relevant in this case. However, the scalar energy $M_S$ is spread over a large volume because the cloud peaks at $r_{\rm{cloud}}\sim \frac{1}{(M\mu)^2}M$. Thus, the scalar density --~which is the quantity directly coupled to the geometry through Einstein's equations~-- is always negligible. Figure~\ref{fig:density} shows the energy-density profile of the scalar cloud during the evolution corresponding to the right panel of Fig.~\ref{fig:evolution}. As in the case of the disk, also the density of the scalar cloud is orders of magnitude smaller than the energy-density associated to the BH horizon, $\sim 1/M^2$, so that the gravitational pull of the cloud produces a negligible effect on the background geometry. Furthermore, the corrections vanish near the BH horizon, so that also the superradiant energy extraction is unaffected\footnote{In principle, superradiant extraction from the BH horizon could be also affected by external perturbers, e.g. other compact objects in the vicinity of the BH. While our analysis already indicates that this correction is negligible, the fact that the near-horizon geometry of a BH cannot be easily deformed by, e.g., tidal forces~\cite{Binnington:2009bb} gives further support that considering isolated BHs in the context of superradiant instabilities is a reliable approximation (see also Ref.~\cite{Barausse:2014tra} for an analysis of environmental effects in gravitational-wave physics).}.

This discussion is in agreement with the results obtained for the superradiant instability in modified Kerr geometries, for example the Kerr-de Sitter metric studied in Ref.~\cite{Zhang:2014kna}. In such case, a value of the cosmological constant comparable to that of the cloud density, $\Lambda\sim 10^{-8}/M^2$, has no impact on the instability.
 
%%%%%%%%%%%%%%%%%%%%%%%%%%%%%%%%%%%%%%%%%%%%%%%%%%
\subsection{Hairy BHs: do they ever form?}\label{sec:hairy}
%%%%%%%%%%%%%%%%%%%%%%%%%%%%%%%%%%%%%%%%%%%%%%%%%%

%
\begin{figure}[ht]
\begin{center}
\begin{tabular}{c}
\epsfig{file=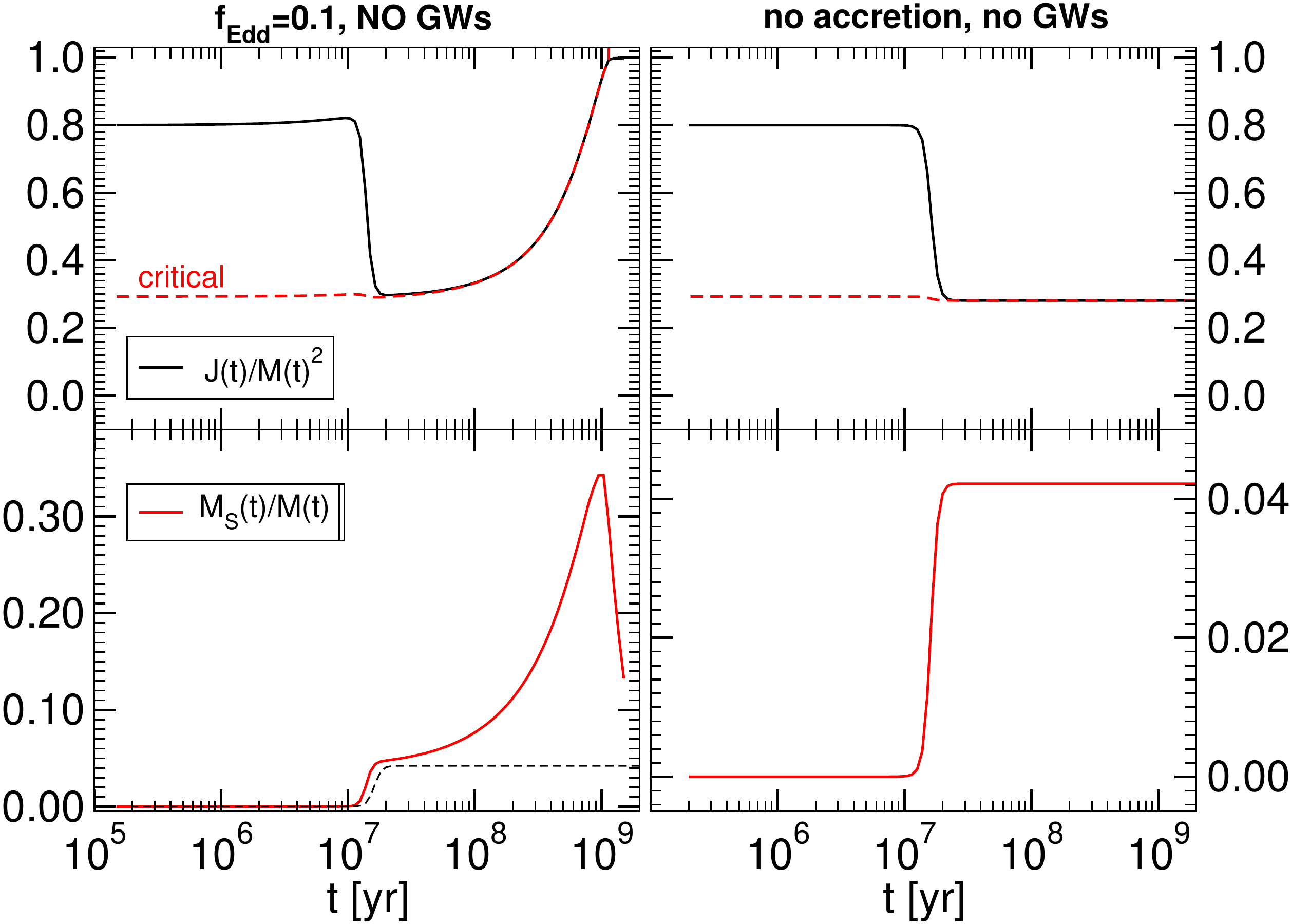,width=0.6\textwidth,angle=0,clip=true}
\end{tabular}
\end{center}
\caption{\label{fig:NOGWs}
Evolution with the same initial conditions as in the right set of Fig.~\ref{fig:evolution} but turning off the emission of gravitational waves as in the case of a complex scalar field. The left panels show the case of mass accretion at the rate $f_{\rm{Edd}}=0.1$, whereas the right panels show the case in which also accretion is turned off. For comparison, the scalar mass in the right bottom panel is also shown in the left bottom panel by a dashed black curve. When accretion is effective, the scalar cloud can become heavier.
}
\end{figure}

The most plausible formation scenarios for BHs involve gravitational collapse of matter, and are likely to form -- on free-fall time scales -- a geometry which is well-described
by the Kerr metric. Our results then have two important consequences: the BH evolves through accretion, gravitational-wave emission and superradiance,
but at late times it will not populate the region defined by Eq.~\ref{region}. In addition, the backreaction of the scalar condensate on the geometry is always small, i.e.,
even in the presence of a scalar cloud the geometry it is that of a Kerr BH to very good approximation. This is relevant for electromagnetic tests of the Kerr hypothesis, which are ultimately based on geodesic motion and would likely not be able to detect the effects of the cloud directly. On the other hand, during the evolution the system emits a nearly monochromatic gravitational-wave signal, which is an interesting source for next-generation gravitational-wave detectors~\cite{Arvanitaki:2010sy,Yoshino:2012kn,Yoshino:2013ofa}.

Our results apply equally to real and complex scalars, and despite recent works finding hairy BH solutions which depart significantly from the Kerr metric~\cite{Herdeiro:2014goa}.
The reason is that our formation scenario starts from a Kerr BH. Thus superradiance can only extract a finite amount of mass from the BH (in fact, at most 29\% of the initial BH mass, cf. e.g.~\cite{Begelman:2014aea}),
and therefore can only grow to a limited value. We find that this value is never sufficient to impart significant changes to the geometry.
By contrast, Refs.~\cite{Herdeiro:2014goa,Herdeiro:2014jaa,Herdeiro:2015gia} find that generically, stationary hairy BHs are smoothly connected to boson stars (see Part~\ref{part:BFstar} for more details on boson stars), and that therefore arbitrarily small BHs (or arbitrarily large ratios $M_S/M$)
are possible. What our results show is that these configurations do not arise from the evolution of initially isolated Kerr BHs; however, we have not ruled out that such solutions -- representing observationally large deviations from the Kerr geometry -- may arise as the end-state of some other initial conditions, most likely involving a large scalar field environment.

For completeness, in Fig.~\ref{fig:NOGWs} we show an evolution starting with the same initial conditions as in the right panel of Fig.~\ref{fig:evolution} but turning off gravitational-wave emission, which corresponds to taking a stationary, complex-scalar cloud in place of Eq.~\ref{scalar}. In the left panels we consider the case of mass accretion at the rate $f_{\rm{Edd}}=0.1$, whereas in the right panels also accretion has been turned off. In the latter case, the scalar mass and the BH angular momentum saturate after the superradiant extraction and the system would never leave the plateau configuration with a scalar mass $M_S\sim 0.04 M$ and a reduced spin $J/M^2\sim0.3$. However, when accretion is turned on (left panels), the scalar mass can attain more than $30\%$ of the BH mass during the evolution. This is due to the fact that the BH mass and angular momentum grow through accretion when superradiance is still effective and can therefore continue feeding the scalar cloud. This process lasts until angular-momentum accretion becomes inefficient at $J/M^2\sim 0.998$. Nonetheless, even in this most favorable case for the growth of the scalar cloud, the energy-density of the scalar field is negligible and the geometry is very well described by the Kerr metric. 

%%%%%%%%%%%%%%%%%%%%%%%%%%%%%%%%%%%%%%%
\subsection{Higher-\texorpdfstring{$l$}{l} modes}
%%%%%%%%%%%%%%%%%%%%%%%%%%%%%%%%%%%%%%%

So far we have neglected the superradiant growth of higher multipoles with $l>1$. This is justified by the fact that the instability time scale increases with $l$ (cf. Eq.~\ref{omega}) and the emission of gravitational waves for increasing $l$ becomes even more negligible~\cite{Yoshino:2013ofa}. The spin-down process is always dominated by the lowest possible superradiant mode $l=m=1$. For example, for the  evolution shown in Fig.~\ref{fig:evolution} the spin down due to the growth of the multipole $l=2$ would only affect the evolution of the BH after time scales of the order $t\sim 10^{13}{\rm{yr}}$.

However, because the superradiance condition depends on the azimuthal number $m$, for certain parameters it might occur that the modes with $l=m=1$ are stable, whereas the modes with $l=m=2$ are unstable, possibly with a superradiant extraction stronger than accretion. When this is the case, our previous analysis confirms that the depleted region of the Regge plane is the union of various holes, as predicted in Ref.~\cite{Arvanitaki:2010sy} by using a linearized analysis.

In the case where axion nonlinearities are taken into account further spin-down due to higher multipoles is expected to be damped due to the axion self-interactions either through the mixing of superradiant with nonsuperradiant levels or through the occurrence of explosive nonlinear effects, such as the bosenova collapse of the axion cloud~\cite{Arvanitaki:2010sy,Yoshino:2012kn}.

%%%%%%%%%%%%%%%%%%%%%%%%%%%%%%%%%%%%%%%%%%%%%%%%%%%%%%%%%%%%%%%%%%%%%%%%%%%%%%%%
\section{Bounds on the mass of bosonic fields from gaps in the Regge plane}\label{sec:bounds_RP}
%%%%%%%%%%%%%%%%%%%%%%%%%%%%%%%%%%%%%%%%%%%%%%%%%%%%%%%%%%%%%%%%%%%%%%%%%%%%%%%%%

%
\begin{figure}
\begin{center}
\epsfig{file=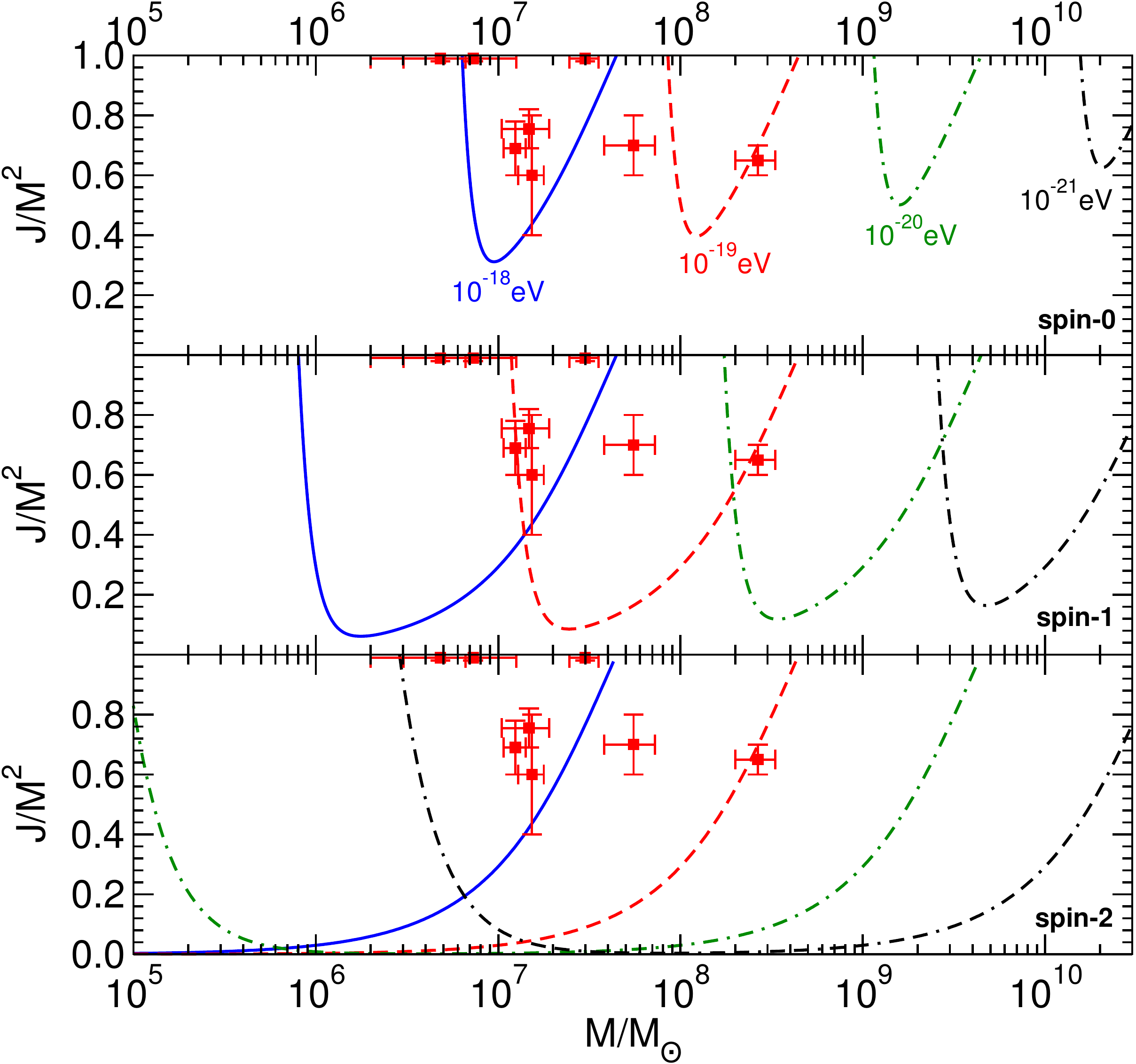,width=0.65\textwidth,angle=0,clip=true}
\caption{Contour plots in the BH Regge plane~\cite{Arvanitaki:2010sy}
  corresponding to an instability time scale shorter than $\tau_{\rm
    {Salpeter}}$ for different values of the boson field mass
  ${{\mu}}\hbar$ and for the most unstable modes. Top, middle and bottom panels show the case of scalar (spin-0), vector (spin-1) and tensor (spin-2) massive fields.
  The experimental points (with error bars) refer to the supermassive BHs
  listed~\cite{Brenneman:2011wz}. Supermassive BHs lying above each
  of these curves would be unstable on an observable time scale, and
  therefore each point rules out a range of the boson field
  masses. Note that the rightmost part of each curve is universal, $a\sim a_{\rm{crit}}$ (cf. Eq.~\eqref{acrit}), i.e. it does not depend on the spin of the field.\label{fig:bound2}
\label{fig:ReggePlane}}
\end{center}
\end{figure}

As shown in the previous Section, a very generic and solid prediction of BH superradiant instabilities is the existence of holes in the Regge plane.
Thus, the estimates for the instability time scale, together with reliable spin measurements for massive BHs, can be used to impose stringent constraints on the allowed mass range
of ultralight bosons~\cite{Arvanitaki:2010sy,Pani:2012vp,Brito:2013wya,Arvanitaki:2014wva}. 
These bounds follow from the requirement that astrophysical spinning BHs should be stable, in the sense that
the superradiant instability time scale $\tau$ should be larger than some observational threshold. For isolated BHs we can take the
observational threshold to be the age of the Universe, $\tau_{\rm{Hubble}}=1.38\times 10^{10}$~yr. However, as we discussed in this Chapter, for supermassive BHs we may
worry about possible spin growth due to mergers with other BHs and/or accretion. As discussed in the previous sections, the most likely mechanism to produce fastly-spinning BHs is
prolonged accretion~\cite{Berti:2008af}. Our results confirm that a conservative assumption to estimate the astrophysical consequences of the
instability is to compare the superradiance time scale, obtained within the linearized analysis, to the minimum time scale over which accretion could spin up the BH. 
For simplicity we assume that mass growth occurs via accretion at the Eddington limit, so that the
BH mass grows exponentially with $e$-folding time given by the
Salpeter time scale, $\tau_{\rm{Salpeter}}\sim 4.5\times 10^7$ years.

In order to quantify the dependence of the boson mass bounds on the mass and spin of supermassive BHs, in Fig.~\ref{fig:ReggePlane} we show exclusion regions
in the BH Regge plane. More precisely, using the results shown in Section~\ref{sec:massive_unified}, we plot contours corresponding to an instability
time scale of the order of the Salpeter time for four different masses
of the bosonic field and considering the unstable mode with the largest growth rate. From top to bottom, the three panels refer to a spin-0, spin-1 and spin-2 field, respectively.
The plot shows that observations of supermassive BHs with $10^5M_\odot\lesssim
M\lesssim 10^{10}M_\odot$ spinning above a certain threshold would exclude a wide range of boson-field masses. Because superradiance is stronger for bosonic fields with spin, the exclusion windows are wider as the spin of the field increases, and they also extend almost down to $J\sim0$ in the case of spin-1 and spin-2 bosons. This latter feature is important because current spin measurements might be affected by large systematics.

Nonetheless, it is clear from Fig.~\ref{fig:ReggePlane} that almost any supermassive BH spin measurement would exclude a
considerable range of masses. Similar exclusion plots exist in the region $M_\odot\lesssim M\lesssim 10^5 M_\odot$ for larger values of $\mu$. Indeed, the only parameter that regulates the instability is the combination $\mu M$. Thus, the best bound comes from the most massive BHs for which spin measurements are reliable, e.g. the BH candidate Fairall~9~\cite{Schmoll:2009gq}.

Using these arguments, from the analysis of Refs.~\cite{Arvanitaki:2010sy,Pani:2012vp,Brito:2013wya} we can obtain the following bounds\footnote{These bounds were obtained using the linearized analysis, summarized in Section~\ref{sec:massive_unified}. By including the effects of gravitational-wave emission and gas accretion, the results of Section~\ref{sec:evolution} show that the linearized prediction should be corrected by Eq.~\eqref{region}, cf. Fig.~\ref{fig:ReggeMC} and discussion in Sec.~\ref{sec:evolution}. Nonetheless, such corrections would not affect the order of magnitude of these constraints. In Ref.~\cite{Arvanitaki:2014wva}, the authors estimate the statistical and systematic errors affecting these bounds, finding exclusions regions at approximately $2\sigma$ and $1\sigma$ for stellar-mass and supermassive BHs, respectively.}:
%%%%
\begin{eqnarray}
m_S \lesssim 5\times 10^{-20} {\rm{eV}}  & \lor  & m_S \gtrsim 10^{-11} {\rm{eV}}   \,, \label{bound_spin0}\\
m_V \lesssim 5\times 10^{-21} {\rm{eV}}  & \lor  & m_V \gtrsim 10^{-11} {\rm{eV}}   \,, \label{bound_spin1}\\
m_T \lesssim 5\times 10^{-23} {\rm{eV}}  & \lor  & m_T \gtrsim 10^{-11} {\rm{eV}}   \,, \label{bound_spin2}
\end{eqnarray}
%%%%
for the mass of ultralight scalar, vector and tensor fields, respectively. Note that, for a single BH observation, superradiant instabilities can only exclude a \emph{window} in the mass range of the fields, as shown in Fig.~\ref{fig:ReggePlane}. Nonetheless, by combining different BH observations in a wide range of BH masses and assuming\footnote{Recently, the first detection of intermediate-mass BHs was reported~\cite{2014Natur.513...74P}, suggesting the BH mass spectrum might be populated continuously from few solar masses to billions of solar masses.} that spinning BHs exist in the entire mass range $M_\odot\lesssim M\lesssim 10^9 M_\odot$, one is able to constrain the range above, where the lower bound comes from the lightest massive BHs (with $M\approx 5M_\odot$), whereas the upper bound comes from the heaviest supermassive BHs for which spin measurements are reliable. If the largest known supermassive BHs with $M\simeq 2\times 10^{10}
M_\odot$ \cite{McConnell:2011mu,2012arXiv1203.1620M} were confirmed to have nonzero spin, we could get even more stringent bounds.

For each BH observation, the upper limit comes from the fact that when $M\mu\ll1$ the time scale grows with some power of $1/(\mu M)$ and eventually the instability is ineffective on astrophysical time scales. The lower limit comes from the fact that the instability exists only when the superradiant condition is satisfied, and this imposes a constraint on $\mu$ for a given azimuthal number $m$\footnote{As $m$ increases, larger values of $\mu$ are allowed in the instability region and virtually any value of $\mu$ gives some unstable mode in the eikonal ($l,m\gg1$) limit. However, the instability is highly suppressed as $l$ increases so that, in practice, only the first few allowed values of $l=m$ correspond to an effective instability.}. 
Indeed, the rightmost part of the curves shown in Fig.~\ref{fig:ReggePlane} for fixed $\mu$ is universal and arises from saturation of the superradiant condition, $a\sim a_{\rm{crit}}$, where $a_{\rm{crit}}$ is given in Eq.~\eqref{region}. Such condition does not depend on the spin of the field, and this explains why the upper bounds in Eqs.~\eqref{bound_spin0}--\eqref{bound_spin2} are the same for scalar, vector and tensor fields.

%%%%%%%%%%%%%%%%%%%%%%%%%%%%%%%%%%
\section{Discussion}
%%%%%%%%%%%%%%%%%%%%%%%%%%%%%%%%%%
If ultralight bosonic degrees of freedom exist in nature, massive BHs should have a maximum spin lower than the Kerr bound and should be endowed with large dipolar bosonic clouds. Thus, observations of highly-spinning BHs can be used to constrain such fields, for example to put bounds on axions or on massive gravitons. Such predictions are based on a linearized analysis which neglects the effects of backreaction and other competitive effects such as accretion. In this Chapter, we have extended such analysis by including the emission of gravitational waves from the cloud and the most conservative case of gas accretion. By adopting an adiabatic approximation, we have simulated the evolution of the scalar condensate around a spinning BH. 

Our results show that the effects of gravitational-wave emission are always too small to affect the evolution of the BH mass and spin, but they contribute to dissipate the scalar condensate. Indeed, we have shown that the scalar condensates are eventually re-absorbed by the BH and dissipated through quadrupolar gravitational waves, in accord to the BH no-hair theorems~\cite{HawkingBook,Heusler:1995qj,Sotiriou:2011dz,Graham:2014ina}. Nonetheless, the mass of the cloud remains a sizeable fraction of the BH total mass over cosmological times, so that such systems can be considered as (quasi)-stationary hairy BHs for any astrophysical purpose. 
The energy-density in the scalar field is negligible and the geometry is very well described by the Kerr metric during the entire evolution. Thus, the prospects of imagining deviations from Kerr due to superradiantly-produced bosonic clouds in the electromagnetic band~\cite{Lu:2014zja,GRAVITY} are low (however, see Ref.~\cite{Cunha:2015yba} for nice results on the shadow of these BHs), but such systems are a primary source for observations aiming at testing the Kerr hypothesis through gravitational-wave detection~\cite{LIGO,VIRGO,KAGRA,ET,ELISA}.

The role of gas accretion is twofold. On the one hand, accretion competes against superradiant extraction of mass and angular momentum. On the other hand accretion might produce the optimal conditions for superradiance, for example by increasing the BH spin before the instability becomes effective or by ``pushing'' the BH into the instability region in the Regge plane.

Our Monte Carlo simulations confirm that a very generic prediction of BH superradiant instabilities is the existence of holes in the Regge plane. For mass accretion near the Eddington rate, such depleted regions are very well described by Eq.~\ref{region}, which corrects the estimate obtained just by comparing the instability time scale against a typical accretion time scale. A more sophisticated analysis --~including radiative effects~\cite{Thorne:1974ve}, the geometry of the disk, the effect of mergers on the evolution~\cite{Barausse:2012fy} and also accretion by tidal disruption~\cite{2012ApJ...749L..42B}~-- would be important to refine the bounds previously derived~\cite{Arvanitaki:2010sy,Kodama:2011zc,Pani:2012vp,Brito:2013wya} (summarized in Section~\ref{sec:bounds_RP}). 

The main limitation of our analysis is the assumption of adiabatic evolution, which is nonetheless well motivated given the large difference in the time scales of the problem. For the same reason, exact numerical simulations --~although important to test our results~-- would be extremely difficult to perform.  Some of our formulas were derived in the small $\mu M$ limit. Although they provide reliable results also when $\mu M\sim {\cal O}(1)$, our analysis could be extended using the exact numerical results derived in Ref.~\cite{Yoshino:2013ofa}. 

It would also be interesting to include scalar self-interactions, which are relevant for axions. Assuming an axion sine-Gordon potential, $V(\Phi)=f_a^2 \mu^2[1-\cos(\Phi/f_a)]$, the results of our analysis would remain valid provided $f_a$ is sufficiently large. In this regime it is reasonable to expect that the self-interactions would not change the evolution shown in Fig.~\ref{fig:evolution} considerably. On the other hand, if the axion decay constant $f_a$ corresponds to the GUT scale, $f_a\approx 10^{16}{\rm{GeV}}$, the bosenova occurs when $M_S\gtrsim 0.16 M$~\cite{Yoshino:2012kn}. In this case, assuming the initial evolution would not be affected considerably, our results suggest that the scalar cloud might reach the threshold for the bosenova condensation.

Finally, we have focused on the scalar case but it is likely that similar results can be derived also for massive vector and tensor fields because in such cases the superradiant instability is stronger.

%%%%%%%%%%%%%%%%%%%%%%%%%%%%%%%%%%%%%%%%%%%%%%%%%%%%%%%%%%%%%%%%%%%%
%%%%%%%%%%%%%%%%%%%%%%%%%%%%%%%%%%%%%%%%%%%%%%%%%%%%%%%%%%%%%%%%%%%%

%%%%%%%%%%%%%%%%%%%%%%%%%%%%%%%%%%%%%%%%%%%%%%%%%%%%%%%%%%%%%%%%%%%%
\part{Interaction between bosonic dark matter and stars}\label{part:BFstar}
\chapter{Bosonic dark matter and stars}\label{chapter:BFstar}
\emph{This part is based on Refs.~\cite{Brito:2015yga,Brito:2015yfh}}.
%%%%%%%%%%%%%%%%%%%%%%%%%%%%%%%%%%%%%%%%%%%%%%%
\section{Introduction}
%%%%%%%%%%%%%%%%%%%%%%%%%%%%%%%%%%%%%%%%%%%%%%%
The evidence for DM in observations is overwhelming, starting with galaxy rotation curves,
gravitational lensing and the cosmic microwave background~\cite{Bertone:2004pz}.
While carefully concocted modified theories of gravity can perhaps explain almost all observations,
the most attractive and accepted explanation lies in DM being composed mostly of cold, collisionless particles~\cite{Bertone:2004pz,Klasen:2015uma,Marsh:2015xka}.

Several candidates for dark matter have been proposed~\cite{Bertone:2004pz,Klasen:2015uma,Marsh:2015xka}, of which ultralight bosonic fields, such as axions, axion-like candidates~\cite{Marsh:2015wka,2014NatPh..10..496S,2014PhRvL.113z1302S} or ``hidden photons''~\cite{Ackerman:mha} are an attractive possibility. Axions were originally devised to solve the strong-CP problem, but recently a plethora of other, even lighter fields with masses $10^{-10}-10^{-33}\,{\rm{eV}}/c^2$, have also become an interesting possibility, in what is commonly known as the axiverse scenario~\cite{Arvanitaki:2009fg}. On the other hand massive vector fields arise in the so-called hidden $U(1)$ sector~\cite{Holdom:1985ag,ArkaniHamed:2008qn,Pospelov:2008jd,Goodsell:2009xc}. This family of candidates are part of what is now referred to as weakly interacting slim particles (WISPs)~\cite{Arias:2012az} in some of the DM literature.

It is appropriate to emphasize that DM has not been seen nor detected through any of the known standard model interactions.
The only evidence for DM is through its gravitational effect. Not surprisingly, the quest for DM is one of the most active fields of research of this century.
Because DM can only interact feebly with Standard Model particles, and thanks to the equivalence principle, the most promising channel to look for DM imprints consists of gravitational interactions.
 
In particular, regions where gravity is strong, such as compact stars, might be a good place to look for signals of DM. Since massive bosonic fields are viable DM candidates~\cite{Bertone:2004pz,Klasen:2015uma}, a natural question is whether they can be accreted inside stars and lead to stable configurations with observable effects. Although not originally framed in the context of DM capture, such solutions, made of both a perfect fluid and a massive complex scalar field, exist~\cite{Henriques:1989ar,Henriques:1989ez,Lopes:1992np,Henriques:2003yr,Sakamoto:1998aj,deSousa:2000eq,deSousa:1995ye,Pisano:1995yk} and can model the effect of bosonic DM accretion by compact stars. Complementary to these studies, accretion of fermionic DM has also been considered, by modeling the DM core with a perfect fluid and constructing a physically motivated equation of state~\cite{Leung:2011zz,Leung:2013pra,Tolos:2015qra}.

%%%%%%%%%%%%%%%%%%%%%%%%%%%%%%%%%%%%%%%%%%%%%%%%%%%%%%%%%%%%%%%%%
\section{Stars may survive dark matter capture}
%%%%%%%%%%%%%%%%%%%%%%%%%%%%%%%%%%%%%%%%%%%%%%%%%%%%%%%%%%%%%%%%%

The work presented in this Part may be useful on several fronts. Firstly, it shows that stable DM cores inside compact stars {\it are possible}.
The question of whether these cores are actually formed through dynamical processes in DM environments is harder to answer.
The standard lore -- which the results we will present do not support -- is that the dynamics happen in two stages (see e.g., Refs.~\cite{Goldman:1989nd,Kouvaris:2011fi,Bramante:2014zca,Bramante:2015cua,Kurita:2015vga}):

\noindent {\bf 1.} {\it Accumulation stage,} where DM is captured by the star due to gravitational deflection and a non-vanishing
cross-section for collision with the star material~\cite{Press:1985ug,Gould:1989gw,Goldman:1989nd,Bertone:2007ae}. 
The DM material eventually thermalizes with the star, and accumulates inside a sphere of r.m.s. radius $r_{th}\sim \left(k_BT/\rho_c m_{DM}\right)^{1/2}$, with $k_B$ Boltzmann's constant, $T,\,\rho_c$ the temperature and density of the star, and $m_{DM}$ the mass of the DM particles. The high-density of compact stars provide the ideal environment for the star to accumulate a considerable amount of DM.

\noindent {\bf 2.} {\it BH formation,} after the DM core becomes self-gravitating. The newly formed BH eventually eats the host star~\cite{Gould:1989gw,Goldman:1989nd,Kouvaris:2011fi,Bramante:2014zca,Bramante:2015cua,Kurita:2015vga}. 

As we will detail in Chapter~\ref{sec:fluid}, stable, self-gravitating and self-interacting bosonic DM cores exist and can be explicitly constructed~\cite{Brito:2015yga} . In fact, a similar construction was recently performed for fermionic DM models~\cite{Leung:2013pra}. Thus, the scenario above cannot be generic. In fact, all of these works {\it assume} that gravitational collapse ensues once DM becomes self-gravitating. The stability and evolution of stars is a far more complex affair, and in particular DM dispersion or an increase in the star temperature can easily rule out the collapse scenario. We show that the gravitational cooling mechanism~\cite{Seidel:1993zk} not only disperses bosonic condensates, but it prevents -- generically -- gravitational collapse to occur. 

We do not take into account non-gravitational interactions between the star and DM; however, there are strong reasons to suspect that some of the main features are independent, at the qualitative level, of the nature of the interaction. For example, although our results are formally only valid for zero-temperature bosonic condensates, finite temperature effects are expected to be negligible for bosonic masses much larger than the central temperature of the host star~\cite{Bilic:2000ef,Latifah:2014ima}, such as bosonic fields with masses $\gtrsim$ keV inside old neutron stars or white dwarfs. In fact, Refs.~\cite{Bilic:2000ef,Latifah:2014ima} showed that stable bosonic stars exist for temperatures below a critical temperature which scales linearly with the boson field mass. Finite temperatures tend to increase the radius of the boson star (comparing a star with the same total mass), but do not significantly affect the star's maximum stable mass. In addition, for axionic-type couplings, for instance, all or most of the core energy will be dissipated away under electromagnetic radiation, on relatively small timescales~\cite{Iwazaki:1998eg,Iwazaki:1999my,Iwazaki:2014wka}. 

Finally, our study predicts that bosonic DM cores drive the star to 
vibrate at a frequency dictated by the scalar field mass, $f=2.5\times 10^{14}\,\left(m_{B}c^2/eV\right)\,{\rm{Hz}}$~\cite{Brito:2015yga}, providing a clear means to identify the presence of dark matter in stars, provided these modes are excited to measurable amplitudes.
Helioseismology, developed to the level of a precision science, can now measure individual modes each with an amplitude of $\sim 10\,{\rm{cm}} \,\rm{s}^{-1}$~\cite{ChristensenDalsgaard:2002ur}. Thus, provided efficient mechanisms exist to gather sufficient DM at the cores of stars, these oscillations will be a smoking gun for DM.

%%%%%%%%%%%%%%%%%%%%%%%%%%%%%%%%%%%%%%%%%%%%%%%
\section{General Framework}
%%%%%%%%%%%%%%%%%%%%%%%%%%%%%%%%%%%%%%%%%%%%%%%
In this last part of the thesis we will be interested in a massive scalar $\phi$ or vector $A_{\mu}$ minimally coupled to gravity, and described by the action
\be
S=\int d^4x \sqrt{-g} \left( \frac{R}{\kappa} - \frac{1}{4}F^{\mu\nu}\bar{F}_{\mu\nu}- \frac{\mu_V^2}{2}A_{\nu}\bar{A}^{\nu}
-\frac{1}{2}g^{\mu\nu}\bar{\phi}^{}_{,\mu}\phi^{}_{,\nu} -\frac{\mu_S^2\bar{\phi}\phi}{2}
+\mathcal{L}_{\rm{matter}}\,
\right)
\,.\label{eq:MFaction}
\ee
We take $\kappa=16\pi$, $F_{\mu\nu} \equiv \nabla_{\mu}A_{\nu} - \nabla_{\nu} A_{\mu}$ is the Maxwell tensor and $\mathcal{L}_{\rm{matter}}$ describes additional matter fields, that we consider to be described by a perfect fluid. We focus on massive, non self-interacting fields, but our results are easily generalized. In fact, we discuss in Section~\ref{sec:collision} how our results 
generalize to a quartic self-interaction term. The mass $m_B$ of the boson under consideration is related to the mass parameter above through $\mu_{S,V}=m_{B}/\hbar$, and the theory is controlled by the dimensionless coupling
\begin{equation}
\frac{G}{c\hbar} M_T\mu_{S,\,V} = 7.5\cdot 10^{9} \left(\frac{M_T}{M_{\odot}}\right) \left(\frac{m_{B}c^2}{eV}\right)\,,\label{dimensionless_massparameter}
\end{equation}
where $M_T$ is the total mass of the bosonic configuration.

Varying the action~\eqref{eq:MFaction}, the resulting equations of motion are
\begin{subequations}
\label{eq:MFEoMgen}
\begin{eqnarray}
  \label{eq:MFEoMScalar}
  &&\nabla_{\mu}\nabla^{\mu}\phi =\mu_S^2\phi
			\,,\\
  \label{eq:MFEoMVector}
  &&\nabla_{\mu} F^{\mu\nu} =
      \mu_V^2A^\nu\,,\\
  \label{eq:MFEoMTensor}
  &&\frac{1}{\kappa} \left(R^{\mu \nu} - \frac{1}{2}g^{\mu\nu}R\right)=
      \frac{1}{4\pi}\left(\frac{1}{2}F^{(\mu}_{\,\,\alpha}\bar{F}^{\nu)\alpha}- \frac{1}{8}\bar{F}^{\alpha\beta}F_{\alpha\beta}g^{\mu\nu}\right.\nonumber\\
   &&  \left. - \frac{1}{4}\mu_V^2A_{\alpha}\bar{A}^{\alpha}g^{\mu\nu}+\frac{\mu_V^2}{2}A^{(\mu}A^{\nu)} \right)
    -\frac{1}{4}g^{\mu\nu}\left( \bar{\phi}^{}_{,\alpha}\phi^{,\alpha}+{\mu_S^2}\bar{\phi} \phi\right)   \nonumber\\
   &&   +\frac{1}{4}\bar{\phi}^{,\mu}\phi^{,\nu}+\frac{1}{4}\phi^{,\mu}\bar{\phi}^{,\nu}+\frac{1}{2}T_{\rm{fluid}}^{\mu\nu}\,.
			%-i\frac{q}{2}A^{\mu}\left(\Psi\nabla^{\nu}\Psi^{\ast}-\Psi^{\ast}\nabla^{\nu}\Psi\right) \nonumber\\
%
 %  &&-\frac{q^2}{4}g^{\mu\nu}\Psi\Psi^{\ast}A_{\alpha}A^{\alpha}+\frac{q^2}{2} \Psi\Psi^{\ast}A^{\mu}A^{\nu}+i\frac{q}{4}g^{\mu\nu}A_{\alpha}\left(\Psi\nabla^{\alpha}\Psi^{\ast}-\Psi^{\ast}\nabla^{\alpha}\Psi\right)\,.
\end{eqnarray}
\end{subequations}
Here, the stress-energy tensor for the perfect fluid is given by~\cite{rezzolla2013relativistic}
\be\label{stress_energy_PF}
T_{\rm{fluid}}^{\mu\nu}=\left(\rho_F+P\right)u^{\mu}u^{\nu}+P g^{\mu\nu}\,,
\ee
with $u^{\mu}$ the fluid's four-velocity, $\rho_F$ its total energy density in the fluid frame and $P$ its pressure.
The massive vector field equations~\eqref{eq:MFEoMVector} imply that the vector field must satisfy the constraint,
\be\label{divA}
\mu^2_V\nabla_{\mu}A^{\mu}=0\,,
\ee
while from the Bianchi identities, it follows that the fluid must satisfy the conservation equations
\be\label{divT}
\nabla_{\mu}T_{\rm{fluid}}^{\mu\nu}=0\,.
\ee
In addition we impose conservation of the baryonic number~\cite{rezzolla2013relativistic}: 
\be\label{baryonic}
\nabla_{\mu}\left(n_F u^{\mu}\right)=0\,,
\ee
where $n_F$ is the baryonic number density in the fluid frame and $m_N n_F$ is the fluid's rest-mass density for baryons of mass $m_N$.
To close this system of equations we also need to complement the system with an equation of state relating $n_F$, $\rho_F$ and $P$. Specific equations of state will be discussed in Chapter~\ref{sec:fluid}.

In the next Chapters we consider only everywhere regular solutions of the system~\eqref{eq:MFEoMScalar}--\eqref{eq:MFEoMTensor}.

%%%%%%%%%%%%%%%%%%%%%%%%%%%%%%%%%%%%%%%%%%%%%%%%%%%%%%%%%%%%%%%%%%%%%%%%%%%%%%%
\section{Self-gravitating bosonic stars}
%%%%%%%%%%%%%%%%%%%%%%%%%%%%%%%%%%%%%%%%%%%%%%%%%%%%%%%%%%%%%%%%%%%%%%%%%%%%%%%
Compact solutions of the system~\eqref{eq:MFEoMScalar}--\eqref{eq:MFEoMTensor}, without including the perfect fluid, exist for both real and complex fields. Scalar fields have been extensively studied in the literature, while similar solutions for vector fields will be constructed in Chapter~\ref{sec:osci}.

The existence of these solutions is not something that one would trivially guess. In fact, in four-dimensional flat spacetime, Derrick's theorem shows that no stable static non-topological compact scalar field solutions exist for any scalar field potential~\cite{Derrick:1964ww}\footnote{Non-topological solitons have an associated Noether charge, unlike topological solitons.}. For complex fields, a way out of this theorem is to consider a harmonic behavior for the field:
\be\label{BS}
\phi(t,r)=\phi(r)e^{i\omega t}\,.
\ee
In general this is not enough to guarantee the existence of solutions. For some non-linear potentials, localized flat space solutions of the form~\eqref{BS} exist~\cite{Coleman:1985ki,Lee:1991ax} (the so-called Q-balls). However, for the most general renormalizable potential, namely $V(\phi)=\mu_S^2|\phi|^2+\lambda |\phi|^4/2$, solutions do not exist in flat space. For solutions with this potential to exist, one must couple the scalar field to gravity. These are the so-called boson stars~\cite{Kaup:1968zz,Ruffini:1969qy}. Although the field oscillates with a frequency $\omega$, the stress-energy tensor of this field will not depend on time, as can easily be checked by plugging in the ansatz~\eqref{BS} into the field equations~\eqref{eq:MFEoMgen}. Thus the metric of these configurations is static.

These arguments only apply for complex fields, for which there is an associated Noether charge. One could expect that for real fields, due to the absence of a conserved current, stable compact solutions do not exist. However Ref.~\cite{Seidel:1991zh} showed that stable compact configurations, similar to boson stars, also exist in this case. Unlike boson stars, for real scalar fields, the stress-energy tensor must itself be time-dependent. For these objects, called oscillatons, both the metric and the scalar field oscillate periodically in time. 
Oscillatons are in fact not truly periodic solutions of the field equations, as they decay through quantum and classical processes. However, their lifetime $T_{\rm{decay}}$ is extremely large for all masses of interest~\cite{Page:2003rd,Grandclement:2011wz},
\be
T_{\rm{decay}}\sim 10^{324}\left(\frac{1\,{\rm{meV}}}{m_B{\rm{c^2}}}\right)^{11}\, {\rm{yr}}\,.
\ee

Both boson stars and oscillatons, can be thought of as being a macroscopic collection of bosonic particles held together by gravity. Assuming that the kinetic energy of these particles is sufficiently low for them to be gravitationally bound, two ingredients must be in place~\cite{Liebling:2012fv}: (i) an attractive interaction to hold the particles together; (ii) a counterbalancing pressure to avoid the field to collapse to a BH. The first ingredient is obviously provided by gravity. The non-zero stress-energy momentum of the field curves the geometry and consequently the field must interact gravitationally with itself. The second ingredient is less obvious from the physical point of view, but one can argue that the dispersive nature of the Klein-Gordon equation, the same dispersion that is encoded in Heisenberg's uncertainty principle, provides the necessary counterbalancing pressure.

\begin{figure}[htb]
\begin{center}
\begin{tabular}{c}
\epsfig{file=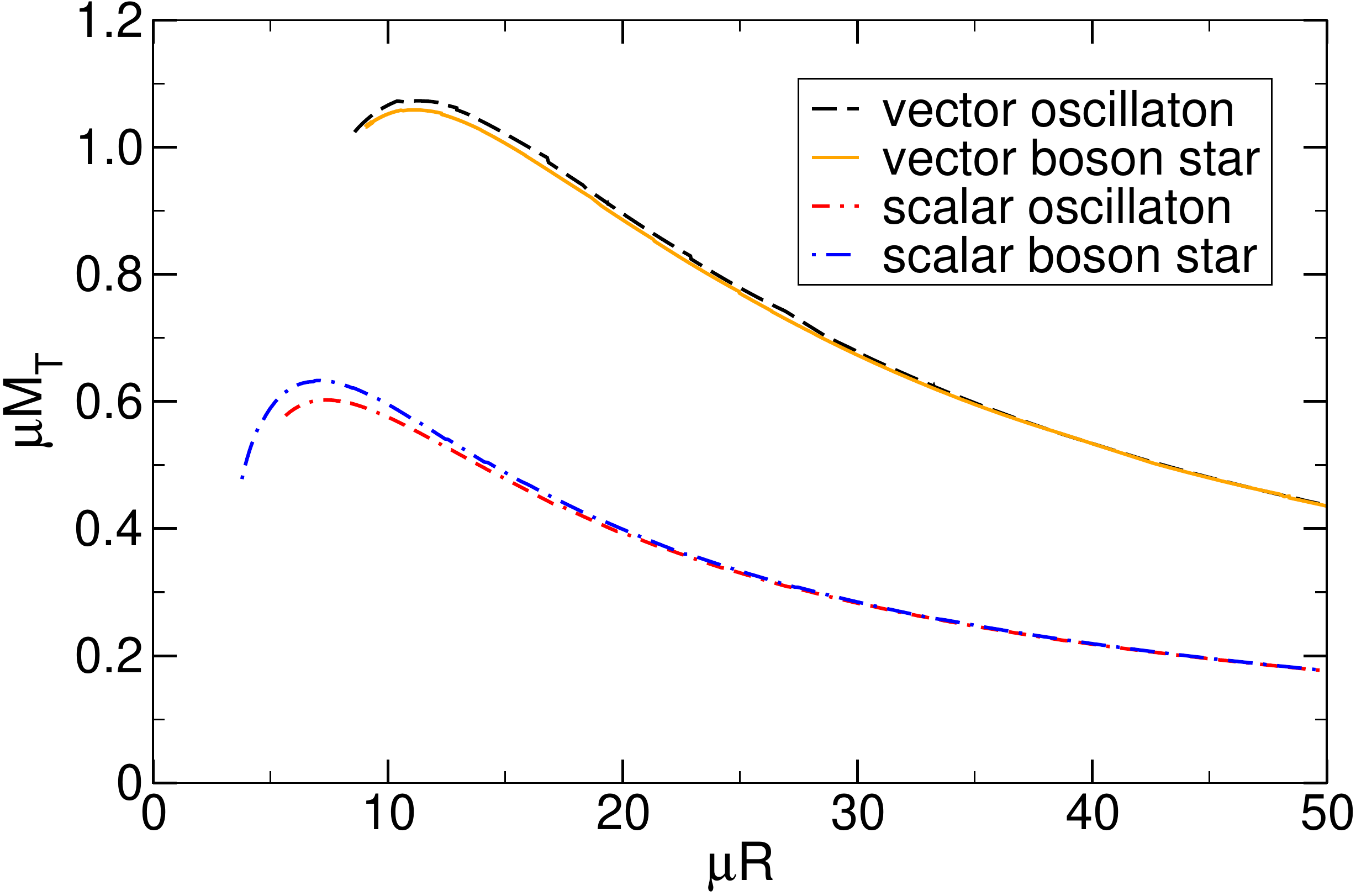,width=10cm,angle=0,clip=true}
\end{tabular}
\caption{Comparison between the total mass of a boson star ({\it complex} scalar or vector fields) and an oscillaton ({\it real} scalar or vector fields), as a function of their radius $R$. $R$ is defined as the radius containing 98\% of the total mass. The procedure to find the diagram is outlined in Chapter~\ref{sec:osci}.\label{MvsR}}
\end{center}
\end{figure}

Indeed, assuming that the bosonic star is a macroscopic object satisfying the uncertainty principle, one can give a good estimate of its maximum possible mass~\cite{Liebling:2012fv}. 
Consider a boson star or oscillaton in its ground state, satisfying the uncertainty principle, $\Delta p\Delta x\geq \hbar$. Assuming that the star is confined within a finite radius $\Delta x=R$, and taking the uncertainty in the momentum to be $\Delta p=m_B c$, with $m_B$ the mass of the bosonic particle, we get
\begin{equation}
m_B\,c\,R\geq \hbar.
\end{equation}
In fact, one could have guessed this result, by assuming that the star is spread out within the Compton wavelength of the bosonic field $\lambda_c=\hbar/(c m_B)$. The maximum mass of the star saturates this bound, while its radius can be compared with its Schwarzschild radius $R_S=2G M_{\rm{max}}/c^2$, where $M_{\rm{max}}$ is the maximum mass of the star. Substituting yields
\begin{equation}
M_{{\rm{max}}}\sim \frac{1}{2}\frac{\hbar c}{G m_B}=0.5 M_{\rm{P}}^2/m_B\,,
\end{equation}
where the Planck's mass is given by $M_{\rm{P}}\equiv\sqrt{\hbar c/G}$. As shown in Fig.~\ref{MvsR} this is a very good estimate of $M_{{\rm{max}}}$ and correctly predicts that the mass of a boson star or oscillaton is inversely proportional to the mass of the bosonic field~\footnote{This is only true when the field's potential is a simple mass term. Adding self-interaction terms adds another contribution to the effective pressure, changing this picture~\cite{Colpi:1986ye}.}

As a final note, we should emphasize that bosonic stars share common features with neutron stars, such as their mass versus radius curves (see Fig.~\ref{MvsR}). This similarity can be used to understand the physics of compact stars in different scenarios, using a simple but very robust model~\cite{Liebling:2012fv}.

%

%%%%%%%%%%%%%%%%%%%%%%%%%%%%%%%%%%%%%%%%%%%%%%%%%%%%%%%%%%%%%%%%%%%%%%%%%%
\subsection{Brief overview of solutions}
%%%%%%%%%%%%%%%%%%%%%%%%%%%%%%%%%%%%%%%%%%%%%%%%%%%%%%%%%%%%%%%%%%%%%%%%%%

Let us give a brief summary of the most popular solutions which have been studied so far. For more detailed reviews on the subject see Refs.~\cite{Jetzer:1991jr,Schunck:2003kk,Liebling:2012fv,Macedo:2013jja}.

\paragraph{Boson stars.}
Boson stars are regular compact solutions of the Einstein-Klein-Gordon equations for a {\it complex} massive scalar field. 
Some of these solutions have been claimed in the literature as possible candidates for supermassive horizonless BH mimickers~\cite{Torres:2000dw}.
They can be classified according to the scalar potential in the Klein-Gordon Lagrangian~\eqref{eq:MFaction}~\cite{Schunck:2003kk}:
\begin{itemize}
\item {Mini boson stars}: the scalar field potential is given by $V(\phi)=\mu_S^2|\phi |^2$, where $\mu_S$ is the scalar field mass. 
For non-rotating boson stars the maximum mass is $M_{{\rm{max}}}\approx 0.633 M_{\rm{P}}^2/\mu_S$, with $M_{\rm{P}}$ being the Planck mass~\cite{Kaup:1968zz,Ruffini:1969qy}. 
Considering values of $\mu_S$ typically found within the Standard Model, this mass limit is much smaller than the Chandrasekhar limit for a fermion star, approximately $M_{\rm{P}}^3/\mu^2$. For ultralight boson masses $\mu_S$, as those motivated by string axiverse scenarios~\cite{Arvanitaki:2009fg}, and relevant in the DM context, mini boson stars may have a total mass compatible with that observed in active galactic nuclei~\cite{Schunck:2003kk}.
\item {Massive boson stars}: the scalar potential has an additional quartic scalar field term, $V(\phi)=\mu_S^2|\phi|^2+\lambda |\phi|^4/2$~\cite{Colpi:1986ye} (see also Ref.~\cite{Eby:2015hsq} for a detailed study of these solutions in the context of DM physics). In this case the maximum mass can be comparable to the Chandrasekhar limit and for $\lambda \gg \mu_S^2/M_{\rm{P}}^2$ one can estimate $M_{{\rm{max}}}\approx 0.062\lambda^{1/2}M_{\rm{P}}^3/\mu_S^2$. 
\end{itemize}
One can also find other types of boson stars by changing the scalar potential or by considering non-minimal couplings as done in Ref.~\cite{vanderBij:1987gi}. For some non-linear potentials, similar solutions, generically called Q-balls, exist even in flat space~\cite{Coleman:1985ki,Lee:1991ax} (when coupled to gravity some of these solutions can give rise to very heavy boson stars~\cite{Friedberg:1986tq}. See also Ref.~\cite{Schunck:1999zu} for potentials with more generic self-interaction terms). Finally, boson stars can also be found in alternative theories of gravity, such as scalar-tensor theories~\cite{Torres:1997np,Torres:1998xw,Whinnett:1999ma}. A more detailed list of solutions can be found in Ref.~\cite{Schunck:2003kk}.

\paragraph{Oscillatons.}

For a {\it real} massive scalar field minimally coupled to gravity, with $V(\phi)=\mu_S^2\phi^2$, compact configurations were first shown to exist in Ref.~\cite{Seidel:1991zh}, while the generalization for a scalar field with a quartic interaction was considered in~\cite{UrenaLopez:2012zz}. Interestingly, for some non-linear potentials, such as the Higgs double well potential or the axionic sine-gordon potential, a real scalar field counterpart of the Q-balls exist and are dubbed oscillons~\cite{Bogolyubsky:1976nx,Gleiser:1993pt,Copeland:1995fq,Kolb:1993hw,Gleiser:2006te} (note that oscillons are built in a {\it Minkowski} background).  
For real massive vector fields, Ref~\cite{Garfinkle:2003jf} found convincing indications that the same kind of oscillatory solutions form in the gravitational collapse of a wide set of initial data.

Boson stars and oscillatons share very similar structures, as summarized in Fig.~\ref{MvsR}, where we plot the mass-radius relation for spherically symmetric boson stars and oscillatons (including massive vectors that will be discussed in Chapter~\ref{sec:osci}). Boson stars and oscillatons have a maximum mass $M_{\rm{max}}$, given approximately by
\be
\frac{M_{\rm{max}}}{M_{\odot}}=8\times 10^{-11}\,\left(\frac{\rm{eV}}{m_{B}c^2}\right)\,,\label{max_mass}
\ee
for scalars and slightly larger for vectors. 

\paragraph{Boson-Fermion stars}

The extension to mixed stars, composed both by a complex scalar field and a perfect fluid, was first considered in Ref.~\cite{Henriques:1989ar} and further studied in~\cite{Henriques:1989ez,Lopes:1992np,Henriques:2003yr} (exact solutions in $2+1$-dimensions were also found in~\cite{Sakamoto:1998aj}). The stability of these objects was studied in~\cite{Jetzer:1990xa,Henriques:1990xg,Henriques:1990vx,ValdezAlvarado:2012xc}. Slowly rotating boson-fermion stars were constructed in~\cite{deSousa:2000eq}, while extensions to allow for an interaction between the scalar field and the fermionic fluid were considered in~\cite{deSousa:1995ye,Pisano:1995yk}.

Boson-fermion stars were shown to exist in a wide variety of configurations. For small boson masses they can be either dominated by the bosonic component or the fermionic, or have bosonic and fermionic components of the same order of magnitude~\cite{Henriques:1989ar,Henriques:1989ez,Lopes:1992np,Henriques:2003yr}. For large boson masses only two types of configurations exist, either bosonic dominated or fermionic dominated, with a sharp transition between the two configurations, when one of the configurations reaches $\sim 10\%$ of the total mass~\cite{Henriques:1989ez,Lopes:1992np}.

In Chapter~\eqref{sec:fluid} we will show that similar solutions also exist for real massive fields. However in this case the fluid must itself oscillate with a dominant frequency given by twice the boson mass. 

%%%%%%%%%%%%%%%%%%%%%%%%%%%%%%%%%%%%%%%%%%%%%%%
\section{Outline of Part III}
%%%%%%%%%%%%%%%%%%%%%%%%%%%%%%%%%%%%%%%%%%%%%%%

We start by constructing scalar and vector oscillatons in Chapter~\ref{sec:osci}. There, we first review how scalar oscillatons are constructed and then present for the first time massive vector field oscillatons.

Chapter~\ref{sec:fluid} is devoted to the study of stars with bosonic cores and their growth. We first study
stellar configurations formed by both a perfect fluid and a real massive scalar field. These solutions are a
generalization of the fluid-boson stars found and studied in detail in Refs.~\cite{Henriques:1989ar,Henriques:1989ez,Lopes:1992np,Henriques:2003yr}. 
We show that the presence of the scalar field induces very specific oscillations in the star's material, and argue that for most of the parameter space of interest, these stars are stable against small perturbations.
We finish by discussing how these cores might grow, arguing that collapse to a BH can be avoided whenever gravitational cooling mechanisms are
efficient~\cite{Alcubierre:2003sx,Seidel:1993zk,Guzman:2006yc,Madarassy:2014jfa}. This suggests that previous claims, assuming the collapse of the host star to
a BH above a certain threshold might not always be valid if DM is composed of massive bosonic fields.

%%%%%%%%%%%%%%%%%%%%%%%%%%%%%%%%%%%%%%%%%%%%%%%%%%%%%%%%%%%%%%%%%%%%%%%%%%%%%%%
\chapter{Oscillatons}\label{sec:osci}
%%%%%%%%%%%%%%%%%%%%%%%%%%%%%%%%%%%%%%%%%%%%%%%%%%%%%%%%%%%%%%%%%%%%%%%%%%%%%%%

%%%%%%%%%%%%%%%%%%%%%%%%%%%%%%%%%%%%%%%%%%%%%%%
\section{Introduction}
%%%%%%%%%%%%%%%%%%%%%%%%%%%%%%%%%%%%%%%%%%%%%%%
Massive bosonic fields minimally coupled to gravity can form structures~\cite{Kaup:1968zz,Ruffini:1969qy,Khlopov:1985jw,Seidel:1991zh,Comment:2015_v2,Guth:2014hsa,Brito:2015pxa}. Self-gravitating {\it complex} scalars may give rise to static, spherically-symmetric geometries called boson stars, while the field itself oscillates~~\cite{Kaup:1968zz,Ruffini:1969qy} (for reviews, see Refs.~\cite{Jetzer:1991jr,Schunck:2003kk,Liebling:2012fv,Macedo:2013jja}). Very recently, analogous solutions for complex massive vector fields where also shown to exist~\cite{Brito:2015pxa}. On the other hand, {\it real} scalars may give rise to long-term stable oscillating geometries, but with a non-trivial time-dependent stress-energy tensor, called oscillatons~\cite{Seidel:1991zh}. Both solutions arise naturally as the end-state of gravitational collapse~\cite{Seidel:1991zh,Garfinkle:2003jf,Okawa:2013jba}, and both structures share similar features.

In this Chapter we construct compact solutions of the Einstein field equations, either for a minimally coupled \emph{real} massive scalar or vector field. 
The methods here presented to construct oscillatons are a straightforward generalization to those used to construct solutions for complex fields. Since the latter are easier to construct, and less generic than the oscillatons, we focus here solely on real fields. We first review the formalism introduced in Ref.~\cite{Seidel:1991zh} to construct massive scalar oscillatons and then show how this formalism can be extended to massive vector fields.   

%%%%%%%%%%%%%%%%%%%%%%%%%%%%%%%%%%%%%%%%%%%%%%%%%%%%%%%%%%%%%%%%%%%%%%%%%%%%%%%
\section{Massive scalar field}
%%%%%%%%%%%%%%%%%%%%%%%%%%%%%%%%%%%%%%%%%%%%%%%%%%%%%%%%%%%%%%%%%%%%%%%%%%%%%%%
We start by considering a \emph{real} massive scalar minimally coupled to gravity. In Ref.~\cite{Seidel:1991zh} it was shown that solutions to the field equations describing spherically symmetric compact configurations exist. We consider a general time-dependent spherically symmetric metric
\be\label{metric}
ds^2=-F(t,r)dt^2+B(t,r)dr^2+r^2d\Omega^2\,.
\ee
Rescaling the scalar field as $\phi\to \phi/\sqrt{8\pi}$ and defining the function $C(t,r)=B(t,r)/F(t,r)$, the field equations~\eqref{eq:MFEoMScalar} and~\eqref{eq:MFEoMTensor} lead to a system of PDEs given by
\beq\label{eqs_scalar}
\frac{\dot{B}}{B}&=&r \dot{\phi}\phi'\,,\label{eqs_scalar1}\\
\frac{B'}{B}&=&\frac{r}{2}\left(C\dot{\phi}^2+(\phi')^2+B\mu_S^2\phi^2\right)+\frac{1}{r}\left(1-B\right)\label{eqs_scalar2}\,,\\
\frac{C'}{C}&=&\frac{2}{r}\left[1+B\left(\frac{1}{2}r^2\mu_S^2\phi^2-1\right)\right]\,,\label{eqs_scalar3}\\
\ddot{\phi}C&=&-\frac{1}{2}\dot{C}\dot{\phi}+\phi''+\phi'\left(\frac{2}{r}-\frac{C'}{2C}\right)-B\mu_S^2\phi\label{eqs_scalar4}\,,
\eeq
where an overdot denotes $\partial/\partial t$ and a prime denotes $\partial/\partial r$.
These equations suggest the following periodic expansion
\beq\label{series_scalar}
B(t,r)&=&\sum_{j=0}^{\infty} B_{2j}(r)\,\cos\left(2j\omega t\right)\,,\nonumber\\
C(t,r)&=&\sum_{j=0}^{\infty} C_{2j}(r)\,\cos\left(2j\omega t\right)\,,\nonumber\\
\phi(t,r)&=&\sum_{j=0}^{\infty} \phi_{2j+1}(r)\,\cos\left[\left(2j+1\right)\omega t\right]\,.
\eeq
Inserting this expansion into the system~\eqref{eqs_scalar2}--\eqref{eqs_scalar4} and truncating the series at a given $j$, yields a set of ordinary differential equations for the radial Fourier components of the metric functions and the scalar field. We note that out of the four  Eqs.~\eqref{eqs_scalar1}--~\eqref{eqs_scalar4} we only need to use three. The remaining one can be checked to be satisfied a posteriori. In practice we only compute the Fourier expansion~\eqref{series_scalar} up to $j=j_{\rm{max}}$. This introduces a certain error in the accuracy at which the full system of equations is satisfied. In general the larger $j_{\rm{max}}$ is, the smaller the error~\cite{Fodor:2009kg,Grandclement:2011wz}, and we explicitly checked that this was the case. 

We impose regular boundary conditions at $r=0$, i.e., $\phi_{2j+1}'(0)=0$,  $B_{0}(0)=1$, $B_{2j}(r)=0$ for $j\geq 1$, while $\phi_{2j+1}(0)$ and $C_{2j}(0)$ are free parameters. At infinity $r\to\infty$, asymptotic flatness requires $\phi_{2j+1}\to 0$, $C_{0}=B_{0}\to 1$ and $C_{2j}=B_{2j}\to 0$ for $j\geq1$\footnote{For asymptotically flat spactimes the metric function $C(t,r)\to B^2(r)/\alpha(t)$ when $r\to\infty$, for some arbitrary function $\alpha(t)$. Thus we can always rescale $t$ such that $C(t,r)\to B^2(r)$ at $r\to \infty$.}. 
The system~\eqref{eqs_scalar2}--\eqref{eqs_scalar4} supplemented with this set of boundary conditions is an eigenvalue problem for the frequency $\omega$. Fixing one of the free parameters, e.g, $\phi_{1}(0)$, one can shoot for the other remaining free parameters, requiring that the boundary conditions are satisfied. For each choice of $\phi_{1}(0)$ there will be a unique family of solutions satisfying the above boundary conditions, characterized by the number of nodes in the scalar field profile.

\begin{figure*}[htb]
\begin{center}
\begin{tabular}{cc}
\epsfig{file=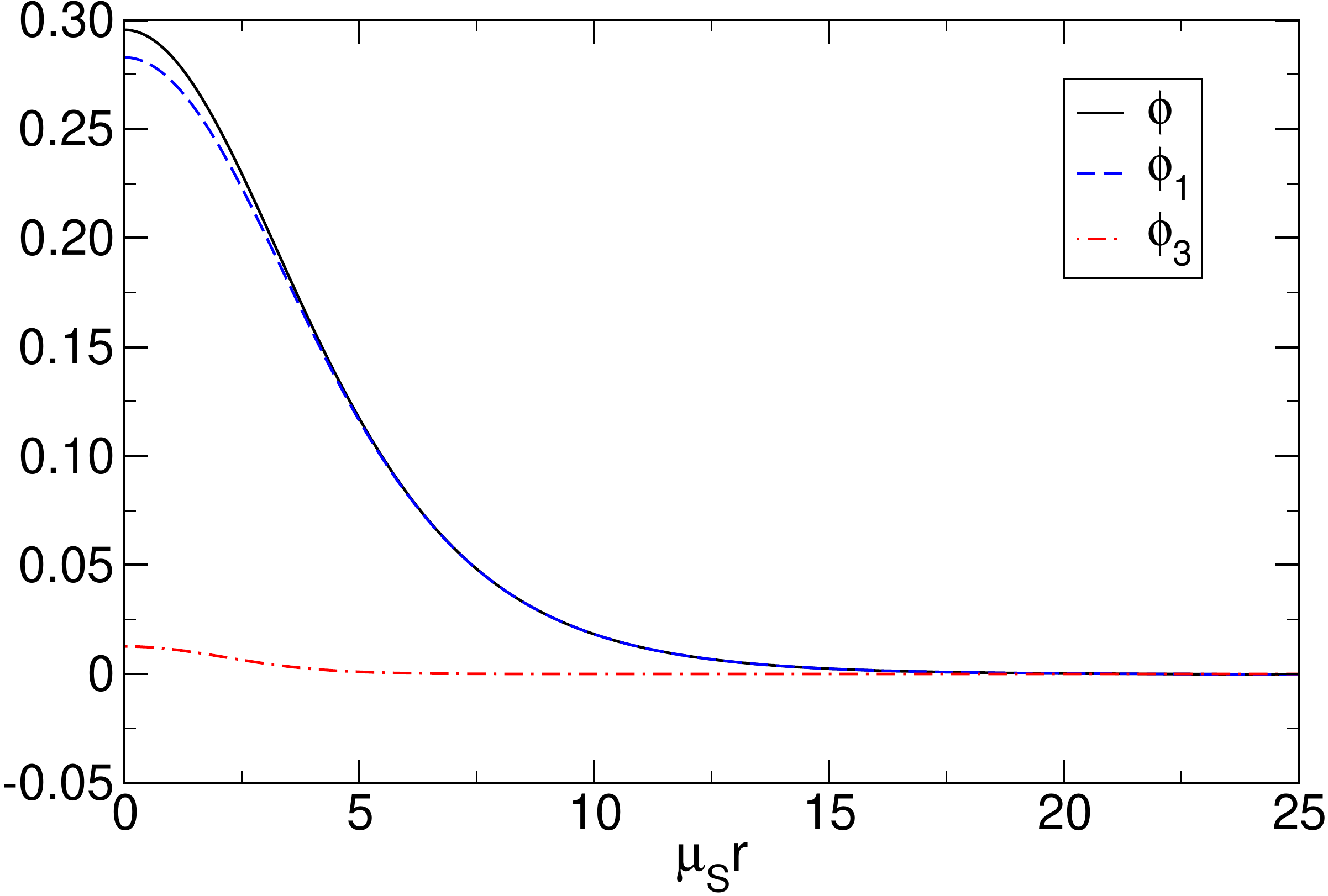,width=7.3cm,angle=0,clip=true}&
\epsfig{file=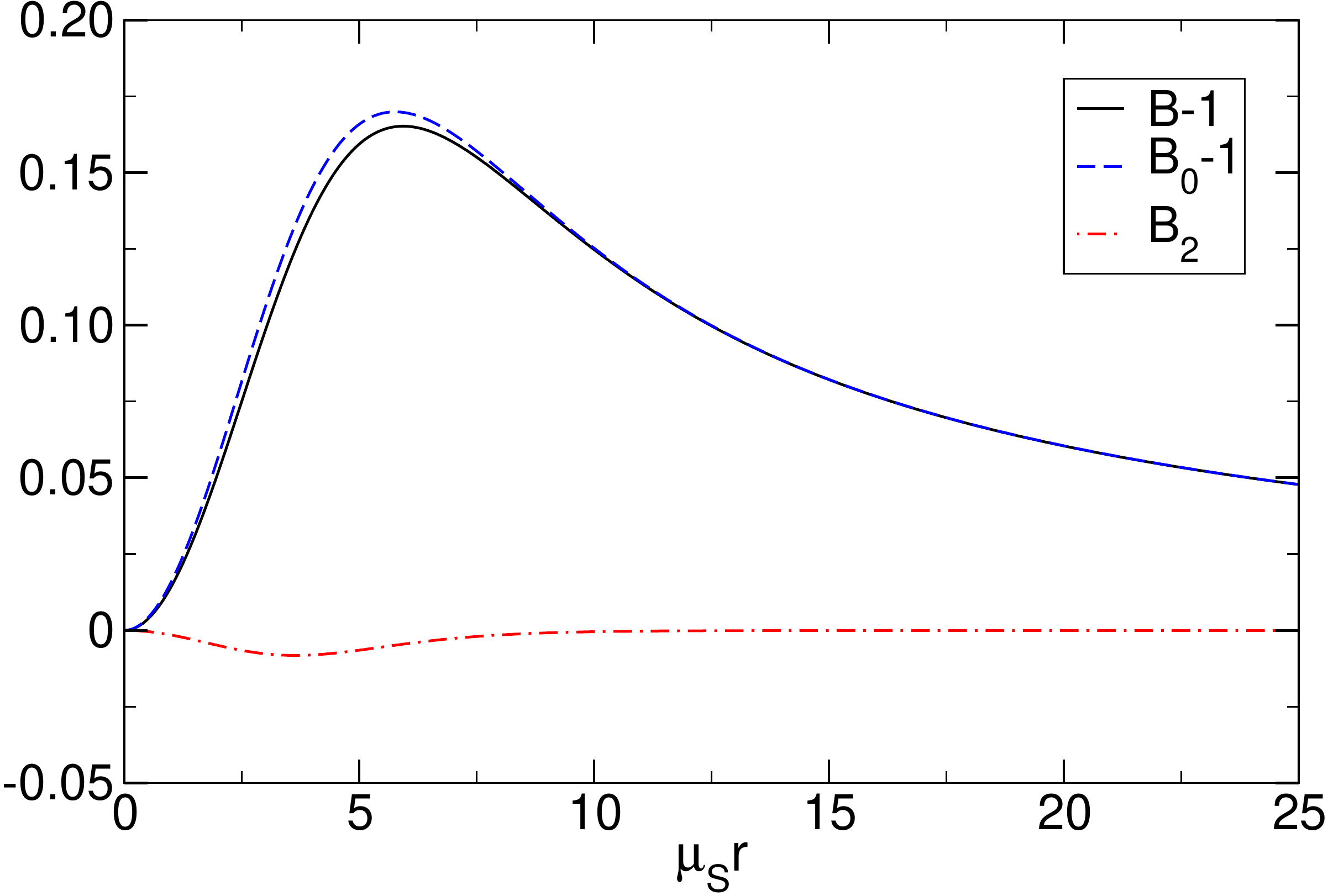,width=7.3cm,angle=0,clip=true}
\end{tabular}
\caption{Left: Scalar field configuration $\phi$ at $\omega t=0$ for $\phi_1(0)=0.2828$ and its first Fourier components with $j_{\rm{max}}=1$. Right: Corresponding radial metric coefficient $B$ at $\omega t=0$ and its first Fourier components. This configuration has a total mass $M_T\approx 0.57/\mu_S$ and fundamental frequency $\omega\approx 0.912\mu_S$.\label{scalar_osci}}
\end{center}
\end{figure*}

Due to the presence of a mass term in the scalar potential, the scalar field decays in a Yukawa-like fashion $e^{-r\sqrt{\mu_S^2-\omega^2}}/r$ at large distances. Thus, at infinity the metric asymptotically approaches the Schwarzschild solution and the total mass of a given configuration can be computed using
\be\label{ADM_mass}
M_T=\lim_{r\to\infty} m(r)=\lim_{r\to\infty} \frac{(B-1)r}{2B}\,.
\ee
We should note that oscillatons are not truly stable configurations, but decay on very long time-scales due to a radiative tail. 
However, since the amplitude of this tail is exponentially suppressed, for our purposes it is enough to compute the mass at some finite radius and consider it to be the mass of the oscillaton.
Although these stars do not possess a well-defined surface where the field vanishes, the configuration is exponentially suppressed at a radius $r\sim 1/\mu_S$. Thus, one can define an effective radius inside which much of the mass is localized. We will define the radius $R$ of the oscillaton as being the radius such that $m(R)$ is $98\%$ of the total mass $M_T$.  

\begin{figure}[htb]
\begin{center}
\begin{tabular}{c}
\epsfig{file=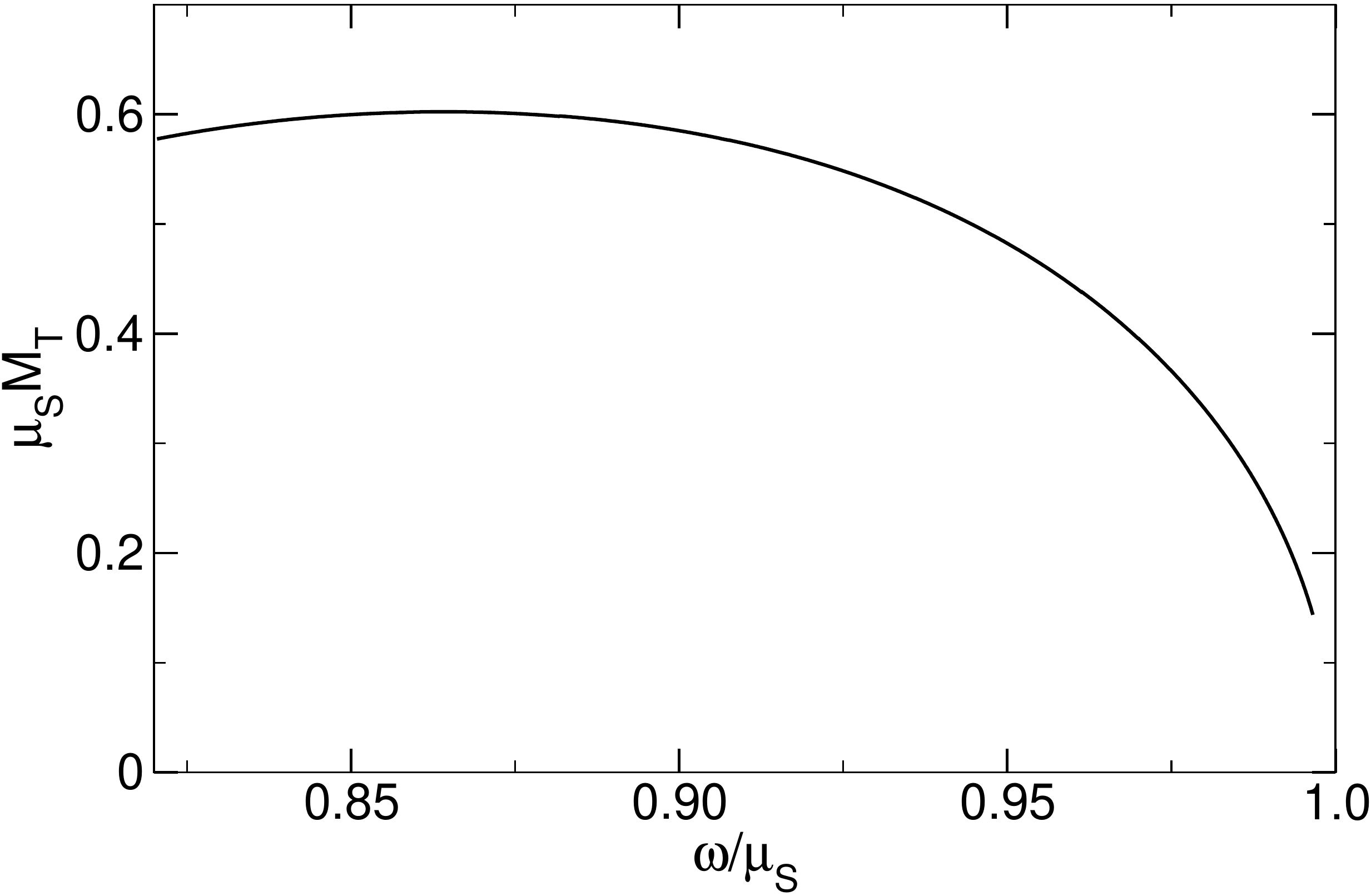,width=10cm,angle=0,clip=true}
\end{tabular}
\caption{Total mass $M_T$ of the scalar oscillaton as a function of the fundamental frequency $\omega$. The maximum mass is $M_T\sim 0.6/ \mu_S$ for $\omega \sim 0.864\mu_S$. This point marks the threshold between stable and unstable configurations. Stars to the right of the maximum are stable while those to the left are unstable.\label{scalar_Mvsw}}
\end{center}
\end{figure}

In Fig.~\ref{scalar_osci} we show an example of a configuration. The profile is smooth everywhere and we find that the series~\eqref{series_scalar} typically converges already for $j=2$. The fundamental frequency satisfies $\omega\lesssim \mu_S$ as shown in Fig.~\ref{scalar_Mvsw}, where we plot the mass $M_T$ as a function of $\omega$. In the Newtonian limit $M_T\to 0$, the solutions become spatially diluted with $\omega\to \mu_S$ (cf. Fig.~\ref{MvsR}). For smaller $\omega$ the star becomes more compact, with a maximum mass given by $M_T\sim 0.6/\mu_S$ for $\omega \sim 0.864\mu_S$, in agreement with previous studies~\cite{Seidel:1991zh,Alcubierre:2003sx}. 

Due to the time-dependence of these solutions, a perturbative analysis of their linear stability is very challenging. However the close similarity between oscillatons and boson stars, suggest that the solutions are stable from $\omega=\mu_S$ down to the maximal mass~\cite{Gleiser:1988ih,Lee:1988av}. The results of Refs.~\cite{Seidel:1991zh,Alcubierre:2003sx}, where Numerical Relativity techniques were used to study how oscillatons behave when slightly perturbed, suggest that this is indeed the case. Since scalar oscillatons have been widely discussed in the literature, we will not discuss them further and instead show that similar configurations exist for massive vector fields. 

%%%%%%%%%%%%%%%%%%%%%%%%%%%%%%%%%%%%%%%%%%%%%%%%%%%%%%%%%%%%%%%%%%%%%%%%%%%%%%%
\section{Massive vector field}
%%%%%%%%%%%%%%%%%%%%%%%%%%%%%%%%%%%%%%%%%%%%%%%%%%%%%%%%%%%%%%%%%%%%%%%%%%%%%%%
%
\begin{figure*}[htb]
%\begin{center}
\begin{tabular}{cc}
\epsfig{file=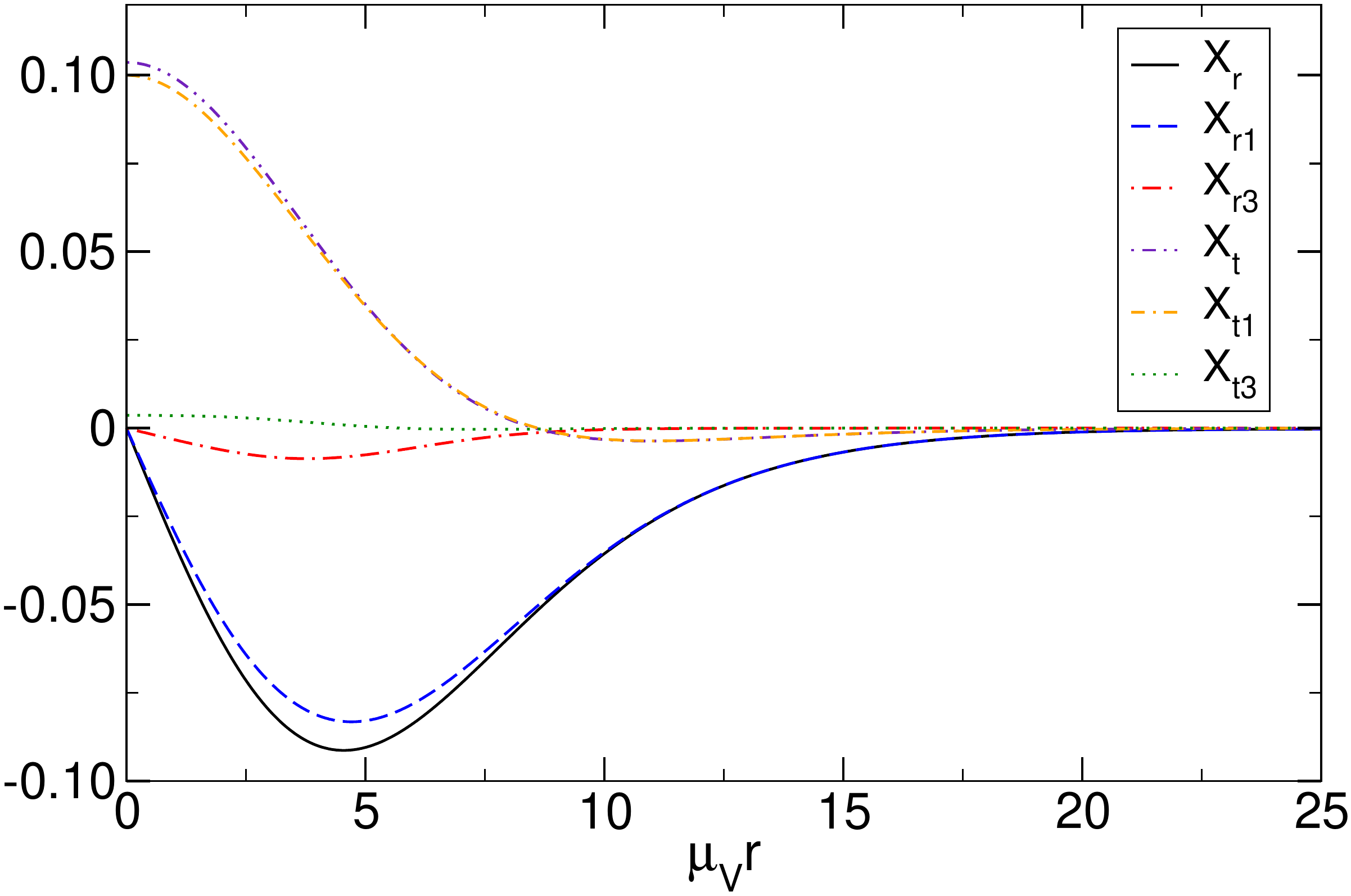,width=7.3cm,angle=0,clip=true}&
\epsfig{file=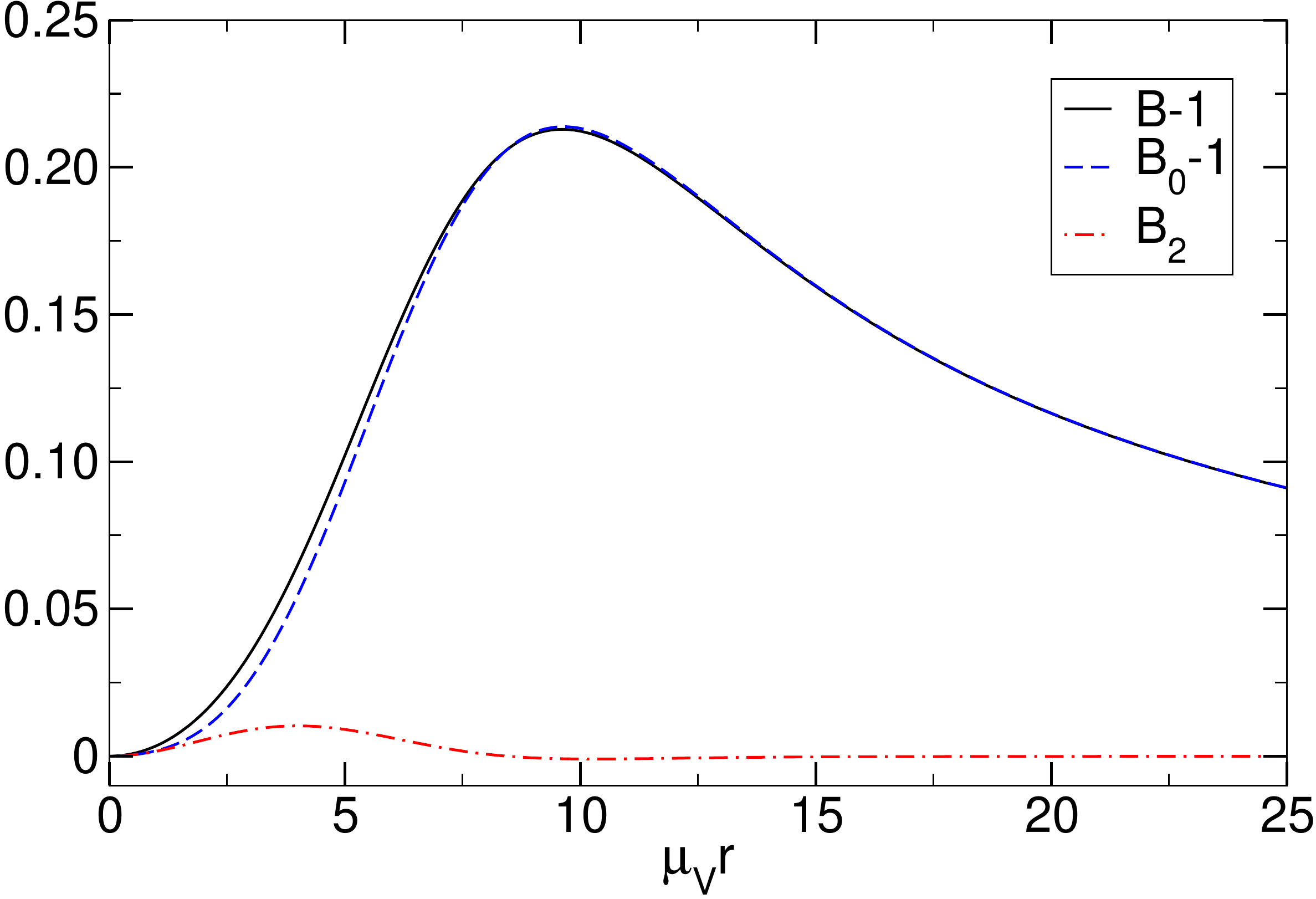,width=7.3cm,angle=0,clip=true}
\end{tabular}
\caption{Left: Vector field profiles at their peak values for $X_{t\,1}(0)=0.1$ and its first Fourier components with $j_{\rm{max}}=1$. Right: Corresponding radial metric coefficient $B$ at $\omega t=0$ and its first Fourier components. This configuration has a total mass $M_T\approx 1.044/\mu_V$ and fundamental frequency $\omega\approx 0.902\mu_V$.\label{vector_osci}}
%\end{center}
\end{figure*}

Very recently, the Einstein-Proca field equations~\eqref{eq:MFEoMVector} and~\eqref{eq:MFEoMTensor} have been shown to admit compact configurations, similar to scalar boson stars, for a complex massive vector field~\cite{Brito:2015pxa}. On the other hand, for real vector fields, Ref.~\cite{Garfinkle:2003jf} found strong numerical evidences that vector oscillatons can form in the gravitational collapse of a real massive vector field. Using the formalism introduced above, we can show that real massive vector fields can indeed form oscillating compact structures when minimally coupled to gravity. This is, as far as we are aware, the first time that these solutions are explicitly constructed. 

We consider the metric~\eqref{metric} and a spherically symmetric vector field
\be
A_{\mu}dx^{\mu}=A_t(t,r) dt+A_r(t,r) dr\,.
\ee
To simplify the equations we define the following functions
\be
X_t\equiv \sqrt{C} A_t\,,\quad
X_r\equiv \frac{1}{\sqrt{C}} A_r\,,\quad
W\equiv \frac{\sqrt{C}}{B}\left(\dot{A}_r-A'_t\right),
\ee
where again $C\equiv B/F$.
From eqs.~\eqref{eq:MFEoMVector} and~\eqref{divA} we find
\begin{align}
&X'_t=\frac{X_r\dot{C}}{2}+C\dot{X}_r-\frac{X_t}{r}\left(B-1\right)
-BW\left(1-r W X_t\right)\,,\\
&\dot{X}_t=\frac{1}{r^2}\partial_r\left(r^2 X_r\right)\label{EQMax_2}\,,\\
&X_t=-\frac{1}{\mu_V^2r^2}\partial_r\left(r^2W\right)\label{EQMax_3}\,,\\
&\dot{W}=-\mu_V^2 X_r\label{algebraic_W}\,,
\end{align}
while from Einstein's field equations~\eqref{eq:MFEoMTensor} we have
\beq
\frac{\dot{B}}{B}&=&2 r \mu^2_V X_r X_t \label{eq_vector1}\,,\\
\frac{B'}{B}&=&r\left(\mu_V^2 C X_r^2+\mu_V^2 X_t^2+BW^2\right)+\frac{1}{r}\left(1-B\right)\,,\\
\frac{C'}{C}&=&\frac{2}{r}\left[1+B\left(r^2W^2-1\right)\right]\,.
\eeq
These equations suggest the following Fourier expansions
\beq\label{series_vector}
%B(t,r)&=&\sum_{j=0}^{\infty} B_{2j}(r)\,\cos\left(2j\omega t\right)\,,\nonumber\\
%C(t,r)&=&\sum_{j=0}^{\infty} C_{2j}(r)\,\cos\left(2j\omega t\right)\,,\nonumber\\
X_t(t,r)&=&\sum_{j=0}^{\infty} X_{t\,2j+1}(r)\,\cos\left[\left(2j+1\right)\omega t\right]\,,\nonumber\\
X_r(t,r)&=&\sum_{j=0}^{\infty} X_{r\,2j+1}(r)\,\sin\left[\left(2j+1\right)\omega t\right]\,,\nonumber\\
W(t,r)&=&\sum_{j=0}^{\infty} W_{2j+1}(r)\,\cos\left[\left(2j+1\right)\omega t\right]\,,
\eeq
while the metric functions are expanded as in~\eqref{series_scalar}. Once more, eq.~\eqref{eq_vector1} will not be used to find the solutions. On the other hand, from Eq.~\eqref{algebraic_W}, one can find $W_{2j+1}$ algebraically, which greatly simplifies the equations (note that $W$ is just an auxiliary function that we introduced to simplify the equations, and so one can easily check that, after finding $W_{2j+1}$ from eq.~\eqref{algebraic_W} and $X'_{r\,2j+1}$ from eq.~\eqref{EQMax_2}, eq.~\eqref{EQMax_3} is automatically satisfied).
Similarly to the scalar case, we can find a set of ordinary differential equations by truncating the series at some $j$ and then solve the eigenvalue problem, imposing regular boundary conditions at $r=0$ and asymptotic flatness. This imposes $X_{r\, 2j+1}(0)=0$,  $B_{0}(0)=1$, $B_{2j}(r)=0$ for $j\geq 1$, while $X_{t\, 2j+1}(0)$ and $C_{2j}(0)$ are free parameters. At infinity, besides the usual conditions for the metric functions, we require $X_{t\,2j+1}=X_{r\,2j+1}\to 0$. We can then find solutions by fixing $X_{t\,1}(0)$ and use the same method as for the scalar case.

\begin{figure}[htb]
\begin{center}
\begin{tabular}{c}
\epsfig{file=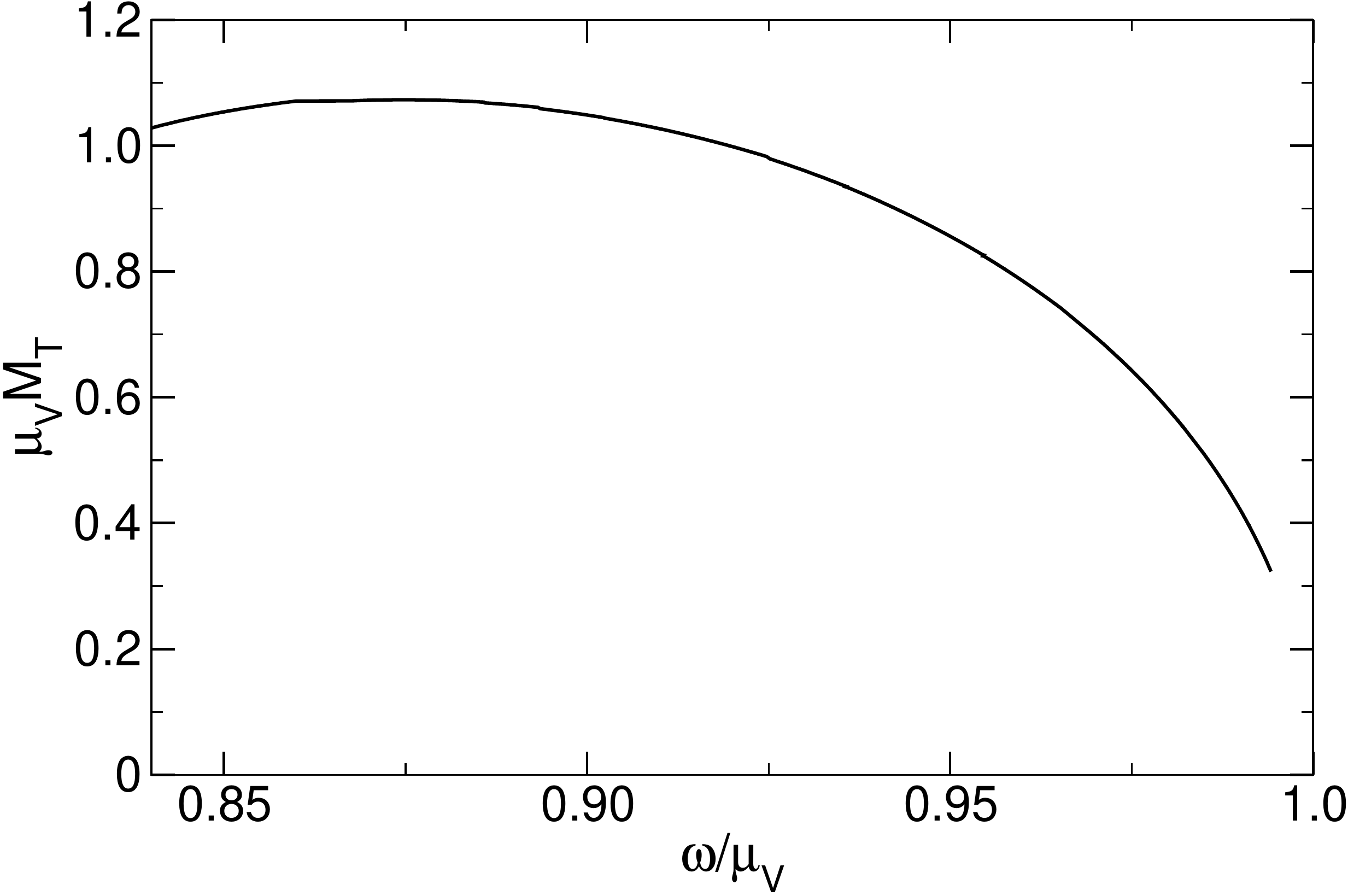,width=10cm,angle=0,clip=true}
\end{tabular}
\caption{Total mass $M_T$ of the vector oscillaton as a function of the fundamental frequency $\omega$. The maximum mass is $M_T\approx 1.07/ \mu_V$ for $\omega \approx 0.875\mu_V$. This point marks the threshold between stable and unstable configurations.\label{vector_Mvsw}}
\end{center}
\end{figure}

Our results are summarized in Figs.~\ref{vector_osci}--\ref{vector_Mvsw}. The overall behavior is analogous to the scalar case. 
The series~\eqref{series_vector} converges rapidly and already for $j=2$ one gets an accuracy for the ADM mass better than $\sim 0.2\%$. The total mass $M_T$ as a function of frequency $\omega$ is shown in Fig.~\ref{vector_Mvsw}. The behavior is analogous to the one found in the scalar case, although the maximum is slightly larger, $M_T\sim 1.07/ \mu_V$ (cf. Fig~\ref{MvsR}), and occurs at $\omega \sim 0.875\mu_V$. Not surprisingly, the overall behavior is almost identical to the one found for Proca stars (i.e. complex vector field boson stars)~\cite{Brito:2015pxa}, as can be seen in Fig.~\ref{MvsR}. 

By considering radial perturbations of Proca stars, it was shown in Ref.~\cite{Brito:2015pxa} that the maximum mass also corresponds to a branching point separating unstable from stable solutions. Although full numerical simulations are needed, vector oscillatons should also follow the same pattern. In particular, similar conclusions should hold: configurations which reach the unstable branch will either quickly collapse to BHs, migrate back to the stable branch via mass ejection, a phenomenon known as the
gravitational cooling mechanism~\cite{Alcubierre:2003sx,Seidel:1993zk,Guzman:2006yc} (see also Section~\ref{sec:collision} in the next Chapter), or simply completely disperse.

As a final word, we should note that although we only discussed fundamental states, characterized by $X_t$ having one node and $X_r$ being nodeless, excited states~--~solutions with more nodes~--~also exist. Since those are expected to be unstable~\cite{Balakrishna:1997ej,Balakrishna:2007mr} we will not discuss them here.

%

%%%%%%%%%%%%%%%%%%%%%%%%%%%%%%%%%%%%%%%%%%%%%%%%%%%%%%%%%%%%%%%%%%%%%%%%%%%%%%%
\section{Conclusions}
%%%%%%%%%%%%%%%%%%%%%%%%%%%%%%%%%%%%%%%%%%%%%%%%%%%%%%%%%%%%%%%%%%%%%%%%%%%%%%%

The picture we discussed in this Chapter is quite generic and shows that any self-gravitating, massive bosonic field can form compact structures, either in the form of boson stars or oscillatons. In particular, we showed for the first time that real self-gravitating massive vector fields can form oscillatons. 
It is reasonable to admit that DM could be composed of different kinds of fundamental entities, but which, when gravitationally clustered into macroscopic lumps, display some universality. Within this context, in the next Chapter we will show how these structures could leave potential imprints in stars.

%%%%%%%%%%%%%%%%%%%%%%%%%%%%%%%%%%%%%%%%%%%%%%%%%%%%%%%%%%%%%%%%%%%%%%%%%%%%%%%
\chapter{Stars with dark matter cores}\label{sec:fluid}
%%%%%%%%%%%%%%%%%%%%%%%%%%%%%%%%%%%%%%%%%%%%%%%%%%%%%%%%%%%%%%%%%%%%%%%%%%%%%%%
%%%%%%%%%%%%%%%%%%%%%%%%%%%%%%%%%%%%%%%%%%%%%%%%%%%%%%%%%%%%%%%%%%%%%%%%%%%%%%%
\section{Introduction}
%%%%%%%%%%%%%%%%%%%%%%%%%%%%%%%%%%%%%%%%%%%%%%%%%%%%%%%%%%%%%%%%%%%%%%%%%%%%%%%
In the previous Chapter we showed that self-gravitating massive bosonic fields can form compact configurations. 
An interesting possibility is that these structures might be accreted by stars, forming a bosonic core in their interior. For complex bosonic fields, such solutions were already considered in the literature~\cite{Henriques:1989ar,Henriques:1989ez,Lopes:1992np,Henriques:2003yr,Sakamoto:1998aj,deSousa:2000eq,deSousa:1995ye,Pisano:1995yk}. We will show that real bosonic fields can also cluster inside stars and give rise to oscillating configurations, where both the star's material and the field oscillate.

%%%%%%%%%%%%%%%%%%%%%%%%%%%%%%%%%%%%%%%%%%%%%%%%%%%%%%%%%%%%%%%%%%%%%%%%%%%%%%%
\section{Setting and Fourier-expansion}\label{subsec:setting}
%%%%%%%%%%%%%%%%%%%%%%%%%%%%%%%%%%%%%%%%%%%%%%%%%%%%%%%%%%%%%%%%%%%%%%%%%%%%%%%

We consider the system~\eqref{eq:MFEoMScalar},~\eqref{eq:MFEoMTensor}, and for simplicity we focus only on the scalar case. Similar configurations should also exist for vector fields.
Due to the presence of the scalar field the fluid will, in general, oscillate with a radial velocity:
\be
dr/dt=u^r/u^0=V(t,r)\,,
\ee
where $u^0=\Gamma/\sqrt{-g_{tt}}$ and $\Gamma=\left(1-U^2\right)^{1/2}$ is the Lorentz factor connecting the fluid's comoving frame with the frame of the observer at rest with respect to a spacelike hypersurface of constant $t$~\cite{1991A&A...252..651G}. Using the normalization condition $u^{\mu}u_{\mu}=-1$ one has $U=V\sqrt{-g_{rr}/g_{tt}}$.\footnote{Not assuming conservation of the baryonic number, given by Eq.~\eqref{baryonic}, is equivalent to neglecting the radial velocity $V\sim 0$. As we show below, this is indeed a good approximation for small bosonic cores. Physically, non conservation of the baryonic number leads to a conversion of bosonic matter to baryonic matter and vice-versa. This mimics a fundamental interaction between the scalar field and star's material for which the baryon number is not conserved. On the other hand, whether a conversion between fundamental fields can indeed occur indirectly through their gravitational coupling is an interesting discussion that we leave for future work. We should also note that for \emph{complex} fields, which give rise to the boson-fermion stars first studied in Refs.~\cite{Henriques:1989ar,Henriques:1989ez}, the fluid and the metric are static, and so for this case we set $V=0$.}

Once more, we will work with a rescaled scalar field, $\phi\to \phi/\sqrt{8\pi}$ and consider a general time-dependent, spherically symmetric metric, as in Eq.~\eqref{metric}.
The field equations, together with the conservation equations~\eqref{divT} and~\eqref{baryonic} lead to a system of PDEs given by
\beq
&&\dot{B}/B=r\left[\dot{\phi}\phi'-\frac{8\pi B V(P+\rho_F)}{1-U^2}\right]\,,\label{cont_eq1}\\
&&\dot{\rho}_{F}=\left(P+\rho_F\right)\frac{\dot{n}_B+Vn_F'}{n_F}-V\rho'_{F} \,,\label{cont_eq12}\\
&&B'/B=\frac{r}{2}\left[C\dot{\phi}^2+(\phi')^2+B\left(\mu_S^2\phi^2+16\pi\frac{\rho_{F}+PU^2}{1-U^2}\right)\right]
+(1-B)/r\,,\\
&&C'/C=2/r+Br\left(\mu_S^2\phi^2+8\pi\rho_{F}-8\pi P\right)-2B/r\,,\\
&&\ddot{\phi}C=-\dot{C}\dot{\phi}/2+\phi''+2\phi'/r-C'\phi'/(2C)-B\mu_S^2\phi\,,\\
&&2P'=-\left(1-U^2\right)\left(P+\rho_{F}\right)\left(CB'-BC'\right)/(BC)\nn\\
&&+V\left[\left(P+\rho_F\right)\left(4CV-r\dot{C}\right)-2rC\dot{P}\right]/r\nn\\
&&-2\left(P+\rho_{F}\right)C\dot{V}+2CV\left(P+\rho_F\right)\frac{\dot{n}_B+Vn_F'}{n_F}\,,\\
&&V'=-\left(1-U^2\right)\left[\left(\dot{B}+VB'\right)/(2B)+\left(\dot{n}_B+Vn_F'\right)/n_F\right]\nn\\
&&-V\left[4+C\left(2r\dot{V}-4V^2\right)+rV\left(\dot{C}+VC'\right)\right]/(2r)
\,.\label{cont_eq2}
\eeq
where we recall that $C\equiv B/F$ and $U\equiv\sqrt{C}V$. We will not make use of the conservation equations~\eqref{cont_eq1} and ~\eqref{cont_eq12}. One can check a posteriori that these equations are satisfied up to a certain error introduced by the ansatz we use.
%
%\beq
%&&\dot{B}/B=r\dot{\phi}\phi'\,,\qquad 2\dot{\rho_{F}}= r\left(P+\rho_{F} \right)\dot{\phi}\phi'\,,\label{cont_eq1}\\
%
%&&B'/B=(r/2)\left[C\dot{\phi}^2+(\phi')^2+B\left(\mu_S^2\phi^2+16\pi\rho_{F}\right)\right]\nn\\
%&&+(1-B)/r\,,\\
%
%&&C'/C=2/r+Br\left(\mu_S^2\phi^2+8\pi\rho_{F}-8\pi P\right)-2B/r\,,\\
%
%&&\ddot{\phi}C=-\dot{C}\dot{\phi}/2+\phi''+2\phi'/r-C'\phi'/(2C)-B\mu_S^2\phi\,,\\
%
%&&2P'=-\left(P+\rho_{F}\right)\left(CB'-BC'\right)/(BC)\,.\label{cont_eq2}
%\eeq
%

Employing the periodic expansion~\eqref{series_scalar}, one can easily see that the fluid's energy density, rest-mass density, pressure and radial velocity can be consistently expanded as 
\beq\label{series_fluid}
%B(t,r)&=&\sum_{j=0}^{\infty} B_{2j}(r)\,\cos\left(2j\omega t\right)\,,\nonumber\\
%C(t,r)&=&\sum_{j=0}^{\infty} C_{2j}(r)\,\cos\left(2j\omega t\right)\,,\nonumber\\
%\phi(t,r)&=&\sum_{j=0}^{\infty} \phi_{2j+1}(r)\,\cos\left[\left(2j+1\right)\omega t\right]\,,\nonumber\\
\rho_F(t,r)&=&\sum_{j=0}^{\infty} \rho_{F\,2j}(r)\,\cos\left(2j\omega t\right)\,,\nonumber\\
n_F(t,r)&=&\sum_{j=0}^{\infty} n_{F\,2j}(r)\,\cos\left(2j\omega t\right)\,,\nonumber\\
P(t,r)&=&\sum_{j=0}^{\infty} P_{2j}(r)\,\cos\left(2j\omega t\right)\,,\nonumber\\
V(t,r)&=&\sum_{j=1}^{\infty} V_{{2j}}(r)\,\sin\left(2j\omega t\right)\,.
\eeq
The equations of motion need to be supplemented by an equation of
state. We will focus on an ideal fluid and polytropic equation of state~\cite{rezzolla2013relativistic}:
\be
P=K \left(m_N n_F\right)^{\gamma},\quad \rho_F(P)=\left(P/K\right)^{1/\gamma}+P/\left(\gamma-1\right)\,,
\ee
where we take $K=100/\mu_S^2$ and $\gamma=2$, which can mimic neutron stars~\cite{1965ApJ...142.1541T,ValdezAlvarado:2012xc}. For this choice, the star is also isentropic, i.e. the fluid's specific entropy is constant along the star~\cite{rezzolla2013relativistic}.
We will also consider the equation of state $P=K \rho_F^{\gamma}$, which is equivalent to the previous one when the fluid's internal energy density is much smaller than the fluid's rest-mass density. This is a good model for cold and old neutron stars~\cite{rezzolla2013relativistic}. In the following, we will compare the results obtained in both models. Although our results can be generalized to other equations of state, we should note that generic equations of state do not allow for a straightforward expansion such as~\eqref{series_fluid}. A possibility is to consider the oscillating components to be a small perturbation of a static star, along the lines of what is usually done to construct slowly-rotating stars~\cite{Hartle:1967he,Hartle:1968si}~\footnote{In this case, at lowest order, the expansion~\eqref{series_fluid} would be given by $\rho_F(t,r)=\rho_{F\,0}(r)+\epsilon^2 P_{2}(r)\,\partial\rho_{F\,0}/\partial P_{0}\cos\left(2\omega t\right)$, $n_F(t,r)=n_{F\,0}(r)+\epsilon^2 P_{2}(r)\,\partial n_{F\,0}/\partial P_{0}\cos\left(2\omega t\right)$, $P(t,r)=P_{0}(r)+\epsilon^2 P_{2}(r)\cos\left(2\omega t\right)$, with $\epsilon$ a small bookkeeping parameter and we assume an equation of state in the absence of the scalar field of the form $\rho_{F\,0}\equiv \rho_{F\,0}(P_{0})$ and $n_{F\,0}=n_{F\,0}(P_{0})$.}. The construction here presented can then be straightforwardly applied. We have explicitly checked that using this approach one can generalize our results to a generic equation of state.

%%%%%%%%%%%%%%%%%%%%%%%%%%%%%%%%%%%%%%%%%%%%%%%%%%%%%%%%%%%%%%%%%%%%%%%%%%%%%%%
\subsection{Small note on units}
%%%%%%%%%%%%%%%%%%%%%%%%%%%%%%%%%%%%%%%%%%%%%%%%%%%%%%%%%%%%%%%%%%%%%%%%%%%%%%%
For generic polytropic equations of state and in geometrical units  $G=c=1$, the constant $K$ has dimensions $[L]^{2\left(\gamma-1\right)}$, where $[L]$ denotes dimensions of length, while in non-geometric units it has dimensions $[L]^{3\gamma-1}[M]^{1-\gamma}[T]^{-2}$, where $[M]$ denotes dimensions of mass and $[T]$ dimensions of time.
In geometric units all quantities have dimensions of length, so two quantities $\hat{L}$ and $L$ with dimensions of length, obtained using $\{\hat{K},\hat{\gamma}\}$ and $\{K,\gamma\}$, respectively, correspond to the same solution (in the absence of the scalar field) if they are related by~\cite{1964ApJ...140..434T}
\be\label{rescale}
\hat{L}/L=\hat{K}^{1/2(\hat{\gamma}-1)}/K^{1/2(\gamma-1)}\,.
\ee

To transform a dimensionless variable $X$ to a dimensionfull quantity $\bar{X}$ with dimensions $[L]^l [M]^m [T]^t$, one can use the following equation~\cite{Noble:2015anf}:
\be
\bar{X}=\bar{K}^x c^y G^z X\,,
\ee
where $x=(l+m+t)/[2(\gamma-1)]$, $y=[(\gamma-2)l+(3\gamma-4)m-t]/(\gamma-1)$ and $z=-(l+3m+t)/2$, and $\bar{K}$ is the value of the constant $K$ in non-geometric units. For example, for the mass of a star we get:
\be
\bar{M}=\bar{K}^{1/2(\gamma-1)}c^3 c^{-1/(\gamma-1)}G^{-3/2} M(K=1)\,.
\ee
where $M(K=1)$ denotes the dimensionless mass for units with $K=1$.

For our choice $\gamma=2$, $\sqrt{K}$ has units of length and can be used to set the length-scale of the problem in the absence of the scalar field. Without loss of generality we will also set $m_N=1$. The choice $K=100$ was considered in e.g. Ref.~\cite{ValdezAlvarado:2012xc}, which we used to check the accuracy of our code. Other values of $K$ can be obtained by fixing $\mu_S$ and rescale all the quantities using Eq.~\eqref{rescale}. For example, fixing $\mu_S=1$, mass and radius are measured in units of $\sqrt{K}$.

%%%%%%%%%%%%%%%%%%%%%%%%%%%%%%%%%%%%%%%%%%%%%%%%%%%%%%%%%%%%%%%%%%%%%%%%%%%%%%%
\section{Numerical procedure}
%%%%%%%%%%%%%%%%%%%%%%%%%%%%%%%%%%%%%%%%%%%%%%%%%%%%%%%%%%%%%%%%%%%%%%%%%%%%%%%
To construct the stars we employ the same method used in the previous Chapter, the difference being that, due to the presence of the fluid, the solutions are now parametrized by 
two parameters, e.g., $n_{F\,0}(0)$ and $\phi_1(0)$, while for the radial velocity we impose $V_{2j}(0)=0$. Additionally, we also need to impose boundary conditions at the star's radius. We define the radius $R$ of the star to be the location where the pressure drops to zero, $P(R)=0$.
For high scalar field central densities, first-order terms $j=1$ in the density might become of the order of the zeroth-order term, making it difficult to find these configurations with good accuracy and impose the boundary condition at the star's radius. However, as explained below, for a given $n_{F\,0}(0)$, we expect these configurations to become unstable at some threshold $\phi_1(0)>\phi^c_{1}(0)$. To avoid these numerical difficulties, we will mostly focus on small $\phi_1(0)$.
  
Due to the different length scales present in the problem, the solutions can also be characterized by the mass coupling $\mu_S M_0$, where $M_0$ is the mass of the \emph{static} star for vanishing scalar field, corresponding to the same value of central rest-mass density $n_{F\,0}(0)$. Depending on the numerical value of $\mu_S M_0$, we employ different numerical strategies. For small $\mu_S M_0$, the scalar field density profile extends beyond the star's radius. For this case we compute the profile inside the star and at the star's radius impose the matching with the outer solution. The full solution is then found by imposing asymptotically flat boundary conditions. For large $\mu_S M_0$, the scalar field is exponentially suppressed inside the star. To prevent numerical errors from spoiling the full solution, we perform tree integrations: we first find the radius at which the scalar field drops to zero and then compute the remaining solution by imposing the scalar field to be zero after this radius. The solution outside the star is then found by matching it with the inner solution.

A useful quantity to describe scalar-fluid stars is the scalar field's energy density, given by
\be\label{scalar_density}
2\rho_{\phi}=-2T_0^{\phantom{0}0}=-\dot{\phi}^{2}/g_{tt}+\phi^{'2}/g_{rr}+\mu_S^2\phi^2\,,
\ee
and the energy density measured by an observer at rest with respect to a spacelike hypersurface of constant $t$, given by
\be\label{fluid_density}
\rho_{\mathcal{F}}=-T_0^{\phantom{0}0}=\Gamma^2\left(\rho_{F}+P\right)-P\,.
\ee
Note that, for our solutions, the contribution from the fluid's kinetic energy to $\rho_{\mathcal{F}}$ is negligible, and so we have in general $\Gamma\sim 1$ and $\rho_{\mathcal{F}}\sim\rho_F$.
With this, we define the time-average total mass in the fluid and bosons as
\be
M_{F,\,B}=\int_0^{\infty} 4\pi \left<\sqrt{B}\,\rho_{{\mathcal{F}},\,\phi}\right> r^2 dr\,,\label{fermion_mass}
\ee
where $<>$ denotes a temporal average. The total mass $M_T$ can be found in the usual way through the metric component $g_{rr}$ which asymptotically approaches the Schwarzschild solution at infinity (cf. Eq.~\eqref{ADM_mass}).  

%%%%%%%%%%%%%%%%%%%%%%%%%%%%%%%%%%%%%%%%%%%%%%%%%%%%%%%%%%%%%%%%%%%%%%%%%%%%%%%
\section{Results}\label{BF_results}
%%%%%%%%%%%%%%%%%%%%%%%%%%%%%%%%%%%%%%%%%%%%%%%%%%%%%%%%%%%%%%%%%%%%%%%%%%%%%%%
We will discuss our results assuming baryon conservation during DM accretion and dynamics,
but we will also discuss stars for which the DM-baryon cross section is so large that
conversion between one and the other is extreme, to the point where solutions with zero fluid velocity
are allowed. These solutions conserve baryon number on the average, but not instantaneously.
Additionally, this also serves as a model for stars composed of fields for which there is no conserved current, such as Majorana fermions or real bosonic fields.
%%%%%%%%%%%%%%%%%%%%%%%%%%%%%%%%%%%%%%%%%%%%%%%%%%%%%%%%%%%%%%%%%%%%%%%%%%%%%%%
\subsection{Conserved baryon number}
%%%%%%%%%%%%%%%%%%%%%%%%%%%%%%%%%%%%%%%%%%%%%%%%%%%%%%%%%%%%%%%%%%%%%%%%%%%%%%%
%
\begin{figure}[ht]
\begin{center}
\begin{tabular}{c}
\epsfig{file=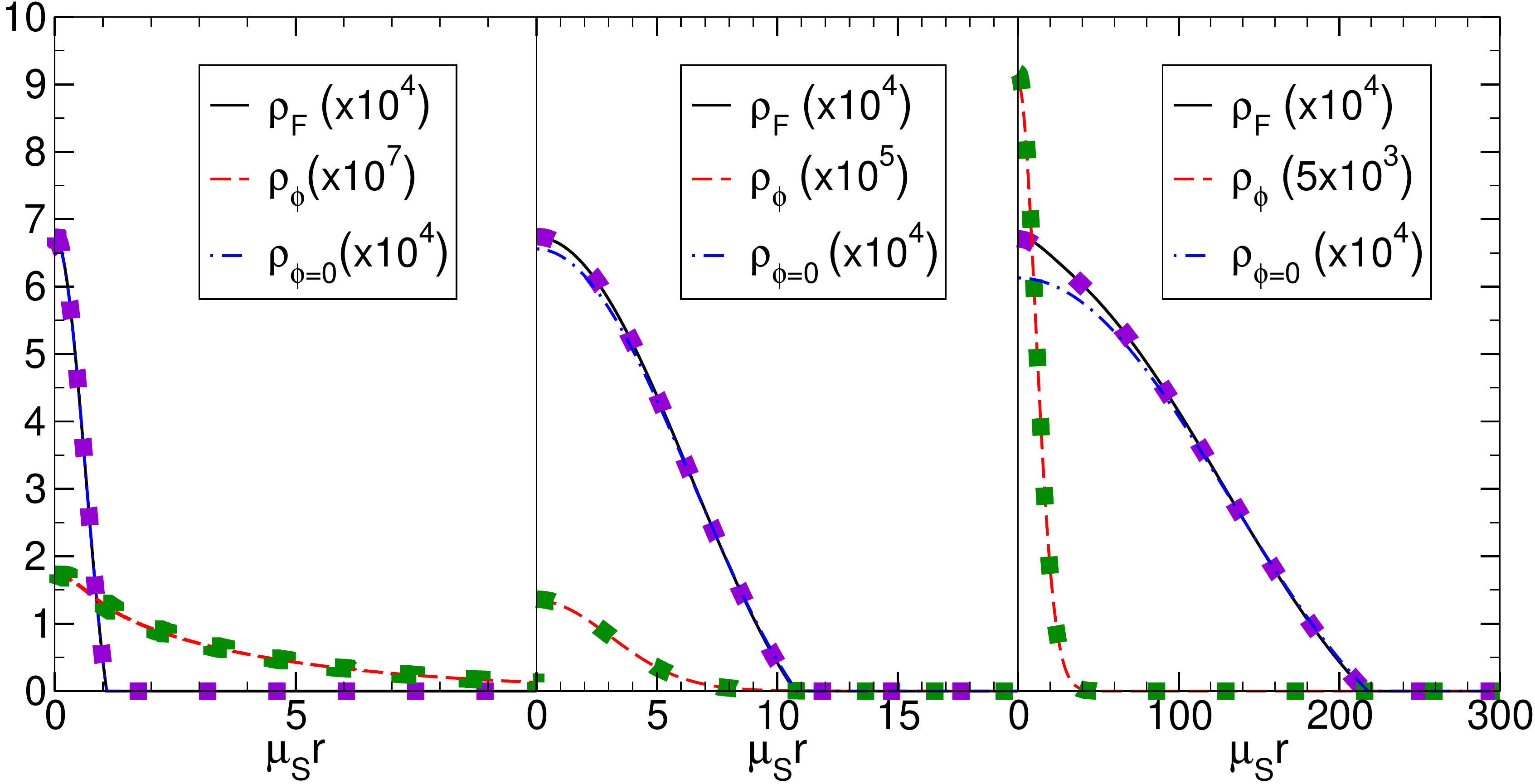,width=10cm,angle=0,clip=true}
\end{tabular}
\caption{
Comparison between the (time average) energy density of the scalar field $\rho_{\phi}$ and the fluid $\rho_{F}$ for mixed scalar oscillatons and baryon fluids, for scenarios where the total baryon number is conserved (fluid velocity $V\neq 0$). 
We fix $\rho_{F\,0}(0)=0.0006332$, and have
from left to right, $\mu_S M_0=0.1,\,M_B/M_T\approx 21\%$, corresponding to $\phi_1(0)=0.026$ and $\omega M_0\approx 0.0993$,
%Middle
$\mu_S M_0=1,\,M_B/M_T\approx 0.66\%$, corresponding to $\phi_1(0)=0.025$ and $\omega M_0\approx 0.863$,
%Right: 
and $\mu_S M_0=20,\,M_B/M_T\approx 0.54\%$, corresponding to $\phi_1(0)=0.015$ and $\omega M_0\approx 16.221$.
Squares denote the corresponding quantities for complex fields (i.e. mixed boson-fluid stars for the same $M_F$ and $M_B$). The overlap is nearly complete.
Here $M_0$ and $\rho_{\phi=0}$ are the total mass of the star (for the same $\rho_{F\,0}(0)$) and the energy density of the fluid (for the same $M_F$), respectively, when the scalar field vanishes everywhere. In the left panel, the $\rho_{\phi=0}$ and the $\rho_{F}$ lines are indistinguishable, because light fields have a negligible influence on the fluid distribution. For $\rho_{F\,0}(0)=0.0006332$, we have, in our units, $M_0=1$ which corresponds to a star with $M_0\sim M_{\odot}$. Solutions with larger $\rho_{F\,0}(0)$ can also be obtained and the qualitative picture remains the same. 
\label{scalarfluid_osci}}
\end{center}
\end{figure}
\begin{figure*}[ht]
\begin{center}
\begin{tabular}{cc}
\epsfig{file=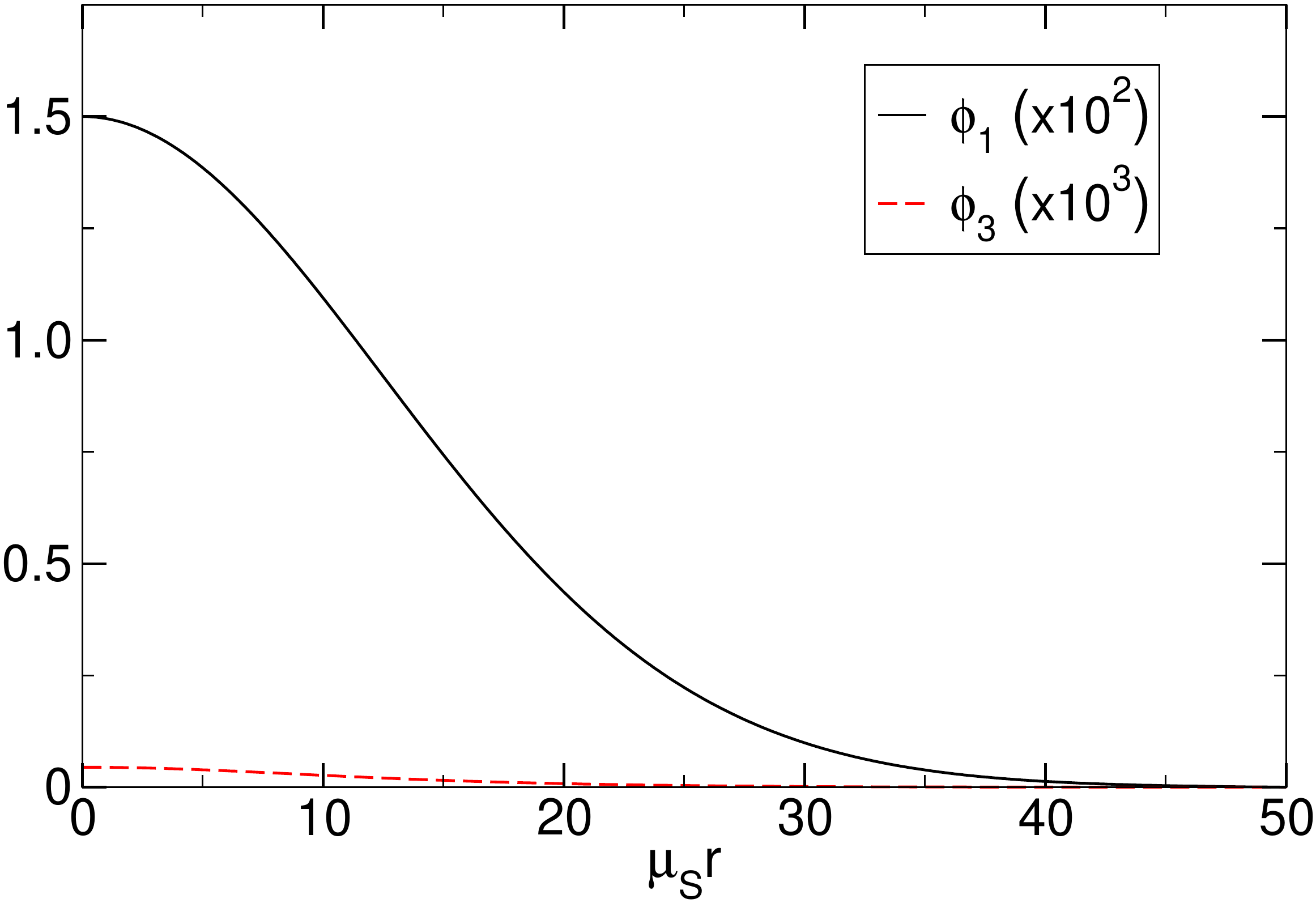,width=7cm,angle=0,clip=true}&
\epsfig{file=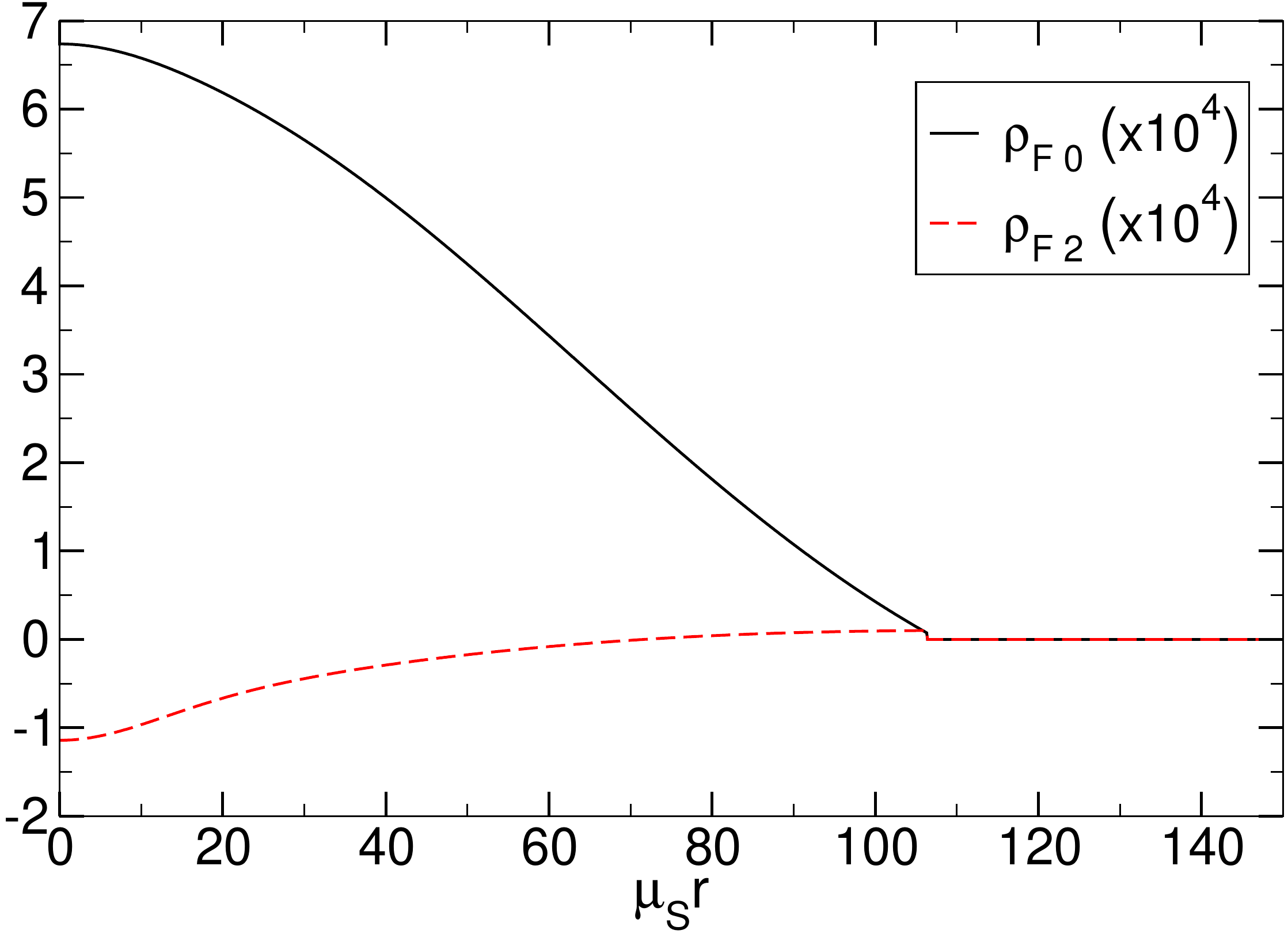,width=7cm,angle=0,clip=true}\\
\epsfig{file=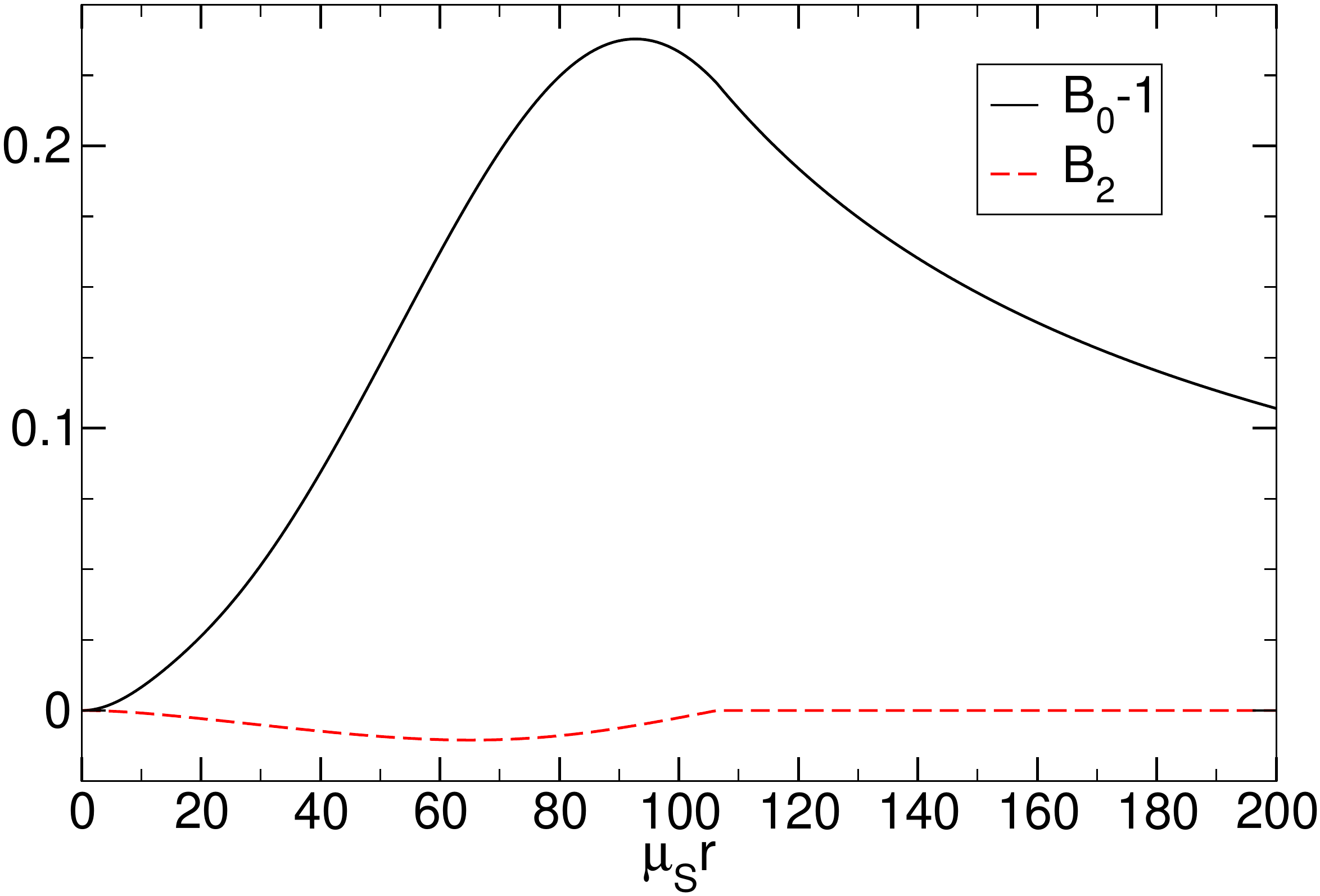,width=7cm,angle=0,clip=true}&
\epsfig{file=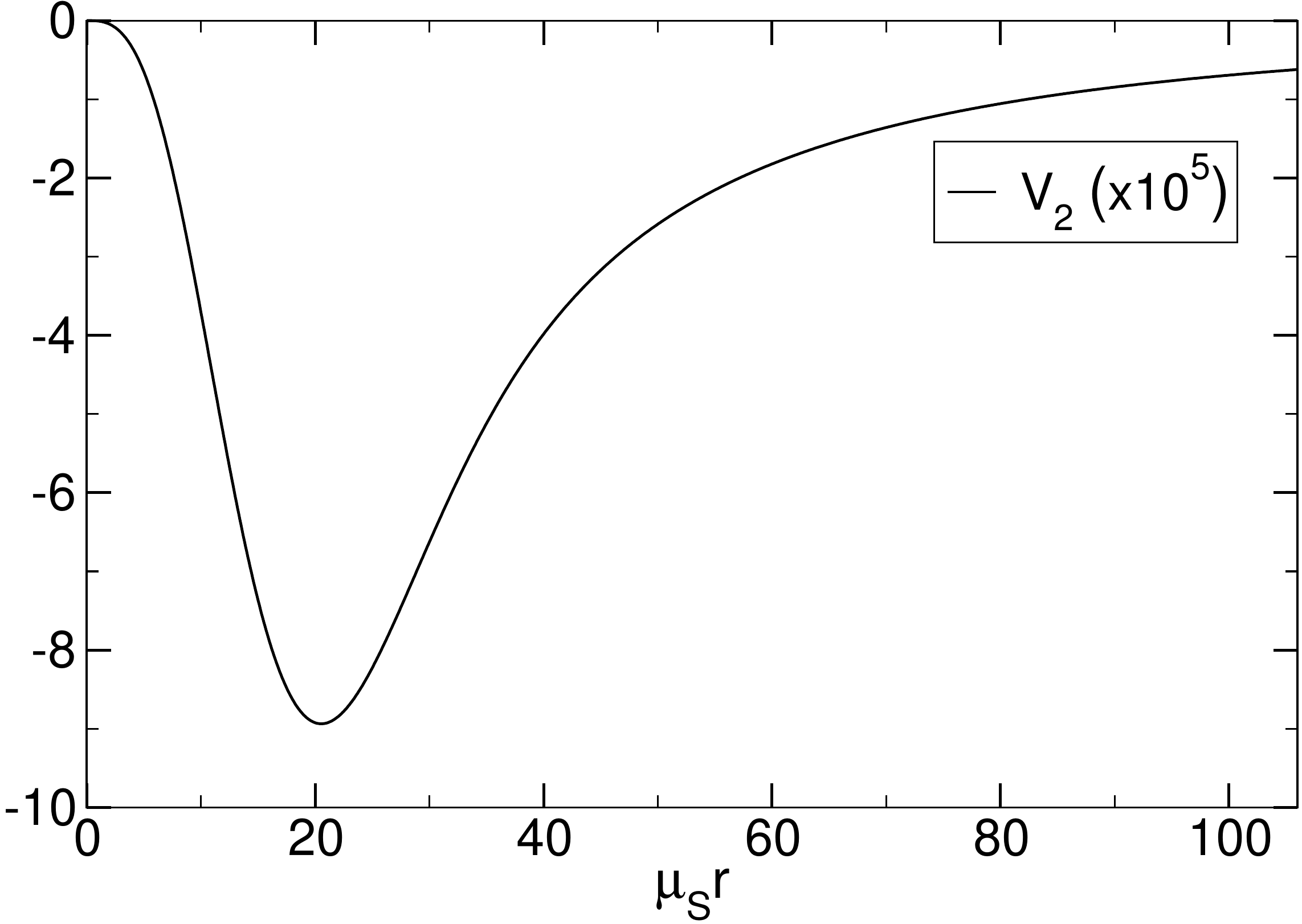,width=7cm,angle=0,clip=true}
\end{tabular}
\caption{Scalar field (top left) configuration, density profile of the fluid (top right), and corresponding metric component $g_{rr}$ (bottom left) for $\mu_S M_0=10$ and $M_B/M_T\approx 0.54\%$, corresponding to $\phi_1(0)=0.015$, $n_{F\,0}(0)=0.0006332$ and $\omega M_0\approx 8.118$. We plot the first Fourier components for $j_{\rm{max}}=1$. In the bottom-right panel, we also plot the velocity profile of the fluid (here $V_2$ is the dominant time-dependent component of the velocity profile, cf. equation~\eqref{series_fluid}).
\label{scalarfluid_solution_v2}}
\end{center}
\end{figure*}
\begin{figure}[htb]
\begin{center}
\begin{tabular}{c}
\epsfig{file=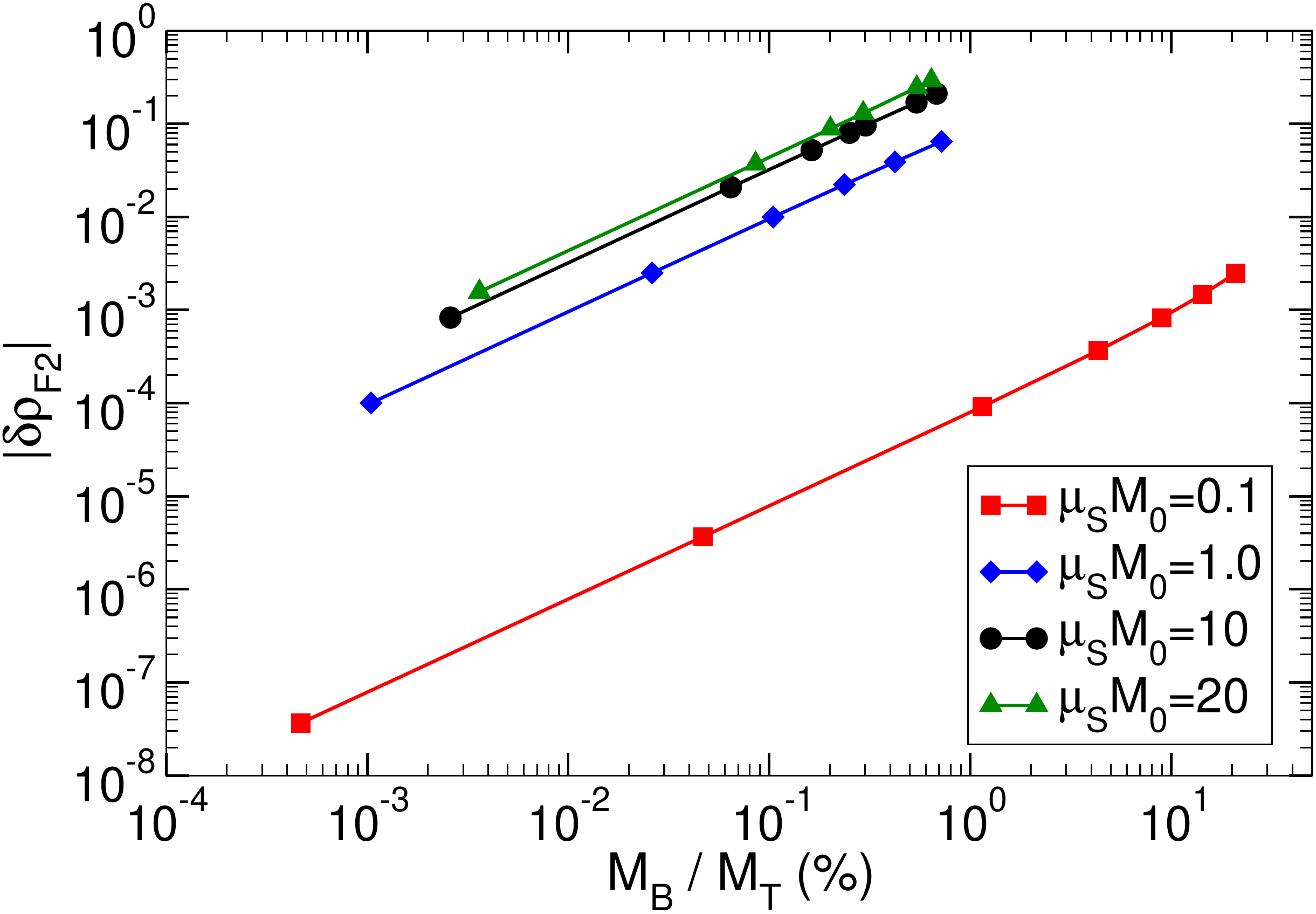,width=8cm,angle=0,clip=true}
\end{tabular}
\caption{
Amplitude of the oscillations $\delta\rho_{F\,2}\equiv\rho_{F\,2}(0)/\rho_{F\,0}(0)$ as a function of $M_B/M_T$. The symbols denote actual solutions that we computed. We find that for small mass ratios $M_B/M_T$ and large $\mu M_0$ the amplitude is well fitted by $\delta\rho_{F\,2}\sim 10(\mu M_0)^{1/2} M_B/M_T$.
%$\rho_{F\,2}/\rho_{F\,0}\propto 0.286 M_B/M_T$, $\rho_{F\,2}/\rho_{F\,0}\propto 0.0833 M_B/M_T$ and $\rho_{F\,2}/\rho_{F\,0}\sim 0.00007 M_B/M_T$, for $\mu_S M_0=10$, $\mu_S M_0=1$ and $\mu_S M_0=0.1$, respectively. 
\label{osci_linear}}
\end{center}
\end{figure}

Our results for composite stars with conserved total baryon numbers are summarized in Figs.~\ref{scalarfluid_osci}-\ref{osci_linear}. The overall behavior and global structure of these DM-cored stars is dependent on the new mass scale introduced by the scalar field mass. For very small $\mu M_0$, the Compton wavelength of the scalar is very large and the scalar field spreads throughout the spacetime. This is shown in the left panel of Fig.~\ref{scalarfluid_osci} for $\mu_SM_0=0.1$.
For very large scalar field masses, on the other hand, the scalar is confined to a small region inside the star, as seen in the right panel of Fig.~\ref{scalarfluid_osci} for $\mu_SM_0=20$. In fact, when $\mu_S M_0$ is extremely large, as happens for many DM models (c.f. eq.~\eqref{dimensionless_massparameter}), the scalar core hardly knows about the existence of the star outside, and behaves, to a very good precision, exactly like the pure oscillatons we described in the previous Chapter.
Notice also that for large mass couplings, one can have a large density, small oscillaton inside a fluid star.
As we discuss below, our argument then indicates that the oscillaton can be in the stable branch, indicating that the whole configuration is stable.
In other words, stable, self-gravitating bosonic DM cores inside stars are possible. These results complement similar recent findings for fermionic DM cores~\cite{Leung:2013pra}.

Accordingly, the detailed structure of these stars will also depend on the dimensionless coupling $\mu_S M_0$.
For large couplings, one can think of these composite stars as a regular fluid star, where at the center sits a small
pulsating oscillaton. It is then natural to expect that the oscillations in the oscillaton density will {\it induce oscillations} in the fluid material. Indeed, this is a generic feature borne out of our results.
A typical star structure is shown in Fig.~\ref{scalarfluid_solution_v2} for a 0.54\% scalar composition
and $\mu M_0=10$, corresponding to a scalar core well inside the baryon fluid. A general feature of these stars is
that they {\it oscillate}, driven by the scalar field, with a frequency
\be
f=2.5\times 10^{14}\,\,\frac{m_{B}c^2}{\rm{eV}}\,{\rm{Hz}}\,.
\ee
In particular, the local density is a periodic function of time with a period dictated by the scalar field. These oscillations are signalled by a nonzero $j=1$ component of the baryon density expansion \eqref{series_fluid}, and are driven by the time-varying component of the oscillaton's density; therefore, the {\it amplitude} of the fluid's oscillations is expected to scale with the mass of the oscillaton. As shown in Fig.~\ref{osci_linear}, for small mass ratios $M_B/M_T$ and large $\mu M_0$ we find that the amplitude of these oscillations is described by the approximate relation
\be
 \delta\rho_{F\,2}\equiv\rho_{F\,2}(0)/\rho_{F\,0}(0)\sim 10 (\mu M_0)^{1/2} M_B/M_T\,.
\ee
Even for $M_B/M_T=0.01$, and for $\mu_S M_0=10$ the oscillations are of the order of $30\%$ of the static component.
For this particular setup, where the baryon number is conserved, the fluid velocity is nonzero, as it is shown in the bottom-right panel of Fig.~\ref{scalarfluid_solution_v2}.

%%%%%%%%%%%%%%%%%%%%%%%%%%%%%%%%%%%%%%%%%%%%%%%%%%%%%%%%%%%%%%%%%%%%%%%%%%%%%%%
\subsection{Non-conservation of baryon number}
%%%%%%%%%%%%%%%%%%%%%%%%%%%%%%%%%%%%%%%%%%%%%%%%%%%%%%%%%%%%%%%%%%%%%%%%%%%%%%%
%
\begin{figure}[htb]
\begin{center}
\begin{tabular}{c}
\epsfig{file=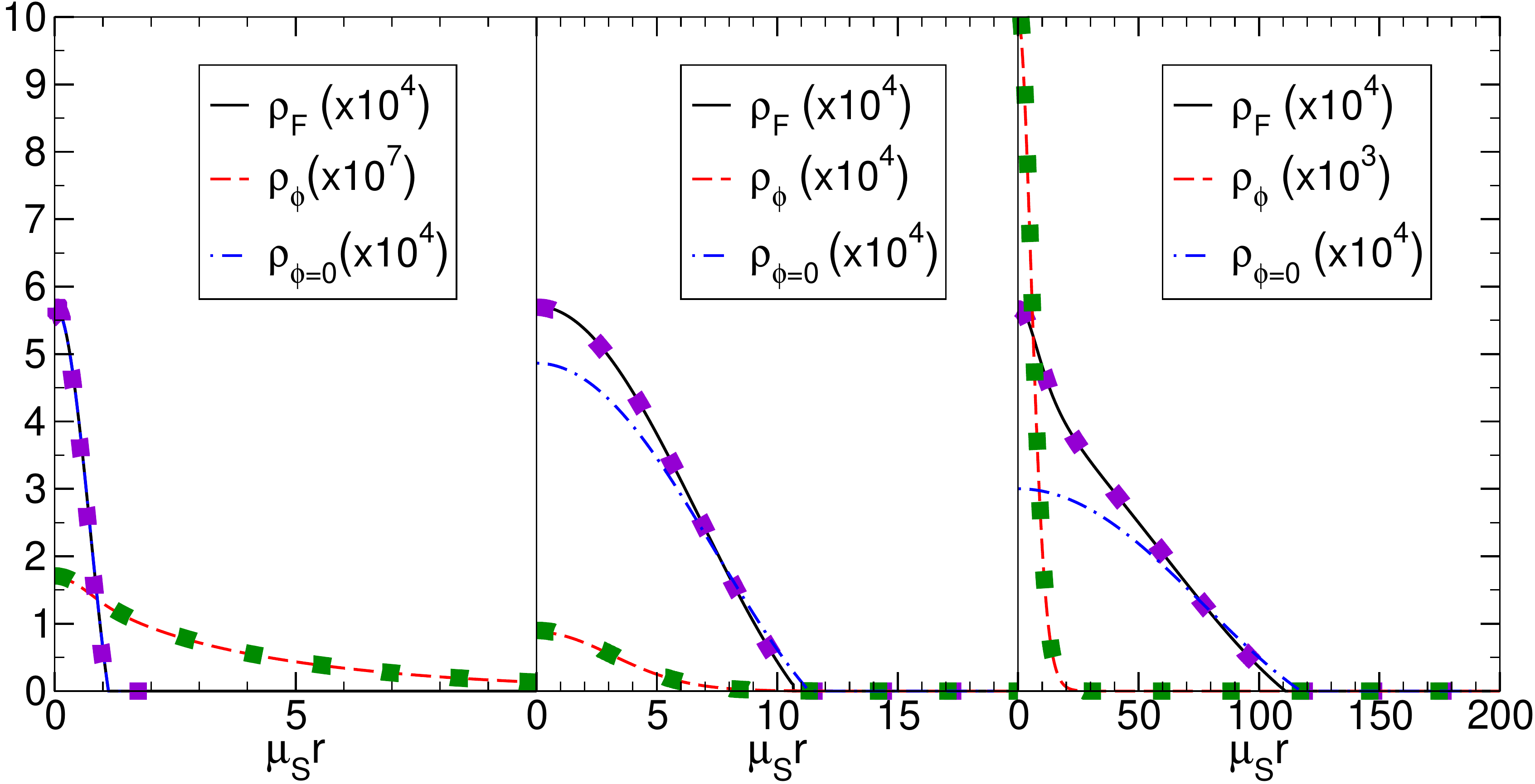,width=10cm,angle=0,clip=true}
\end{tabular}
\caption{Same as Fig.~\ref{scalarfluid_osci} but for $V=0$ and $P=K\rho_F^{\gamma}$.
We fix $\rho_{F\,0}(0)=0.00057092$, and have from left to right, $\mu_S M_0=0.1,\,M_B/M_T\approx 21\%$, corresponding to $\phi_1(0)=0.026$ and $\omega M_0\approx 0.0993$,
%Middle: 
$\mu_S M_0=1,\,M_B/M_T\approx 5\%$, corresponding to $\phi_1(0)=0.064$ and $\omega M_0\approx 0.873$,
%Right: 
and $\mu_S M_0=10,\,M_B/M_T\approx 5\%$, corresponding to $\phi_1(0)=0.06982$ and $\omega M_0\approx 8.629$.
Once more, squares denote the corresponding quantities for complex fields (i.e. mixed boson-fluid stars for the same $M_F$ and $M_B$).
\label{scalarfluid_osci_2}}
\end{center}
\end{figure}
\begin{figure*}[htb]
\begin{center}
\begin{tabular}{ccc}
\epsfig{file=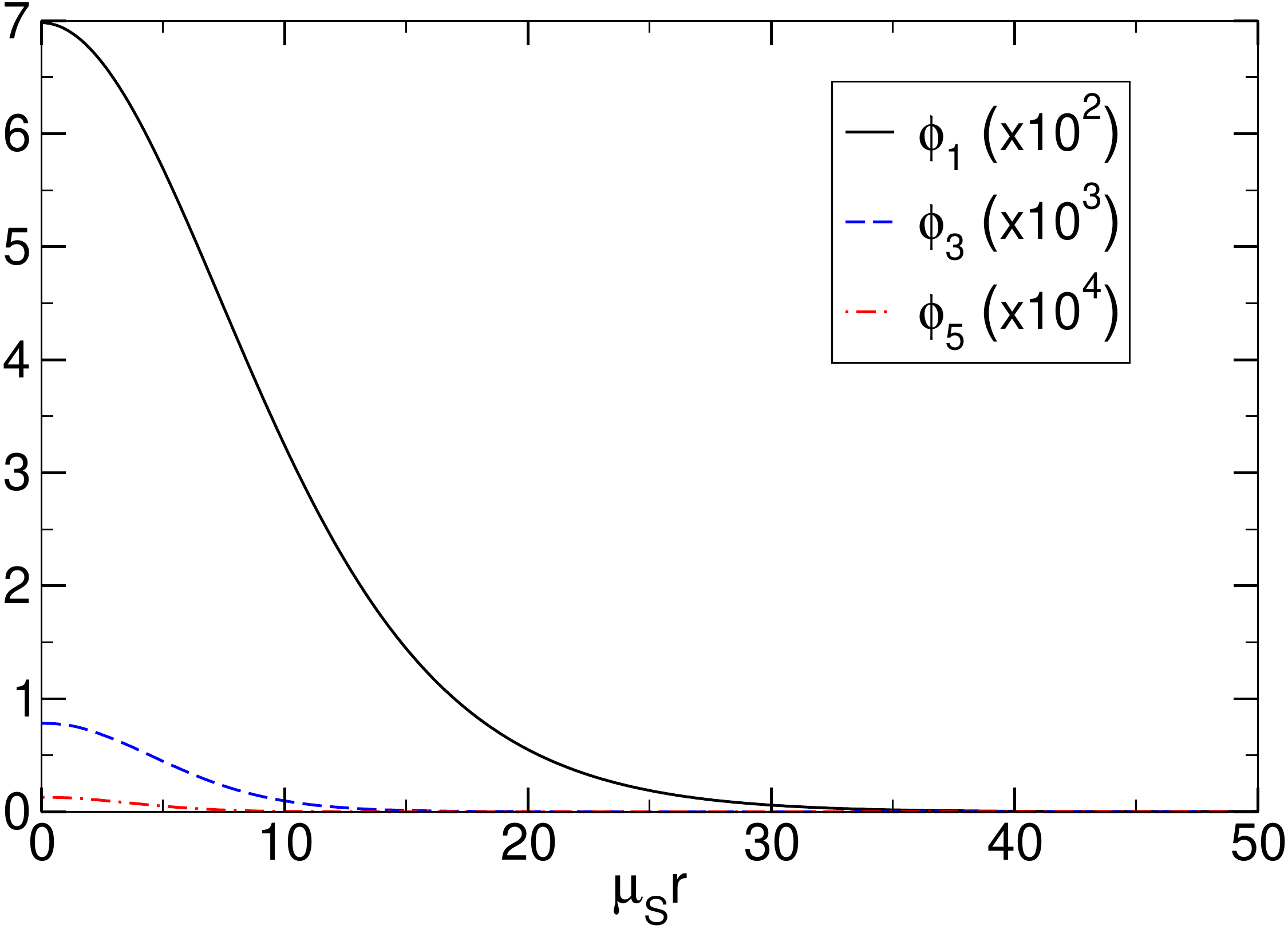,width=5cm,angle=0,clip=true}
\epsfig{file=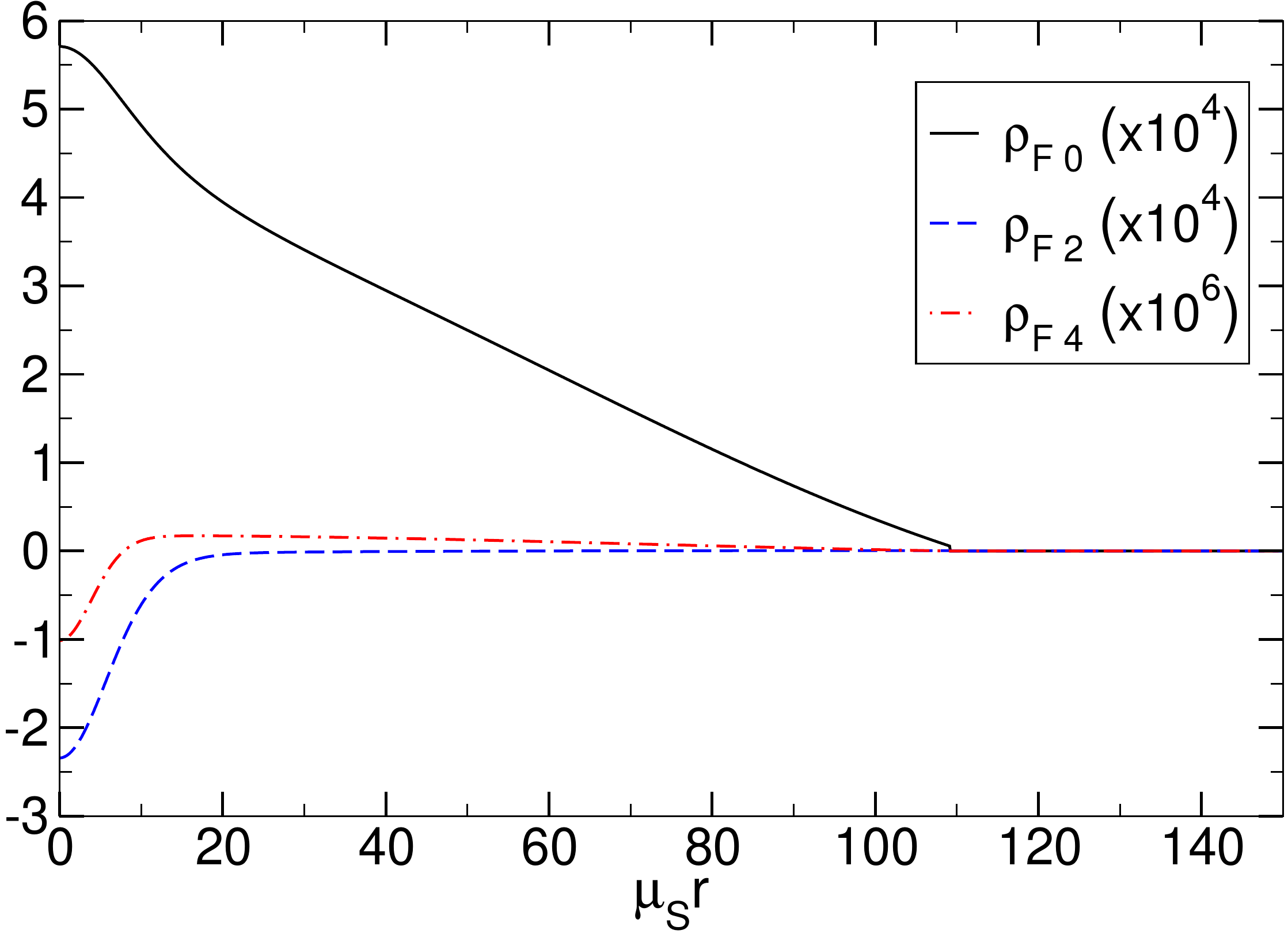,width=5cm,angle=0,clip=true}
\epsfig{file=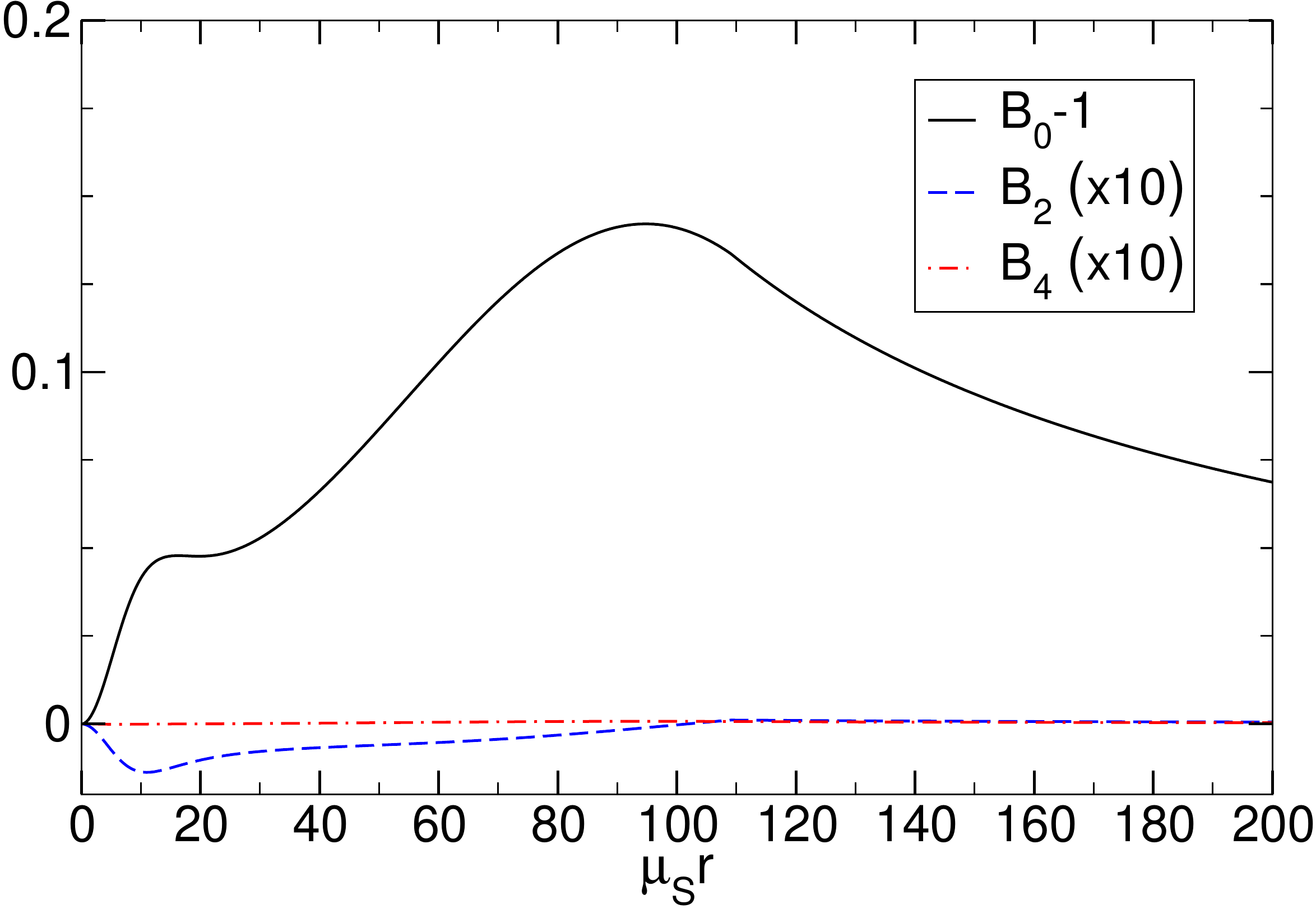,width=5cm,angle=0,clip=true}
\end{tabular}
\caption{Scalar field (left) configuration, density profile of the fluid (center) and corresponding metric component $g_{rr}$ (right) for $\mu_S M_0=10$ and $M_B/M_T\approx 5\%$, corresponding to $\phi_1(0)=0.06982$, $\rho_{F\,0}(0)=0.00057092$ and $\omega M_0\approx 8.629$. Here we take $V=0$ and consider $P=K\rho_F^{\Gamma}$. We plot the first Fourier components for $j_{\rm{max}}=2$.
\label{scalarfluid_solution}}
\end{center}
\end{figure*}
\begin{figure}[htb]
\begin{center}
\begin{tabular}{cc}
\epsfig{file=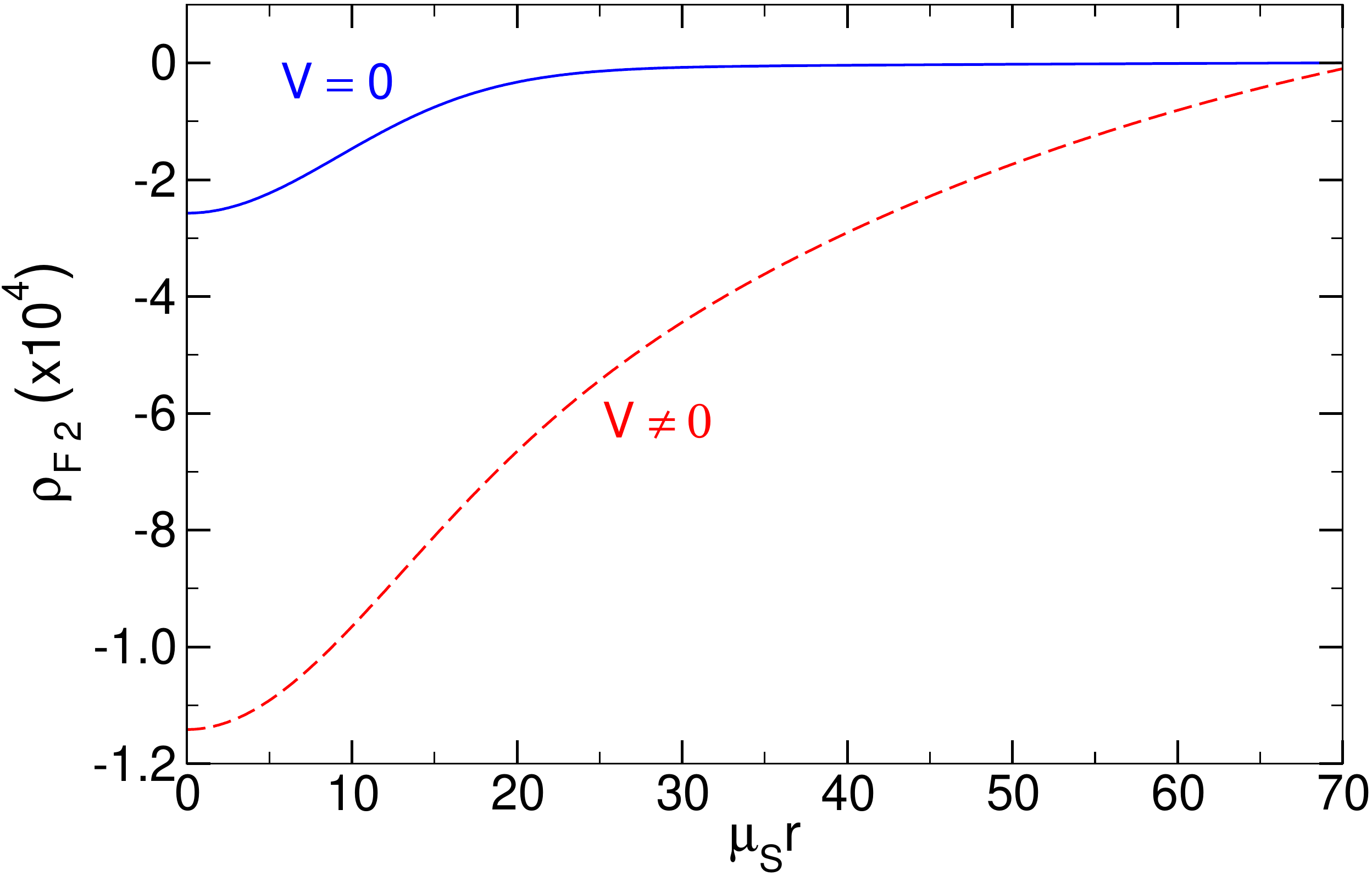,width=8cm,angle=0,clip=true}
\end{tabular}
\caption{For the sake of comparison, we show here the amplitude of the oscillations ($\rho_{F2}$ is the dominant time-dependent component of the fluid density, cf. equation (\ref{series_fluid})) for $\mu_S M_0=10$ and $M_B/M_T\approx 0.54\%$, corresponding to the solution of Fig.~\ref{scalarfluid_solution_v2}, computed using $V=0$ with the equation of state $P=K\rho_F^{\gamma}$ and $V\neq 0$ with the equation of state $P=K \left(m_Nn_F\right)^{\gamma}$, $\rho_F(P)=\left(P/K\right)^{1/\gamma}+P/\left(\gamma-1\right)$.\label{scalar_fluid_comparison}}
\end{center}
\end{figure}
The previous results can be compared and contrasted with the extreme case where DM and baryonic matter can convert into one another.
Under this assumption, we find that there are solutions, summarized in Figs.~\ref{scalarfluid_osci_2}-\ref{scalar_fluid_comparison}, that allow a star to have zero fluid velocity. We find that despite this, the overall qualitative behavior is the same as those of baryon-conserving stars.

The density distribution of baryon non-conserving stars is shown in Fig.~\ref{scalarfluid_osci_2} for different dimensionless mass couplings. 
Again, the mass coupling changes drastically the global behavior of these stars; large mass couplings result in a small bosonic DM core
which is oblivious of the baryons surrounding it.

The structure of a star is shown in Fig.~\ref{scalarfluid_solution}. Because the equations are technically less challenging to handle in this case, we can accurately compute their $j=2$ Fourier components and consider larger values of $M_B/M_T$.
As we said, the qualitative behavior is similar, and in particular these composite stars also oscillate in density, driven by the density-varying oscillaton sitting at their center.
Since the effect of the radial velocity is negligible for small $M_B/M_T$, taking $V\sim 0$ describes with very good accuracy the main properties of these stars. For large $\mu M_0$, the only noticeable effect of $V$ on the solution is to slightly increase the amplitude of the oscillations in comparison to the $V=0$ case. A comparison for the solution of Fig.~\ref{scalarfluid_solution_v2}, where this difference is noticeable, is shown in Fig.~\ref{scalar_fluid_comparison}. In general, the smaller $\mu M_0$ and $M_B/M_T$ the better the agreement between the two cases.

We stress that the overall behavior might have been anticipated from an analysis of Fig.~\ref{MvsR}: for light fields, $\mu_S M_0<1$ the scalar profile is extended and the pure oscillaton solution is broad and light. As such, the scalar has a negligible influence on the fluid distribution (as can be seen from the fact that the zero-scalar line $\rho_{\phi=0}$ overlaps with the fluid line in Figs.~\ref{scalarfluid_osci} and~\ref{scalarfluid_osci_2}), and these stars simply have an extended scalar condensate protruding away from them.
In fact, our results are compatible with a decoupling between the boson and fluid for large $\mu_S M_B$. For this case, Fig.~\ref{MvsR} alone is enough to interpret the bosonic distribution.
For example, for $\mu_S M_0=10, M_B/M_T=5\%$, we get $\mu_S M_B\sim 0.3$, which would imply from Fig.~\ref{MvsR} that $\mu_S R\sim 20$ for the scalar field distribution. This is indeed apparent from Fig.~\ref{scalarfluid_osci_2}.
Similar conclusions were reached when studying mixed fermion fluid/boson stars with complex fields~\cite{Lopes:1992np}. In fact, the structure of mixed oscillatons and fluid stars is almost identical to that of boson stars and fluids, as can be seen from Figs.~\ref{scalarfluid_osci} and~\ref{scalarfluid_osci_2}, where we overplot with dotted lines the complex field case.

Overall, our results are consistent with what was previously found for boson-fermion fluid stars~\cite{Henriques:1989ar,Henriques:1989ez}. We expect that field configurations with high $M_B/M_T$ for large $\mu_S M_0$ should follow the same kind of behavior as that found in boson-fermion stars. In particular we expect that bosonic dominated stars should also be possible when increasing $M_B/M_T$~\cite{Henriques:1989ar,Henriques:1989ez}.

Finally, as we discuss below, a careful stability analysis shows that, for sufficiently small $\phi_1(0)$ and for stars which are stable in the absence of scalars,
composite stars are dynamically stable. On the other hand, our results show that these configurations can be understood well from the mass-radius relation of oscillatons.
The maximum mass supported is (\ref{max_mass}), $M_{\rm{max}}/M_{\odot}=8\times 10^{-11}\,\rm{eV}/(m_{B}c^2)$, which for a neutron star and an axion field of mass $10^{-5}$ eV falls well within the stability regime~\cite{Gleiser:1988}.

%%%%%%%%%%%%%%%%%%%%%%%%%%%%%%%%%%%%%%%%%%%%%%%%%%%%%%%%%%%%%%%%%%%%%%%%%%%%%%
\subsection{Stability of fluid-boson stars}
%%%%%%%%%%%%%%%%%%%%%%%%%%%%%%%%%%%%%%%%%%%%%%%%%%%%%%%%%%%%%%%%%%%%%%%%%%%%%%

As was mentioned before, the stability properties of fluid-boson stars are not expected to depend on the details of the scalar field description. Therefore, we
assume that the results for the well-studied case of a fermionic star with a complex scalar field, holds in more generic cases (in particular, when the scalar field is real).
Since fermion-boson star solutions depend on two parameters (i.e., for instance $\{n_{F}(0), \phi_1(0)\}$), the stability theorems for single parameter solutions can not be directly applied. This implies that the change in stability of these solutions cannot easily be inferred from the extremes of a mass versus radius diagram, and requires a more careful analysis. Nevertheless, one can argue that a {\it necessary} condition for stability is that the binding energy $M_T-m_N N_F-m_B N_B$ be negative~\cite{Henriques:1989ez}, where $N_B$ is the number of bosons
(associated to the conservation of the Noether charge) and $N_F$ is the number of fermions
(associated to the conservation of the baryonic number) defined, respectively, by
\beq
N_{F}=\int_0^{\infty} 4\pi \sqrt{B}\,n_{F} r^2 dr\,,\\
N_{B}=\int_0^{\infty} 4\pi \sqrt{C}\,\omega|\phi|^2 r^2 dr\,.
\eeq

For the solutions shown in Fig.~\ref{osci_linear}, the binding energy defined in terms of the masses $M_T-M_F-M_B$ is always negative. Although negative values do not necessarily imply stability, they do give strong support to the claim that these configurations are stable. A more careful stability analysis shows that for sufficiently small $\phi_1(0)$ and for stars which are stable in the absence of scalars, the negativity of the binding energy is a good criterion for stability~\cite{Henriques:1990xg}.

More strict stability criteria to find the critical point --which separates the stable from the unstable configurations-- can be obtained either through a rather involved dynamical analysis or using alternative approaches proposed in Refs.~\cite{Henriques:1990xg,ValdezAlvarado:2012xc} (in particular, see e.g. Fig. 1 of Ref.~\cite{Henriques:1990xg} and Fig. 2 of Ref.~\cite{ValdezAlvarado:2012xc}). All these methods showed that there is a wide region in the parameter space for which these solutions are stable. These results were validated by numerical evolutions, showing that unstable stars, depending on the initial perturbation, either migrate to a stable star or collapse to a BH.

These studies were extended to fermion-boson stars with self-gravitating scalar fields by allowing for solutions with comparable number of bosonic and fermionic particles~\cite{Brito:2015yfh}, and also for this case the picture remains the same: under small perturbations, the stable configurations just oscillate with the quasi-normal modes, while unstable configurations either migrate to a stable star or collapse to a BH.
This confirms that there is nothing special regarding fermion-boson stars with self-gravitating scalar fields, in contrast to some claims in the literature~\cite{Goldman:1989nd,Kouvaris:2011fi,Bramante:2014zca,Bramante:2015cua,Kurita:2015vga}; these configurations can be either stable or unstable, depending on the parameters of the system.

%%%%%%%%%%%%%%%%%%%%%%%%%%%%%%%%%%%%%%%%%%%%%%%%%%%%%%%%%%%%%%%%%%%%%%%%%%%%%%%%%%%%%
\section{Stars with scalar cores in scalar-tensor theories}\label{sec:ST}
%%%%%%%%%%%%%%%%%%%%%%%%%%%%%%%%%%%%%%%%%%%%%%%%%%%%%%%%%%%%%%%%%%%%%%%%%%%%%%%%%%%%%
Scalar fields are a fundamental component of scalar-tensor theories of gravity, one of the most natural and extensively studied extensions of GR~\cite{Fujii:2003pa}. These theories are normally characterized by a non-minimal coupling between a scalar field and gravity. Very compact stars with vanishing scalar in scalar-tensor theories can be unstable towards the growth of a scalar field, a phenomena called {\it spontaneous scalarization}~\cite{Damour:1993hw,Pani:2010vc,Palenzuela:2013hsa,Berti:2015itd}. The final state is a static star with a non-vanishing but static scalar profile.

A natural consequence of our work is that in these theories stars can also have scalar cores if the scalar field is massive and time-dependent. Given that massive scalar-tensor theories are poorly constrained, this opens the way to improve current bounds on these theories from binary pulsar experiments. We will leave this for future work and focus instead on massless scalar-tensor theories. We argue that even for a massless scalar field, some theories might admit stars with long-lived scalar cores.

Let us focus on the simplest possible case, that of a complex scalar-tensor theory. This is conceptually easier to handle because it allows for the existence of spherically symmetric solutions with a static metric. This theory is formally equivalent to a tensor-multi-scalar theory with two real scalar fields~\cite{Damour:1992we,Horbatsch:2015bua}. 
Our results also apply to single scalar-tensor theories with nonminimally coupled real scalar fields, the difference being that in these theories the geometry must also oscillate. 

In the physical (Jordan) frame the scalar is non-minimally coupled to the Ricci scalar~\cite{Fujii:2003pa}. By performing a conformal transformation, one can write the theory in the Einstein frame as~\cite{Damour:1992we,Horbatsch:2015bua}
\beq\label{ST_action}
S=\int d^4 x\sqrt{-g}\left[\frac{R}{16\pi}-g^{\mu\nu}\partial_{\mu}\bar{\phi}\partial_{\nu}\phi
%-V(\phi,\bar{\phi})
\right]
+S_m\left[A^2\left(\phi,\bar{\phi}\right)g_{\mu\nu};\Phi\right]\,,
\eeq
where $\bar{\phi}$ denotes the complex conjugate of $\phi$, $A^2\left(\phi,\bar{\phi}\right)$ is a generic function of the scalar field, and $S_m$ denotes the matter action. The matter fields, denoted collectively by $\Phi$, are minimally coupled to the Jordan frame metric $\tilde{g}_{\mu\nu}=A^2\left(\phi,\bar{\phi}\right)g_{\mu\nu}$, where the tilde denote quantities computed in the Jordan frame. This guarantees that the weak equivalence principle holds. 
By varying the action~\eqref{ST_action}, one obtains the following scalar field equation (apart from the Einstein-Klein-Gordon equations minimally coupled to the matter fields):
\be
%R_{\mu\nu}&=&4\nabla_{(\mu}\bar{\phi}\nabla_{\nu)}\phi+2V(\phi,\bar{\phi})g_{\mu\nu}+8\pi G_{\star}\left(T_{\mu\nu}-\frac{1}{2}T g_{\mu\nu}\right),\\
\Box \phi=
%\frac{\partial V}{\partial \bar{\phi}}
-2\frac{\partial \log A}{\partial\bar{\phi}} T\,,\label{ST_scalar}
\ee
where $T$ denotes the trace of the matter fields' stress-energy tensor. The physical stress-energy tensor (written in the Jordan frame) is related to the Einstein-frame stress energy tensor by
\be
T_{\nu}^{\mu}=A^4\tilde{T}_{\nu}^{\mu},\quad T_{\mu\nu}=A^2\tilde{T}_{\mu\nu}, \quad T=A^4\tilde{T}\,,
\ee
where $\tilde{T}^{\mu\nu}$ is the physical stress-tensor in the Jordan frame, given by Eq.~\eqref{stress_energy_PF} (for details see e.g. Ref.~\cite{Pani:2014jra}).

We will assume that at spatial infinity the scalar field vanishes and that the function $A$ can be expanded as
\be
A\approx 1+\alpha\phi+\bar{\alpha}\bar{\phi}+\frac{1}{2}\beta\phi\bar{\phi}+\frac{1}{4}\beta_1\phi^2+\frac{1}{4}\bar{\beta}_1\bar{\phi}^2+\ldots\,,
\ee
where $\beta$ is a real constant, while $\alpha$ and $\beta_1$ are complex numbers. Without loss of generality we set $\alpha=\beta_1=0$. 
Applying this expansion to Eq.~\eqref{ST_scalar} one immediately sees that the field acquires an effective position-dependent mass term given by $\mu_{\rm{eff}}^2=-\beta T$~\cite{Cardoso:2013fwa,Cardoso:2011xi}.
By taking the ansatz
\be
\phi=\frac{1}{\sqrt{16\pi}}\phi(r)e^{-i \omega t}\,,
\ee
and expanding the equations of motion around $\phi_0=0$, we find
\beq
&&B'/B=(r/4)\left[C \omega^2\phi^2+(\phi')^2+B\left(4\beta\rho_{F}\phi^2+32\pi\rho_{F}\right)\right]
+(1-B)/r\,,\label{ST_eq1}\\
&&C'/C=2/r+(Br)/2\left[2\beta\phi^2(\rho_{F}-P)+16\pi\rho_{F}-16\pi P\right]
-2B/r\,,\\
&&\phi''=\beta B (\rho_{F}-3P)\phi-C\omega^2\phi-2\phi'/r+C'\phi'/(2C)\,,\\
&&2P'=-\left(P+\rho_{F}\right)\left(CB'-BC'\right)/(BC)
-\beta\phi\phi'\left(P+\rho_{F}\right)/(16\pi)\,.\label{ST_eq2}
\eeq
Here, we consider the matter fields to be described by a perfect fluid. Note that in all the equations we are only considering terms up to order $\phi^2$. The method to find compact stars is the same as described in the previous Sections, so we will not dwell on it further. For the perfect fluid we will consider the polytropic equation of state $P=K\rho_F^{\gamma}$, with the parameters used in the previous Sections.

\begin{figure}[htb]
\begin{center}
\begin{tabular}{cc}
\epsfig{file=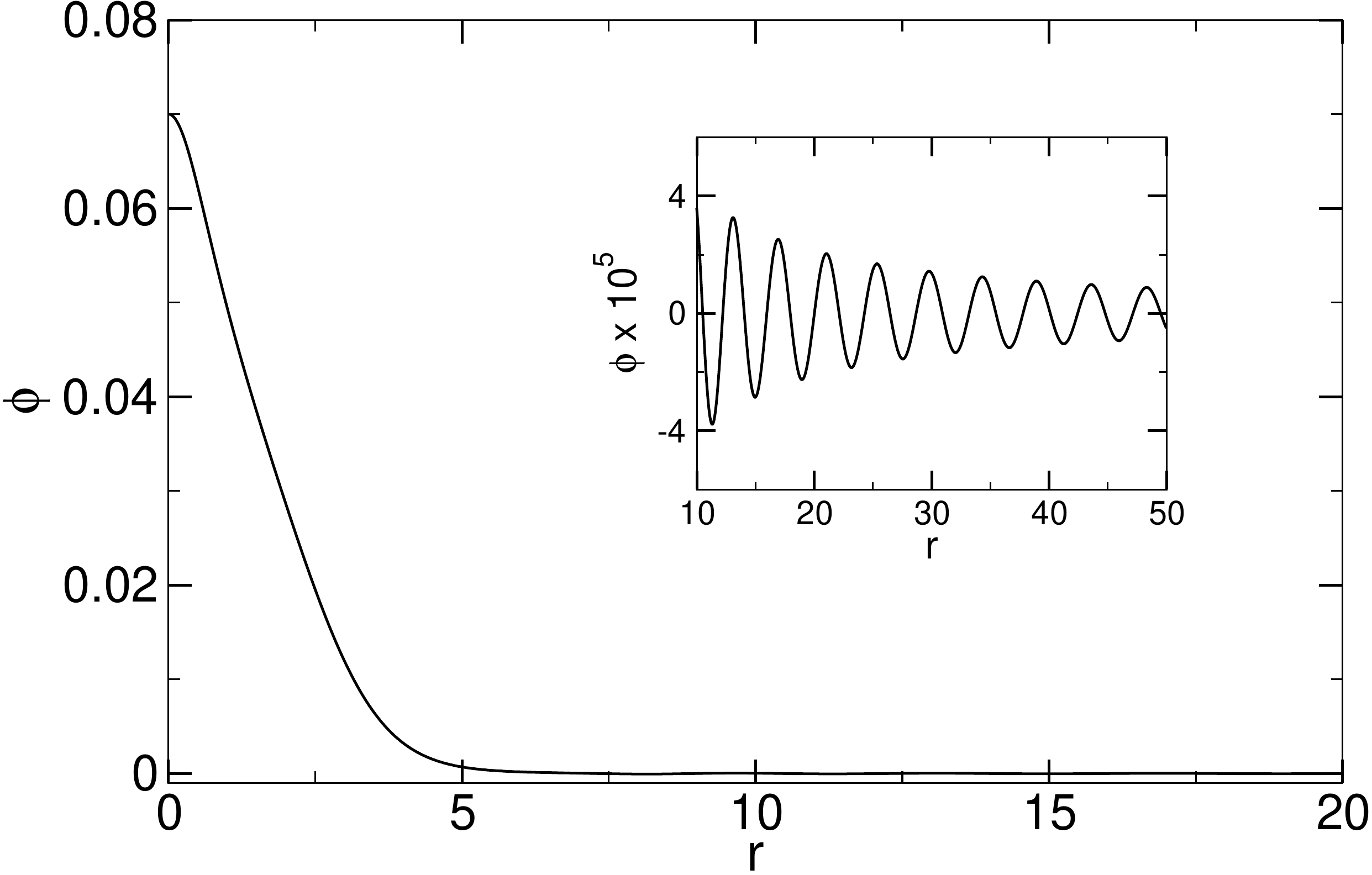,width=7.4cm,angle=0,clip=true}&
\epsfig{file=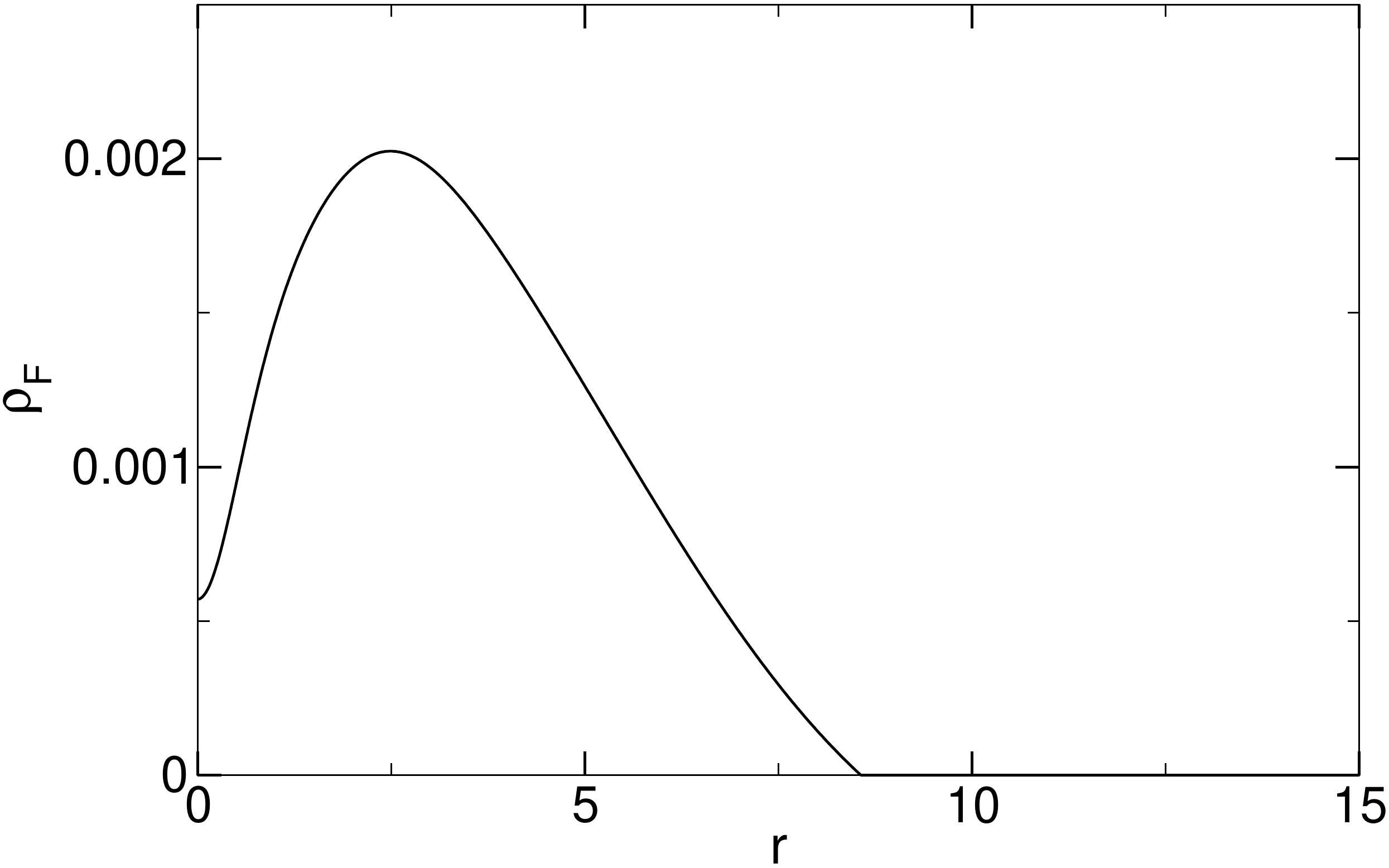,width=7.5cm,angle=0,clip=true}
\end{tabular}
\caption{Scalar field (top) and fluid's density (bottom) profiles for $\beta=7000$ and $\phi(0)=0.07$, by expanding the system~\eqref{ST_eq1}--~\eqref{ST_eq2} up to order $\phi^2$. For this solution we get $\omega=1.22$.
% and $M_T=1.89755$.
The inset of the left panel shows a zoom of the scalar field at large distances.\label{scalartensor_solution}}
\end{center}
\end{figure}
\begin{figure}[htb]
\begin{center}
\begin{tabular}{cc}
\epsfig{file=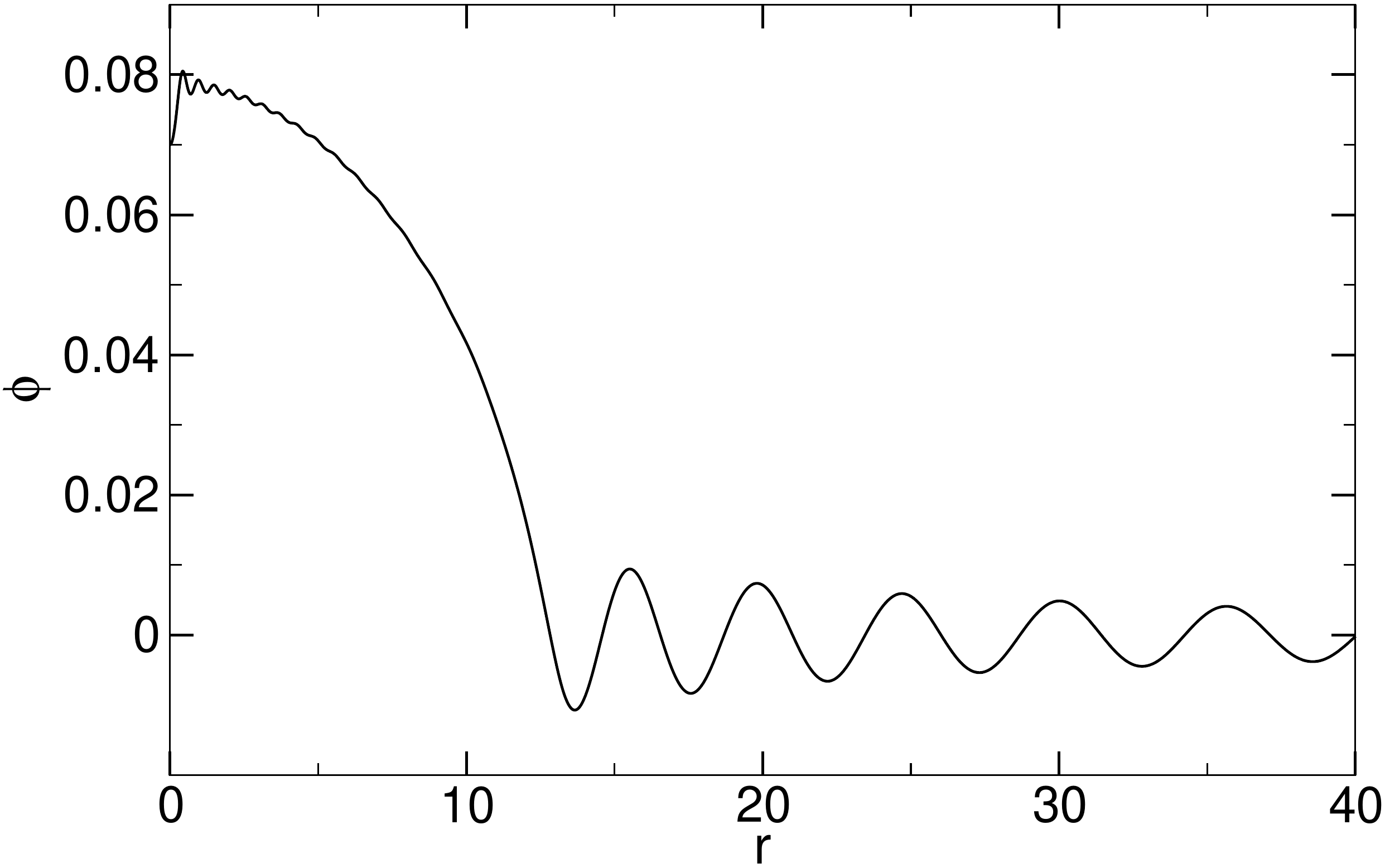,width=7.4cm,angle=0,clip=true}&
\epsfig{file=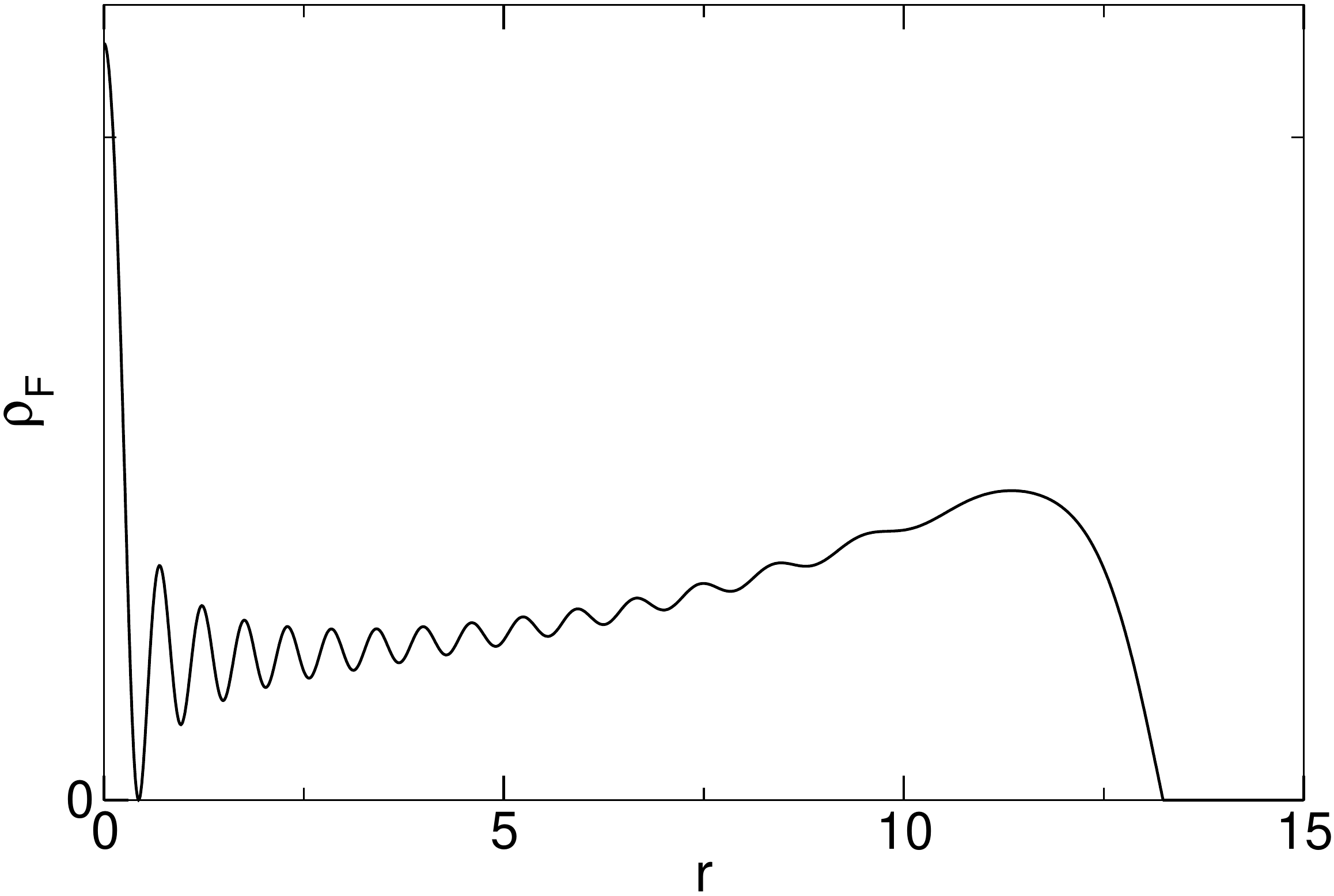,width=7.4cm,angle=0,clip=true}
\end{tabular}
\caption{Scalar field (top) and fluid's density (bottom) profiles for $A=e^{\beta|\phi|^2/2}$, $\beta=7000$ and $\phi(0)=0.07$. For some specific frequencies the scalar field acquires a non-trivial profile inside the star. However the amplitude of the radiating tail is non-negligible. 
\label{ST_total}}
\end{center}
\end{figure}
A solution is shown in Fig.~\ref{scalartensor_solution}. We have not been able to find solutions for which the scalar decays exponentially at infinity~\footnote{and thus truly stationary solutions. In other words, a time-varying scalar that decays as $1/r$ at large distances leads to a non-zero flux of energy at infinity}. However, we find that for some specific frequencies $\omega$ the scalar field is exponentially suppressed inside the star. Outside the star these solutions display an oscillating tail, indicating that they are not truly stable solutions but are instead long-lived solutions, slowly decaying through the emission of scalar radiation. This is very similar to what happens for oscillatons~\cite{Fodor:2009kg,Grandclement:2011wz}. We have only been able to find such solutions in the range $\beta\gg 1$. Negative values of $\beta$ are highly constrained by binary pulsar experiments~\cite{Damour:1996ke}, however positive values of $\beta$ remain unconstrained~\footnote{However see Refs.~\cite{Mendes:2014ufa,Palenzuela:2015ima} for a recent proposal to constrain positive values of $\beta$.}.

Although a careful analysis is out of the scope of this work, our results make it possible that some massless scalar-tensor theories allow for the existence of stars with long-lived scalar cores. We would like to emphasize that our results are formally only valid up to order $\phi^2$. In the regime where such solutions exist, higher-order terms are in general important and should be taken into consideration. For the specific cases we tried, in particular $A=e^{\beta|\phi|^2/2}$, higher-order terms change drastically the solution as shown in Fig.~\ref{ST_total}. Although, for some specific frequencies, some solutions display a non-trivial profile inside the star, the amplitude of the radiative tail is non-negligible (and thus these solutions will dissipate over smaller time-scales). However full dynamical studies are needed to accurately compute the time-scale over which these configurations disperse.

%%%%%%%%%%%%%%%%%%%%%%%%%%%%%%%%%%%%%%%%%%%%%%%%%%%%%%%%%%%%%%%%%%%%%%%%%%%%%%%
\section{Accretion and growth of dark matter cores}\label{sec:collision}
%%%%%%%%%%%%%%%%%%%%%%%%%%%%%%%%%%%%%%%%%%%%%%%%%%%%%%%%%%%%%%%%%%%%%%%%%%%%%%%

We have shown that pulsating stars with DM cores exist as solutions of the field equations, even when the scalar DM core is self-gravitating. Although more studies are required, we also argued that these equilibrium solutions are stable. Do they form dynamically?
Pulsating purely bosonic states certainly do, through collapse of generic initial data~\cite{Seidel:1991zh,Garfinkle:2003jf,Okawa:2013jba}. There are two different channels
for formation of composite fluid/boson stars. One is through gravitational collapse in a bosonic environment, through which the star is born already with a DM core. The second process consists of capture and accretion of DM into the core of compact stars. A careful analysis for WIMPS has been done some time ago, showing that a significant amount of DM can be captured during the star's lifetime~\cite{Press:1985ug,Gould:1989gw}. The capture rate calculation for bosonic condensates follows through, {\it if} the condensate is small enough that it can be considered pointlike (we recall that bosonic condensates have a size determined by its total mass; very light condensates are spatially broad). If the condensate is ultralight and macroscopically-sized, interactions with the star are likely to be enhanced.

%%%%%%%%%%%%%%%%%%%%%%%%%%%%%%%%%%%%%%%%%%%%%%%%%%%%%%%%%%%%%%%%%%%%%%%%%%%%%%%
\subsection{Growth of DM cores}
%%%%%%%%%%%%%%%%%%%%%%%%%%%%%%%%%%%%%%%%%%%%%%%%%%%%%%%%%%%%%%%%%%%%%%%%%%%%%%%

Once the scalar is captured it will interact with the boson core. Interactions between complex fields have shown that
equal mass collisions at low energies form a bound configuration~\cite{Bernal:2006ci,Palenzuela:2006wp}. In other words, two bosonic cores composed of complex fields interact and form a more massive core at the center. This new bound configuration is in general asymmetric and will decay on large timescales~\cite{Macedo:2013jja}, the final state being spherical~\cite{Bernal:2006it}. 

\begin{figure}[htb]
\begin{center}
 \epsfig{file=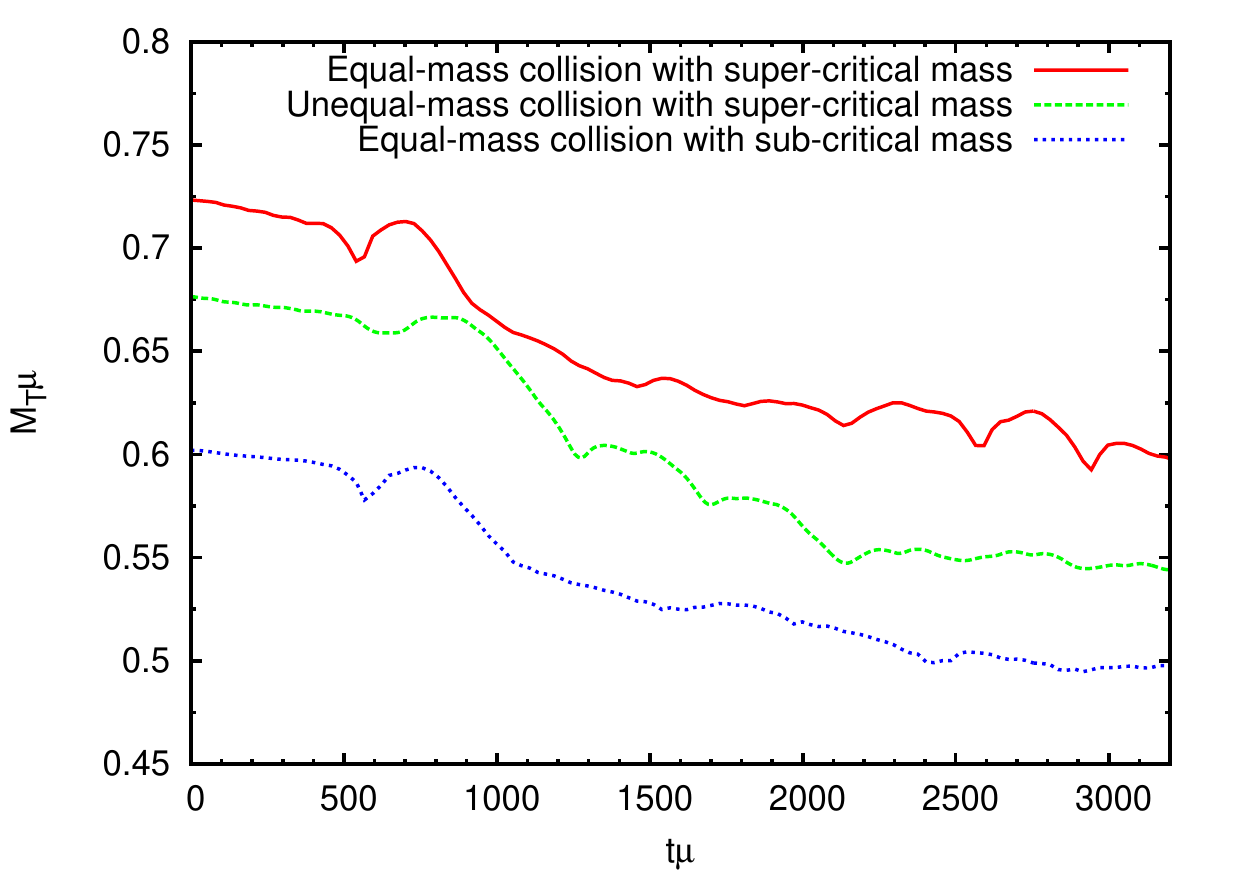,width=10.cm,angle=0,clip=true}
 \caption{Total mass of final object for the collision of scalar
 condensates, with total mass below and above the critical threshold
 $M_{\rm{max}}$. No BH formation is observed.}
 \label{fig:collision_totalmass}
\end{center}
\end{figure}

The analysis of Section~\ref{BF_results} makes it clear that for large and small boson masses, the boson and fluid behave as decoupled entities. As such, accretion by the DM core is well approximated by considering the collision between oscillatons. 
Using Numerical Relativity simulations~\footnote{These simulations and Fig.~\ref{fig:collision_totalmass} are courtesy of Dr. Hirotada Okawa. See Ref.~\cite{Brito:2015yfh} for details.}, we considered collisions between two scalar oscillatons for three different cases:

{\bf (i)} Two equal-mass oscillatons colliding at sufficiently small energies. 
Note that the total mass is larger than the peak value and one would naively predict gravitational collapse to a BH.
Instead, the final result is an oscillating object below the critical mass as shown by the red-solid curve in
Fig.~\ref{fig:collision_totalmass}.

{\bf (ii)} Unequal-mass oscillatons for a total mass above the peak value.
Also in this case the final configuration relaxes to a perturbed configuration
which oscillates around a stable configuration on the curve of Fig.~\ref{MvsR}.
The time evolution of the total mass is depicted by the green-dashed curve in Fig~\ref{fig:collision_totalmass},
where one can see that the total mass gradually decreases after the collision.

{\bf (iii)} Two equal-mass oscillatons but for a total mass below the peak value.
Even for this case, the final outcome is also an oscillating object below the critical mass
whose mass as a function of time is shown by the blue-dotted curve in Fig.~\ref{fig:collision_totalmass}.
It is also important to highlight that the final total mass is {\it higher} than those of the individual stars, showing that
cores {\it can} grow.

All these collisions share common features. In particular, one of the main conclusions that one can draw, is that collapse to a BH is generically avoided for the cases where the total mass is larger than the critical stable mass \footnote{Collapse to BHs {\it can} occur for certain special initial conditions, such as high-energy collisions of boson stars or even spherically symmetric collapse~\cite{Choptuik:2009ww,Palenzuela:2007dm,Okawa:2013jba}. These results are however not in contradiction with our statement and findings that, for accretion-related problems collapse to BHs seems to be avoided.}. This is a general feature of what has been termed the ``gravitational cooling mechanism'': a very efficient (dissipationless) mechanism that stops them from growing past the unstable point, through the ejection of mass~\cite{Seidel:1993zk,Alcubierre:2003sx,Guzman:2006yc}. Such features have been observed in the past in other setups, such as spherically symmetric gravitational collapse~\cite{Seidel:1993zk}
(see Fig. 2 in Ref.~\cite{Okawa:2013jba}), slightly perturbed oscillatons~\cite{Alcubierre:2003sx} or fields with a quartic self-interacting term~\cite{Guzman:2006yc}. Gravitational cooling provides a counter-example to an often-used {\it assumption} in the literature, that stars accreting DM will grow past the Chandrasekhar limit for the DM core and will collapse to a BH~\cite{Gould:1989gw,Bertone:2007ae,McDermott:2011jp,Graham:2015apa,Kouvaris:2013kra,Bramante:2014zca,Bramante:2015cua,Kouvaris:2011fi}. These results show that this need not be the case, if the DM core is prevented from growing by a self-regulatory mechanism, such as gravitational cooling. In fact, avoidance of the BH final state has been seen in collisions of super-critical neutron stars as well~\cite{Kellermann:2010rt,Noble:2015anf}, which means that the phenomena is not exclusive of scalar fields.

The previous results concerned exclusively non-interacting fields. Extending these calculations to quartic self-interactions show that the same qualitative features arise also for self-interacting fields, in particular strong gravitational-cooling effects~\cite{Brito:2015yfh}.

Thus, even though other more detailed simulations are still needed, the likely scenario for evolution would comprise a core growth through minor mergers, slowing down close to the mass-radius peak (see Fig.~\ref{MvsR}), at which point it stops absorbing any extra bosons~\cite{Alcubierre:2003sx,Okawa:2013jba}. In other words, the unstable branch is never reached. This phenomenology is specially interesting, as it would also provide a capture mechanism for these fields which is independent of any putative nucleon-axion interaction cross-section: as we discussed, the bosonic core grows (in mass) through accretion until its peak value. At its maximum, it has a size $R_B/M_{\odot}\sim M_B/M_{\odot}$. This is the bosonic core {\it minimum size}, as described by Fig.~\ref{MvsR}. In other words, even for $M_B=0.01M_T$ the boson core has a non-negligible size and is able to capture and trap other low-energy oscillatons.

%%%%%%%%%%%%%%%%%%%%%%%%%%%%%%%%%%%%%%%%%%%%%%%%%%%%%%%%%%%%%%%%%%%%%%%%%%%%%%%%%%%
\section{Conclusions}\label{sec:conclusion_osci}
%%%%%%%%%%%%%%%%%%%%%%%%%%%%%%%%%%%%%%%%%%%%%%%%%%%%%%%%%%%%%%%%%%%%%%%%%%%%%%%%%%%

The main purpose of this Chapter was to understand how DM might affect the structure of compact stars, using a fully relativistic setup. 
Self-gravitating, massive bosonic fields can form compact structures, which can cluster inside stars, leading to oscillating configurations with distinctive imprints. 
In particular, since the fundamental frequency is $\omega\sim \mu_{S,\,V}$ for non-compact stars (cf. Chapter~\ref{sec:fluid}), these oscillations imply that both the bosonic field and the fluid density (which is coupled to it gravitationally), vary periodically with a frequency
\be
f=2.5\times 10^{14}\,\left(\frac{m_{B}c^2}{eV}\right)\,{\rm{Hz}}\,,
\ee
or multiples thereof. For axion-like particles with masses $\sim 10^{-5} \,eV/c^2$, these stars would emit in the microwave band.
These oscillations are driven by the boson core and might have observable consequences; it is in principle even possible that resonances occur when the frequency of the scalar is equal to the oscillation frequency of the unperturbed star. 
The joint oscillation of the fluid and the boson might be called a {\it global thermalization of the star}, and is expected to occur also
for boson-star-like cores (which give rise to static boson cores), once the scalar is allowed to have non-zero couplings with the star material. Such couplings where recently considered
in Ref.\cite{Arvanitaki:2015iga}, who showed that the oscillations could leave imprints in the Earth's breathing modes and possibly be observable with Earth-based detectors. Although further work is needed, our analysis shows that stars could also work as good DM detectors. Additional signatures could also occur in the presence of a DM core. For example, bosonic cores are intrinsically anisotropic. A certain degree of anisotropy in a neutron star might leave important imprints that could potentially be measured~\cite{Silva:2014fca} (some effective field theories going beyond the mean-field approximation also predict that inside neutron stars, nuclear matter may become anisotropic, see e.g~\cite{Adam:2010fg,Nelmes:2012uf,Adam:2014dqa,Adam:2015lpa}. The interplay between these two effects is also an interesting subject for future research). 

We argued that composite structures can be stable, even when the DM core is self-gravitating.
We should stress that previous works on the subject of DM accretion by stars have implicitly assumed that
the DM core is able to grow without bound and eventually collapse to BHs~\cite{Gould:1989gw,Bertone:2007ae,McDermott:2011jp,Graham:2015apa,Kouvaris:2013kra}.
Our arguments, coming from full nonlinear simulations of the field equations~\cite{Brito:2015yfh}, show that the core may stop growing when it reaches a peak value, at the threshold
of stability, if DM is composed of light massive fields. Gravitational cooling quenches the core growth for massive cores and the core growth halts, close to the peak value (c.f. Fig.~\ref{MvsR}). Similar mechanisms have also been shown to be effective in collisions of super-critical neutron stars~\cite{Kellermann:2010rt} (see also~\cite{Noble:2015anf} and references therein), and thus this phenomena is a very generic feature of self-gravitating solutions, and not only of bosonic fields. These studies show that BH formation depends very sensitively on the initial conditions of the system and cannot be solely inferred from the linear stability of the stars.

Our argument focused on core growth through lump accretion, and does not address other forms of growth, in particular more continuous processes like spherical accretion or wind accretion. Partial results in the literature indicate that gravitational cooling mechanisms are also active in these setups~\cite{Hawley:2000dt}. To study BH formation in these different scenarios, a complete scan of the parameter space would be necessary. This could be done along the lines of Refs.~\cite{Hawley:2000dt,Noble:2015anf}.

Future works should consider more realistic equations of state and possibly include viscosity in the star's fluid and local thermalization. Viscous timescales for neutron star oscillations can be shown to be large compared to the star dynamical timescale $R$,
but small when compared to the (inverse of) the accretion rate likely to be found in any realistic configuration~\cite{1990ApJ...363..603C}. As such, we expect that viscosity will damp global oscillations of the star, eventually leading to a depletion of the scalar field core. A similar effect will occur with local thermalization of the scalar with the star material if the central temperature of the star is much larger than the mass of the bosonic field~\cite{Bilic:2000ef,Latifah:2014ima}. On the other hand, although more detailed studies of these effects are still necessary, the results of Refs.~\cite{Bilic:2000ef,Latifah:2014ima} suggest that, for bosonic fields with masses $\gtrsim$ keV inside old neutron stars or white dwarfs, local thermalization should not significantly affect our results.

We also argued that in theories where a scalar field acquires an effective mass due to the presence of matter, long-lived oscillating configurations might form inside stars, in a region of the parameter space which remains unconstrained. Whether these configurations actually form and whether they have peculiar observable imprints, can only be accessed through a fully dynamical analysis. On the other hand, our results apply directly to massive scalar-tensor theories (see e.g. Ref.~\cite{Chen:2015zmx}). Further work is still needed, but our results raise the interesting possibility that DM cores could be used to further constrain these theories.

There are a number of other setups where similar results may hold. For example, minimally coupled, multiple (real) scalars, interacting only gravitationally, were also shown to give rise to similar configurations~\cite{Hawley:2002zn}. In higher dimensions, one may ask if purely gravitational oscillatons exist. Such solutions could arise due to the compactification of extra dimensions, which effectively give rise to massive bosonic fields. A natural extension of our work would be to look for such solutions in theories with massive spin-2 fields~\cite{deRham:2014zqa}.
Another outstanding open problem concerns the construction of rotating oscillatons. Rotating boson stars were obtained for both complex scalar fields~\cite{Yoshida:1997qf} and more recently for complex vector fields~\cite{Brito:2015pxa}. For real fields, the time-dependence of the metric makes the explicit construction of rotating oscillatons an highly intricate task. A possibility would be to use Numerical Relativity methods to construct such solutions. We hope to possibly solve some of these problems and further develop this subject in the near-future. 
%%%%%%%%%%%%%%%%%%%%%%%%%%%%%%%%%%%%%%%%%%%%%%%%%%%%%%%%%%%%%%%%%%%%

%%%%%%%%%%%%%%%%%%%%%%%%%%%%%%%%%%%%%%%%%%%%%%%%%%%%%%%%%%%%%%%%%%%%%%%%%%%%%%
\appendix
%%%%%%%%%%%%%%%%%%%%%%%%%%%%%%%%%%%%%%%%%%%%%%%%%%%%%%%%%%%%%%%%%%%%%%%%%%%%%%%%%
\chapter{Quasinormal modes and quasibound states: Numerical methods}\label{app:modes}
%%%%%%%%%%%%%%%%%%%%%%%%%%%%%%%%%%%%%%%%%%%%%%%%%%%%%%%%%%%%%%%%%%%%%%%%%%%%%%%%%
This appendix details the numerical computation of BH eigenfrequencies for massive perturbations.
To have a well-defined problem we need to define boundary conditions, and these
determine an eigenvalue problem for the frequency $\omega$, which can be solved using 
several different tools~\cite{Berti:2009kk,Pani:2013pma}. 
At the horizon we must impose regular boundary conditions, which correspond to purely ingoing waves
\be
\label{BC_hor}
\Phi_j(r)\sim e^{-i\omega r_*}\,,\qquad j=1,2,\ldots\,,
\ee
as $r_*\to -\infty$, where $\Phi_j(r)$ is any of the radial perturbative functions of the massive field. 
On the other hand, the asymptotic behavior of the solution at infinity is given by
\be
\label{BC_inf}
\Phi_j(r)\sim B_j e^{- k_{\infty} r}r^{-\frac{M(\mu^2-2\omega^2)}{k_{\infty}}}+C_j e^{ k_{\infty} r}r^{\frac{M(\mu^2-2\omega^2)}{k_{\infty}}}\,,
\ee
where $k_{\infty}=\sqrt{\mu^2-\omega^2}$, such that Re$(k_{\infty})>0$. For massive fields we have to consider two kinds of modes: (i) the QNM, which corresponds to purely outgoing waves at infinity, i.e., they are defined by $B_j=0$; (ii) quasibound states, defined by $C_j=0$ and correspond to modes spatially localized within the vicinity of the BH and that decay exponentially at spatial infinity.

%%%%%%%%%%%%%%%%%%%%%%%%%%%%%%%%%%%%%%%%%
\section{Continued-fraction method}
%%%%%%%%%%%%%%%%%%%%%%%%%%%%%%%%%%%%%%%%%
The use of the continued fraction method requires a suitable \emph{ansatz}. For a linear perturbation on a Schwarzschild background, we take it to be
\begin{equation}
\label{ansatz}
\Phi_j(\omega,r)=f(r)^{-2iM\omega}r^{\nu}e^{-qr}\sum_n{a^{(j)}_n}f(r)^n\,,
\end{equation} 
where $\nu$ and $q$ are defined as below Eq.~\eqref{ansatz2}.

%%%%%%%%%%%%%%%%%%%%%%%%%%%%
\subsection{Massive spin-2 field on a Schwarzschild background: Axial dipole}
%%%%%%%%%%%%%%%%%%%%%%%%%%%%
Inserting \eqref{ansatz} into \eqref{oddl1} leads to a three-term recurrence relation of the form 
\beq
\alpha_0 a_1+\beta_0 a_0&=&0\,,\nonumber\\
\alpha_n a_{n+1}+\beta_n a_n+\gamma_n a_{n-1}&=&0\,,\qquad n>0\,,
\eeq
where,
\begin{align}
\alpha_n &= (n+1)(n+1-4i\omega)\,,\\
\beta_n &= -2 \left(n^2+n-1\right)+\frac{\omega ^2 (2 n-4 i \omega +1)}{q}\nn\\
&-3 q (2 n-4 i \omega +1)+4 i (2 n+1) \omega -4 q^2+12 \omega ^2\,,\\
\gamma_n &= q^{-2}\left(nq+q^2-3q-2 i q \omega -\omega ^2\right)\nn\\
&\times\left(nq+q^2+3q-2 i q \omega -\omega ^2\right)\,.
\end{align}
The QNM or quasibound-state frequencies can be obtained solving numerically the continued fraction equation
\be
\beta_0-\frac{\alpha_0 \gamma_1}{\beta_1-\frac{\alpha_1 \gamma_2}{\beta_2-\frac{\alpha_2 \gamma_3}{\beta_3-\ldots}}}=0\,.
\ee
This method has been extensively used and described in detail elsewhere~\cite{Leaver:1985ax,Berti:2009kk,Pani:2012bp}, some routines are freely available \cite{webpage} so we will not discuss it any further. 
%%%%%%%%%%%%%%%%%%%%%%%%%%%%%%%%%%%%%%
\subsection{Massive spin-2 field on a Schwarzschild background: Axial modes with \texorpdfstring{$l\geq 2$}{lg2}}
%%%%%%%%%%%%%%%%%%%%%%%%%%%%%%%%%%%%%%
For $l\geq 2$ the axial modes satisfy a pair of coupled differential equations, Eqs.~\eqref{axial_bi1} and \eqref{axial_bi2}. Inserting \eqref{ansatz} into these equations leads to a three-term matrix-valued recurrence relation,
\begin{eqnarray}
\label{recrelation}
\bm{\alpha}_0 \mathbf{U}_1+\bm{\beta}_0 \mathbf{U}_0&=&0\,,\nonumber\\
\bm{\alpha}_n \mathbf{U}_{n+1}+\bm{\beta}_n \mathbf{U}_n+\bm{\gamma}_n \mathbf{U}_{n-1}&=&0\,,\qquad n>0\,,
\end{eqnarray}
The quantity $\mathbf{U}_n=\left(a_n^{(1)},a_n^{(2)}\right)$ is a two-dimensional vectorial coefficient and $\bm{\alpha}_n$, $\bm{\beta}_n$, $\bm{\gamma}_n$ are $2\times 2$ matrices whose form reads,
\beq
\bm{\alpha}_n &=&
\begin{pmatrix}
\alpha_n & 0 \\
   0     & \alpha_n 
\end{pmatrix}\,, \qquad
\bm{\beta}_n =
\begin{pmatrix}
\beta_n & \lambda-2 \\
   -2   & \beta_n-3 
 \end{pmatrix}\,, \nonumber\\
\bm{\gamma}_n &=&
\begin{pmatrix}
  \gamma_n & 6-3\lambda \\
   0 & \gamma_n+9 
\end{pmatrix}\,, \nonumber
\eeq
with
\begin{align}
\alpha_n &= (n+1)(n+1-4i\omega)\,,\\
\beta_n &= 2-\lambda -2 \left(n^2+n-1\right)+\frac{\omega ^2 (2 n-4 i \omega +1)}{q}\nn\\
&-3 q (2 n-4 i \omega +1)+4 i (2 n+1) \omega -4 q^2+12 \omega ^2\,,\\
\gamma_n &= q^{-2}\left[q^2 \left(n^2-4 i n \omega -6 \omega ^2-9\right)+2 q^3 (n-2 i \omega )\right.\nn\\
&\left.-2 q \omega ^2 (n-2 i \omega )+q^4+\omega ^4\right]\,.
\end{align}
The matrix-valued three-term recurrence relation can be solved using matrix-valued continued fractions~\cite{Rosa:2011my,Pani:2012bp}. The QNM or quasibound frequencies are roots of the equation $\mathbf{M}\mathbf{U}_0=0$, where
\be
\mathbf{M}\equiv \bm{\beta}_0+\bm{\alpha}_0 \mathbf{R}^{\dagger}_0\,,
\ee
with $\mathbf{U}_{n+1}=\mathbf{R}^{\dagger}_n \mathbf{U}_n$ and
\be
\mathbf{R}^{\dagger}_n=-\left(\bm{\beta}_{n+1}+\bm{\alpha}_{n+1}\mathbf{R}^{\dagger}_{n+1}\right)^{-1} \bm{\gamma}_{n+1}\,.
\ee
For nontrivial solutions we then solve numerically
\be
\det |\mathbf{M}|=0\,.
\ee
%
%When solving the equation we expect to find two independent solutions, each one corresponding to a different eigenvector. 
%%%%%%%%%%%%%%%%%%%%%%%%%%%%%%%%%%%%%%%%%%%%%%%%%%%
\section{Direct integration for quasibound states}
%%%%%%%%%%%%%%%%%%%%%%%%%%%%%%%%%%%%%%%%%%%%%%%%%%%
To compute the spectrum of quasibound states a direct integration approach is often possible, since the solutions asymptotically vanish at spatial infinity, and desirable because it converges faster. We start with a series expansion close to the horizon of the form
\be
\label{bc_di}
\Phi_j(\omega,r)=e^{-i\omega r_*}\sum_n{b^{(j)}_n}(r-r_H)^n\,,
\ee
where the coefficients $b^{(j)}_n$ for $n\geq 1$ can be found in terms of $b^{(j)}_0$ by solving the near-horizon equations order by order.
We then integrate outward up to infinity where the condition $C_j=0$ in Eq.~\eqref{BC_inf} is imposed. This allow us to obtain the frequency spectrum using a shooting method. This method can be extended to solve systems of coupled equations~\cite{Rosa:2011my,Pani:2012bp}. Consider a system of $N$ coupled equations. Imposing the ingoing wave boundary condition at the horizon~\eqref{bc_di} we may obtain a family of solutions at infinity characterized by $N$ parameters, corresponding to the $N$-dimensional vector of the coefficients $\bm{b_0}=\{b^{(j)}_0\}$, with $j=1,\ldots,N$. Note that all the solutions of the system of coupled equations must have the form~\eqref{bc_di} near the horizon. We may then compute the bound-state spectrum by choosing a suitable orthogonal basis for the space of initial coefficients $b^{(j)}_0$. To do so we perform $N$ integrations from the horizon to infinity and construct the $N\times N$ matrix 
\be
\label{detS}
\bm{S_m} (\omega)=\lim_{r\to \infty}
 \begin{pmatrix}
  \Phi_{(1)}^{(1)} & \Phi_{(1)}^{(2)} & \ldots & \Phi_{(1)}^{(N)} \\
  \Phi_{(2)}^{(1)} & \Phi_{(2)}^{(2)} & \ldots & \ldots \\
  \ldots  & \ldots  & \ldots & \ldots  \\
  \Phi_{(N)}^{(1)} & \ldots & \ldots & \Phi_{(N)}^{(N)}
 \end{pmatrix}\,,
\ee
where the superscripts denote a particular vector of the chosen basis, for example, $\Phi_j^{(1)}$ corresponds to $\bm{b_0}=\{1,0,\ldots,0\}$, $\Phi_j^{(2)}$ corresponds to $\bm{b_0}=\{0,1,\ldots,0\}$, and $\Phi_j^{(N)}$ corresponds to $\bm{b_0}=\{0,0,\ldots,1\}$. The bound-state frequency $\omega_0=\omega_R+i\omega_I$ will then correspond to the solutions of
\be
\det|\bm{S_m}(\omega_0)|=0\,,
\ee
which in practice corresponds to minimizing $\det\bm{S_m}$ in the complex plane at arbitrarily large distances. 

%%%%%%%%%%%%%%%%%%%%%%%%%%%%%%%%%%%%%%%%%%%%%%%%%%%%%%%%%%%%%%%%%%%%%%
\chapter{Green's function approach to compute waveforms}\label{app:GF}
%%%%%%%%%%%%%%%%%%%%%%%%%%%%%%%%%%%%%%%%%%%%%%%%%%%%%%%%%%%%%%%%%%%%%%

Consider a wave equation with a source given by:
\be
\frac{d^2 \tilde{Z}_g}{dr_{g*}^2}+\left(\omega^2-V_g\right)  \tilde{Z}_g=\left(1-\frac{r_g}{r}\right)S\,,\label{axialZg_2}\\
\ee
The Green's function $G_{l\omega}$ of this equation is defined by
\be
\frac{d^2 G_{l\omega}}{dr_{g*}^2}+\left(\omega^2-V_g\right)  G_{l\omega}=\delta(r_{g*}-r'_{g*})\,.
\ee
To construct the Green's function we choose two independent solutions of the homogeneous equation associated with Eq.~\eqref{axialZg_2}, $\tilde{Z}_g^{H}$ and $\tilde{Z}_g^{\infty}$, which satisfy the following boundary conditions:
\begin{equation} \label{boundinf}
\left\{
 \begin{array}{l}
 \tilde{Z}_g^{\infty}\sim e^{i\omega r_{g*}}\,,\\
\tilde{Z}_g^{H}\sim A_{\rm{out}}e^{i\omega r_{g*}}+A_{\rm{in}}e^{-i\omega r_{g*}}\,,  
\end{array}\right.
 \quad r_{g*}\to +\infty
\end{equation}
\begin{equation}\label{boundhor}
\left\{
 \begin{array}{l}
\tilde{Z}_g^{\infty}\sim B_{\rm{out}}e^{i\omega r_{g*}}+B_{\rm{in}} e^{-i\omega r_{g*}}\,,\\
\tilde{Z}_g^{H}\sim e^{-i\omega r_{g*}}\,,
\end{array}\right.
 \quad r_{g*}\to -\infty\,,
\end{equation}
where $\{A,B\}_{\rm{in},\rm{out}}$ are constants. 
By imposing wave-like ingoing boundary conditions at the horizon and outgoing boundary at infinity (see e.g. the discussion in Sec.~\ref{sec:QNMs} of Chapter~\ref{chapter:nonbi}), the Green's function reads
\be
G_{l\omega}(r'_{g*},r_{g*})=\frac{1}{W}\left\{
 \begin{array}{l}
\tilde{Z}_g^{H}(r_{g*})\tilde{Z}_g^{\infty}(r'_{g*})\,, \quad r_{g*}<r'_{g*}\,,\\
\tilde{Z}_g^{\infty}(r_{g*})\tilde{Z}_g^{H}(r'_{g*})\,, \quad r_{g*}>r'_{g*}\,,
\end{array}\right.
\ee
where $W$ is the Wronskian of these two linearly independent solutions, and it is constant by virtue of the field equation~\eqref{axialZg_2}. Evaluating $W$ at infinity one gets,
\begin{equation} \label{wronskian}
W=\tilde{Z}_g^{H}\frac{d\tilde{Z}_g^{\infty}}{dr_{g*}}-\tilde{Z}_g^{\infty}\frac{d\tilde{Z}_g^{H}}{dr_{g*}}=2i\omega A_{\rm{in}}\,.
\end{equation}

The solution to Eq.~\eqref{axialZg_2} with appropriate boundary conditions is then given by
\beq
&&\tilde{Z}_g(r_{g*})=\int_{-\infty}^{+\infty} dr'_{g*}\,\,G_{l\omega}(r'_{g*},r_{g*})\, S(r'_{g*})=\nn\\
&&=\frac{\tilde{Z}_g^{\infty}(r_{g*})}{W}\int_{-\infty}^{r} dr'_{g*}\,\,\tilde{Z}_g^{H}(r'_{g*})\, S(r'_{g*})+
\frac{\tilde{Z}_g^{H}(r_{g*})}{W}\int_{r}^{+\infty} dr'_{g*}\,\,\tilde{Z}_g^{\infty}(r'_{g*})\, S(r'_{g*})
\,.
\eeq

Evaluating this expression at $r_{g*}\to +\infty$ we find
\beq
&&\tilde{Z}_g(r_{g*}\to\infty)=\frac{\tilde{Z}_g^{\infty}(r_{g*})}{W}\int_{-\infty}^{+\infty} dr'_{g*}\,\,\tilde{Z}_g^{H}(r'_{g*})\, S(r'_{g*})=\nn\\
&&=\frac{e^{i\omega r_{g*}}}{2i\omega A_{\rm{in}}}\int_{r_g}^{+\infty} dr'\,\,\tilde{Z}_g^{H}(r')\, S(r'_{g*})\left(1-\frac{r_g}{r'}\right)^{-1}\,.
\eeq
This integral can be computed numerically by first integrating the homogeneous part of Eq.~\eqref{axialZg_2} with the boundary condition~\eqref{boundhor} to get $\tilde{Z}_g^{H}$ and then compute $A_{\rm{in}}$ by equating the solution obtained numerically to~\eqref{boundinf}. The waveform in the time-domain is then obtained performing the integral:
\be
Z_g(t,r)=\frac{1}{\sqrt{2\pi}}\int_{-\infty}^{+\infty} e^{-i\omega t}\tilde{Z}_g(\omega,r)d\omega\,.
\ee
For more details on the numerical procedure see, e.g., Ref.~\cite{Cardoso:2002ay}.

%%%%%%%%%%%%%%%%%%%%%%%%%%%%%%%%%%%%%%%%%%%%%%%%%%%%%%%%%%%%%%%%%%%%%%%%%%%%
\chapter{Massless fields around Kerr black holes}\label{app:Teu_eqs}
%%%%%%%%%%%%%%%%%%%%%%%%%%%%%%%%%%%%%%%%%%%%%%%%%%%%%%%%%%%%%%%%%%%%%%%%%%%%

%
\begin{table}[htb]
% \scriptsize
\centering \caption{Wavefunction $\psi$ for each value of the spin weight-$s$. The spin coefficient is given by $\rho\equiv-1/(r-ia\cos\theta)$. The quantities $\phi_0$, $\phi_2$, $\Psi_0$ and $\Psi_4$ are Newman-Penrose scalars~\cite{Newman:1961qr} describing electromagnetic and gravitational perturbations, respectively. The quantities $\chi_0$ and $\chi_1$ denote components of the Dirac spinor along dyad legs.} 
\vskip 12pt
\begin{tabular}{|c|| c c c c |}
\hline%
$s$      & 0  & ($1/2$, $-1/2$) & ($1$, $-1$) & ($2$, $-2$)\\
$\psi$  & $\Phi$  & ($\chi_0$,$\rho^{-1}\chi_1$)	&($\phi_0$,$\rho^{-2}\phi_2$) &($\Psi_0$,$\rho^{-4}\Psi_4$)\\
\hline
\end{tabular}
\label{tab:NP_scalars}
\end{table}
The wave equation for linearized fluctuations around the Kerr geometry~\eqref{Kerr} was studied by Teukolsky, Press and collaborators in great detail~\cite{Teukolsky:1972my,Teukolsky:1973ha,Teukolsky:1974yv,Press:1973zz}. 
Following Carter's unexpected result on the separability of the Hamilton-Jacobi equation for the geodesics in a Kerr geometry~\cite{Carter:1968rr}, he also noted that the analogue scalar field equation was separable~\cite{Carter:1968ks}, as was explicitly shown in Ref.~\cite{Brill:1972xj}. In a breakthrough work (see Ref.~\cite{Teukolsky:2014vca} for a first-person historical account), it was shown that linearized perturbations of the Kerr geometry could be described with a single master equation, describing ``probe'' scalar ($s=0$), massless Dirac ($s=\pm 1/2$), electromagnetic ($s=\pm 1$) and gravitational ($s=\pm 2$) fields in a Kerr background~\cite{Teukolsky:1972my}. The master equation reads
\begin{eqnarray}
&&\left[\frac{\left(r^2+a^2\right)^2}{\Delta}-a^2\sin^2\theta\right]\frac{\partial\psi^2}{\partial t^2}+\frac{4Mar}{\Delta}\frac{\partial\psi^2}{\partial t\partial\phi}+
\left[\frac{a^2}{\Delta}-\frac{1}{\sin^2\theta}\right]\frac{\partial\psi^2}{\partial\phi^2}\nn\\
&&-\Delta^{-s}\frac{\partial}{\partial r}\left(\Delta^{s+1} \frac{\partial \psi}{\partial r}\right)-\frac{1}{\sin\theta}\frac{\partial}{\partial \theta}\left(\sin\theta\frac{\partial\psi}{\partial \theta}\right)-2s\left[\frac{a(r-M)}{\Delta}+\frac{i\cos\theta}{\sin^2\theta}\right]\frac{\partial\psi}{\partial\phi}\nn\\
&&-2s\left[\frac{M(r^2-a^2)}{\Delta}-r-ia\cos\theta\right]\frac{\partial\psi}{\partial t}+\left(s^2\cot^2\theta-s\right)\psi=0\,,
\end{eqnarray}
where $s$ is the field's spin weight, and the field quantity $\psi$ is directly related to Newman-Penrose quantities as shown in Table~\ref{tab:NP_scalars}.
By Fourier transforming $\psi(t,r,\theta,\phi)$ and using the ansatz
\be\label{teu_eigen}
\psi=\frac{1}{2\pi}\int d\omega e^{-i\omega t}e^{im\phi}S(\theta)R(r)\,,
\ee
Teukolsky found separated ODE's for the radial and angular part, which read, respectively
\be\label{teu_radial}
\Delta^{-s}\frac{d}{dr}\left(\Delta^{s+1}\frac{dR}{dr}\right)+\left(\frac{K^2-2is(r-M)K}{\Delta}+4is\omega r-\lambda\right)R=0\,,
\ee
and
{\small
\begin{align}
\frac{1}{\sin\theta}\frac{d}{d\theta}&\left(\sin\theta\frac{dS}{d\theta}\right)\nn\\
&+\left(a^2\omega^2\cos^2\theta-\frac{m^2}{\sin^2\theta}-2a\omega s\cos\theta
-\frac{2m s\cos\theta}{\sin^2\theta}-s^2\cot^2\theta+s+A_{slm}\right)S=0\,, \label{spheroidal}
\end{align}}
where $K\equiv (r^2+a^2)\omega-am$ and $\lambda\equiv A_{slm}+a^2\omega^2-2am\omega$. Together with the orthonormality condition
\be\label{sphe_norm}
\int_0^{\pi}|S|^2\sin\theta d\theta=1\,,
\ee
the solutions to the angular equation~\eqref{spheroidal} are known as spin-weighted spheroidal harmonics $e^{im\phi}S\equiv S_{slm}(a\omega,\theta,\phi)$. When $a\omega=0$ they reduce to the spin-weighted spherical harmonics $Y_{slm}(\theta,\phi)$~\cite{Goldberg1967}.
For small $a\omega$ the angular eigenvalues are (cf. Ref.~\cite{Berti:2005gp} for higher-order terms)
\be
A_{slm}=l(l+1)-s(s+1)+\mathcal{O}(a^2\omega^2)\,.
\ee
The computation of the eigenvalues for generic spin can only be done numerically~\cite{Berti:2005gp}. 

Besides these equations, to have complete information about the gravitational and electromagnetic fluctuations, we need to find the relative normalization between $\phi_0$ and $\phi_2$ for electromagnetic fields and between $\Psi_0$ and $\Psi_4$ for gravitational perturbations. This was done in Refs.~\cite{Teukolsky:1974yv,1973ZhETF..65....3S,Staro2} assuming the normalization condition~\eqref{sphe_norm} and using what is now known as the Teukolsky-Starobinsky identities (see also~\cite{Chandra} for details).

Defining the tortoise coordinate $r_*$ as $dr/dr_*=\Delta/(r^2+a^2)$, Eq.~\eqref{teu_radial} has the following asymptotic solutions
\be\label{bc_kerr}
R_{slm}\sim \mathcal{T}\Delta^{-s} e^{-ik_H r_*}+\mathcal{O} e^{ik_H r_*}\,, {\rm{as}}\quad r \to r_+\,,\quad 
R_{slm}\sim \mathcal{I}\frac{e^{-i\omega r}}{r}+\mathcal{R}\frac{e^{i\omega r}}{r^{2s+1}}\,, {\rm{as}} \quad r \to \infty\,,
\ee
where $k_H=\omega-m\Omega_{\rm{H}}$ and $\Omega_{\rm{H}}=a/(2M r_+)$ is the angular velocity of the BH horizon. Regularity at the horizon requires purely ingoing boundary conditions, i.e., $\mathcal{O}=0$ (see Section 3 in Ref.~\cite{Berti:2009kk} for a careful discussion of boundary conditions).

%%%%%%%%%%%%%%%%%%%%%%%%%%%%%%%%%%%%%%%%%%%%%%%%%%%%%%%%%%%%%%%%%%%%%%%%%%%%%%%%%%%%%%
\chapter{Further details on the  magnetized Kerr-Newman black hole background}\label{app:KN}
%%%%%%%%%%%%%%%%%%%%%%%%%%%%%%%%%%%%%%%%%%%%%%%%%%%%%%%%%%%%%%%%%%%%%%%%%%%%%%%%%%%%%%

The full magnetized Kerr--Newman solution can be found in Refs.~\cite{Ernst:1976:KBH,Diaz:1985xt,Aliev:1989wz,Gibbons:2013yq}.
For $q=-2\tilde{a}M^2 B$ and at second order in the spin, the solution reads
\beq\label{KN_Ernst}
ds^2=H\left[-Fdt^2+\Sigma\left(\frac{dr^2}{\Delta}+d\theta^2\right)\right]
+\frac{A\sin^2\theta}{\Sigma H}\left(H_0 d\phi-\varpi dt\right)^2\,,
\eeq
where $F={\Sigma\Delta}/{A}$, $H_0$ is introduced to remove the conical singularity~\cite{Hiscock:1981np} and 
%
% \begin{widetext}
%
\begin{align}
&\Delta=r^2-2Mr+M^2\tilde{a}^2+q^2\,,\\
&\Sigma=r^2+\tilde{a}^2M^2\cos^2\theta\,,\\
&A=r^4+M^2 r \tilde{a}^2 \left[\sin ^2\theta (2 M-r)+2 r\right]\,,\\
&H=1+\frac{1}{2}B^2 r^2 \sin ^2\theta+\frac{1}{16} B^4 r^4 \sin ^4\theta\nn\\
&+\tilde{a}^2\left[\frac{1}{8} B^6 M^4 r^2 \sin ^2 2\theta
+\frac{1}{8} B^4 M^2\left(2 M r \sin ^6\theta\right.\right.\nn\\
&\left.\left.+2
   M \cos ^2\theta \left(M \cos ^4\theta+2 \cos ^2\theta(M-2 r)+9 M+8 r\right)\right.\right. \nn\\
&\left.\left.-8 M r+r^2 \sin ^4\theta\right)+\frac{B^2 M^2}{2 r}
 \sin ^2\theta (r-M(7+\cos 2\theta))\right]\,,\\
&H_0\equiv H(r,\theta=0)=1+3 B^4 M^4 \tilde{a}^2\,,\\
&\varpi=\frac{M^2 \tilde{a}}{64 r^3} \left[-B^4 r^3 (12 \cos 2\theta+\cos 4 \theta) (r-2 M)\right.\nn\\
&\left.+B^2 r^2 \left(256-B^2 r (154 M+51   r)\right)+128\right]\,.
\end{align}
%
% \end{widetext}
This solution reduces to the Ernst metric (cf. Eq.~\eqref{Ernst} in Chapter~\ref{chapter:magnetic}) when $\tilde{a}=0$. To second order in $\tilde{a}$, the event horizon is located at
\be
r_+=2M-\tilde{a}^2\left(\frac{M}{2}+2B^2M^3\right)\,, \label{rp}
\ee
and the vector potential of the magnetic field is given by
\be
A=\Phi_0 dt+\Phi_3\left(H_0 d\phi-\varpi dt\right)\,.
\ee
The explicit form of the functions $\Phi_0$ and $\Phi_3$ is not important here, so we refer the reader to Ref.~\cite{Gibbons:2013yq}.

Interestingly, these solutions incorporate Wald's result for the charge induction~\cite{Wald:1974np} in the small-$B$ limit. This allows us to understand the Wald's charge as being the one needed to have a vanishing total electric charge at infinity. Indeed the total physical charge of the solution is given by~\cite{Aliev:1988wy,Gibbons:2013yq}
\be\label{charge}
Q=q\left(1-\frac{1}{4}q^2B^2\right)+2\tilde{a} M^2 B\,.
\ee
Due to the vacuum polarization and accretion of particles of opposite charge, BHs have a tendency to quickly lose their charge~\cite{Gibbons:1975kk}. In order to be neutral, a BH must then satisfy $q\left(1-\frac{1}{4}q^2B^2\right)=-2\tilde{a} M^2 B$.
Solving for $q$ and expanding in the small-$B$ limit we find
\be
q_{\mathrm neutral}/M=-2\tilde{a}BM+\mathcal{O}\left[\tilde{a}^3(BM)^5\right]\,.
\ee
The result above reduces to Wald's results to first order in $BM$ and also in the small-rotating limit.

Note that $q$ and $\tilde{a}$ do not have a direct physical meaning for the exact geometry of the magnetized BH. The conserved electric charge of the magnetized BH is given by $Q$~\eqref{charge}, while the true conserved angular momentum of the exact magnetized BH solutions can be evaluated from thermodynamic considerations, as it was done in Ref.~\cite{Gibbons:2013dna}. Although this quantity can be quite complicated, expanding in the small-$\tilde{a}$ limit and considering a BH with Wald's charge, one recovers the standard relation for the angular momentum of a Kerr BH,
\be
J=\tilde{a}M^2+\mathcal{O}\left(\tilde{a}^3\right)\,.
\ee

For a BH with charge $q=-2\tilde{a}M^2B+\mathcal{O}\left(\tilde{a}^3\right)$, the horizon's angular velocity $\Omega_{\rm{H}}$ is given by
\be\label{angular_vel}
\Omega_{\rm{H}}=\frac{\tilde{a}}{4M}+2\tilde{a}M B^2\left(1-2B^2M^2\right)+\mathcal{O}\left(\tilde{a}^3\right)\,.
\ee
Note that $\Omega_{\rm{H}}$ is slightly different from the case of a magnetized BH with $q=0$. Indeed, when $q\neq 0$ a charged BH has a gyromagnetic ratio $q/M$~\cite{Carter:1968rr}, so it can acquire an angular momentum when immersed in a uniform magnetic field. The extra term proportional to $B$ in~\eqref{angular_vel} is related to this effect. This can be seen by computing $\Omega_{\rm{H}}$ for a BH with $\tilde{a}=0$,
\beq\label{angular_charged}
\Omega^{(\tilde{a}=0)}_H&=&-\frac{8 qB \left[B^2 \left(q^2-4 M \sqrt{M^2-q^2}-4 M^2\right)+4\right]}{\left(B^4 q^4+24 B^2 q^2+16\right)
   \left(\sqrt{M^2-q^2}+M\right)}\nn\\
	&\sim&-\frac{qB}{M}+2B^3M q+\mathcal{O}\left(q^3M^3\right)\,,
\eeq
where in the last step we linearized in $q$. Taking $q=q_{\mathrm neutral}$ we get the extra term proportional to $B$ in~\eqref{angular_vel}.

For a bosonic wave with frequency $\omega$ and azimuthal number $m$, superradiant scattering is possible whenever $\omega_R<m\Omega_{\rm{H}}$~\cite{Teukolsky:1974yv} or (to second order in rotation):
\be\label{superwald}
\tilde{a}>\frac{4M\omega}{m\left(1+8B^2M^2-16B^4M^4\right)}\,,
\ee
The effect of the charge induced by the magnetic field is to change the superradiant threshold which, for a BH with $q=0$, is given by
\be\label{supernocharge}
\tilde{a}>\frac{4M\omega}{m}\,.
\ee

\newpage
\section{Scalar field on a magnetized Kerr-Newman background}

We show here the coefficients appearing in Eq.~\eqref{KG_KN} of the main text:
%
%\begin{widetext}
\begin{align}
\mathcal{V}_0&=\frac{3 B^{12} M^4 m^2}{128}  (r-2M) r^5+\frac{1}{64} B^{10} M^4 m^2 r^3 (23 r-48M)\nn\\
&-\frac{B^8M^2 m^2}{128}  \left(r^4-968 M^4+136 M^3 r-280 M^2 r^2-14 M r^3\right) \nn\\
&-\frac{B^6 M^2
   m^2 \left(544 M^3+48 M^2 r-20 M r^2+r^3\right)}{16 r}+\frac{B^4 M^3m^2 (9 r^2-46 M^2+10 M r)}{2 r^3}\nn\\
&+\frac{B^2 M^2\left(r \left(-4 l (l+1)M^2+m^2 r
   (r+4M)+8M^2\right)-24M^3\right)}{r^5}\nn\\
&	+\frac{M^2}{r^5} \left[l(l+1)(r-4M)+r \left(m^2-(r-2M) r \omega ^2-1\right)+12M\right]-24M^4\,,\\
\mathcal{V}_2&=-\frac{9}{128} B^{12} M^4 m^2 (r-2M) r^5+\frac{1}{64} B^{10} M^4 m^2 (104M-49 r) r^3\nn\\
&+\frac{B^8 M^2 m^2}{256}  \left(-704 M^4+1744 M^3 r-424 M^2 r^2-76 M r^3+5 r^4\right)\nn\\
&+\frac{B^6 M^2 m^2 (r+2M) ((r-36M) r+84M^2)}{16 r}\nn\\
&-\frac{B^4 M^2 m^2 \left[8 M^3 + r (-76 M^2 + 3 r (8 M + r))\right]}{8 r^3}\nn\\
&-\frac{B^2M^2 m^2
   (r-2M)}{r^3}+\frac{(r-2M) M^2\omega ^2}{r^3}\,,\\
\mathcal{V}_4&=\frac{9}{128} B^{12}M^4 m^2 (r-2M) r^5+\frac{1}{64} B^{10}M^4 m^2 r^3 (29 r-64M)\nn\\
&-\frac{1}{256} B^8 M^2 m^2 \left[288 M^4+r \left(336 M^3+r \left(3 r (r-20 M)-56 M^2\right)\right)\right]\nn\\
&+\frac{B^6M^2
   m^2 \left[48 M^3+r (r-4 M) (12 M+r)\right]}{16 r}+\frac{3 B^4M^2 m^2 (r-2M)^2}{8 r^2}\,,\\
\mathcal{V}_6&=-\frac{3}{128} B^{12}M^4 m^2 (r-2M) r^5+\frac{1}{64} B^{10}M^4 m^2 (8M-3 r) r^3\nn\\
&-\frac{1}{256} B^8 M^2 m^2 \left[-64 M^4+80 M^3 r-40 M^2 r^2+4 M r^3+r^4\right]-\frac{B^6 M^2 m^2
   (r-2M)^3}{16 r}\,,\\
\mathcal{V}_8&=\frac{1}{256}  M^2B^8 m^2 (r-2M)^4\,.
\end{align}
%\end{widetext}
% 

%%%%%%%%%%%%%%%%%%%%%%%%%%%%%%%%%%%%%%%%%%%%%%%%%%%%%%%%%%%%%%%%
\chapter{Quasibound states of a massive spin-2 field around a Kerr BH: Analytical results}\label{app:ana}
%%%%%%%%%%%%%%%%%%%%%%%%%%%%%%%%%%%%%%%%%%%%%%%%%%%%%%%%%%%%%%%%
In this appendix we generalize Detweiler's analytical calculations~\cite{Detweiler:1980uk} for the unstable scalar modes of a Kerr BH in the small-mass limit to the case of the massive spin-2 axial dipole, to first-order in the rotation. 

Defining $R(r)=Q/r$ the axial dipole equation~\eqref{axialdi_kerr} can be rewritten as
\begin{align}
&r^2 f\frac{d}{dr}\left(r^2 f\frac{dR}{dr}\right)+\Big[r^4\omega^2-4\tilde{a}m M^2 r\omega-r^2 f \Big( j(j+1)\nn\\
&\left.\left.+\mu^2r^2-\frac{2M s'^2}{r}-\tilde{a}m M^2\frac{12 (4 r-9 M)}{r^4 \omega}\right)\right]R=0\,,
\end{align}
where we have defined $j=l+S=2$ and $s'=3$. From now on we consider $j$ and $s'$ to be generic integers and we replace their specific values only in the final result~\eqref{finalana} below. The latter is valid for any $j$ and $s'$ provided $j<s'$. To use the method of matching asymptotics we start by writing this equation in terms of the dimensionless variable $z=(r-r_+)/r_+$, 
\begin{align}
\label{di_ana}
&Z\frac{d}{dz}\left(Z\frac{dR}{dz}\right)+\Big[4M^2\omega^2(1+z)^4-2\tilde{a}m M \omega(1+z)-j(j+1)Z\nn\\
&\left.-4M^2\mu^2 z(1+z)^3+s'^2 z-\tilde{a}m\frac{3z(1-8z)}{4M \omega(1+z)^3}\right]R=0\,,
\end{align}
where $Z=z(z+1)$.

We first expand the equation above for $z\gg 1$. For this we define the variable $x=4Mk_{\infty}z$ and get the equation   
\be
\label{di_zgg1}
\frac{d^2}{dx^2}\left(x R\right)+\left[-\frac{1}{4}+\frac{\nu}{x}-\frac{j(j+1)}{x^2}\right]xR=0\,,
\ee
where we have defined , $k^2_{\infty}=\mu^2-\omega^2$, $\nu=M\mu^2/k_{\infty}$ and have considered $\omega\sim \mu$.
For quasibound states the solution of this equation with the correct boundary condition at infinity is given by
\be
R_{\infty}(x)\approx C_1 e^{-x/2}x^j U(1+j-\nu,2j+2,x)\,,
\ee
where $C_1$ is a constant and $U(p,q,x)$ is one of the confluent hypergeometric functions~\cite{handmath}. For $z\ll 1$, at leading order, the behavior of the solution reads
\be
R_{\infty}(r)\approx C_1 \left[(2k_{\infty}r)^j\frac{\Gamma[-1-2j]}{\Gamma[-j-\nu]}
+(2k_{\infty}r)^{-j-1}\frac{\Gamma[1+2j]}{\Gamma[1+j-\nu]}\right]\,.
\ee 
Equation~\eqref{di_ana} can also be solved in the region where $r\ll$ max$(j/\omega,j/\mu)$. In this limit,
\be
Z\frac{d}{dz}\left(Z \frac{dR}{dz}\right)+\left[P^2-j(j+1)Z+\bar{s}^2 z\right]R=0\,,
\ee
where we have defined  $\epsilon=2M\mu$, $\bar{s}^2=s'^2-\frac{3\tilde{a}m}{2\epsilon}$, $P=-2Mk_H=-2M(\omega-m\Omega_{\rm{H}})$ and neglect $\mathcal{O}(\tilde{a}^2)$ terms in $P^2$. Note that in order to solve the equation analytically, we neglect terms $\mathcal{O}(\frac{\tilde{a}z^2}{\epsilon})$, so the approximation is valid only if $\tilde{a}\ll j\, M\mu$. 

The solution of the equation above is given in terms of hypergeometric functions. Imposing ingoing waves at the horizon we get that the general solution is given by
\be
R_{H}(r)=C_2 e^{-2P\pi}(-1)^{2iP}z^{iP}(1+z)^{\sigma}
_2F_1(-j+i P+\sigma,1+j+i P+\sigma,1+2i P,-z)\,,
\ee
where $_2F_1(a,b,c,z)$ is the hypergeometric function~\cite{handmath} and $\sigma=\sqrt{\bar{s}^2-P^2}$. Using the asymptotic properties of the hypergeometric function~\cite{handmath} we can derive the large-distance limit $z\gg 1$ of this solution
\begin{align}
&R_{H}(r)\approx C_2 \Gamma[1+2iP]\nn\\
&\times\left[\frac{(2M)^{1+j}\Gamma[-1-2j]}{\Gamma[-j+i P-\sigma]\Gamma[-j+i P+\sigma]}r^{-j-1}\right.\nn\\
&\left.+\frac{(2M)^{-j}\Gamma[1+2j]}{\Gamma[1+j+i P-\sigma]\Gamma[1+j+i P+\sigma]}r^{j}\right]\,.
\end{align} 
The near- and far-region solutions have an overlapping region when $M\omega\ll j$ and $M\mu\ll j$ and one can find a matching condition equating the coefficients of $r^j$ and $r^{-j-1}$:
\begin{align}
&\frac{\Gamma [2 j+1] \Gamma [-j-\nu]}{\Gamma [-2 j-1] \Gamma [j-\nu
   +1]}=(4k_{\infty} M)^{2 j+1}\nn\\
	\times&\frac{\Gamma [-2 j-1]\Gamma \left[j+i
   P-\sigma+1\right] \Gamma \left[j+i
   P+\sigma+1\right]}{\Gamma [2 j+1] \Gamma \left[-j+i
   P-\sigma\right] \Gamma \left[-j+i P+\sigma\right]}\,.
\end{align}
At leading order for $M k_{\infty}$ the right hand side vanishes. In the left hand side this corresponds to the poles of $\Gamma[j+1-\nu]$, which are given by $\nu^{(0)}=j+1+n$ for a non-negative integer $n$, yielding the expected hydrogen-like quasibound states. We obtain, to lowest order in $M\mu$,
\be
\label{ana_real}
k_{\infty}^2=\mu^2-\omega_R^2\approx\mu^2\left(\frac{M\mu}{j+n+1}\right)^2\,.
\ee
In order to get the imaginary part of the spectrum, we expand around this value to get the next-to-leading order correction. Writing $\nu\equiv \nu^{(0)}+\delta\nu$ and assuming $\delta\nu\ll 1$ we get (for details see e.g.~\cite{Furuhashi:2004jk})
\be	
\label{ana_kerr}
\delta\nu\approx -\frac{(4 k_{\infty} M)^{2 j+1}\Gamma [-2 j-1] \Gamma [2 j+n+2]}{\Gamma [1+2 j]^2\Gamma [2 j+2]
   \Gamma [n+1] }
	\frac{\Gamma [j+iP-\sigma +1] \Gamma [j+i P+\sigma +1]}{\Gamma [-j+i P-\sigma ] \Gamma [-j+i P+\sigma ]}\,.
\ee
Since there is a pole in one of the $\Gamma$-functions we take to lowest order in $P$ and $\tilde{a}/\epsilon$, $\Gamma[-j+i P-\sigma]\approx \Gamma[-j-s']$.
%and use $\Gamma[-p]/\Gamma[-q]=(-1)^{p-q}\Gamma[q+1]/\Gamma[p+1]$.
% To cancel the pole we use 
%
%\be
%\lim_{\tilde{\epsilon}\to 0}\frac{\Gamma[-1-2j-2\tilde{\epsilon}]}{\Gamma[-j-s'-\tilde{\epsilon}]}=\frac{(-1)^{1+j-s'}}{2}\frac{\Gamma[j+s'+1]}{\Gamma[2j+2]}\,,
%\ee
%
%where we have taken into account that the angular eigenvalue $l$ receives higher-order corrections in the rotating case~\cite{Rosa:2012uz}. 
We then get in this limit
\begin{align}
\delta\nu\approx (-1)^{j-s'}\frac{(4 k_{\infty} M)^{2 j+1} \Gamma [2 j+n+2]\Gamma[j+s'+1]^2}{2\Gamma [1+2 j]^2\Gamma [2 j+2]^2
   \Gamma [n+1] \Gamma [-j+s']}\Gamma [j+iP-\sigma +1] \,,
\end{align}
where the factor $2$ in the denominator comes from a specific limit of the $\Gamma$ functions and it is related to the fact that $l(l+1)$ is not the exact angular eigenvalue in a rotating background. In the nonrotating limit, the result above must be multiplied by a factor $2$. [see the discussion of Appendix C2 in Ref.~\cite{Pani:2012vp} for details].
The imaginary part of the bound-mode frequency reads
\be
i\omega_I=\frac{\delta\nu}{M}\left(\frac{M\mu}{j+n+1}\right)^3\,.
\ee
To understand how this scales with $M\mu$ in the small-mass limit, we note that for $\tilde{a}\ll M\mu$  and at first-order in $P$ we have  $\Gamma [j+iP-\sigma +1]\sim -i P/{P^2}\sim -\frac{iP}{4M^2\mu^2}$. Finally we get 
\begin{align}
&M\omega_I\approx (-1)^{j+1-s'}(\tilde{a}m-2r_+\mu)(M\mu)^{4j+3}\times\nn\\
&\frac{4^{2j-1} \Gamma [2 j+n+2]\Gamma[j+s'+1]^2}{(j+1+n)^{2j+4}\Gamma [1+2 j]^2\Gamma [2+2 j]^2
   \Gamma [n+1] \Gamma [-j+s']}\,. \label{finalana}
\end{align}
The fundamental mode, $n=0$, for the axial dipole ($j=2\,,s'=3$) reads 
\be
M\omega_I\approx (\tilde{a}-2r_+\mu)\frac{40 (M\mu) ^{11}}{19683}\,.
\ee
The formula above is valid when $0\neq\tilde{a}\ll M\mu$ whereas, in the nonrotating case, it must be multiplied by a factor $2$ as explained above. A comparison with the numerical results for the non-rotating case and for the rotating case is shown in Fig.~\ref{fig:axial_ana} and in Fig.~\ref{fig:ana_vs_num}, respectively.

We note the importance of the factor $\Gamma [j+i P-\sigma +1]$, which takes the form $\Gamma [j-s'+1]$ at lowest order in $P$ and $\tilde{a}/\epsilon$ and diverges because $s'>j$ $(s'=3,j=2)$. This is not the case in the axial perturbations of the Proca field $(s'=1,j\geq 1)$ and the perturbations of the scalar field $(s'=0,j\geq 0)$~\cite{Detweiler:1980uk,Furuhashi:2004jk,Rosa:2011my,Pani:2012bp}.  It is this factor that contributes with a term $(M\mu)^{-2S}$ for the imaginary part of the quasibound frequency, resulting in a power-law of the form $\omega_I/\mu\propto -(M\mu)^{4j-2S+5}=-(M\mu)^{4l+2S+5}$.

%%%%%%%%%%%%%%%%%%%%%%%%%%%%%%%%%%%%%%%%%%%%%%%%%%%%%%%%%%%%%%%%%%%%%%%%%%%%%%%%%%%%%%%
\section{Note on the monopole of Proca and massive spin-2 field}
\begin{figure}[htb]
\begin{center}
% \begin{tabular}{c}
\epsfig{file=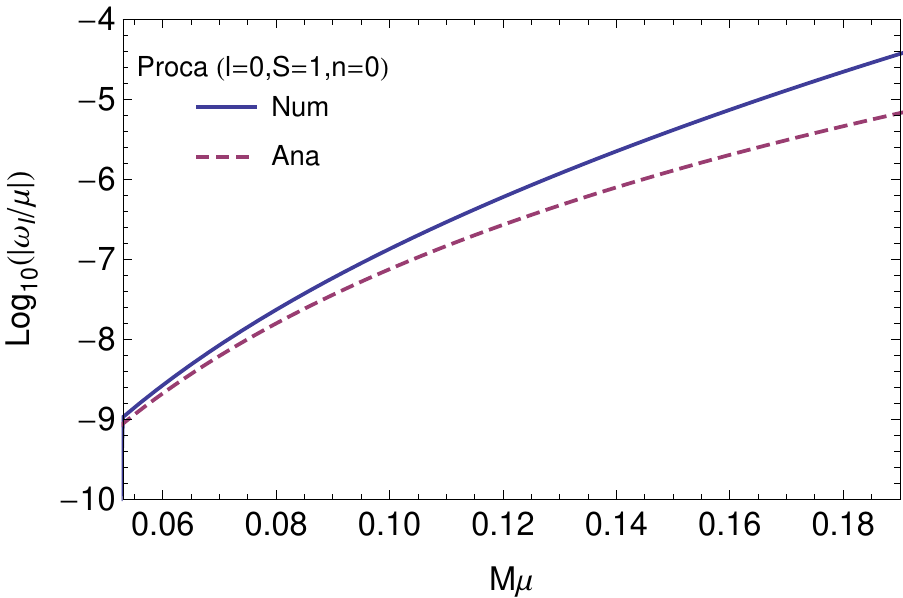,width=10cm,angle=0,clip=true}
% \end{tabular}
\caption{Comparison between the numerical and analytical results for the Proca field mode $l=0$, $n=0$ as a function of the mass coupling $M\mu$. The solid line shows the numerical data and the dashed shows the analytical formula $\omega_I/\mu\approx -\frac{3}{4}(M\mu)^{7}$.\label{fig:proca_mono}}
\end{center}
\end{figure}

%%%%%%%%%%%%%%%%%%%%%%%%%%%%%%%%%%%%%%%%%%%%%%%%%%%%%%%%%%%%%%%%%%%%%%%%%%%%%%%%%%%%%%%
The monopole equation for the Proca field~\cite{Rosa:2011my} is given by
\be
\frac{d^2 u_{(2)}}{dr_*^2}+\left[\omega^2-f(r)\left(\mu^2+\frac{2}{r^2}-\frac{6M}{r^3}\right)\right]u_{(2)}=0\,.
\ee
This can be written in the form~\eqref{di_ana} taking $\tilde{a}=0$, $j=l+S=1$, and $s'=2$. We can then solve analytically this equation in the same way as we did for the axial dipole and all the formulas apply. 
%However to cancel the pole in~\eqref{ana_kerr} we use
%
%\be
%\frac{\Gamma[-1-2j]}{\Gamma[-j-s']}=(-1)^{1+j-s'}\frac{\Gamma[j+s'+1]}{\Gamma[2j+2]}\,,
%\ee
%
%since in the nonrotating case $j$ is an integer. 
We then find that for this mode
\be
\frac{\omega_I}{\mu}\approx -\frac{8 (M\mu) ^7(n+1) (n+3)}{(n+2)^5} \,,
\ee
in agreement with the numerical results of Rosa and Dolan~\cite{Rosa:2011my}. In Fig.~\ref{fig:proca_mono} we compare the numerical results and the analytical formula in the small-mass limit.

Another interesting behavior that we can infer comparing with the axial dipole is that it seems that $s'$ is simply given by the sum of the spin projection $S$ and the spin of the field, i.e., $s'=s+S=1+1=2$ for the monopole of the Proca field and $s'=s+S=2+1=3$ for the massive spin-2 field. 

Unfortunately the monopole equation for the massive spin-2 field~\eqref{evenl0} does not have a simple and understandable form in the limit $z\ll 1$ due to the complex form of the potential. However in the limit $z\gg 1$ we can deduce the equation 
\be
\frac{d^2}{dx^2}\left(x R_0\right)+\left[-\frac{1}{4}+\frac{\nu}{x}-\frac{6}{x^2}\right]xR_0=0\,,
\ee
where $R_0=\varphi_0/r$. This looks exactly like the axial dipole equation in the same limit~\eqref{di_zgg1}. By comparison we can see that the monopole acquires a centrifugal term with $j=l+S=2$ in agreement with the numerical results presented in Sec.~\ref{subsec:quasibound}.

%%%%%%%%%%%%%%%%%%%%%%%%%%%%%%%%%%%%%%%%%%%%%%%%%%%%%%%%%%%%%%%%%%%%%%%%%%%%%%

%%%%%%%%%%%%%%%%%%%%%%%%%%%%%%%%%%%%%%%%%%%%%%%%%%%%%%%%%%%%%%%%%%%%%%%%%%%%%%%%%%%%%%%%%%%%%%%%%%%%%%%%%
%\clearpage
%\newpage

\cleardoublepage
\phantomsection
\bibliographystyle{apsrev4-1_mod}
%\bibliography{ref}


\addcontentsline{toc}{chapter}{\bibname}
%\cleardoublepage

\end{document}